\documentclass[
headinclude=true, 
headsepline=true,
twoside=true,
BCOR=10mm, 
    fontsize=10 pt,
paper=17cm:24cm, 
pagesize, 
DIV=14, 
headings=normal, 
appendixprefix=true,
draft=false,
numbers=noendperiod, 
toc=bibliography, 
parskip=false, 
captions=bottombeside,
version=last 
]{scrbook}

\usepackage{ifdraft}
\ifdraft{
	\usepackage[notref,notcite]{showkeys}
	\usepackage[obeyspaces,hyphens,spaces]{url}
	\renewcommand*{\showkeyslabelformat}[1]{%
		\fbox{\parbox{1.15cm}{\normalfont\small\ttfamily\url{#1}}}}
}{}

\addtokomafont{caption}{\small}

\renewenvironment{thebibliography}[1]{
  \begin{oldthebibliography}{#1}
    \setlength{\itemsep}{.3em}
    \setlength{\parskip}{0em}
}
{
  \end{oldthebibliography}
}
\usepackage[backend=biber,style=ieee,isbn=false,citestyle=numeric-comp,dashed=false,url=false]{biblatex}
\DefineBibliographyStrings{english}{phdthesis = {Ph.D. thesis}}
\makeatletter
\DeclareRobustCommand*{\bfseries}{%
   \not@math@alphabet\bfseries\mathbf
   \fontseries\bfdefault\selectfont
   \boldmath
}
\makeatother

\usepackage{tabularx}

\makeatletter
\renewcommand{\@chapapp}{}

\makeatother

\usepackage{subfiles}
\usepackage[dutch,american]{babel}
\usepackage[T1]{fontenc}
\usepackage{lmodern, microtype}
\usepackage{wrapfig}
\usepackage{pdfpages}

\setcapindent{0pt}
\usepackage[labelfont=bf]{caption}

\usepackage{xcolor}
\usepackage{graphicx}
\usepackage{amssymb}
\usepackage{amsmath}
\usepackage{verbatim}
\usepackage{subcaption}
\usepackage{csquotes}
\usepackage[all]{xy}
\usepackage{changepage}
\usepackage{mdframed}
\usepackage{appendix}
\usepackage{adjustbox}

\definecolor{arxivc}{RGB}{22,72,145}
\def\arxivtourl#1{http://arxiv.org/abs/#1}      
\DeclareFieldFormat{eprint:arxiv}{%
    \ifhyperref
        {\color{arxivc}\href{\arxivtourl#1}{[arxiv:\nolinkurl{#1}]}}
        {\nolinkurl{#1}}}

\mathchardef\mhyphen="2D
\newcommand{\onlyinsubfile}[1]{#1}
\newcommand{\notinsubfile}[1]{}

\AtBeginEnvironment{subappendices}{%
\section*{Appendices}
\counterwithin{figure}{chapter}
\counterwithin{table}{chapter}
}

\usepackage[pdftex,hypertexnames=false]{hyperref}

\graphicspath{{figures/curv-profs/}{figures/acknowledgements/}{figures/ricci2d/}{figures/defects/}{figures/monte-carlo/}{figures/slice3d/}{figures/}}

\newcommand{\R}{\mathbb{R}}
\newcommand{\dbar}{\bar{d}}
\newcommand{\dbav}{\dbar_\textrm{av}}
\newcommand{\delmax}{\delta_\textrm{max}}

\newcommand{\be}{\begin{equation}}
\newcommand{\ee}{\end{equation}}
\renewcommand\[{\begin{equation}}
\renewcommand\]{\end{equation}}

\newcommand{\mo}{\mathcal{O}}
\newcommand{\code}[1]{\texttt{#1}}

\def\d{\mathrm{d}}
\def\D{\mathrm{D}}
\def\pd{\partial}
\def\e{\textrm e}
\def\i{\imath}

\newcommand{\tr}{\mathrm{Tr}}
\newcommand{\U}{\textrm{U}} 
\newcommand{\vd}{\Delta}  
\newcommand{\amp}{\mathcal{A}}

\newcommand{\dx}{\frac{\d}{\d x}}
\newcommand{\dxi}[1]{\frac{\d^{#1}}{\d x^{#1}}}
\newcommand{\dy}{\frac{\d}{\d y}}
\newcommand{\dyi}[1]{\frac{\d^{#1}}{\d y^{#1}}}
\newcommand{\x}{\mathbf{x}}
\newcommand{\y}{\mathbf{y}}
\newcommand{\bp}{\mathbf{p}}
\newcommand{\cZ}{\mathcal{Z}}
\newcommand{\ba}{\mathbf{a}}
\newcommand{\bn}{\mathbf{n}}
\newcommand{\bb}{\mathbf{b}}
\newcommand{\bfm}{{\mathbf{m}}}
\newcommand{\bH}{\mathbb{H}}
\newcommand{\bR}{\mathbb{R}}

\renewcommand{\c}{\alpha} 

\newcommand{\lambdac}{\tilde{\lambda}}
\newcommand{\kg}{k_0}
\newcommand{\kcc}{\tilde{k}_3}

\newcommand{\targn}{\tilde{N}_{31}}

\newcommand{\vthet}{\vec{\theta}}
\newcommand{\vthetm}{\vec{\theta}_\textrm{min}}
\newcommand{\seh}{S^\textrm{EH}}
\newcommand{\ttot}{t_\textrm{tot}}
\newcommand{\optwo}{O_2}

\makeatletter
\newcommand*\bigcdot{\mathpalette\bigcdot@{.5}}
\newcommand*\bigcdot@[2]{\mathbin{\vcenter{\hbox{\scalebox{#2}{$\m@th#1\bullet$}}}}}
\makeatother

\newcommand{\diagblock}[2]{\begin{array}{c} #1 \\ \eqref{#2} \end{array}}

\def\blfootnote{\xdef\@thefnmark{}\@footnotetext}
\long\def\symbolfootnote[#1]#2{\begingroup%
\def\thefootnote{\fnsymbol{footnote}}\footnote[#1]{#2}\endgroup}

\begin{document}
\renewcommand{\onlyinsubfile}[1]{}
\renewcommand{\notinsubfile}[1]{#1}

\hypersetup{pageanchor=false}

\includepdf[pages=-,fitpaper=true]{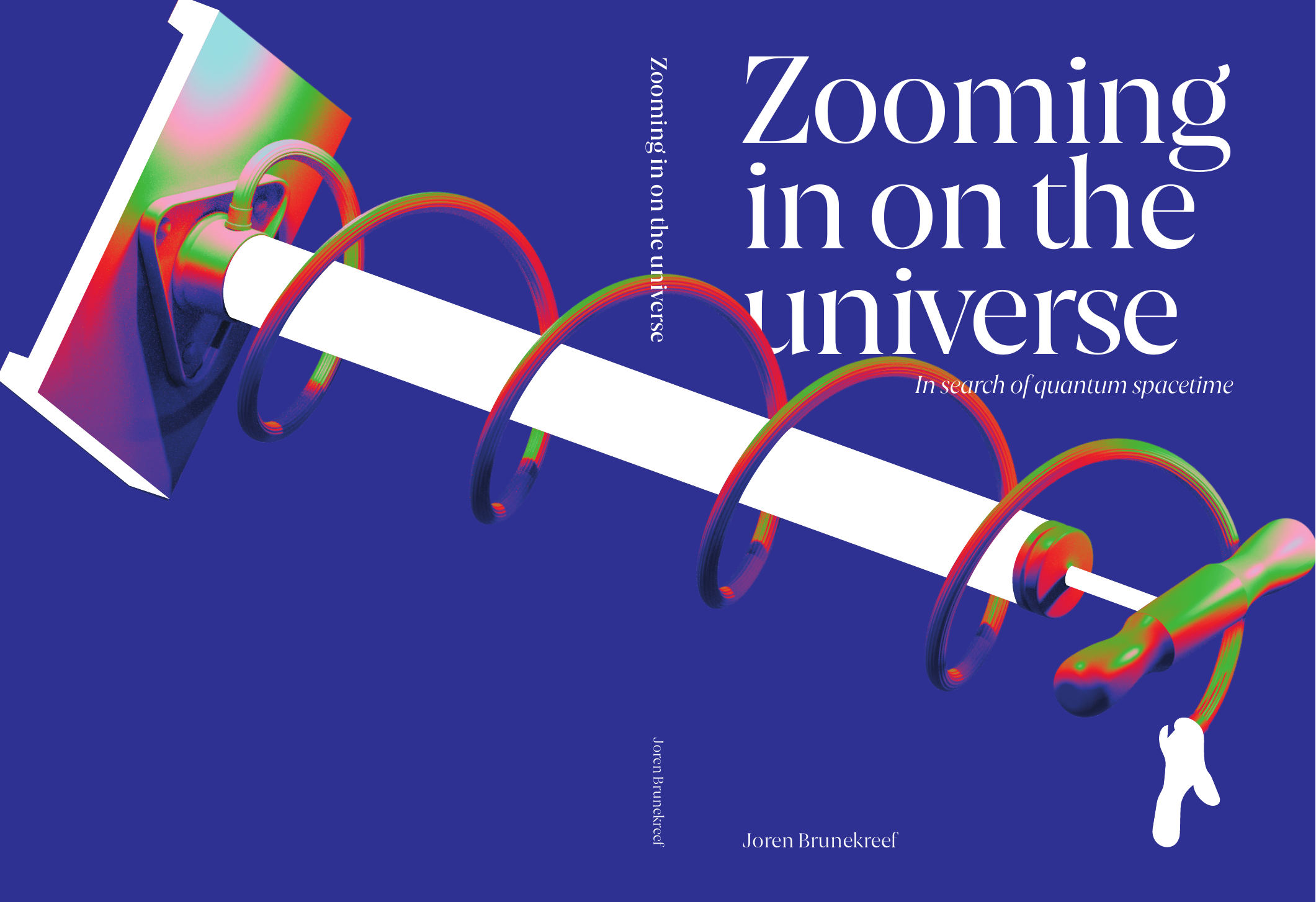}
\thispagestyle{empty}
\vspace*{3.6cm}

{\centering
{\Huge Zooming in on the Universe}\vspace{.2cm} \\
{\Large In Search of Quantum Spacetime}\vspace{1.45cm}\\

{\Huge Het Universum onder de Loep}\vspace{.2cm} \\
{\Large Op Zoek naar Kwantumruimtetijd}\vspace{1.45cm}\\

{\LARGE Joren Brunekreef}\\

}
\newpage~\thispagestyle{empty}
\vfill
\noindent 
Ph.D. thesis, Radboud University Nijmegen, September 2023\\
ISBN: 978-94-6483-444-4 \\
\vspace{0.4cm}

\noindent
Cover design by Dani\"el Roozendaal --- www.danielroozendaal.com

\newpage
\thispagestyle{empty}
\vspace*{0.2cm}
{\centering
{\Huge Zooming in on the Universe}\vspace{.2cm} \\
{\Large In Search of Quantum Spacetime}\vspace{2.45cm}\\


 
{\LARGE Proefschrift}\vspace{0.8cm}\\
\par} 
{\large{\centering{ \noindent ter verkrijging van de graad van doctor \\aan de Radboud Universiteit Nijmegen \\ op gezag van de rector magnificus prof. dr. J.M. Sanders, \\ volgens besluit van het college voor promoties \\ in het openbaar te verdedigen op \\ \vspace{0.4cm}

woensdag 8 november 2023 \\ 
om 16.30 uur precies\\}}}\vspace{0.4cm}
{\centering
{\large door}\vspace{0.8cm}\\
{\LARGE Joren Willem Brunekreef}\vspace{.6cm}\\
{\large geboren op 9 april 1990 \\ te Utrecht}\par}

 \newpage \thispagestyle{empty}
{\large \noindent Promotor: Prof. dr. R. Loll\\
\\
Manuscriptcommissie: \\
Prof. dr. R.H.P. Kleiss (voorzitter) \\
Prof. dr. J. Ambj\o rn (K\o benhavns Universitet, Denemarken) \\
Dr. T.G. Budd \\
Prof. dr. D. Johnston (Heriot-Watt University, Verenigd Koninkrijk) \\
Prof. dr. E.P. Verlinde (Universiteit van Amsterdam) \\
}
 \vfill
\newpage


\

\frontmatter
\chapter*{Publications}
\textbf{Part \ref{part:curv}} of this thesis is based on the following publications:
\begin{itemize}
    \item [\cite{brunekreef2021quantum}] \fullcite{brunekreef2021quantum}
    \item [\cite{brunekreef2021curvature}] \fullcite{brunekreef2021curvature}
\end{itemize}

\noindent \textbf{Part \ref{part:cdt3d}} of this thesis is based on the following publication:
\begin{itemize}
    \item [\cite{brunekreef2023nature}] \fullcite{brunekreef2023nature}
\end{itemize}

\noindent \textbf{Part \ref{part:mat}} of this thesis is based on the following publications:
\begin{itemize}
    \item [\cite{brunekreef2022onematrix}] \fullcite{brunekreef2022onematrix}
    \item [\cite{brunekreef2023simulating}] \fullcite{brunekreef2023simulating}
\end{itemize}

\newpage

\noindent Other work published during the author's Ph.D. research, but not included in this thesis:
\begin{itemize}
    \item [\cite{brunekreef2022phase}] \fullcite{brunekreef2022phase}
    \item [\cite{brunekreef2021approximate}] \fullcite{brunekreef2021approximate}
\end{itemize}



\hypersetup{pageanchor=true}
\tableofcontents

\mainmatter






\chapter{Introduction}
\label{ch:intro}
There is an apocryphal story about Lord Kelvin, the great physicist and mathematician, addressing the British Association for the Advancement of Science at the turn of the nineteenth century. In his speech, he supposedly claimed that he and many of his colleagues would soon be irrelevant:
\begin{quote}
There is nothing new to be discovered in physics now. All that remains is more and more precise measurement.
\end{quote}
Although it has been questioned whether Lord Kelvin himself spoke these words, it seems that the underlying sentiment was a common one in those days. Steven Weinberg notes, in his 1992 book ``Dreams of a Final Theory'', that
\begin{quote}
[...] there is plenty of other evidence for a widespread, though not universal, sense of scientific complacency in the late nineteenth century. When the young Max Planck entered the University of Munich in 1875, the professor of physics, Jolly, urged him against studying science. In Jolly’s view there was nothing left to be discovered. Millikan received similar advice: ``In 1894,'' he recalled, ``I lived in a fifth-floor flat on Sixty-fourth Street, a block west of Broadway, with four other Columbia graduate students, one a medic and the other three working in sociology and political science, and I was ragged continuously by all of them for sticking to a `finished,' yes, a `dead subject,' like physics, when the new, `live' field of the social sciences was just opening up.''
\end{quote}
Millikan's flatmates could hardly have been more wrong. Physics would soon live through not just one, but multiple upheavals that revolutionized our fundamental understanding of nature. At the time of writing this thesis, nearly 130 years later, theoretical physicists (including me) are still attempting to find answers to important questions that arose as a result of those upheavals in the early twentieth century.

Science is inherently a gradual process, in which new ideas originate in different places at different times. Some ideas are largely ignored, some catch on, some are subsequently forgotten again. Even the great ideas that we take for granted these days often took many years to gain a firm footing in the scientific community. The term `upheaval' may therefore be more appropriate in hindsight, since those present during its occurrence likely experienced it as a slow and drawn-out process. Distinct and incompatible schools of thought can exist simultaneously, and it is perhaps better to view the state of science at any point in time as a complex mixture of varying viewpoints. Still, we can make a rough distinction between physics in the pre-1900 and post-1900 stages. These stages are commonly referred to as \emph{classical} physics and \emph{modern} physics, respectively.\footnote{One could argue that we have reached a \emph{postmodern} stage of physics in recent years. I leave it to the reader to decide whether that is a good or a bad thing.}

\section{Classical and modern physics}
Classical and modern physics differ in a number of distinct fundamental aspects. In this Section, we briefly discuss some of the major overhauls that the field of theoretical physics underwent around the turn of the 20th century.

\subsection{Relativity}
The first major development that taught physicists to think about the world in a fundamentally different way was the formulation of the theory of special relativity by Albert Einstein in 1905 \cite{einstein1905zur}. We highlight one of the problems that this theory set out to solve, since its basic premise is simple to understand. Furthermore, it serves as a thought-provoking starting point where physical reality and everyday intuition grow apart. Experimentalists had determined that the speed of light in vacuum is the same for every observer, regardless of their velocity or the direction in which the light is emitted \cite{michelson1887relative}. This can be contrasted with a baseball pitcher throwing a ball: spectators will observe different throw velocities depending on whether the pitcher is standing still with respect to them, or is throwing the ball from a moving platform. Light rays turn out to behave differently: even if a ray is emitted from a spaceship traveling with a velocity close to the speed of light (for example, with respect to the surface of the earth), both travelers on the spaceship and outside observers (e.g. on earth) will find the ray to travel with the exact same speed. Furthermore, it was found that the outcomes of physical experiments do not depend on the choice of inertial reference frame with respect to which they are measured. This suggests that there is no absolute notion of being ``at rest'' or ``in motion with constant velocity''.

Special relativity's interpretation of these observations is that \emph{all motion is relative}, and that there simply is no unambiguous notion of being ``at rest''. The speed of light is considered a universal constant $c = 299 792 458\textrm{ m/s}$ for any observer, regardless of their velocity. This counterintuitive behavior is resolved in special relativity by introducing a notion of space and time that depends on the choice of reference frame. The result of this merger is a four-dimensional entity called \emph{spacetime}.

A subsequent major advancement in theoretical physics was the development of the general theory of relativity \cite{einstein1915feldgleichungen}, again largely due to Einstein. This promoted spacetime to an even more prominent role in theoretical physics: Einstein realized that the force of gravity can be described in terms of the \emph{curvature} of spacetime. In general relativity, spacetime curvature is described by a field called the \emph{metric tensor}, denoted $g_{\mu \nu}$. Furthermore, the curvature of spacetime is determined by the distribution of matter and energy contained in it. We therefore find ourselves in the following interesting situation: spacetime curvature dictates how objects move, and these objects in turn dictate the curvature of spacetime. In general relativity, spacetime is not just a static canvas on which the structure and dynamics of our universe are painted, but rather constitutes an active fabric that stretches and adapts to the picture being painted on it. Curvature is a central concept in general relativity, and we will return to it on several occasions throughout this thesis.


\subsection{Quantum theory}
Also in the early years of the 20th century, physicists encountered numerous mismatches between experiments and expectations in a very different setting: that of (sub-)atomic physics. The details of these experiments require too much of a digression to discuss here, but it was found that many of these difficulties could be resolved by assuming that, colloquially speaking, \emph{energy comes in discrete bits}. The framework for describing physics with discrete energy levels was called ``quantum theory'', forming the second crossroads where modern physics and classical physics started diverging.

Quantum theory was further developed in the early 20th century, which led to surprising insights that would violently divorce fundamental physics from everyday experience. An often quoted example is the \emph{wave-particle} duality, which entails that matter can have both a wave-like and a particle-like nature under different circumstances. A further surprising feature of quantum theory that was discovered in 1927 by Werner Heisenberg is the \emph{uncertainty principle} \cite{heisenberg1927ueber}, which states that there is a fundamental limit to the accuracy with which a particle's position and momentum can be measured simultaneously. Although one could imagine that this is a consequence of practical issues in the measurement procedure, the uncertainty principle is rather a statement about quantities that cannot be known simultaneously even in principle. Clearly, this feature of quantum theory is difficult to reconcile with a common sense perspective.

The counterintuitive nature of quantum theory has given it the reputation of being a deeply mysterious aspect of reality that defies understanding on many levels. However, the framework of quantum theory has also led to a unification of concepts that were previously not known to be so intimately related. These unifying insights started to appear when physicists attempted to apply quantum laws to the theory of electromagnetism, the most well-known example of a so-called \emph{classical field theory}. It was already known that light rays can be understood as electromagnetic waves, and in the quantum picture these waves can similarly be understood as particles --- called ``photons'' --- under the aforementioned wave-particle duality. These particles were subsequently found to be carriers of the electromagnetic force between charged objects. The unifying language that emerged from this research goes by the name of \emph{relativistic quantum field theory}, which combines special relativity, quantum mechanics, and classical field theory into a single overarching framework. Over the following decades, this led to the development of the Standard Model of particle physics, a specific instance of a quantum field theory. The Standard Model classifies all known elementary particles of nature by viewing them as quantized waves of a specific set of quantum fields. Some of these particles are force-carrying particles, similar to the photon in the case of electromagnetism. In this way, it describes the electromagnetic, weak, and strong interactions between all the elementary particles. The Standard Model of particle physics provides high-precision predictions that have been verified in state-of-the-art particle colliders. We refer the interested reader to \cite{schweber1994qed,riordan1987hunting} for extensive and fascinating treatments of the historical development of this line of research.

However, as impressive as the Standard Model is, it does not constitute a final and complete theory of nature. The fourth fundamental force, gravity, is notoriously missing from this set-up. It is somewhat ironic that the force most familiar to us in daily life is the one most difficult to reconcile with quantum theory, but it has defied all efforts towards its quantization for nearly a century. We previously mentioned that gravity is described by general relativity, in which it manifests itself through the curvature of the spacetime we inhabit. Therefore, it is natural to work towards a quantum theory of gravity by quantizing general relativity. However, even with all the sophisticated machinery developed in the framework of quantum field theory, physicists have so far not succeeded in constructing a quantum incarnation of general relativity that is valid at all energy scales. It is possible to construct an effective quantum field theory in the so-called \emph{perturbative} regime of quantum gravity, where the gravitational field is weak. This might result in a theory valid up to a certain energy scale, but it fails as one approaches the so-called \emph{Planck scale} due to the perturbative nonrenormalizability of gravity in four dimensions \cite{goroff1985quantum}. In order to attain a more complete understanding of nature at the fundamental level, we also need to understand quantum gravity at and beyond the Planck scale, where its quantum behavior dominates.

Over the years, physicists have constructed a whole zoo of different strategies to improve our understanding of quantum gravity. One candidate solution to the problem is superstring theory \cite{green1988superstring,johnson2002dbranes}, which may unify all fundamental interactions into a single quantum theory and simultaneously provide answers to unsolved problems like the black hole information paradox \cite{raju2021lessons}. In this framework, the fundamental constituents of the universe are one-dimensional objects called \emph{strings}, and all elementary particles and forces arise as consequences of the vibrations and mutual interactions of these strings. Superstring theory is a fascinating and vast subject, and it has permeated into many areas of theoretical physics like cosmology, nuclear physics and condensed matter physics, while even spurring on the development of new branches in pure mathematics. However, it has so far not met its original expectations as a complete ``Theory of Everything''.

Many other approaches to quantum gravity exist, such as spin foam models, asymptotic safety, and causal set theory. An excellent overview of the main modern lines of research can be found in \cite{oriti2009approaches}. Each approach comes with particular merits and drawbacks, and only time will tell which of them, if any, will survive further scrutiny. 

In this thesis, we focus our attention on \emph{nonperturbative} approaches to quantum gravity, with a leading role reserved for the framework of Causal Dynamical Triangulations (CDT). In contrast to the aforementioned perturbative approaches, nonperturbative models of quantum gravity do not assume the gravitational field to be weak. One class of methods for studying quantum gravity nonperturbatively is formed so-called lattice approaches, in which one discretizes spacetime by introducing an ultraviolet cutoff. This is analogous to the methods used in lattice field theory \cite{montvay1994quantum}, where one investigates a quantum field theory placed on a discretized spacetime lattice. The continuum behavior of such a theory may be investigated by letting the lattice spacing approach zero. In the next section, we give a brief introduction to lattice models of quantum gravity. We start by presenting a simpler example of lattice methods applied to a quantum mechanical particle, and subsequently explain how these ideas can be extended to gravity. The discussion will progressively become more technical, eventually culminating in a description of the aforementioned model of Causal Dynamical Triangulations.

\section{Quantum gravity from triangles}
\label{intro-sec:cdt}
\subsection{Path integrals}
\begin{figure}[ht!]
	\centering
	\includegraphics[width=0.7\textwidth]{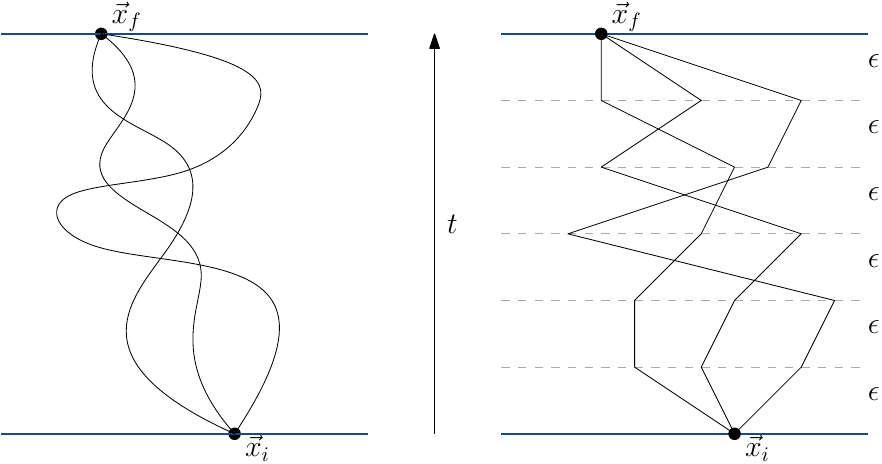}
	\caption{Schematic illustration of the path integral approach in the context of a quantum mechanical particle. On the left, we show three examples of smooth paths a particle can take from $\vec{x}_i$ to $\vec{x}_f$. On the right, we divide the paths into straight line segments for discrete, equal time intervals of duration $\epsilon$. }
	\label{intro-fig:qmpi}
\end{figure}

The lattice approaches that we consider in this thesis are based on the \emph{Feynman path integral} formulation of quantum mechanics. The intuition behind path integral methods is best understood by the example of a single quantum mechanical free particle. In classical physics, a free particle moves in a straight line with a constant velocity. In the context of quantum mechanics we are interested in the \emph{probability amplitude} for the particle to propagate from an initial point $\vec{x}_i$ to a final point $\vec{x}_f$ in a certain time $T$. According to the path integral formulation of quantum mechanics, such a particle in fact takes \emph{all} possible paths $\vec{x}(t)$ for which $\vec{x}(0) = \vec{x}_i$ and $\vec{x}(T) = \vec{x}_f$. Each path is assigned a probability amplitude depending on the path's \emph{action} $S[\vec{x}(t)]$. For a free particle of mass $m$, the action of a path $\vec{x}(t)$ takes the form
\begin{equation}
	S[\vec{x}(t)] = \int dt \frac{m}{2} \dot{\vec{x}}^2(t),
\end{equation}
which we recognize as the line integral of the particle's kinetic energy along the path. The probability amplitude assigned to such a path is the complex exponential of the action
\begin{equation}
	e^{i S[\vec{x}(t)]}.
\end{equation}
Summing (or `integrating') over all the path probability amplitudes then gives us the total probability amplitude for the particle to propagate from $\vec{x}_i$ to $\vec{x}_f$.\footnote{This raises the following interesting question: to what extent should we view the path integral as an accurate picture of reality? Or, specified to our example: does the particle \emph{in fact} take an infinite number of paths from $\vec{x}_i$ to $\vec{x}_f$, or is this formulation merely a useful instrument to compute probability amplitudes? We do not address this controversial question here, but we refer the interested reader to \cite{deutsch2011fabric} and Part I of \cite{wallace2012emergent} for thought-provoking discussions on this topic.} In order to compute this sum over paths, it is convenient to make use of a discretization scheme. Instead of attempting to sum over all smooth curves $\vec{x}(t)$ for which $\vec{x}(0) = \vec{x}_i$ and $\vec{x}(T) = \vec{x}_f$, we break up the path into a number of pieces of equal time duration, where each piece is a straight line. We call this a \emph{piecewise straight} path, for obvious reasons. By breaking up the path into ever smaller time intervals \emph{ad infinitum}, we can approach a limit in which we effectively sum over all possible paths for the particle to move from $\vec{x}_i$ to $\vec{x}_f$ in time $T$. We call this the \emph{continuum limit} of the discretized theory. As a warm-up for what's soon to come, we present here the formal expression of the continuum path integral for the quantum mechanical free particle:
\begin{equation}
	Z_\textrm{qm} = \int_{\vec{x}(0) = \vec{x}_i}^{\vec{x}(T)=\vec{x}_f} \mathcal{D} \vec{x}(t) \exp\left[ i \int_0^T dt \frac{m}{2} \dot{\vec{x}}^2(t) \right]
	\label{intro-eq:qm-pi}
\end{equation}
The measure $\mathcal{D} \vec{x}(t)$ indicates that we are integrating over paths $\vec{x}(t)$. This path integral can be computed mathematically rigorously, resulting in the continuum propagator of the particle. We illustrate the construction of the continuum path integral (left) and its time-discretized counterpart (right) in Fig.\ \ref{intro-fig:qmpi}. We consider the special case of a particle located along a one-dimensional line, with time increasing in the upwards direction. In both the continuum and discrete settings, we present three possible paths the particle can take. In the discretized picture, the particle jumps sideways in every time interval $\epsilon = T / N$, starting at $\vec{x}_i$ and ending at $\vec{x}_f$. Letting the time interval $\epsilon$ approach zero, these piecewise straight paths can approximate any continuous path to arbitrary accuracy. 

We now make a conceptual leap\footnote{A fairly substantial one, to put it mildly.} from the quantum mechanical free particle to quantum gravity, starting by introducing the formal continuum path integral for gravity. When studying a quantum mechanical particle, the dynamical variable was the particle's position, and the objects we summed over in the path integral \eqref{intro-eq:qm-pi} were all possible paths the particle could take. For gravity, the dynamical variable is the metric tensor field. In the quantum gravitational path integral we therefore sum over all possible ways in which spacetime can be curved. This idea is captured in the following formal expression:\footnote{Note that we consider pure gravity here, i.e. a universe without matter fields.}
\begin{equation}
	Z = \int \displaylimits_{\textrm{geometries }} \hspace{-1.5em} \mathcal{D} \left[g_{\mu \nu} \right] e^{i S_\textrm{EH}[g_{\mu \nu}]}.
	\label{intro-eq:gravpi}
\end{equation}
We call $Z$ the path integral for quantum gravity. The integration is over all possible equivalence classes $[g_{\mu \nu}]$ of Lorentzian spacetime metrics $g_{\mu \nu}$ modulo diffeomorphisms, i.e. all Lorentzian \emph{geometries}. The path integral is weighted by the complex exponential of the so-called Einstein-Hilbert action $S_\textrm{EH} [g_{\mu \nu}]$. This action (for a $d$-dimensional geometry) takes the form
\begin{equation}
	S_\textrm{EH}[g_{\mu \nu}] = \frac{1}{16 \pi G} \int_M d^d x \sqrt{-g} \left(R-2 \Lambda\right),
	\label{intro-eq:eh}
\end{equation}
where $G$ is the Newton gravitational constant and $\Lambda$ is the cosmological constant. Furthermore, the integration is over the manifold $M$ on which the metric $g_{\mu \nu}$ is defined, $g$ is the determinant of this metric tensor, and $R$ is the associated Ricci scalar. The trouble is that the expression \eqref{intro-eq:gravpi} is purely formal, and there is no straightforward prescription for integrating over the space of all possible spacetime metrics. Furthermore, one still has to define the appropriate class of geometries to integrate over.

This is a difficult question to answer \emph{a priori}, and there is no fundamental reason to choose one such class of spacetimes over another. It turns out that the choice one makes here can leave a strong imprint on the resulting properties of the model. 
One such choice is to define the class of spacetimes as the continuum limit of triangulated geometries with a well-defined causal structure.\footnote{The motivation for this choice can be traced back to Teitelboim \cite{teitelboim1983causality}.}  This is the approach advocated in the framework of \emph{Causal Dynamical Triangulations} (CDT), the main model of interest in this thesis. We now proceed to outline the most important features of this model, starting with the prescription it uses to make sense of the formal path integral \eqref{intro-eq:gravpi}.

\subsection{Summing over triangulations}
We first provide some motivation for the CDT implementation of the quantum gravitational path integral by drawing parallels to the method we sketched for defining the path integral of the quantum mechanical point particle. We subsequently list the technical details involved in this procedure. Recall from the treatment of a quantum mechanical particle that we used a discretization procedure to properly define the sum over all paths. The framework of CDT uses a similar idea to define the sum over all spacetimes: whereas we previously constructed arbitrary paths from a sequence of straight line segments, we now construct arbitrary spacetimes by gluing a collection of flat building blocks together. Again we aim to take a continuum limit where the number of building blocks grows to infinity while their typical size shrinks to zero, effectively summing over all possible spacetime geometries. This is an infinite sum over piecewise flat ``triangulated'' geometries, which should be understood as the \emph{definition} of the purely formal path integration as it appears in Eq.\ \eqref{intro-eq:gravpi} in the context of CDT. It takes the form
\begin{equation}
	Z = \sum_{T \in \mathcal{T}} \frac{1}{C_T} e^{i S_R[T]},
	\label{intro-eq:cdt-sum}
\end{equation}
where $\mathcal{T}$ is an ensemble of triangulations of fixed topology. The factor $C_T$ is the order of the automorphism group of $T \in \mathcal{T}$, implying that triangulations with a nontrivial discrete symmetry have a smaller weight in the partition function. Finally, $S_R$ denotes a discretized form of the Einstein-Hilbert action, called the \emph{Regge action}. We turn our attention to the functional form of the Regge action later in this section. 

We stated earlier that CDT aims to construct the path integral over Lorentzian spacetimes, understood as the continuum limit of a sum over triangulations with a well-defined causal structure. In what follows, we discuss the technical definition of CDT triangulations in $d$ spacetime dimensions.

\subsection{Causal Lorentzian triangulations}
The elementary building blocks of CDT are $d$-dimensional \emph{simplices}, generalizations of triangles to arbitrary dimension. These $d$-simplices are pieces of flat $d$-dimensional Minkowski space. We choose the edges of the simplices to be either spacelike with squared length $\ell_\textrm{sl}^2 = a^2$, or timelike with squared length $\ell_\textrm{tl}^2 = -\alpha a^2$, where $a$, having units of length, parametrizes the ``lattice spacing'' and $\alpha$ is a positive real number. This asymmetry factor $\alpha$ allows for distinct finite rescaling of spacelike and timelike directions, and it takes on the role of a  coupling parameter of the theory in the continuum limit. By gluing together the simplicial building blocks in an appropriate way, we obtain \emph{simplicial manifolds} \cite{ambjorn1997geometry} that serve as approximations of spacetime. We furthermore require that these geometries can be foliated with respect to an integer proper time parameter. The leaves of this foliation are $(d-1)$-dimensional simplicial hypersurfaces of the geometry, in which all links (or 1-simplices) are spacelike and of equal length. We call such hypersurfaces the spatial slices of the geometry. In between two neighboring spatial slices, we have a $d$-dimensional slab geometry consisting of $d$-simplices with both spacelike and timelike links, where the timelike links connect the two distinct slices. The topology of all the spatial slices is taken identical, emulating the situation in classical Einstein gravity where every spacetime is causal.

\begin{figure}[ht!]
	\centering
	\includegraphics[width=0.8\textwidth]{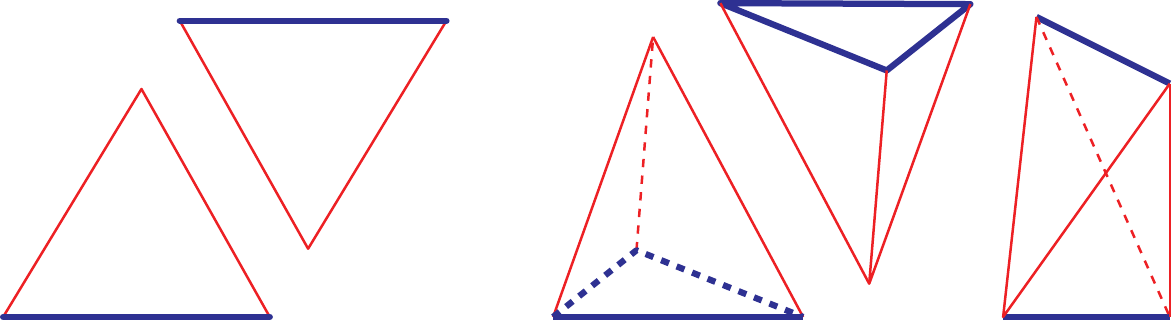}
	\caption{Elementary simplicial building blocks of CDT geometries in two (left) and three (right) dimensions. Time is increasing in the upwards direction. Blue edges are spacelike, red edges are timelike.}
	\label{intro-fig:cdt-blocks}
\end{figure}

We can characterize the $d$-simplices by the integer time parameter of the slices in which their vertices reside. We denote a $d$-simplex with $p$ vertices in the slice labeled $t$ and $q$ vertices in the slice labeled $(t+1)$ as a $(p,q)$-simplex. We illustrate the distinct building blocks of CDT in two and three dimensions in Fig.\ \ref{intro-fig:cdt-blocks}. The spacelike links are drawn in blue and the timelike links in red. In cases where no confusion can arise, we will omit the brackets and write $pq$-simplex in place of $(p,q)$-simplex. In two dimensions, there are only two types of simplices, namely, $(1,2)$ and $(2,1)$. The three basic building block types in three dimensions are $(3,1)$, $(1,3)$, and $(2,2)$.

\pagebreak
\subsection{The Regge action in CDT}
\label{intro-sec:regge-cdt}
What we have accomplished so far is to attach a meaning to the concept of a regularized sum over geometries. The continuum Einstein-Hilbert action \eqref{intro-eq:eh} should be adapted accordingly so that it can be implemented on the triangulations we sum over. In fact, a simplicial analogue of the Einstein-Hilbert action was already constructed by Tullio Regge in 1961 \cite{regge1961general}, well before the advent of triangulated quantum gravity. Regge's aim was to study general relativity in a coordinate-free language. He recognized that spacetime can be triangulated by simplices with edges of variable length, and that the Einstein-Hilbert action associated to a spacetime geometry can be written as a function of these edge lengths. This prescription can be applied to the CDT setting where all edges are of fixed squared length $\ell_\textrm{sl}^2$ or $\ell_\textrm{tl}^2$. 

Consider again the expression \eqref{intro-eq:eh} for the continuum action. We see that the second term is proportional to the volume of spacetime. Since a CDT geometry is constructed from a set of basic $d$-dimensional simplicial building blocks $\sigma^{(d)}$, the total volume of such a geometry can be expressed as a sum over the volumes $V_{\sigma^{(d)}}$ of these building blocks:
\begin{equation}
	V[T] = \sum_{\sigma^{(d)} \in T} V_{\sigma^{(d)}}.
	\label{intro-eq:total-vol}
\end{equation}

The first term in the continuum action \eqref{intro-eq:eh} is proportional to the integrated Ricci scalar curvature of the spacetime. The triangulations under consideration are piecewise flat, and the curvature is concentrated at the $(d-2)$-dimensional ``hinges'' of these simplicial manifolds. In order to make this more precise, we first define the notion of a \emph{deficit angle} associated to a hinge. Denote by $\sigma^{(d)} \succ \sigma^{(d-2)}$ all $d$-dimensional simplices $\sigma^{(d)}$ that share a $(d-2)$-dimensional hinge $\sigma^{(d-2)}$. Every such $\sigma^{(d)}$ has an internal angle $\delta\left(\sigma^{(d)}, \sigma^{(d-2)}\right)$ between the ``faces'' associated to $\sigma^{(d-2)}$. If the space perpendicular to the hinge is flat, these internal angles sum up to $2\pi$. We therefore define the deficit angle $\epsilon\left(\sigma^{(d-2)}\right)$ associated to a hinge as follows:
\begin{equation}
	\epsilon\left(\sigma^{(d-2)}\right) = 2 \pi - \sum_{\sigma^{(d)} \succ \sigma^{(d-2)}} \delta\left(\sigma^{(d)}, \sigma^{(d-2)}\right).
\end{equation}

\begin{figure}[ht!]
	\centering
	\begin{subfigure}[t]{0.45\textwidth}
	\centering
	\includegraphics[height=0.8\linewidth]{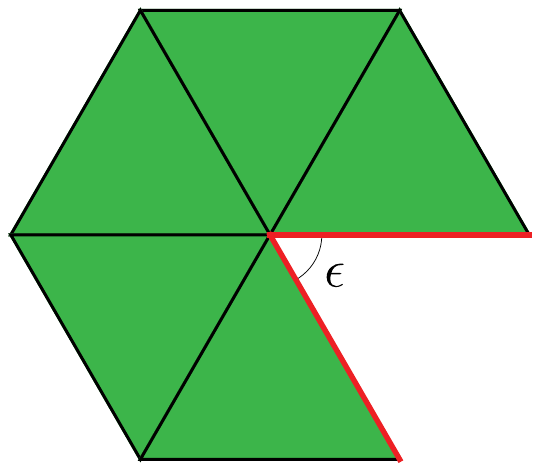}
	\end{subfigure}
	\hspace{0.01\textwidth}
	\begin{subfigure}[t]{0.45\textwidth}
	\centering
	\includegraphics[height=0.8\linewidth]{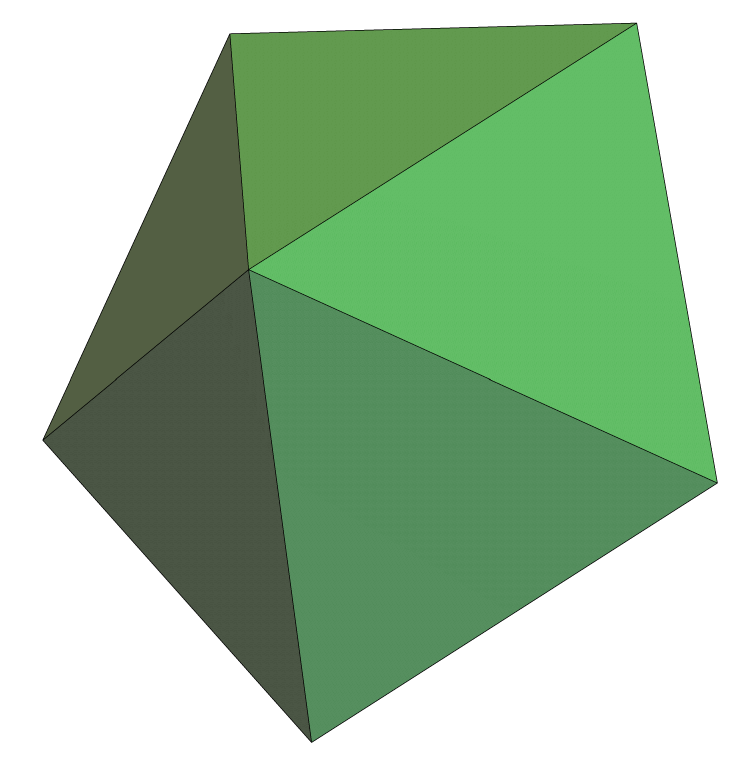}
	\end{subfigure}
	\caption{Gluing together five equilateral triangles (left) along the red edge, we obtain a surface (right) with positive curvature concentrated at the center vertex. The missing angle $\epsilon = \frac{2\pi}{6}$ is the \emph{deficit angle}.}
	\label{intro-fig:angle-deficit}
\end{figure}

\pagebreak
A positive value for the deficit angle indicates the presence of positive Gaussian curvature, whereas a negative deficit angle indicates negative Gaussian curvature. We illustrate this in Fig.\ \ref{intro-fig:angle-deficit} by an example from two dimensions, in which we glue five equilateral triangles together along their edges. The positive deficit angle $\epsilon$ translates into positive curvature concentrated at the center vertex after the gluing. In the language of the previous paragraph, the triangles $\sigma^{(2)}$ are the faces associated to the center ``hinge'' vertex $\sigma^{(0)}$.

We are now in a position to formulate a discretized version of the Einstein-Hilbert action, which goes by the name of the \emph{Regge action} $S_R$. It reads
\begin{equation}
	S_R = \frac{1}{8 \pi G} \sum_{\textrm{hinges } \sigma^{(d-2)}} V_{\sigma^{(d-2)}} \,\epsilon\left(\sigma^{(d-2)}\right) - \Lambda \sum_{\substack{d-\textrm{simplices} \\ \sigma^{(d)}}} V_{\sigma^{(d)}},
	\label{intro-eq:regge-cdt}
\end{equation}
where $V_{\sigma^{(d)}}$ is the $d$-dimensional volume of a $d$-simplex $\sigma^{(d)}$ and $V_{\sigma^{(d-2)}}$ is the $(d-2)$-dimensional volume of a hinge $\sigma^{(d-2)}$. In CDT, the basic building blocks are a finite number of types of $d$-dimensional simplices. A consequence of this choice is that the deficit angle at a hinge comes in discrete units, and only depends on which different types of $d$-simplices meet at such a hinge. As a result, the CDT Regge action simplifies, and can be written in terms of the simplex counting variables $N_d$.

The sum in Eq.\ \eqref{intro-eq:cdt-sum} is over Lorentzian triangulations with a complex phase factor, and interference of the phases makes the sum difficult to compute. It turns out that CDT allows for a \emph{Wick rotation}, where the Lorentzian geometries are mapped to triangulations of Euclidean signature, and the weight is rendered purely real by an analytic continuation of the Regge action, denoted $S_E[T]$. The result is a Euclideanized version of the CDT partition function, which we write as
\begin{equation}
	Z=\sum_{T \in \mathcal{T}} \frac{1}{C_T} e^{-S_E[T]}.
\end{equation}
This is the partition function of a \emph{statistical} system, where the exponential of the Euclidean Regge action is given the interpretation of a real Boltzmann weight. A major advantage is that the Euclidean version of the partition function is amenable to sampling by \emph{Monte Carlo simulations}, which we discuss in depth in Chapter \ref{ch:monte-carlo}. Note that attaching a physical interpretation to results obtained in this setting still requires us to rotate back to Lorentzian signature in the end.

As an example, we now present expressions of the Euclidean Regge action for CDT in two and three dimensions \cite{ambjorn2001dynamically,ambjorn2012nonperturbative}. They are formulated in terms of the counting variables $N_x$, where $x$ indicates the type of simplex being counted. As an example: $N_2$ refers to the number of 2-simplices $\sigma^{(2)}$ i.e.\ triangles, and $N_3$ to the number of 3-simplices $\sigma^{(3)}$ i.e.\ tetrahedra in the system.
In these variables, we can write the aforementioned parametrizations of the Regge action in two and three dimensions as (we omit the obvious dependence on the triangulation $T$):
\begin{align}
S_E^{(2)} &=  \lambda N_2, \label{intro-eq:cdt-regge-2d}\\
S_E^{(3)} &= -k_0 N_0 + k_3 N_3. \label{intro-eq:cdt-regge-3d}
\end{align}
The factors $\lambda, k_0, k_3$ are dimensionless bare coupling parameters of the models that carry the dependence on the dimensionful parameters $G, \Lambda, a$ from the continuum action. 
In two dimensions, the curvature term in the action depends only on the topology of the system due to the Gauss-Bonnet theorem, which states that
\begin{equation}
	\frac{2}{a^2} \sum_{\sigma^{(d-2)} \in T} \epsilon\left(\sigma^{(d-2)}\right)  V_{\sigma^{(d-2)}} = 2 \pi \chi,
\end{equation}
where $\chi$ is the Euler characteristic of the triangulation $T$. Therefore, the curvature term only gives rise to an irrelevant global prefactor in the Boltzmann weight, and $\lambda = \frac{2 \Lambda a^2}{16 \pi G}$ is the sole coupling parameter of the model in two dimensions. In three dimensions, there are two independent couplings which (when setting $\alpha = 1$) take the form 
\begin{equation}
	k_0 = \frac{a}{4G}, \quad \quad k_3 = \frac{a^3 \Lambda}{48 \sqrt{2} \pi G} + \frac{a}{4G}\left(\frac{3}{\pi} \arccos{1/3}-1\right).
\end{equation}

\subsection{From lattice to continuum}
\label{intro-subsec:lat-to-cont}
It is important to keep in mind that the goal of the CDT approach is to formulate a \emph{continuum} theory of quantum gravity. The introduction of discrete geometries is purely a technical device for defining and regularizing the path integral, giving us a handle on the sum over spacetimes. In the end, we are interested in the continuum limit of this sum over spacetimes, in which the number of building blocks grows to infinity and their typical size approaches zero. Taking this limit involves a renormalization procedure for the coupling constants of the model, where we express the bare parameters in terms of their renormalized counterparts. Constructing a continuum quantum field theory from a lattice model ordinarily requires the existence of a second-order phase transition, where the diverging correlation length can be used as a yardstick for translating lattice units to physical length dimensions. This is in principle no different in fully-fledged four-dimensional CDT, where the continuum theory should contain two local degrees of freedom at low energies, as seen by performing a Dirac analysis of the constraints in four-dimensional general relativity. 

Second-order phase transitions are indeed present in 4D CDT \cite{ambjorn2012second,ambjorn2017new}, and it may be possible to construct a continuum theory of quantum gravity by approaching these transitions in a specific way. However, in the absence of a fixed background geometry --- and thereby, a simple notion of correlation length --- it is not straightforward to see what quantities should be renormalized, and subsequently, how they should be renormalized \cite{ambjorn2014renormalization,ambjorn2020renormalization}. Additionally, it is difficult to numerically investigate regions of the phase space near phase transitions, further adding to the complexity of the issue.

A simpler notion of continuum limit exists in two- and three-dimensional CDT. For CDT in two dimensions, the curvature term in the action is topological, and as a result only the cosmological coupling $\lambda$ needs to be renormalized in order to obtain a continuum limit. The critical value $\lambdac$ for which the expectation value $\left\langle N_2 \right\rangle$ of the volume diverges depends only on the entropy of the two-dimensional triangulations. This critical value is known to equal $\lambdac = \ln 2$ exactly, as shown analytically in \cite{ambjorn1998nonperturbative}. In three-dimensional CDT, the critical value $\kcc$ of the cosmological coupling also depends on the gravitational coupling $\kg$, raising the possibility that this coupling needs to be fine-tuned to a critical value simultaneously. However, computer simulations of this model have provided evidence that this is not the case \cite{ambjorn2001nonperturbative,ambjorn2001computer,ambjorn20023d}. It was found that an extended geometry emerges for a certain range of $\kg$, where this coupling sets an effective length scale and is not renormalized. The continuum limit coincides with the limit of infinite system volume $\left\langle N_3 \right\rangle$, obtained by fine-tuning only the cosmological coupling $k_3$ to its ($\kg$-dependent) critical value $\kcc(\kg)$.

\subsection{Observables}
\label{intro-sec:observables}
The CDT partition function takes the form of a sum over triangulated geometries. We can probe the properties of a superposition of such geometries (a ground state of ``quantum geometry'') by investigating the quantum expectation value
\begin{equation}
	\langle \mo \rangle = \frac{1}{Z} \sum_{T \in \mathcal{T}} \frac{1}{C_T} \, \mo[T] \, e^{-S_E[T]},
	\label{intro-eq:obs-exp}
\end{equation}
of suitable geometric operators $\mo$ that are defined on the triangulations $T \in \mathcal{T}$. We consider such a geometric operator a quantum gravitational \emph{observable} if it is invariant under relabeling of the simplices, and it has an interpretation in the continuum limit of the model. Straightforward examples of geometric operators of this type are the number of $d$-simplices $N_d[T]$ in the triangulation, and the maximum link distance\footnote{Here we define the link distance $d(v, w)$ between two vertices $v,w$ of the triangulation as the minimum number of edges that connect $v$ to $w$.} $d_\textrm{max}[T]$ between any two of its vertices. On the continuum side, $N_d[T]$ can be interpreted as the effective volume of the resulting quantum spacetime, whereas $d_\textrm{max}[T]$ can be interpreted as its effective diameter. 

Quantum gravitational observables must respect the diffeomorphism symmetry of gravity. One of the strengths of the CDT approach is that it defines the nonperturbative path integral without reference to a coordinate system, and is therefore diffeomorphism-invariant by construction. This invariance carries over to observables formulated on CDT systems as long as such observables are also formulated without reference to coordinates: the role of diffeomorphism-invariance is now taken by invariance under relabeling of the simplices. A strategy for constructing such observables is to integrate local geometric scalar quantities over the full spacetime, and CDT observables are generally indeed of such a global nature.

A minor disclaimer is in place here: the term ``observable'' suggests a relation to quantities that we can in principle measure experimentally in the real world. However, we make broader use of the term, including in settings where we cannot necessarily link the expectation value of an observable to a physical experiment --- for example, in the context of two- and three-dimensional CDT, which are toy models of quantum gravity. A more direct link between quantum gravitational observables and physical experiment outcomes may exist for CDT in four dimensions. There is evidence \cite{ambjorn2011secondorder,ambjorn2012second,ambjorn2017new} that indicates the presence of continuous phase transitions in four-dimensional CDT, making this model a candidate for a continuum theory of four-dimensional quantum gravity. Observables in this theory (in a semi-classical regime) may therefore be matched to properties of the spacetime we inhabit, eventually perhaps allowing us to verify --- or falsify --- whether the theory can form a suitable description of our physical reality.

A small number of quantum gravitational observables have been studied in the context of CDT. Examples are the spectral and Hausdorff dimensions, which provide information about the effective dimensionality of the spacetimes generated by the CDT path integral. A recent addition to the list of CDT observables is the \emph{quantum Ricci curvature} \cite{klitgaard2018introducing}, which allows for investigating the curvature properties of CDT universes. Although one has access to curvature information of CDT simplicial manifolds at the cutoff scale through deficit angles, it turns out there are several obstacles that make this prescription ill-suited for use in the quantum setting. The quantum Ricci curvature is a novel method for defining curvature that addresses this issue. In this thesis, we devote special attention to the quantum Ricci curvature, building upon its definition to construct a global quantum gravitational observable called the \emph{curvature profile}. We subsequently investigate the behavior of the curvature profile in two distinct two-dimensional settings, finding further evidence that the prescription is appropriate for extracting curvature information in a non-classical context.

It is generally difficult to compute expectation values of CDT observables using purely analytical methods. One therefore often makes use of Monte Carlo methods to study CDT observables from a numerical point of view. These methods allow for sampling geometries from CDT ensembles, and we can use such samples to compute numerical approximations to expressions like Eq.\ \ref{intro-eq:obs-exp}. We provide a hands-on introduction to Monte Carlo approaches in the context of CDT quantum gravity in Chapter \ref{ch:monte-carlo} of this thesis.

\subsection{Random geometry from matrix integrals}
We have so far described how the CDT Lorentzian path integral can be studied using ensembles of triangulated geometries. Computing the path integral is tantamount to counting all triangulations with a fixed value of the action, at least in an asymptotic sense. Counting such triangulations beyond the two-dimensional case is an unsolved problem. However, there is an interesting connection between the theory of \emph{random matrices} and ensembles of two-dimensional triangulated Riemann surfaces, which was first observed by 't Hooft in 1974 \cite{hooft1993planar} in the context of Yang-Mills theory with a large number $N$ of color degrees of freedom. This connection allows one to investigate the behavior of the two-dimensional \emph{Euclidean} quantum gravitational path integral using analytical techniques. The formal Euclidean path integral for gravity is written
\begin{equation}
	Z = \int \displaylimits_{\substack{\textrm{Euclidean} \\\textrm{spaces } g_{\mu \nu}}} \hspace{-1em} \mathcal{D} \left[g_{\mu \nu} \right] e^{- S_\textrm{EH}[g_{\mu \nu}]}.
	\label{intro-eq:gravpi-euc}
\end{equation}
The difference with the Lorentzian path integral \eqref{intro-eq:gravpi} is that the integration is now over curved spaces\footnote{Since there is no notion of time in geometries with Euclidean signature, we use the term ``spaces'' instead of ``spacetimes'' here.} of Euclidean signature, and that the weight in the action is real instead of complex. The Euclidean path integral can be discretized following the same line of reasoning as presented in the previous sections. This approach is referred to in the literature as \emph{(Euclidean) Dynamical Triangulations}, or simply (E)DT. In two-dimensional DT, the path integral over curved surfaces of Euclidean signature is defined as the continuum limit of the sum over all gluings of equilateral triangles, where all edges are now spacelike. It is interesting to note here that parts of the CDT approach to Lorentzian quantum gravity can be traced back to this triangulated model of the path integral over two-dimensional surfaces of Euclidean signature. Originally, the motivation for studying such triangulated surfaces was not to quantize two-dimensional gravity, but rather to construct a nonperturbative UV-regularized theory of the worldsheet swept out by a bosonic string.

\begin{figure}[tb]
	\centering
	\includegraphics[width=0.45\textwidth]{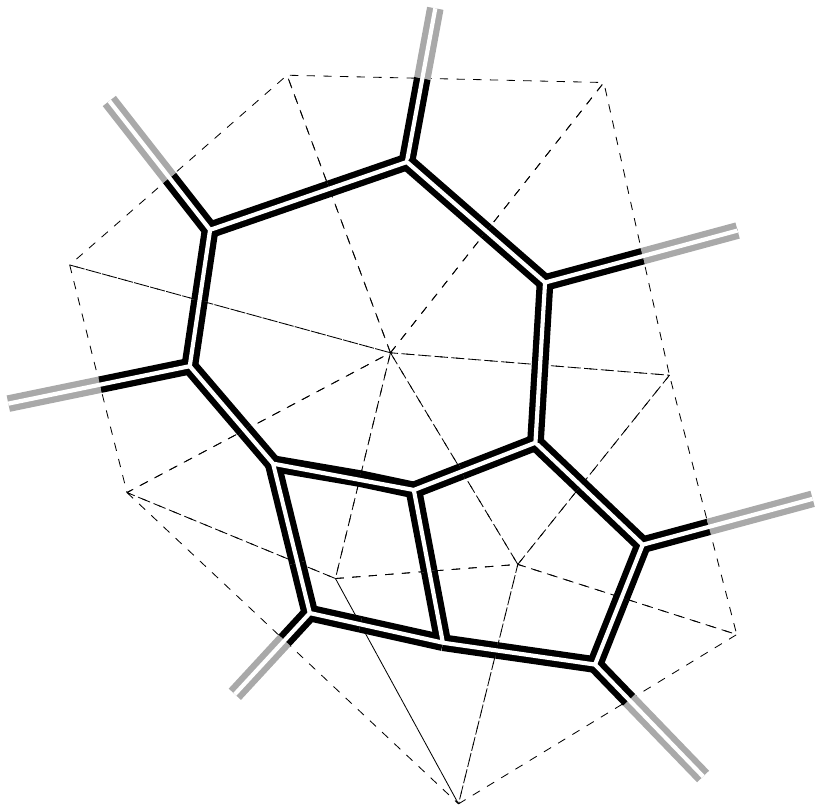}
	\hfill
	\includegraphics[width=0.45\textwidth]{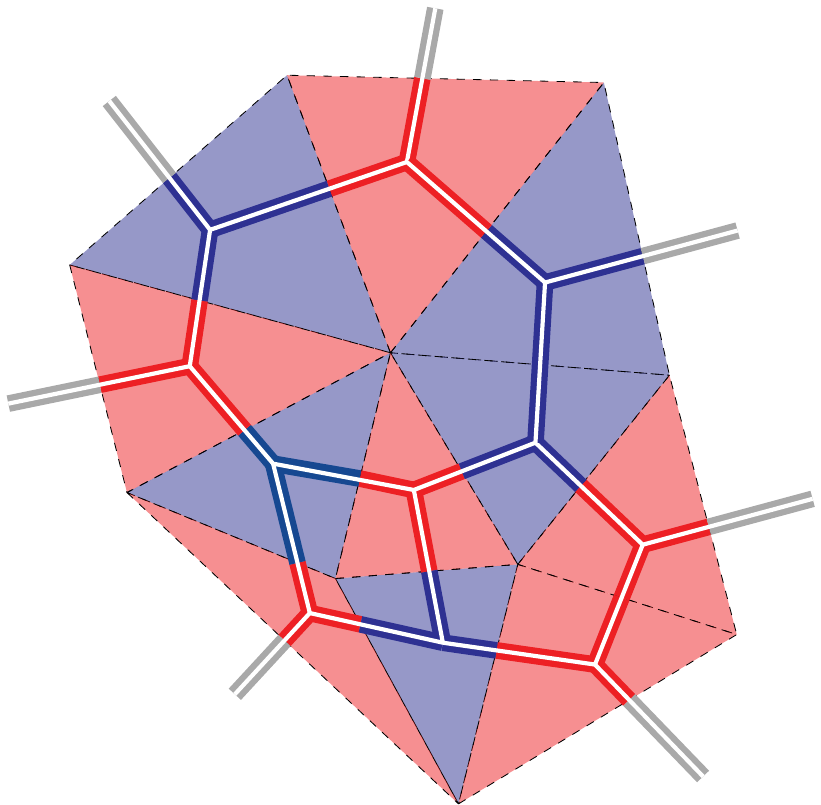}
	\caption{A triangulation and its dual ribbon graph from a one-matrix model (left) and a colored triangulation and its (dually weighted) dual ribbon graph from a two-matrix model (right). The interaction vertices are 3-valent, signifying that the associated matrix model includes at least an $M^3$ interaction term (or $A^3$ and $B^3$ for the two-matrix model).}
	\label{intro-fig:triangulation-dual}
\end{figure}

It is convenient to represent a triangulated surface by its \emph{dual graph}, illustrated in Fig.\ \ref{intro-fig:triangulation-dual}. In this dual graph, the triangles are represented by trivalent vertices, and its edges represent neighboring relations between the triangles. The graphs dual to a triangulated surface correspond to Feynman diagrams in the perturbative expansion of a scalar field theory with a cubic interaction term, also referred to as $\phi^3$ theory. The path integral for this theory is written 
\begin{equation}
	Z_{\phi^3} = \int \mathcal{D} \phi \, e^{- \frac{1}{2}(\partial \phi)^2 + \frac{\lambda}{3!} \phi^3},
\end{equation}
where we point out that this is a \emph{formal} expression that should be understood in terms of its perturbative expansion, since the $\phi^3$ term in the exponential renders this expression ill-defined, regardless of the sign of $\lambda$.
A similar structure emerges when computing the partition function of a probability distribution over random $N \times N$ Hermitian matrices $M_{ij}$. The partition function is defined as
\begin{equation}
	Z_V = \int_{\mathbb{H}_N} dM\, e^{- N \tr \left(\frac{1}{2} M^2 - V(M)\right)},
	\label{intro-eq:mat-int}
\end{equation}
where the measure $dM = \prod_i^N d M_{ii} \prod_{i < j}^N d \Re M_{ij} d \Im M_{ij}$ and $V(M)$ is the interaction potential. We can perturbatively expand the exponential of the interaction potential into a polynomial, so that the full partition function can be computed by the use of Wick's theorem. The partition function then takes the form of a sum over the moments of a Gaussian probability distribution over the matrix ensemble. This perturbative expansion can be represented by \emph{ribbon graphs}, with double lines as propagators instead of the single lines as seen in the diagrams of $\phi^3$ theory. For the specific choice of interaction potential $V(M) = \frac{\lambda}{3!} M^3$, the ribbon graphs can now be viewed as the dual graphs of an oriented triangulated Riemann surface as shown in Fig.\ \ref{intro-fig:triangulation-dual}. Including higher-order interaction terms $M^n$ leads to the appearance of higher-order $n$-gons in the resulting `triangulation'. One of the major insights by 't Hooft in \cite{hooft1993planar} is that the topological genus of such a triangulation determines the power of $N$ with which the corresponding ribbon graph appears in the perturbative expansion of the partition function. This allows one to organize the expansion in terms of the genus. The leading contribution in $N$ comes from the \emph{planar} ribbon graphs, which correspond to triangulations with topological genus zero. 

We can therefore capture the counting of triangulated surfaces in the language of random matrices. This so-called \emph{matrix model} approach has proven to be a powerful analytical tool for investigating the continuum behavior of 2D Euclidean quantum gravity, and we refer to \cite{difrancesco19952d} for a comprehensive overview on the subject. In anticipation of the results presented in Chapter \ref{ch:diff-mat} of this thesis, we highlight one particularly interesting application of matrix models in the context of 2D quantum gravity coupled to Ising spin degrees of freedom. One can couple two separate matrix integrals of the form \eqref{intro-eq:mat-int} with an interaction term $\tr(AB)$ to obtain a two-matrix model, written as
\begin{equation}
	Z_{V_1, V_2} = \int_{\bH_N\times \bH_N}\frac{\d A\, \d B}{c_N} \e^{-N \tr(AB- V_1(A) - V_2(B))}.
	\label{intro-eq:mat-double-int}
\end{equation}
We can represent the so-called ``dually weighted'' ribbon graphs of the associated perturbative expansion by assigning colors to the ribbons associated to the distinct matrix ensembles: blue for $A$, and red for $B$. The interaction term allows for connections between blue and red ribbons. Setting $V_1(M) = V_2(M) = - \frac{1}{2} M^2 + \frac{\lambda}{3!} M^3$, we obtain a diagrammatic expansion into ribbon graphs with both blue and red trivalent ribbon interactions, and ribbons can be connected regardless of their color. The triangulations dual to the graphs generated by the partition function \eqref{intro-eq:mat-double-int} then correspond to all possible \emph{colored} triangulations of closed surfaces (see Fig.\ \ref{intro-fig:triangulation-dual} for an example). We can interpret the two colors as a spin-$\tfrac{1}{2}$ degree of freedom on each triangle, so that the partition function of this two-matrix model corresponds to the partition function of the Ising model coupled to two-dimensional DT quantum gravity. 

Matrix integrals provide an analytical toolbox to describe the behavior of such two-dimensional systems of quantum gravity. For example, it is possible to determine the leading-order growth rate of the number of genus zero triangulations by computing the large-$N$ behavior of the partition function \eqref{intro-eq:mat-int}. This can be achieved by approximating the random matrix integral near saddle points of the action. Furthermore, one can rewrite matrix model partition functions as a determinant of a system of \emph{(bi-)orthogonal polynomials}, which can in some cases give exact results for correlation functions in the model. Both one- and two-matrix models can be recast into a language of matrix differential operators, and we make a detailed investigation of the interesting connections between the integral and differential languages in Chapter \ref{ch:diff-mat} of this thesis.

\section{Thesis overview}
In this thesis, we will investigate aspects of low-dimensional quantum gravity from several distinct viewpoints. Part \ref{part:curv} is dedicated to the study of the \emph{quantum Ricci curvature} (QRC) \cite{klitgaard2018introducing}, a notion of curvature that can be implemented on a much wider class of metric spaces than the (semi-)Riemannian manifolds familiar from classical general relativity. In Chapter \ref{ch:curv-profs}, we discuss the construction of the quantum Ricci curvature as formulated by Klitgaard and Loll, and subsequently average this construction to obtain an observable which we dub the \emph{curvature profile}. The curvature profile captures scale-dependent coarse-grained curvature information, and can be interpreted as a global signature of the space under consideration. We discuss the implementation of the curvature profile on two-dimensional CDT with toroidal topology in Chapter \ref{ch:ricci2d}, and present and interpret our measurement results obtained through Monte Carlo simulations. Chapter \ref{ch:defects} revolves around the application of the curvature profile to surfaces with point defects, where the curvature is maximally concentrated in isolated singularities. By measuring the curvature profiles of the surfaces of Platonic solids, we put the coarse-graining properties of our prescription to the test.

We turn our attention to three-dimensional CDT in Part \ref{part:cdt3d} of this thesis. In particular, we focus on the two-dimensional spatial slice geometries that foliate the triangulated spacetimes in this ensemble. By measuring several of their properties, including the Hausdorff dimension and the entropy exponent, we investigate whether these spatial slices can be modeled as independent systems of two-dimensional random geometry. The slices exhibit a surprising behavior, and we set out to develop a better understanding of their relation to two-dimensional DT quantum gravity.

Part \ref{part:mat} focuses on several technical aspects of discrete quantum gravity research. In Chapter \ref{ch:diff-mat}, we explore a differential reformulation of integrals over random matrices. Such random matrix integrals can be diagonalized for certain special choices of matrix ensembles, and we detail how the analogous procedure can be carried out in the differential language. We detail the parallels between the two formulations, and demonstrate how the differential reformulation might provide a different perspective on the resolution of matrix models in terms of orthogonal polynomials. Chapter \ref{ch:monte-carlo} is based on an unpublished set of notes written by the author, centered around implementing computer simulations of CDT quantum gravity. We explain the general set-up of Markov chain Monte Carlo algorithms and discuss the difficulties one encounters when these are applied to dynamical lattices. Subsequently, we describe the technical aspects of our own open-source C++ implementation of two- and three-dimensional CDT simulations. Constructing these codebases formed an integral part of the author's Ph.D. research, serving as the basis for all simulation results discussed in Chapters \ref{ch:ricci2d} and \ref{ch:slice3d}. We therefore opted to include this chapter in the main body of the thesis, as opposed to relegating it to an appendix.

We end this thesis with a discussion of the lessons learned during the research leading up to this manuscript, and provide a personal perspective on the current and potential future state of CDT research.



\part{Curvature of non-classical spaces}
\label{part:curv}
\chapter{Curvature profiles and the average sphere distance}\label{ch:curv-profs}
We briefly mentioned in the Introduction how curvature plays a central role in Einstein's theory of gravity. The current chapter serves as an introduction to Part \ref{part:curv} of this thesis, in which we go further into detail on the topic of curvature in the context of \emph{quantum} gravity. We start by discussing the difficulties one encounters when attempting to define curvature in the context of simplicial quantum gravity. Subsequently, we review a solution that was proposed in \cite{klitgaard2018introducing}, going by the name of the \emph{quantum Ricci curvature}. We then extend this prescription to formulate a coordinate-invariant observable that can be implemented in both a classical and a quantum setting. This scale-dependent observable, called the \emph{curvature profile}, is the main object of study in the subsequent two chapters. We can use it to coarse-grain local curvature fluctuations occurring on a microscopic length scale in order to obtain information about the distribution of curvature at longer distances.

\section{Riemann curvature}
\begin{figure}[ht!]
	\centering
	\includegraphics[width=0.5\textwidth]{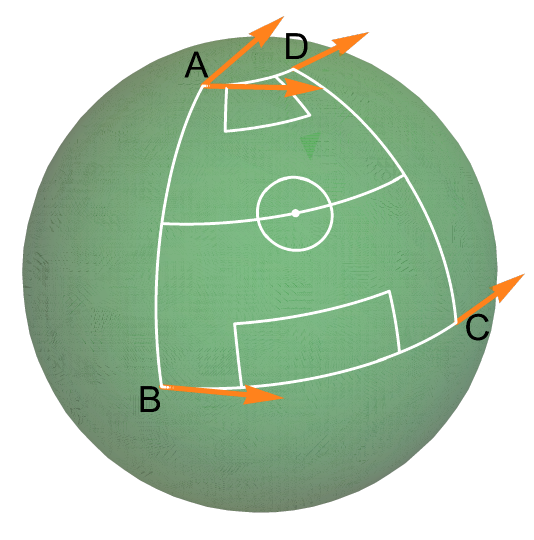}
	\caption{Parallel transport of a vector (`flag') along a closed path (`football pitch') on a curved surface (`planet astroturf'). After traversing the loop $ABCD$ counterclockwise, the vector has rotated with respect to its original orientation.}
	\label{cp-fig:parallel-transport}
\end{figure}

Curvature is a fundamental concept in the field of Riemannian geometry, where it is defined through the \emph{Riemann curvature tensor}. This curvature tensor captures how tangent vectors on a (pseudo-)Riemannian manifold are affected when parallel transporting them around infinitesimally small loops in the geometry. We can illustrate how tangent vectors are affected by parallel transport in a curved space by the following thought experiment. Imagine a linesman on a football pitch, holding a flag pointed straight to the east. We then assign the linesman the task to walk around a closed loop on the pitch, while continuously keeping the flag parallel to its previous position. If the football pitch is perfectly flat\footnote{This is known to be patently false in an experimental setting, where it has delivered my team undeserved goals on multiple occasions.}, the flag will point in the same direction after the linesman returns to his original position, regardless of the shape or size of the loop he traversed. This is not the case, however, when we consider football pitches on a curved surface --- for example, a Siberia-sized football pitch constructed on the surface of the Earth. The effect will be small for short walks, but if the linesman goes on larger trips around this pitch he will find that the flag has rotated appreciably with respect to its original direction upon returning to his starting point. We present a schematic drawing of this thought experiment in Fig. \ref{cp-fig:parallel-transport}.

We now briefly recall the relevant concepts and definitions of continuum (pseudo-) Riemannian geometry. Given a manifold $\mathcal{M}$ equipped with a metric tensor field $g_{\mu\nu}$, the Riemann curvature tensor $R^\mu_{\nu \rho \sigma}$ is defined as follows:
\begin{equation}
	R^\mu_{\nu \rho \sigma} = \partial_\rho \Gamma^\mu_{\nu \sigma} - \partial_\sigma \Gamma^\mu_{\nu \rho} + \Gamma^\mu_{\rho \lambda} \Gamma^\lambda_{\nu \sigma}-\Gamma^\mu_{\sigma \lambda} \Gamma^\lambda_{\nu \rho},
	\label{qp-eq:riem}
\end{equation}
where the $\Gamma^\mu_{\nu \rho}$ are the Christoffel symbols computed from the first derivatives of the metric:
\begin{equation}
	\Gamma^\mu_{\nu \rho} = \frac{1}{2} g^{\mu \lambda} \left( \partial_\nu g_{\rho \lambda} + \partial_\rho g_{\nu \lambda} - \partial_\lambda g_{\nu \rho}\right).
\end{equation}
From the Riemann tensor, we can compute the \emph{Ricci tensor} $R_{\mu \nu}$ by contracting the first and third indices:
\begin{equation}
	R_{\mu \nu} = R^\lambda_{\mu \lambda \nu},
	\label{qp-eq:rict}
\end{equation}
and subsequently the \emph{Ricci scalar} $R$ by performing one further index contraction:
\begin{equation}
	R = R^\mu_\mu.
	\label{qp-eq:rics}
\end{equation}
These last two objects are especially important in the study of gravity, as they appear in the vacuum Einstein field equations for general relativity, along with the metric tensor $g_{\mu \nu}$ and cosmological constant $\Lambda$:
\begin{equation}
	R_{\mu \nu} + \left(\Lambda - \frac{1}{2} R \right) g_{\mu \nu} = 0.
\end{equation}

In this chapter, we describe how one can formulate a notion of Ricci curvature that is well-suited for CDT (and EDT) nonperturbative quantum gravity. The above definition of the Riemann curvature tensor $R^\mu_{\nu \rho \sigma}$ is not directly applicable in this context. The metric tensor $g_{\mu \nu}$ is not twice-differentiable at the $(d-2)$-dimensional hinges of the piecewise-flat geometries encountered in the discretized path integral, and as a result $R^\mu_{\nu \rho \sigma}$ is singular at these hinges (and identically zero everywhere else). 

We discussed in Sec.\ \ref{intro-sec:regge-cdt} how the local Ricci scalar curvature can be formulated on simplicial manifolds using the Regge prescription, sidestepping the need for a Riemann curvature tensor defined in terms of derivatives of the metric tensor. Indeed, this is an appropriate notion of Ricci curvature in the context of classical \emph{Regge calculus} \cite{regge1961general}, which is a discrete approach for finding approximate solutions to the classical Einstein equations.\footnote{Note that one can, in turn, construct simplicial implementations of the Riemann curvature tensor using this prescription, see e.g. \cite{hamber1986simplicial,brewin1988riemann,alsing2011simplicial}.} We can turn this into a diffeomorphism-invariant quantity by integrating the Ricci scalar over the simplicial manifold. However, it was found in simulations that this integrated curvature does not scale properly in the nonperturbative quantum setting for dimensions three and higher, making this approach ill-suited for our purposes.

In what follows, we introduce the so-called \emph{quantum Ricci curvature} \cite{klitgaard2018introducing}, a prescription for studying curvature that addresses some of the aforementioned issues. The quantum Ricci curvature is a quasi-local quantity that can be used to determine curvature information on coarse-grained scales. This property makes it a suitable method for investigating whether CDT geometries resemble any classical curved spaces on any range of scales.

\section{Quantum Ricci curvature}
\label{cp-sec:qrc}
The quantum Ricci curvature was inspired by a notion of Ricci curvature due to Ollivier \cite{ollivier2009ricci}, applicable to more general metric spaces than the smooth, (pseudo-)Riemannian ones we discussed at the start of the previous section.\footnote{We refer to \cite{trugenberger2016random,trugenberger2017combinatorial,kelly2019selfassembly,kelly2021emergence,trugenberger2022emergent,tee2021enhanced,vanderhoorn2020ollivier} for examples of the implementation of the ``Ollivier curvature'' in various random geometric settings.} Its classical starting point is the observation that two sufficiently close and sufficiently small geodesic spheres on a positively curved Riemannian space are closer to each other than their respective centers, 
while the opposite is true on a negatively curved space. This leads to the idea of \textit{defining} curvature on more general
(e.g.\ nonsmooth) metric spaces by comparing the distances of spheres with the distances of their centers.

The quantum Ricci curvature is a particular implementation of this idea, designed specifically for 
use on the ensembles of piecewise flat simplicial configurations of nonperturbative, dynamically triangulated
quantum gravity models. In Chapter \ref{ch:ricci2d} of this thesis, we investigate the quantum Ricci curvature in the context of two-dimensional CDT quantum gravity. It has furthermore been applied to Euclidean dynamical triangulations in two dimensions \cite{klitgaard2018implementing} and more recently to the physically relevant case of CDT in four dimensions \cite{klitgaard2020how}, demonstrating its viability in full 4D quantum gravity. However, the prescription can be implemented in other geometric settings, as long as one has notions of distance and volume. This includes classical continuum spaces, which we focus on in Chapter \ref{ch:defects} as part of an effort to build up a reference catalogue of curvature profiles, for comparison with quantum results. The word ``quantum'' in the ``quantum Ricci curvature'' merely signifies that the motivation to formulate this prescription is applications in nonperturbative quantum gravity, even though the method can be applied in a non-quantum setting.

\begin{figure}[tb]
	\centering
	\includegraphics[width=0.5\textwidth]{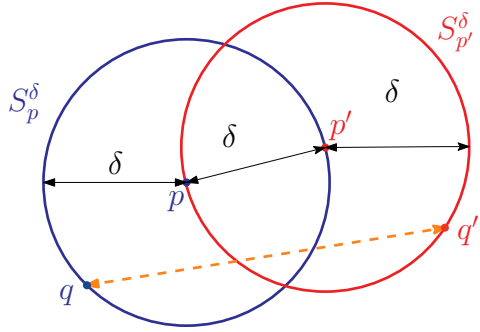}
	\caption{The double-circle configuration involved in computing the average sphere distance $\dbar(S_p^\delta, S_{p'}^\delta)$ of Eq.\ \eqref{cp-sdist} 
	on a two-dimensional Riemannian manifold.
	}
	\label{cp-fig:avg-sphere-dist}
\end{figure}

In what follows, we detail the construction of the quantum Ricci curvature in a smooth two-dimensional Riemannian manifold. This procedure can be extended to the higher-dimensional case.
For a two-dimensional continuum space $M$ with metric $g_{\mu\nu}(x)$ and associated geodesic distance $d_g$,
the quasi-local set-up associated with 
the quantum Ricci curvature $K(p,p')$ consists of two intersecting geodesic circles (one-spheres)
$S_p^\delta$ and $S_{p'}^\delta$ of radius $\delta$, with centers $p$ and $p'$ a distance $\delta$ apart
(Fig.\ \ref{cp-fig:avg-sphere-dist}).

To extract $K(p,p')$ one computes the \textit{average sphere distance}\footnote{In two dimensions, one could also call it the average \textit{circle} 
distance, but we will use the more general term in the remainder of this thesis.} $\bar{d}(S_p^{\delta},S_{p'}^{\delta})$,  
which by definition is given by the normalized double-integral 
\begin{equation}
\bar{d}(S_p^{\delta},S_{p'}^{\delta}):=\frac{1}{\textit{vol}\,(S_p^{\delta})}\frac{1}{\textit{vol}\,(S_{p'}^{\delta})}
\int_{S_p^{\delta}}dq\; \sqrt{h} \int_{S_{p'}^{\delta}}dq'\; \sqrt{h'}\ d_g(q,q')
\label{cp-sdist}
\end{equation}
over all pairs of points $(q,q')\in S_p^\delta \times S_{p'}^\delta$,
that is, by \textit{averaging} the distance $d_g(q,q')$ over the two circles,
where $h$ and $h'$ are the determinants of the induced metrics on $S_p^{\delta}$ and $S_{p'}^{\delta}$,
which are also used to compute the one-dimensional circle volumes $vol(S)$.
From Eq.\ \eqref{cp-sdist}, the quantum Ricci curvature $K(p,p')$ associated with the point pair $(p,p')$ 
is defined by
\begin{equation}
\bar{d}(S_p^{\delta},S_{p'}^{\delta})/\delta=c\, (1 - K(p,p')),\;\;\;\;\;\;\;\; \delta =d_g(p,p'),
\label{cp-qric}
\end{equation}
where the prefactor $c$ is given by $c\! :=\!\lim_{\delta\rightarrow 0} \bar{d}/\delta$, and in general may depend
on the point $p$.
On smooth Riemannian manifolds and for small distances $\delta$, one can expand the quotient (\ref{cp-qric}) 
into a power series in $\delta$, with $c$ a constant that only depends on the dimension. In two dimensions,
one finds \cite{klitgaard2018implementing}
\begin{equation}
\bar{d}/\delta= 1.5746+\delta^2 \left(-0.1440\, \textit{Ric}(v,v)+{\cal O}(\delta)\right),
\label{cp-2dexp}
\end{equation}
where $\mathit{Ric}(v,v)\! =\! R_{ij}v^iv^j$ is the usual Ricci curvature, evaluated on the unit vector $v$ at the point $p$ 
in the direction of $p'$, and the numerical coefficients are rounded to the digits shown.

\section{Classical curvature profiles}
We can probe \emph{global} information about the curvature of a Riemannian manifold $M$ by integrating the average sphere
distance (\ref{cp-sdist}) over all positions $p$ and $p'$ of the two circle centers, 
while keeping their distance $\delta$ fixed. The spatial average over the average sphere distance at
the scale $\delta$ is 
\begin{equation}
\bar{d}_\textrm{av}  (\delta):=\frac{1}{Z_\delta}
\int_{M}d^2x \sqrt{g} \int_M d^2 x' \sqrt{g}\;\,  \bar{d}(S_x^{\delta},S_{x'}^{\delta})\; \delta_D(d_g(x,x'),\delta),
\label{cp-avdist}
\end{equation}   
where $\delta_D$ denotes the Dirac delta function and the normalization factor $Z_\delta$ is given by
\begin{equation}
Z_\delta=\int_{M}d^2x \sqrt{g} \int_M d^2 x' \sqrt{g}\;\; \delta_D(d_g(x,x'),\delta).
\label{cp-zdelta}
\end{equation}
The integration in Eq.\ \eqref{cp-avdist} includes an averaging over directions, which means that it will
allow us to extract an (averaged) \textit{quantum Ricci scalar} $K_\textrm{av}(\delta)$.
The curvature profile is now given by the quotient
\begin{equation}
\bar{d}_\textrm{av}(\delta)/\delta=:c_\textrm{av} (1 - K_\textrm{av}(\delta)),
\label{cp-profile}
\end{equation}
where the constant $c_\textrm{av}$ is defined by $c_\textrm{av}\! :=\!\lim_{\delta\rightarrow 0} \bar{d}_\textrm{av}/\delta$.
Note that for the special case where $M$ is the round two-sphere no spatial averaging is necessary to obtain the curvature
profile \cite{klitgaard2018introducing}, because in that case the average sphere distance \eqref{cp-sdist}
depends only on the distance $\delta$, and not on the locations of $p$ and $p'$. 

The classical curvature profile can be used on any smooth Riemannian geometry, even if it is highly inhomogeneous and anisotropic. The quantum\footnote{Note that we are still in a classical setting at this stage, and the term ``quantum'' indicates that this is a Ricci scalar obtained from the ``quantum Ricci curvature'' prescription.} Ricci scalar obtained detects the scale-dependent distribution of curvature, while washing out all directional information. It can even be implemented on singular geometries where the curvature is concentrated in several isolated points. We treat a specific class of such geometries in Chapter \ref{ch:defects}, where we compute the (partial) curvature profiles of the surfaces of Platonic solids.

\section{Quantum curvature profiles}
As emphasized earlier, the classical curvature profile (\ref{cp-profile}) 
can be translated directly into a quantum observable,
whose expectation value $\langle \bar{d}_\textrm{av}(\delta)/\delta \rangle$ can 
be determined numerically in an approach like CDT quantum gravity. 
In such a lattice approach, all distances are given in dimensionless
lattice units, which can be converted into dimensionful units invoking the lattice spacing $a$, an ultraviolet length cutoff
that is taken to zero in any continuum limit. 
From the dimensionless curvature $K_\textrm{av}(\delta)$ measured in the nonperturbative quantum theory
one can extract a dimensionful, renormalized quantum Ricci scalar 
$K^r\! (\delta_\textrm{ph})$, which depends on a physical coarse-graining or renormalization scale $\delta_\textrm{ph}\! :=\! a\delta$, 
via
\begin{equation}
K_\textrm{av}(\delta)=:\delta^2 a^2 K^r\! (\delta_\textrm{ph})=(\delta_\textrm{ph})^2 K^r\! (\delta_\textrm{ph}), 
\label{cp-qren}
\end{equation}
in the limit $a\rightarrow 0$. In what follows, we provide further detail on how the average sphere distance \eqref{cp-sdist} can be adapted for triangulated geometries, and how we subsequently put it to use in order to formulate the quantum version of the curvature profile. We again restrict to the case of two-dimensional triangulations, but the reader should keep in mind that the set-up can be generalized to the higher-dimensional case.

\begin{figure}[ht]
	\centering
	\includegraphics[width=0.6\textwidth]{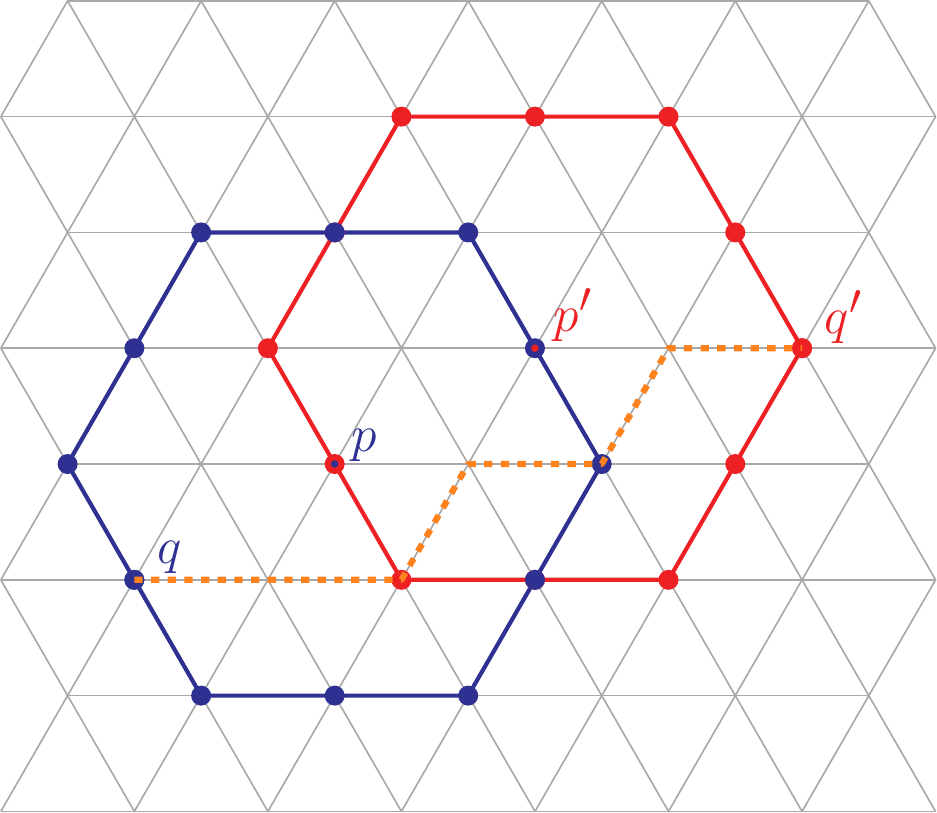}
	\caption{The double-circle configuration involved in computing the average sphere distance $\dbar(S_p^2, S_{p'}^2)$ of Eq.\ \eqref{cp-eq:disc-avg-sph-dst}
	on the two-dimensional regular triangulation of the flat plane, using the link distance.}
	\label{cp-fig:cdt-sphere-pair}
\end{figure}

On a two-dimensional piecewise flat manifold $T$, one can implement formula (\ref{cp-sdist}) using discrete analogues of 
distances and volumes, yielding
\begin{equation}
	\dbar (S^\delta_{p},S^\delta_{p'} ) = \frac{1}{N_1(S^\delta_{p})} \frac{1}{N_1(S^\delta_{p'})} \sum_{q \in S^\delta_{p}} \sum_{q' \in S^\delta_{p'}} d(q,q'), 
	\;\;\; d(p,p')=\delta,
	\label{cp-eq:disc-avg-sph-dst}
\end{equation}
where we continue to use the notation $\dbar$ for the average sphere distance, and $q$, $q'$, $p$, $p'$ now denote vertices of $T$.
The standard notion of distance on an equilateral triangulation $T$ is the integer-valued link distance $d(q,q')$ between pairs of vertices $q$, $q'$ in $T$,
counting the number of links (edges) in the shortest path of contiguous links joining $q$ and $q'$. By definition, a ``sphere'' $S_p^\delta$ centered at the
vertex $p\in T$ is the set of all vertices $q$ at link distance $\delta$ from $p$, and $N_1(S_p^\delta)$ counts the number of vertices
in this set. These sets only loosely resemble the round spheres of the smooth continuum, as is illustrated by Fig.\ \ref{cp-fig:cdt-sphere-pair},
which depicts a pair of spheres on a regular tessellation of the plane in terms of equilateral triangles, forming a hexagonal lattice. 
The dashed orange line indicates a shortest
path between the vertices $q$ and $q'$. As happens generically, this path is not unique. Note that according to our definition, the spheres
consist only of the vertices, and not the links between neighboring vertices, which in Fig.\ \ref{cp-fig:cdt-sphere-pair} 
have merely been included to guide the eye. 
On generic CDT configurations, which are much less regular, one cannot in general link vertices of a ``sphere'' pairwise in a way that
results in a single loop without intersections or overlaps.

Later we will occasionally work with the dual link distance, which is defined on the dual vertices of the triangulation (the centers of the triangles), 
and counts the number of dual links in the shortest path between a pair of dual vertices. The dual links can be thought of as the edges of the trivalent 
graph dual to a given triangulation $T$. By virtue of universality, the link distance and the dual link distance should lead to the same notion of physical distance
in the continuum limit, up to an overall scale due to the difference between the discrete link and dual link units.\footnote{We are not aware of counterexamples
in either DT or CDT models of quantum gravity in any dimension.} One would therefore also expect equivalent physical
results for the average sphere distance (and the quantum Ricci curvature extracted from it) in the continuum limit, independent of the choice of
the discrete distance function. However, since the actual simulations always take place on finite lattices, it can be convenient to make a specific choice.
For example, in the recent investigation of the quantum Ricci curvature in CDT quantum gravity in four dimensions, in order to obtain measurements
with a sufficiently fine distance resolution at the given lattice size, it turned out to be important to use the dual link distance \cite{klitgaard2020how}.    

From the average sphere distance $\bar{d}(S_p^{\delta},S_{p'}^{\delta})$, we can now obtain the quantum Ricci curvature\footnote{The ``$q$'' in $K_q$ stands for ``quantum'' and does not denote a
point or vertex.} $K_q(p,p')$
associated with a point pair $(p,p')$ at distance $\delta$ in essentially the same way as \eqref{cp-qric}. 
Note that such a point pair can be thought of as a generalization of the notion of a tangent vector at the point $p$, which
``points from $p$ to $p'$'' and reflects the directional dependence of the quantum Ricci curvature. To define the
quantum Ricci curvature, we take the quotient of two distances, namely, the average sphere distance and the distance $\delta$,  
\begin{equation}
\bar{d}(S_p^{\delta},S_{p'}^{\delta})/\delta=c_q\, (1 - K_q(p,p')).
\label{cp-qric-disc}
\end{equation}
Considering this expression as a function of $\delta$ (while keeping the ``direction'' $\overline{pp'}$ fixed), the factor $c_q$ by definition
describes its constant, $\delta$-independent part, while $K_q(p,p')$ captures any nontrivial $\delta$-dependence.

In the context of smooth Riemannian geometry, we defined the factor $c_q$ as the value that \eqref{cp-qric} takes near $\delta=0$. On triangulations, the (dual) link distance $\delta$ is integer-valued and the limit $\delta\!\rightarrow\! 0$ of \eqref{cp-qric-disc} is not defined. Moreover, measurements of the average sphere distance at small values of $\delta$ are affected by the details of the discretization. These so-called ``lattice artifacts'' are unphysical, and should therefore be discarded.
One can therefore define $c_q\! :=\! (\bar{d}/\delta)|_{\delta=\delta_0}$, where $\delta_0$ is the distance above which lattice artifacts are
negligible. Numerous investigations of the average sphere distance on triangulated and other piecewise flat manifolds (regular lattices
and Delaunay triangulations
\cite{klitgaard2018introducing}, DT \cite{klitgaard2018implementing} and CDT \cite{klitgaard2020how} configurations) have found that $\delta_0\approx 5$ in terms of the link distance, with
larger values of $\delta_0$ when the dual link distance is used. These results also show that the value of $c_q$ is not universal, but
depends on the details of the lattice discretization. This includes the lattice direction in which the quantum Ricci curvature is evaluated, 
and is related to the absence of exact rotational
invariance around a given lattice vertex. It is similar to the ``staircase effect'' on a regular tiling of the two-dimensional flat plane
by identical squares. In this case, the link distance is anisotropic, since walking $\delta$ steps from a vertex $p$ along a shortest path in 
the diagonal direction results in a zigzag path that only covers a distance $\delta/\sqrt{2}$ in the underlying flat plane (assuming the link length is 1), 
while shortest paths of $\delta$ steps in the horizontal or vertical direction cover a continuum distance of $\delta$ \cite{klitgaard2020how,klitgaard2022new}. 

Exploring the quantum Ricci curvature $K_q$ in the full quantum theory, which is based on a nonperturbative path integral over all spacetimes, requires that we 
construct diffeomorphism-invariant observables $\cal O$ depending on $K_q$. We can then determine their expectation
values
\begin{equation}
\langle {\cal O}\rangle =\frac{1}{Z}\,\sum_T \frac{1}{C_T}\, {\cal O}[T]\, {\mathrm e}^{-S [T]}
\label{cp-expect}
\end{equation}
with the help of Monte Carlo simulations, where the partition function $Z$ takes the form
\begin{equation}
	Z = \sum_{T} \frac{1}{C_T} \, {\mathrm e}^{-S[T]}.
\end{equation}
As explained in detail in the Introduction, invariant observables in pure gravity generically involve a spacetime averaging, to
eliminate the dependence on individual points (or other subregions), which are identified by their unphysical coordinate labels. 

We can formulate a suitable quantum observable of this kind by mimicking our procedure for defining the \emph{classical} curvature profile \eqref{cp-profile} in terms of the classical quantum Ricci curvature \eqref{cp-qric}. To this end, we again eliminate the dependence of the quasi-local quantum Ricci curvature $K_q(p,p')$ on its local arguments by summing over the locations of the circle centers $p$ and $p'$ in the average sphere distance, subject to the constraint that the distance
between $p$ and $p'$ is always equal to $\delta$. The normalized average of the average sphere distance is then written \footnote{By a minor abuse of notation, we use the same symbol $\bar{d}_\textrm{av}$ for both the smooth and the piecewise
flat case. This should not lead to any confusion, since Chapters \ref{ch:ricci2d} and \ref{ch:defects} each focus on only one of these two definitions.} 
\begin{equation}
\bar{d}_\textrm{av}  (\delta):	=  \frac{1}{{\cal N}_\delta} \sum_{p \in T}\sum_{p' \in T} \dbar (S_p^\delta, S_{p'}^\delta) \, \delta_K(d(p,p'), \delta),
	\label{cp-avdist-disc}
\end{equation}
where $\delta_K$ denotes a discrete Kronecker delta that enforces the distance constraint between $p$ and $p'$. The normalization factor ${\cal N}_\delta$ is given by
\begin{equation}
	{\cal N}_\delta = \sum_{p \in T}\sum_{p' \in T}  \delta_K(d(p,p'), \delta).
\end{equation}
These two expressions are entirely analogous to \eqref{cp-avdist} and \eqref{cp-zdelta}, respectively.

Just as in \eqref{cp-avdist}, the expression \eqref{cp-avdist-disc} includes an average over directions, again allowing us to extract an (averaged) quantum Ricci scalar $K_\textrm{av}(\delta)$ 
from the curvature profile, similarly defined as the quotient
\begin{equation}
\bar{d}_\textrm{av}(\delta)/\delta=:c_\textrm{av} (1 - K_\textrm{av}(\delta)).
\label{cp-profile-disc}
\end{equation}
The factor $c_\textrm{av}$ can now be defined as the value $c_\textrm{av}\! :=\! (\bar{d}_\textrm{av}/\delta)|_{\delta=\delta_0}$, where $\delta_0$ is again taken to be some fixed integer value as discussed earlier. 

We point out that it is possible to access \emph{tensorial} information of the Ricci curvature if one does not average over directions between the sphere centers. We will see an example of such direction-dependent measurements in Chapter \ref{ch:ricci2d} of this thesis, where we distinguish between timelike and spacelike directions in two-dimensional CDT. A similar investigation was made in the context of four-dimensional CDT in \cite{klitgaard2020how}.  

Finally, in the quantum theory we will be interested in the expectation value of the 
nonlocal curvature profile ${\cal O}(\delta)\! =\! \bar{d}_\textrm{av}(\delta)/\delta$ (and the averaged quantum Ricci curvature
extracted from it) as a function of $\delta$, obtained by taking the ensemble average over CDT geometries $T$,
\begin{equation}
	\langle \bar{d}_\textrm{av}(\delta)/\delta \rangle \equiv \langle \bar{d}_\textrm{av}(\delta)\rangle /\delta 	
	= \frac{1}{\delta}\, \sum_{T } \frac{1}{C_T}\,  \bar{d}_\textrm{av}(\delta)\, e^{- S_\lambda [T]}   .
	\label{cp-eq:curv-prof-cdt}
\end{equation}
This quantum expectation value of the curvature profile can typically only be computed by resorting to Monte Carlo simulations. In order to study the curvature profile in the continuum limit of the lattice theory, we make use of finite-size scaling methods as discussed in the Introduction: we determine curvature profiles of the lattice model at a range of fixed target volumes, and subsequently investigate the scaling behavior of the resulting profiles. A continuum interpretation of the curvature profile can only be given if the profiles at several distinct volumes can be collapsed on top of each other by rescaling them with a single parameter, also called the \emph{scaling dimension} of the observable. In Chapter \ref{ch:ricci2d}, we discuss the implementation of the quantum curvature profile on the two-dimensional CDT ensemble with the topology of a torus.


\chapter{Curvature in two-dimensional CDT}\label{ch:ricci2d}
\section{Introduction}

In this chapter, we will investigate the quantum Ricci curvature of two-dimensional CDT quantum gravity on a torus, describing (1+1)-dimensional
universes whose spatial slices are compact one-spheres or circles of variable length $L$, 
and where for the convenience of the computer simulations, we cyclically identify the time direction. 
This model was solved analytically in \cite{ambjorn1998nonperturbative} and is well studied; it has  
a spectral dimension $d_S$ of at most 2 and a Hausdorff dimension $d_H$ of almost surely 2 \cite{durhuus2010spectral}, the latter in agreement with earlier theoretical \cite{ambjorn1998nonperturbative,ambjorn1999euclidean} and numerical \cite{ambjorn1999new} results. However, as already emphasized in \cite{ambjorn1999new}, the fact that such dimensions have their
classically expected values,
equal to the topological dimension of the microscopic, triangular building blocks, does not imply that the geometry is locally flat. Plotting the
spatial volume of the universe as a function of (discrete) proper time for a typical member of the ensemble of geometries, one finds characteristic, strongly fluctuating ``candlestick'' profiles (Fig.\ \ref{ric2dcdt-fig:candlestick}, see also Fig.\ \ref{ric2dcdt-fig:volume-profiles} below). This behavior is not compatible with local flatness, 
which for the triangulated spacetimes of CDT would imply coordination 
number 6 for all vertices. However, it could still be the case that averaging the curvature over sufficiently large neighborhoods yields an ``effectively flat'' universe
on a larger scale. 

The new quantum Ricci curvature, which depends on a coarse-graining scale $\delta$ (roughly, the diameter of the averaging region), allows
us to establish whether or not such a flat behavior is present, and at what scale. This is the issue we will investigate below. By virtue of the torus
topology, we will be able to distinguish between the (quasi-)local behavior of the curvature, where only the geometry inside a contractible neighborhood contributes,
and the global behavior, which can involve some ``wrapping around'' one or both of the compact torus directions. Lastly, the directional nature of the 
quantum Ricci curvature will allow us to explore the time- and spacelike directions of the quantum geometry separately, as was done previously in CDT
quantum gravity in four dimensions \cite{klitgaard2020how}. However, as already stressed above, unlike in four dimensions, there is no classical limit for quantum gravity in two dimensions, whose pure quantum character prevents us from formulating any straightforward expectation for its curvature properties.

\begin{figure}[t]
\centerline{\scalebox{0.5}{\rotatebox{90}{\includegraphics[trim=2.5cm 1cm 2.5cm 1cm, width=0.95\textwidth, clip]{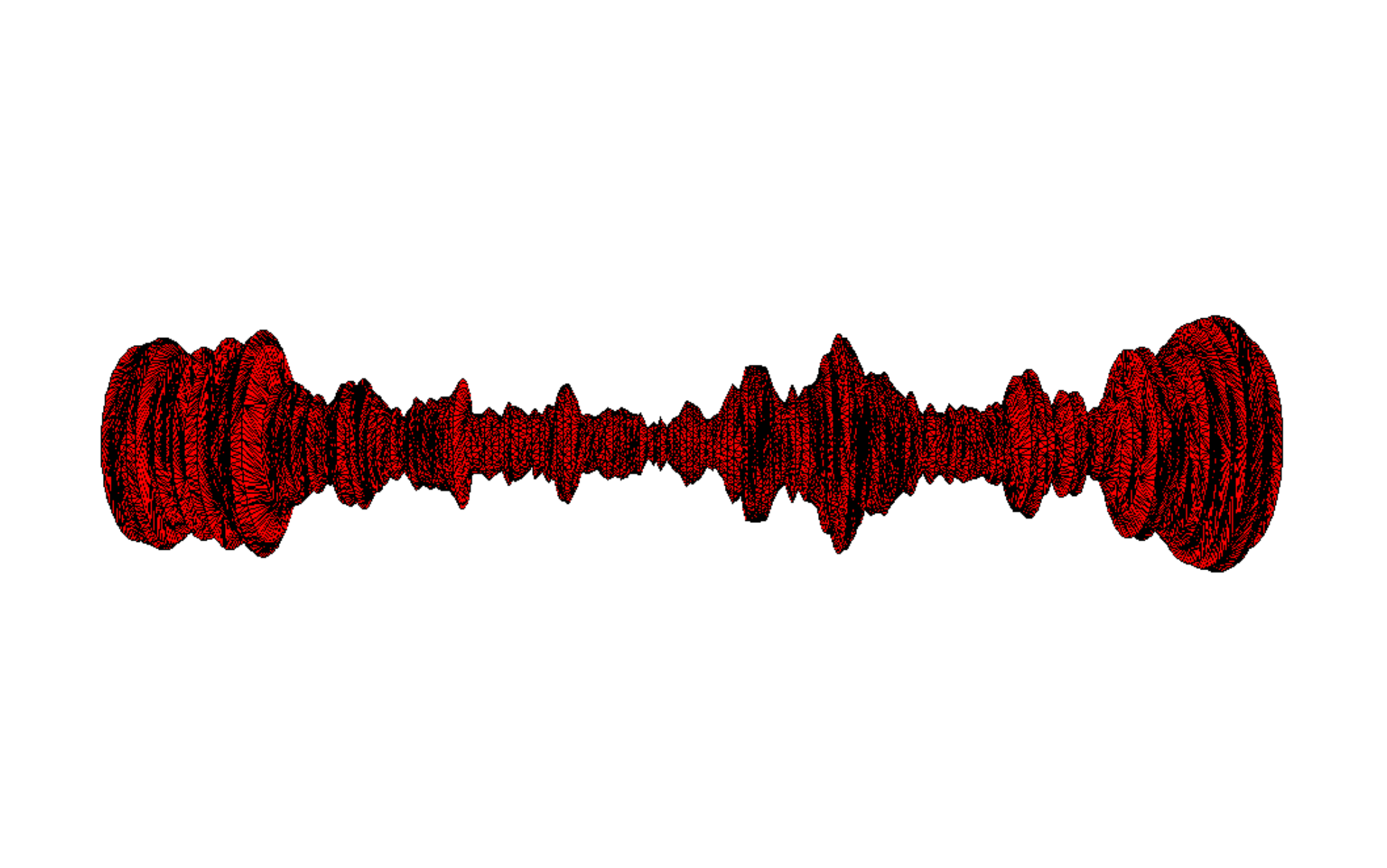}}}}
\caption{Typical configuration contributing to the CDT path integral in two dimensions \cite{ambjorn1999new}, 
with spacetime volume $N_2\! =\! 9408$, illustrating the fluctuating volume of 
the spatial $S^1$-slices as a function of proper time (horizontal direction, cyclically identified in simulations).}
\label{ric2dcdt-fig:candlestick}
\end{figure}

The structure of this chapter is as follows. In Sec.\ \ref{ric2dcdt-sec:2dcdt} we recall some relevant features and ingredients of Causal Dynamical
Triangulations in two spacetime dimensions.
The definition and construction of the quantum Ricci curvature and the associated notion of the curvature profile was discussed in Sec. \ref{cp-sec:qrc} of the introduction to Part \ref{part:curv}, and will not be repeated here.
In Sec.\ \ref{ric2dcdt-sec:numerical}, we describe the set-up of the Monte Carlo simulations and present the curvature measurements.
To help us interpret the outcome, we perform a closer analysis of the influence of the torus topology
in Sec.\ \ref{ric2dcdt-sec:topology}, by determining quantitatively the fraction of data points that is affected by it.
This allows us to separate local and global aspects of the curvature profile and re-evaluate the previous measurements.
Some technical details have been relegated to Appendix \ref{app-sec:intersection-numbers}.
Finally, Sec.\ \ref{ric2dcdt-sec:summary} contains a summary and conclusions.

\section{CDT quantum gravity in two dimensions}
\label{ric2dcdt-sec:2dcdt}

CDT quantum gravity was conceived as a nonperturbative implementation of the Lorentzian gravitational path integral, which involves a
sum over all
four-dimensional spacetime geometries with a physical, Lorentzian signature. It constituted a major advance over previous nonperturbative lattice
formulations\footnote{see \cite{loll1998discrete} for an overall assessment}, which for technical reasons addressed a different problem, 
namely, how to construct a quantum theory 
of gravity for four-dimensional Riemannian metric spaces, which lack a causal structure and with it the distinction between time-, space- and light-like
directions. It is important to realize that beyond perturbation theory on a flat background, where a Wick rotation is available, Lorentzian and Euclidean
quantum gravity are a priori different theories. To the extent they can be shown to exist as fundamental theories, there are no 
compelling arguments for why they should be simply related or even equivalent. 
If for some reason one wants to take unphysical, Riemannian metrics as a starting point, one must deal with the additional
difficulty of showing how Lorentzian concepts like time and causality can be retrieved from them. At this stage, the question is largely academic,
since nonperturbative Euclidean path integrals in four dimensions seem to suffer from generic pathologies and one has not found
evidence of second-order phase transitions, a prerequisite for the existence of a continuum limit of the underlying
regularized models. By contrast, four-dimensional CDT quantum gravity does possess higher-order phase transitions, and the investigation of
several quantum observables (dynamical dimensions, volume and curvature profiles) in CDT has revealed highly nontrivial evidence for the emergence 
of a classical limit from its Planckian quantum dynamics (see \cite{loll2019quantum} for a recent review).  

Since two-dimensional quantum gravity is merely a toy model for the real theory, there is no compelling physical reason to study it in any particular signature.
However, a pivotal aspect of the analytic solution of CDT in two dimensions \cite{ambjorn1998nonperturbative} was to provide a first explicit example of the inequivalence between Lorentzian 
and Euclidean gravitational path integrals, which manifests itself in different values for several universal critical exponents, including fractal dimensions and
the string susceptibility \cite{ambjorn1999euclidean,ambjorn2000lorentzian}. Following the original, purely geometric CDT model, there have been numerous analytical and numerical studies of
coupled Lorentzian gravity-matter systems in two spacetime dimensions, 
involving spin systems \cite{ambjorn1999new,ambjorn2000crossing,ambjorn2009shaken}, dimers \cite{atkin2012analytical,ambjorn2014restricted}, scalar fields 
\cite{ambjorn2012pseudotopological,ambjorn2015spectral}, gauge fields \cite{ambjorn2013twodimensional,candido2021compact} and loop models \cite{durhuus2021critical} 
(see these papers for more extensive bibliographies). Other work in two dimensions has dealt with
the relation between the Euclidean and Lorentzian quantum geometries \cite{ambjorn2000relation}, a possible interpolation between both models 
involving ``locally causal dynamical triangulations'' \cite{loll2015locally}, the universal nature of CDT quantum gravity \cite{ambjorn2013universality},
and a generalized version of CDT, where spatial topology changes are permitted \cite{ambjorn2007putting,ambjorn2008string}. 

\begin{figure}[t]
	\centering

	\includegraphics[width=0.9\textwidth]{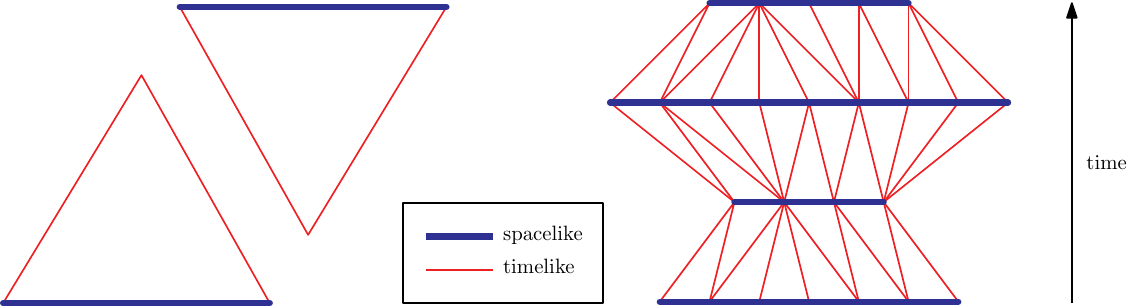}
	\caption{Minkowskian isosceles triangles with one space- and two timelike edges (left) are the elementary building blocks of the 
	piecewise flat lattices of CDT. They are assembled into curved, triangulated spacetimes with a stacked structure (right). Note that the
	graphical representation is not isometric.}
	\label{ric2dcdt-fig:cdt-sample}
\end{figure}

To set the stage for our investigation of the quantum Ricci curvature, we will briefly recap the ingredients of CDT in two dimensions (see \cite{ambjorn2001dynamically,ambjorn2012nonperturbative}
for more comprehensive technical details). The 
gravitational path integral in this approach is defined as the continuum limit of a sum over regularized, curved spacetimes, given in terms of 
simplicial manifolds, the \emph{causal triangulations}. In order to evaluate this nonperturbative path integral, either analytically or numerically, one needs
to analytically continue it, using the Wick rotation that is a key feature of this formulation \cite{ambjorn1998nonperturbative,loll2019quantum}. The CDT path integral after the Wick rotation
has the form of a real-valued partition function
\begin{equation}
	Z = \sum_{T} \frac{1}{C_T} \, {\mathrm e}^{-S_\lambda[T]}, \quad \quad S_\lambda[T] = \lambda\, N_2(T),
	\label{ric2dcdt-eq:partition-sum}
\end{equation}
where the sum is over causal triangulations $T$ of topology $S^1\times S^1$ and $C_T$ denotes the order of the automorphism group of $T$.
Since the topology is fixed, the Euclidean gravitational action $S_\lambda [T]$ consists only of a cosmological-constant term, 
where the bare cosmological constant $\lambda$ multiplies
the volume $N_2(T)$, counting the number of triangles contained in the triangulations $T$. The simple form of $S_\lambda$ comes about because
all triangulations are built from a single type of triangular building block of fixed geometry, a flat Minkowskian isosceles triangle with one spacelike
and two timelike edges, which in the causal triangulations can appear with two different time orientations (Fig.\ \ref{ric2dcdt-fig:cdt-sample}, left). The Wick 
rotation of CDT turns
this into a flat Euclidean triangle, which without loss of generality can be taken to be equilateral. Note that the purely Euclidean path integral in terms of
DT, whose continuum limit is Liouville gravity, also uses flat equilateral triangles and formally looks identical to Eq.\ \eqref{ric2dcdt-eq:partition-sum}, but is defined
on a different configuration space of simplicial manifolds, which do not carry any imprint of the causal structure of CDT. The latter is captured
by causal ``gluing rules'' for the elementary building blocks, which result in a stacked structure, a discrete analogue of global hyperbolicity 
(Fig.\ \ref{ric2dcdt-fig:cdt-sample}, right).\footnote{Note that the ``straightness'' of the one-dimensional simplicial submanifolds of the spacelike edges 
(thick lines) in Fig.\ \ref{ric2dcdt-fig:cdt-sample} is a feature of the graphic representation and does not indicate the absence of extrinsic curvature of
these submanifolds (see \cite{loll2019quantum} for a detailed discussion of the roles of time and causality in CDT).}

Despite being sometimes labelled a ``discrete approach'', CDT quantum gravity is not based on a postulate of fundamental discreteness at the
Planck scale, but uses simplicial building blocks merely as part of a lattice-like regularization of the quantum theory.   
The edge length $a$ of such a simplex or triangle provides a short-distance (UV) cutoff, which in any continuum limit will be sent to zero, accompanied by
a suitable renormalization of the bare coupling constant(s). 
In numerical simulations in two dimensions, the limit $a\rightarrow 0$ is realized by taking $N_2\rightarrow\infty$ for the total number of triangles,
where $a\propto 1/\sqrt{N_2}$. Since the limit of infinite system size cannot be attained with finite computational resources, the behavior of 
an observable in the continuum limit is extrapolated by studying it on
a sequence of systems of large and increasing $N_2$ using finite-size scaling, a standard tool in the analysis of statistical
systems \cite{newman1999monte}. In Monte Carlo studies of any physical quantities, including the quantum Ricci curvature considered below, 
data points at or very near to the cutoff scale are usually discarded, because they are affected by the details of the chosen 
regularization, so-called lattice artifacts. Instead, one is interested in universal, regularization-independent
properties that manifest themselves on scales $\gg a$, and which are characterised by a specific scaling behavior as a
function of the system size.

\section{Measuring curvature profiles}
\label{ric2dcdt-sec:numerical}

Although the partition function \eqref{ric2dcdt-eq:partition-sum} of two-dimensional CDT quantum gravity can be evaluated analytically \cite{ambjorn1998nonperturbative}, 
it is not known at this stage how to compute the expectation value (\ref{cp-eq:curv-prof-cdt}) of the curvature profile exactly. 
To understand its quantitative behavior, we therefore turn to computer simulations. As outlined above, we will measure this observable
at various fixed spacetime volumes, in our case in the range $N_2\in [50k,600k]$, and try to extract its continuum behavior from a finite-size scaling analysis.  

Our procedure used a Markov chain Monte Carlo algorithm (our implementation code can be found at \cite{brunekreef2021jorenb}) to generate a sequence of independent CDT configurations $\{ T_n\}$, $n\! =\! 1,2,\dots ,N$, all 
with torus topology $T^2$
and fixed numbers $N_2$ of triangles and $\tau$ of spatial slices. This sequence serves as a random sample of the probability distribution given 
by the partition function \eqref{ric2dcdt-eq:partition-sum} at fixed volume. 
Since the computation of the exact sphere distance $\bar{d}_{\textrm av}  (\delta)$ of Eq.\ \eqref{cp-avdist-disc} for a given triangulation $T_n$ is very 
costly\footnote{A separate BFS has to be performed for every vertex in $S_p^\delta$, with a maximum depth of $3\delta$,
the upper bound for the distance between any pair of points $(q,q')\! \in\! S_p^\delta\! \times\! S_{p'}^\delta$, corresponding to a path through $p$ and $p'$
(c.f.\ Fig.\ \ref{cp-fig:cdt-sphere-pair}). 
A simple but effective optimization of the BFS is to relabel the spheres such that the ``first'' sphere $S_p^\delta$ is the one containing fewer vertices.}, 
we only compute a sample of sphere distances $\dbar(S_p^\delta, S_{p'}^\delta)$ on each $T_n$, one for each $\delta$ in the chosen 
integer range $\delta\!\in\! [1,\delta_{\textrm max}]$, which will then contribute to the ensemble average (\ref{cp-eq:curv-prof-cdt}) at the given $\delta$-value.

For given $T_n$ and $\delta$, a pair $(S_p^\delta, S_{p'}^\delta)$ of spheres is constructed by first picking a vertex $p\!\in\! T_n$ with uniform probability, 
and finding the set $S_p^\delta$ of all vertices at link distance $\delta$ from the center $p$ by a breadth-first search (BFS) algorithm. 
One then picks uniformly at random a vertex $p'\! \in\! S_p^\delta$, and constructs the set $S_{p'}^\delta$ in the same manner. 
Given such a pair of spheres, the average sphere distance $\dbar(S_p^\delta, S_{p'}^\delta)$ is determined exactly by carrying out a BFS starting from every point 
$q \in S_p^\delta$ until the distances between all pairs $(q,q') \in S_p^\delta \times S_{p'}^\delta$ have been found. 
Repeating this procedure for each triangulation in the sequence $\{ T_n\}$, we obtain $N$ average sphere distance measurements for every 
value of $\delta$, whose average yields a numerical estimate of the expectation value $\langle \bar{d}_{\textrm av}(\delta)\rangle$ and thus of the
curvature profile.

As part of the same Monte Carlo simulation, we also collected average sphere distances using the dual link distance. For each configuration $T_n$
of some sequence $\{ T_n\}$, $n\! =\! 1,2,\dots ,N$, and for each dual distance $\delta\in [1,\delta^*_{\textrm max}]$, 
we randomly picked a dual vertex $p^*$ (unrelated to any vertex $p$
selected for the link-distance measurements), and subsequently a dual vertex $p'^*$ from the $\delta$-sphere $S_{p^*}^\delta$ around $p^*$, etc.,
following a procedure completely analogous to the one just described to extract the average sphere distance $\dbar(S_{p^*}^\delta, S_{p'^*}^\delta)$
with respect to the dual link distance. Since the dual graph is trivalent, the size of the ``dual'' spheres $S_{p^*}^\delta$, based at a dual vertex $p^*$,
grows on average more slowly as a function of $\delta$ than the size of the spheres $S_p^\delta$ based at a vertex $p$. By the same token, one traverses faster
through a given triangulation $T$ along a geodesic when taking steps along links rather than along dual links, a well-known phenomenon that leads to a relative scaling
between the two notions of lattice distance\footnote{roughly speaking, a factor of order 2}, compared with any particular notion of (renormalized) geodesic 
distance in the continuum limit. We therefore expect that the $\delta$-range for which we can determine the curvature profile of triangulations of
a given size $N_2$ is larger when we use the dual link distance, in other words, that $\delta^*_{\textrm max}>\delta_{\textrm max}$, which indeed turns out to be the case.

A prominent feature of the quantum Ricci curvature is its dependence on a length scale $\delta$, captured by the notion of the curvature profile. 
In the context of the nonperturbative CDT quantum theory, and depending on the (finite) lattice size and resolution, this scale allows us in principle to 
distinguish various
regimes: an unphysical regime below the cutoff, a Planckian regime for small $\delta$ above the cutoff, and a (semi-)classical regime for larger $\delta$.
In the region where $\delta$ becomes comparable to the linear system size, the quantum Ricci curvature will by construction ``feel'' the global structure of the 
underlying spacetime, which is determined by our choice of boundary conditions. 

Recall that in nongravitational lattice field theories, one often works with cubic $D$-dimensional lattices, where all directions are identified periodically, yielding a $D$-dimensional torus $T^D$ topologically. If one wants to approximate physics in Minkowski space
(after Wick rotation, in flat Euclidean space), such a compactification does not reflect a physical property, but is chosen for simplicity, and forced upon 
us by the finite size of the system (unless one wants to deal with open boundaries).
When measuring observables, one therefore tries to minimise any dependence on this unphysical global structure. 

The situation in nonperturbative lattice gravity is slightly different. Firstly, since gravity is dynamical, the geometric properties of the lattice are
not fixed a priori. Even if the total spacetime volume is kept fixed, a concept like the ``linear size'' of the spacetime is not
fixed at the outset and at the very least subject to quantum fluctuations. Besides, geometric quantities like lengths, areas and volumes may not scale
``canonically'' in the continuum limit, i.e. not as expected from the topological dimension of the discrete building blocks that were used to
construct the regularized theory. The classic example is two-dimensional Liouville gravity (Euclidean DT), whose quantum geometry is characterised 
by a Hausdorff dimension of four, and not two as one may have expected na\"ively \cite{kawai1993transfer,ambjorn1995fractal,ambjorn1995scaling}. 
Secondly, since gravity is a theory of spacetime, one may be interested in studying specifically the influence of the global spatial or spacetime
topology on the gravitational dynamics, regarding it as a feature rather than a technical necessity. 

Returning to the discussion of the quantum Ricci curvature, its built-in scale dependence means that even in the classical case, it can be used to probe 
not only an infinitesimal, local neighborhood, producing results compatible with a small-$\delta$ expansion like Eq.\ \eqref{cp-2dexp}, but also 
noninfinitesimal $\delta$, giving rise to an averaged or coarse-grained Ricci curvature associated with a macroscopic length scale $\delta$. 
For general metrics, this regime is difficult to access analytically, an issue that is at the heart of the so-called averaging problem in classical
cosmology (see e.g.\ \cite{ellis2012relativistic}). Addressing this problem in terms of the quantum Ricci curvature is of independent interest in the 
classical theory and an issue we will return to elsewhere. 

In the two-dimensional quantum theory we are studying presently, it is clear that
the results for the curvature profile will carry an imprint of the global toroidal topology whenever the sphere radius $\delta$ becomes sufficiently large. 
A feature of the torus we will exploit below is that -- unlike for the two-sphere -- one can formulate sharp criteria for when a particular average sphere distance 
measurement is affected by the nontrivial topology. We already mentioned that the linear extension of the ``double sphere'' $(S_p^\delta, S_{p'}^\delta)$ 
is $3\delta$. If one wants to exclude global topological effects, one should clearly avoid that the area enclosed by the two spheres wraps around the 
torus and self-overlaps in a nontrivial way. However, this is not sufficient. One must also make sure that there are no geodesics between any pair of points
$(q,q')\! \in\! S_p^\delta \times S_{p'}^\delta$ that represent shortcuts by wrapping around the ``back'' of the torus and that would not be present if the
torus was cut open appropriately outside the double sphere region (see Sec.\ \ref{ric2dcdt-sec:topology} for further discussion, as well as related considerations in Chapter \ref{ch:defects}). 
\emph{If} we were on a fixed, flat torus in the continuum,
obtained by identifying opposite sides of a flat rectangle with side lengths $l_1$ and $l_2$, a sufficient condition for this not to happen would be
$6\delta\!\leq\! l_i$. The same condition applies to CDT configurations with respect to the periodically identified time direction. Namely, if we want
to exclude any influence of the compactified time on the quantum Ricci curvature measurements, we should choose $\delta_{\textrm max}$ such that
$6\delta_{\textrm max}\! \leq\! \tau$, where $\tau$ is the (fixed) time extension of a CDT geometry, measured in terms of discrete proper time. 
When using the dual link distance in the construction of the quantum Ricci curvature, the corresponding condition is $3\delta^*_{\textrm max}\!\leq \!\tau$.  
To determine when and how the compactness in the spatial direction on the torus influences the curvature measurements is much more involved, since the
volume of the spatial slices is subject to large quantum fluctuations. 
Results for the curvature profile that are independent of the boundary condition in the time direction will be presented in Sec.\ \ref{ric2dcdt-sub:results},
while the influence of the spatial compactness will be analyzed in detail in Sec.\ \ref{ric2dcdt-sec:topology}.

\subsection{Measurement results}
\label{ric2dcdt-sub:results}

Following the procedure outlined above, we have determined the expectation value $\langle \bar{d}_{\textrm av}(\delta)/\delta \rangle$ 
of the curvature profile on toroidal CDT geometries for two different time extensions $\tau$ and a range of spacetime volumes $N_2$, as a function
of the link distance. 
For configurations with $\tau\! =\! 183$ spatial slices, measurements were performed in the range $\delta\! \in\! [1, \delta_{\textrm max}\! =\! 30]$, 
to avoid any direct influence of the periodic identification in time, a boundary condition chosen merely for the convenience of
the simulations. 
We have collected $50k$ measurements at volumes $N_2\! =\! 50$, 70, 150 and 250$k$, and $30k$ measurements at $N_2\! =\! 350k$, where 
increasing the total volume for fixed $\tau$ results in a larger average size of the spatial slices at integer time.
In addition, we performed $30k$ measurements for configurations with time extension $\tau\! =\! 243$, 
using an extended $\delta$-range with $\delta_{\textrm max}\! =\! 40$ and 
a volume of $N_2\! =\! 600k$. We also determined $\langle \bar{d}_{\textrm av}(\delta)/\delta \rangle$ in terms of the dual link distance, 
for $\tau\! =\! 183$, $\delta^*_{\textrm max}\! =\! 60$ and at volumes 
$N_2\! =\! 100$, 200 and $250k$, based on $50k$ data points. In between measurements, we performed $N_2\!\times\! 1000$ attempted moves. Considering that the equilibrium acceptance ratios are around $10\%$, this implies that every site of the geometry was updated roughly 100 times on average in between two sampled triangulations $T_n$ and $T_{n+1}$.

We used a data blocking procedure to check for autocorrelations in the data collected at subsequent measurement steps, but such correlations were found to be absent. This is perhaps not too surprising, given that a single average sphere distance measurement is quasi-local (as long as $\delta$ does not become too large
compared to the system size), and we have chosen to take only one
data point for every $\delta$ for a given configuration $T_n$, each one from a double-sphere with different initial point $p$.

\begin{figure}[t]
	\centering
	\includegraphics[width=0.7\textwidth]{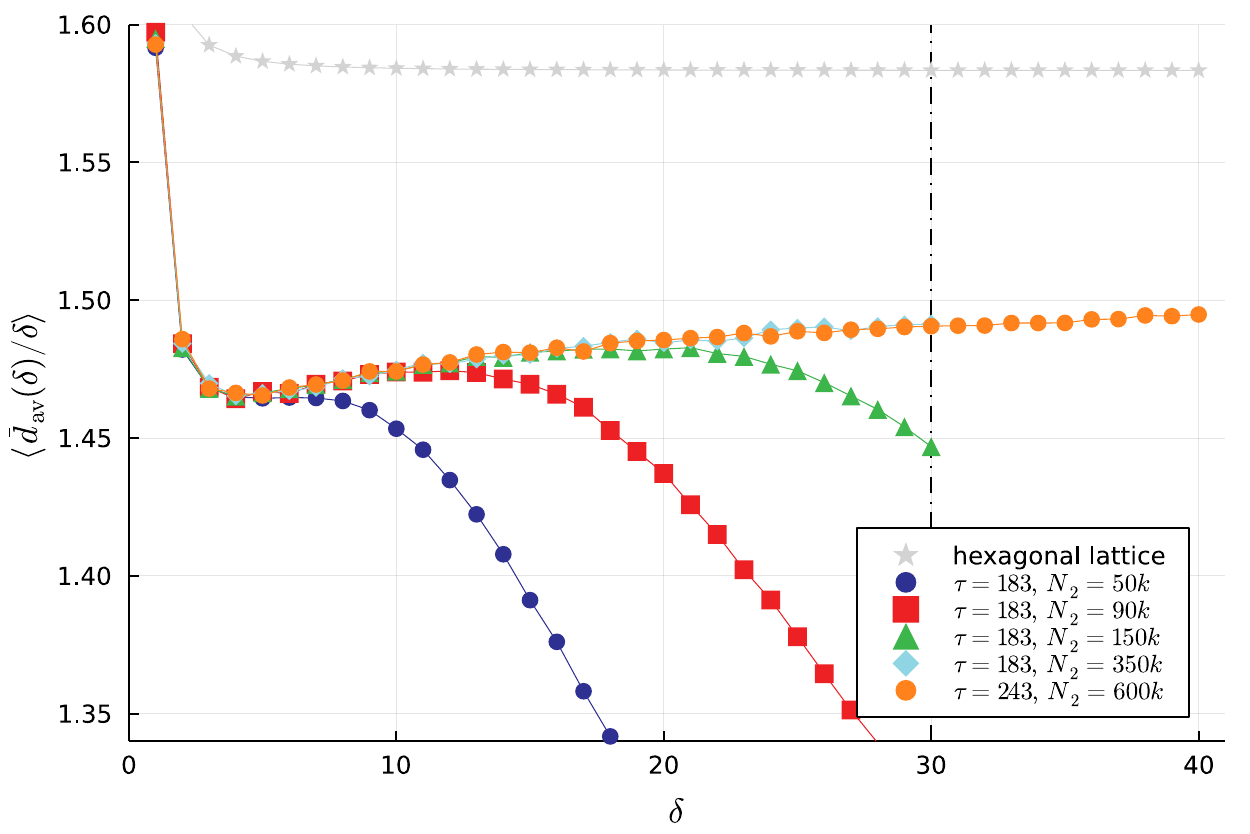}
	\caption{Curvature profile $\langle \bar{d}_{\textrm av}(\delta)/\delta \rangle$ as a function of the link distance $\delta$ for
	CDT quantum gravity on a two-torus, for various volumes $N_2$ and time extensions $\tau$.  Any data for $\tau\! =\! 183$ lying on or to the left of the dash-dotted vertical line at $\delta\! =\! 30$ is not influenced by the periodic identification of time. The classical curvature profile 
	of the flat hexagonal lattice is shown for comparison. (Error bars are smaller than dot size.)}
	\label{ric2dcdt-fig:ric2dcdt}
\end{figure}

Our results for the curvature profile $\langle \bar{d}_{\textrm av}(\delta)/\delta \rangle$ as a function of the link distance 
are shown in Fig.\ \ref{ric2dcdt-fig:ric2dcdt}. The five data sets for the CDT ensemble display an almost identical behavior for $\delta\!\leq\! 5$, namely,
a steeply falling curve due to short-distance lattice artifacts, familiar from previous investigations in \cite{klitgaard2018introducing,klitgaard2018implementing,klitgaard2020how} and Chapter \ref{ch:defects} of this thesis, both in a classical
and a quantum context. Beyond this region, all interpolated curves show a gentle incline, which for the three smaller volumes continues up to some maximum, 
beyond which they decrease rather fast. The location of the maximum appears to be roughly proportional 
to the volume $N_2$, that is, effectively to the volume of the spatial slices. 
We have not included any data points that are potentially affected by the choice of the compactified boundary condition
in the time direction, which for $\tau\! =\! 183$ would lie to the right of $\delta\! =\! 30$, marked by the vertical line in Fig.\ \ref{ric2dcdt-fig:ric2dcdt}. The simulation data
for the two larger volumes do not exhibit any downward slope in the investigated $\delta$-ranges.  

\begin{figure}[t]
	\centering
	\includegraphics[width=0.8\textwidth]{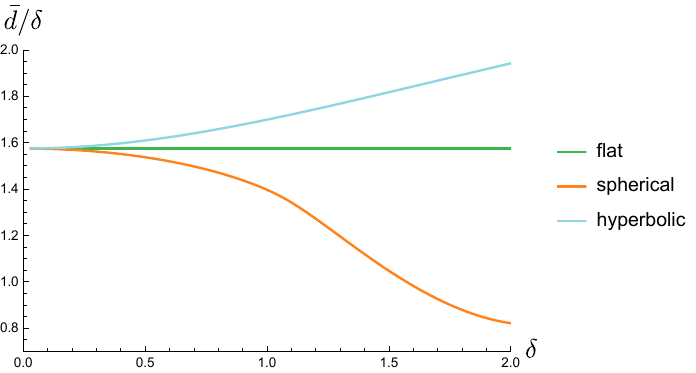}
	\caption{Classical curvature profiles $\bar{d}_{\textrm av}(\delta)/\delta$ as a function of the distance $\delta$ for
	two-dimensional Riemannian manifolds of constant curvature: a hyperboloid of curvature radius 1 (top), flat space (middle) and a sphere of
	curvature radius 1 (bottom).  }
	\label{ric2dcdt-fig:riccishort}
\end{figure}

In trying to interpret these outcomes, let us recall the behavior of the curvature profiles on constantly curved Riemannian spaces \cite{klitgaard2018introducing}, 
which for small $\delta$ is also captured by the expansion (\ref{cp-2dexp}). 
Considering that the quantum geometry we are investigating is presumably homogeneous (``looks the same everywhere'') on sufficiently large
scales, classical spaces of constant Gaussian curvature are natural spaces to compare with. 
As illustrated by Fig.\ \ref{ric2dcdt-fig:riccishort}, 
their curvature profiles increase for negative and decrease for positive curvature as a function of $\delta$, and are constant
(in fact, for all $\delta$) on flat space. However, what we observe in the measurements 
is not an obvious match for any of these cases. The fall-off behavior found for the smaller
volumes could be an indication of positive curvature, of a topological effect due to the compact spatial boundary condition, or of a combination of both. 
Alternatively, if we can demonstrate that
the fall-offs are a purely topological effect, discarding the corresponding data points for large $\delta$ could plausibly lead to a single, universal
enveloping curve, roughly along the data points taken at the largest volume $N_2\! =\! 600k$ in Fig.\ \ref{ric2dcdt-fig:ric2dcdt}, whose
upward slope may indicate the presence of (a small) negative curvature. 

On general grounds, if the impact of the global torus topology can be quantified and removed from our data, the remainder should describe
the pure, quasi-local geometry of a (possibly infinitely extended) contractible region of spacetime, without any topological or boundary effects. 
However, note that such a scenario lacks a distinguished macroscopic scale, which makes it difficult to understand how its curvature behavior 
could possibly resemble that of
a constantly curved space of either positive or negative curvature, both of which are characterised by a curvature radius $\rho$ 
(which in Fig.\ \ref{ric2dcdt-fig:riccishort} is set to 1). 
In terms of a classical interpretation, this only seems to leave flat space. For example, one might envisage a situation
where positive and negative curvatures on microscopic scales average out to produce an effectively
flat space on sufficiently large scales. 
On the other hand, since we are dealing with a pure quantum system, as already emphasized in the introduction, 
there are no compelling arguments for why the quantum geometry should be flat.\footnote{Note that the quantum Ricci curvature for
$\delta\! >\! 1$ is distinct from the deficit-angle prescription of Regge calculus \cite{regge1961general}, summed over a ball of radius $\delta$,
and in particular does not obey a Gauss-Bonnet theorem in two dimensions. 
Regge's definition of local curvature on $D$-dimensional triangulations in terms of the number of $D$-simplices meeting at a
$(D-2)$-simplex is a valid notion for finite simplicial manifolds, but does not give rise to a well-defined renormalized curvature  
in the nonperturbative quantum theory, unlike the quantum Ricci curvature \cite{klitgaard2018introducing}.} 

Let us nevertheless examine the classical flatness hypothesis a bit further. Is it possible that the enveloping curve of Fig.\ \ref{ric2dcdt-fig:ric2dcdt}
would flatten out if we added data for larger volumes? In other words, could there be lattice artifacts 
that mask a ``true'' constant behavior of the data points in the range measured up to now? 
To assess this issue, we have computed the curvature profile 
$\bar{d}_{\textrm av}(\delta)/\delta$ of the regular triangulation of the (infinitely extended) flat plane, corresponding to a hexagonal lattice, as a function of
the link distance, up to $\delta\! =\! 40$.
Using translation invariance, the double sum over points $p$ and $p'$ in Eq.\ \eqref{cp-avdist-disc} reduces to a single sum over points $p'$ at distance $\delta$ 
from a fixed point $p$. The geometry of the situation is illustrated by Fig.\ \ref{cp-fig:cdt-sphere-pair} for $\delta\! =\! 2$, where one should still average over all locations
$p'$ that lie on the blue ``sphere'' $S_p^\delta$. We have included the exactly computed values of the classical curvature profile for the hexagonal lattice for 
$\delta\!\geq\! 3$ in Fig.\ \ref{ric2dcdt-fig:ric2dcdt}, for comparison with the simulation data. After an initial ``overshoot'' caused by lattice artifacts, 
where the points follow a monotonically decreasing curve, they settle to an approximately constant value for
$\delta\! \gtrsim\! 6$, without going through a dip, as the quantum measurements do. (Recall that the offset in the $y$-direction of the curvature profile
is a lattice-dependent, nonuniversal quantity.) This agrees with the expectation that the quantum Ricci curvature vanishes identically for sufficiently
large $\delta$ and is in line with earlier investigations of flat lattices and of approximations of flat space by Delaunay triangulations \cite{klitgaard2018introducing}.\footnote{Our data 
for the hexagonal lattice differ slightly from those presented in \cite{klitgaard2018introducing}, which did not include an average over all directions.}
We conclude that comparison with a classical flat lattice does not provide support to the interpretation that the quantum geometry of the two-torus
is effectively flat. An additional argument against the existence of a $\delta$-value far away from the cutoff scale, above which the curve for
$\langle \bar{d}_{\textrm av}(\delta)/\delta \rangle$ would flatten out, is the absence of a distinguished macroscopic physical scale, as already pointed out
earlier.   

\begin{figure}[t]
	\centering
	\includegraphics[width=0.7\textwidth]{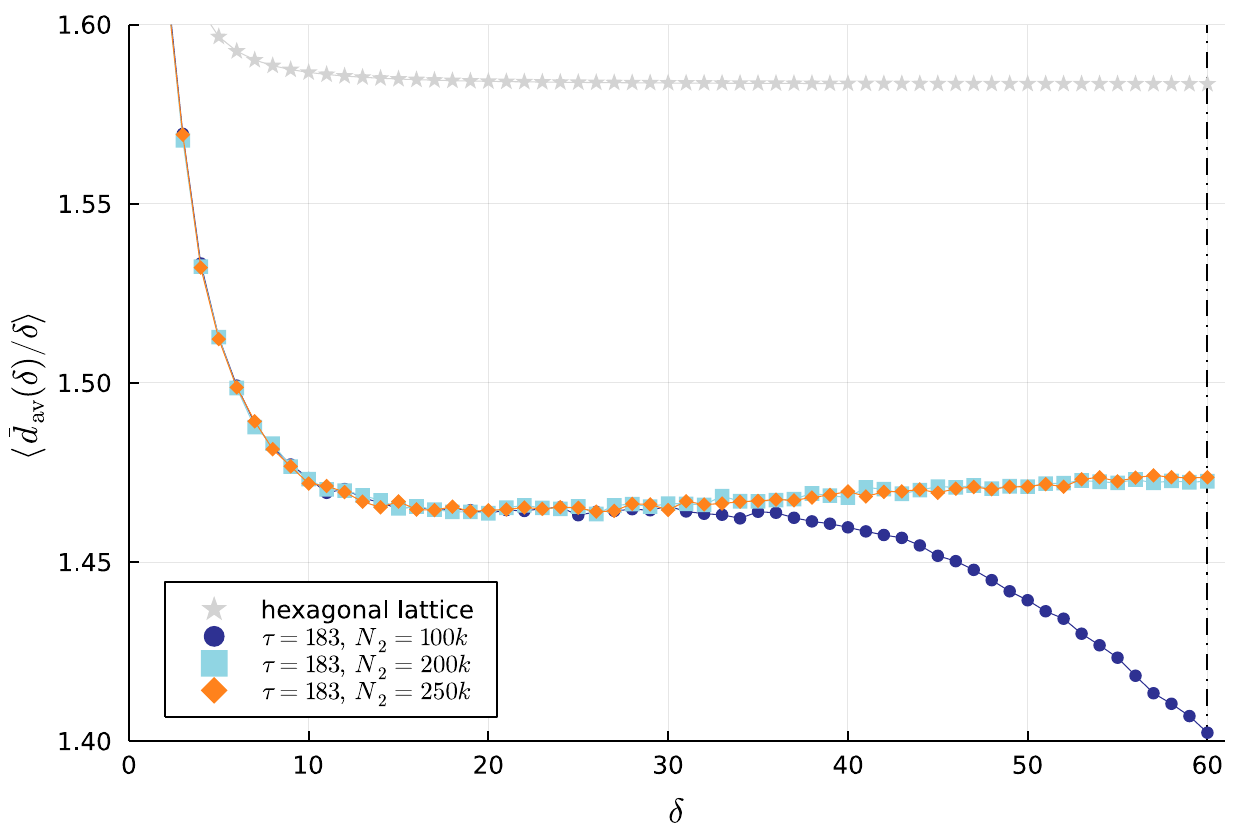}
	\caption{Curvature profile $\langle \bar{d}_{\textrm av}(\delta)/\delta \rangle$ as a function of the dual link distance $\delta$ for
	CDT quantum gravity on a two-torus, for various volumes $N_2$ and time extension $\tau\! =\! 183$. The dash-dotted vertical line at $\delta\! =\! 60$ 
	marks the upper end of the $\delta$-range not influenced by the periodic identification of time. The classical curvature profile 
	of the flat hexagonal lattice as a function of the dual link distance is shown for comparison. (Error bars are on the order of the dot size.)}
	\label{ric2dcdt-fig:dric2dcdt}
\end{figure}

As a cross-check of our measurements in terms of the link distance, 
we also measured the curvature profile $\langle \bar{d}_{\textrm av}(\delta)/\delta \rangle$ using the dual link distance.  
By universality, one would expect to find similar results, up to a finite rescaling of $\delta$ as discussed earlier. This is borne out
by our measurements, which are presented in Fig.\ \ref{ric2dcdt-fig:dric2dcdt}, together with the corresponding classical curvature profile of the 
hexagonal lattice for comparison. The three depicted data sets coincide initially, including the rapid fall-off for small $\delta$ and subsequent
slight dip, before the curve for the smallest volume starts decreasing, while those for the two larger volumes continue on a gently increasing
slope. The behavior is qualitatively very similar to that of Fig.\ \ref{ric2dcdt-fig:ric2dcdt}, up to a rescaling of roughly a factor 2 along the $x$-axis 
(regarding the location of the minimum and the onset of the fall-off of the curves), and
a reduction of the amplitude of the curves in the $y$-direction. 

Before we reach a final conclusion on the interpretation of the measured curvature profiles of Figs.\ \ref{ric2dcdt-fig:ric2dcdt} and \ref{ric2dcdt-fig:dric2dcdt}, 
we still need to understand and quantify to what extent
the periodic identification of the spatial direction of the torus geometries influences the quantum Ricci curvature measurements. 
Because of the fluctuating nature of the spatial slices in the quantum ensemble, this analysis is not completely straightforward. 
It will be the subject of the following section.

\begin{figure}[t]
	\centering
	\includegraphics[width=0.7\textwidth]{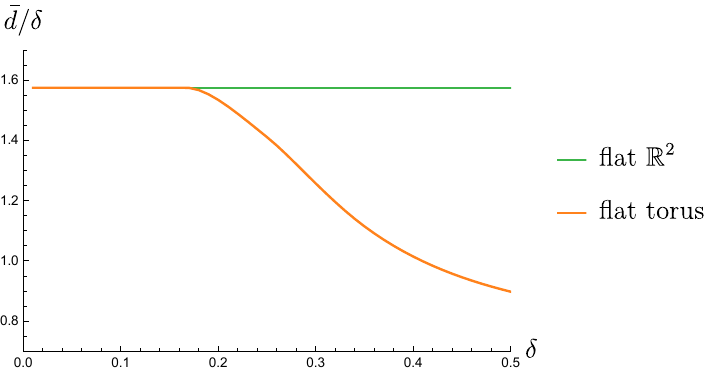}
	\caption{Normalized average sphere distance $\bar{d}/\delta$, measured in one of the directions of a flat classical 
	unit torus.  For comparison, we have included the same quantity on a flat space without compactifications. The two curves coincide initially, but for
	$\delta\! \gtrsim\! 0.17$ the torus data stop following a horizontal curve and switch abruptly to a quasi-linear
	decreasing behavior.}
	\label{ric2dcdt-fig:flattorus}
\end{figure}

\section{The influence of global topology} 
\label{ric2dcdt-sec:topology}

The effect of the global topology on the curvature profile of a smooth classical manifold can be illustrated neatly by considering a flat 
torus \cite{klitgaard2022new}.
The example we consider is a two-torus, obtained by the pairwise identification of the opposite sides of a flat square parametrized by $x\!\in\! [0,1]$
and $y\!\in\! [0,1]$. Fig.\ \ref{ric2dcdt-fig:flattorus} shows the normalized average sphere distance $\bar{d}(S^\delta_p,S^\delta_{p'})/\delta$ corresponding to 
a point pair $(p,p')$, where
$p\! =\! (0.5,0.5)$ and $p'\! =\! (0.5+\delta,0.5)$ have a geodesic distance $\delta\!\in\! [0,0.5]$ purely along the $x$-direction. 
By virtue of the triangle inequality, the largest possible distance between any pair of points $(q,q')\!\in\! S^\delta_p\!\times\! S^\delta_{p'}$ 
is given by $3\delta$, which for sufficiently small $\delta$ is realized by $(q,q')\!=\! ((0.5\! -\!\delta,0.5),(0.5\! +\! 2\delta,0.5))$.
However, if $\delta\! >\! 1/6\!\approx\! 0.1666$, there is a shorter geodesic between $q$ and $q'$ that runs around the torus in the opposite direction, 
leading to the shorter distance $d_g(q,q')\! =\! 1\! -\! 3\delta$. The fact that there is more than one geodesic between a given pair of
points, despite the local flatness of the manifold, is a consequence of the nontrivial topology of the torus. More specifically, there is a
two-parameter family of geodesics between each pair of points, which cannot be smoothly deformed into each other and which differ by their
relative winding numbers around the two torus directions. Of course, only the shortest of these geodesics determines the distance $d_g(q,q')$
that enters into the computation of the average sphere distance. This should be compared with the situation in the flat plane, where there is always a
unique geodesic between two points, determining their Euclidean distance. 

The two cases are contrasted in Fig.\ \ref{ric2dcdt-fig:flattorus},
where the normalized average sphere distance on flat $\R^2$ is a constant $\approx\! 1.5746$ (c.f.\ Eq.\ \eqref{cp-2dexp}). The corresponding curve for the torus, 
computed with the help of Monte Carlo simulations \cite{klitgaard2022new}, coincides with the horizontal curve for $\delta\!  \lesssim\! 1/6$, and beyond
this point starts deviating sharply, entering a regime of quasi-linear decline. In other words, beyond a finite threshold value for $\delta$, there is
a strong influence of the global topology on the average sphere distance, which is systematically lowered due to the presence of ``shortcuts''
between point pairs $(q,q')$ that do not exist for trivial topology. This purely topological effect can be compared to the effect of nontrivial geometry,
which for a sphere of constant curvature also leads to a decreasing curvature profile. However, as illustrated by the corresponding  
curve in Fig.\ \ref{ric2dcdt-fig:riccishort}, the downward slope starts as soon as $\delta\! >\! 0$, as described by Eq.\ \eqref{cp-2dexp}, and without an initial flat plateau. 

\begin{figure}[t]
	\centering
	\includegraphics[width=0.7\textwidth]{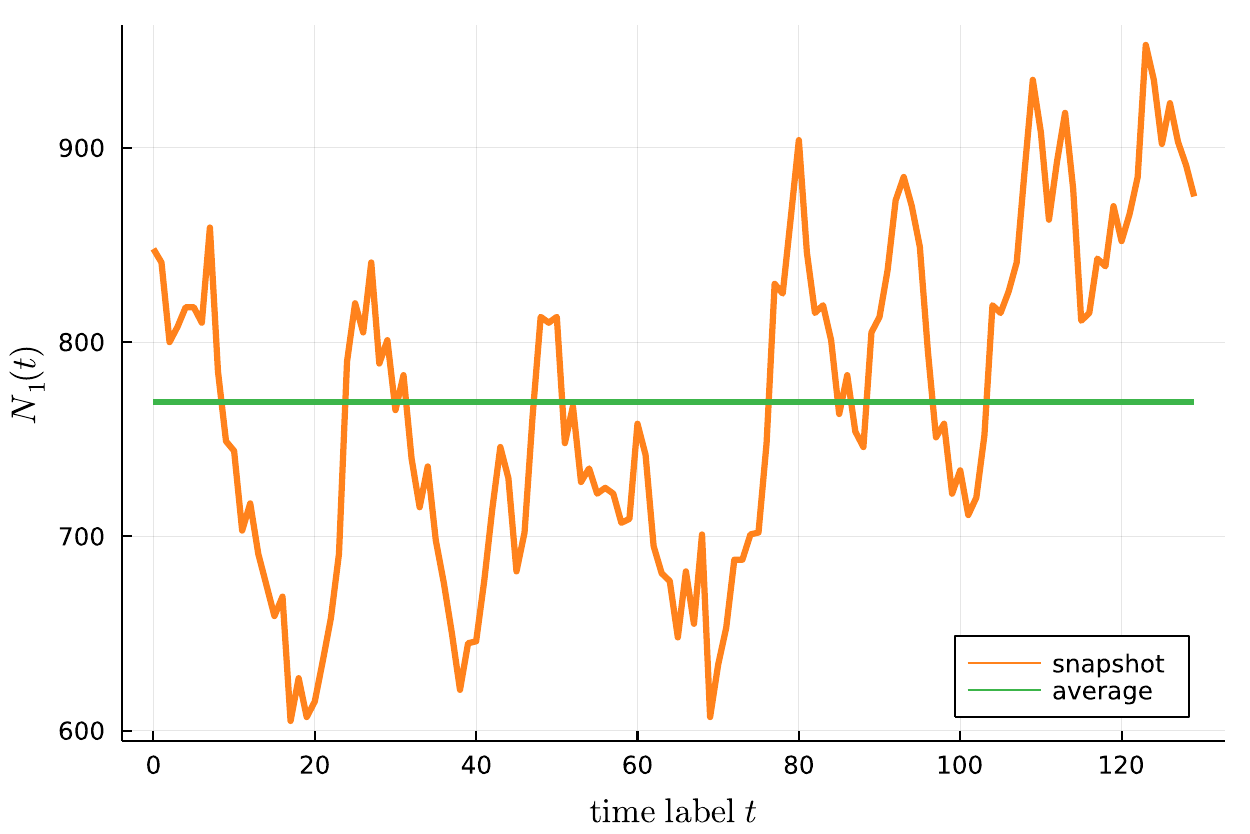}
	\caption{Volume profile $N_1(t)$ of a typical torus geometry in the CDT ensemble, together with the average $\bar{N}_1$ (green), for time
	extension $\tau\! =\! 130$ and volume $N_2\! =\! 200k$.}
	\label{ric2dcdt-fig:volume-profiles}
\end{figure}

Returning to quantum gravity, the sliced structure and fixed time extension of the CDT geometries make it straightforward to deal 
with the periodic boundary condition in the time direction in the same way as one would on a classical torus. 
To avoid any influence of this (unphysical) choice on the measurement of the curvature profile, it is sufficient to observe the bound
$\delta\! \leq \!\tau/6$. The effect of the compactness in the spatial direction is much more difficult to assess because the discrete volume 
(number of edges) $N_1(t)$ of a spatial slice at discrete proper time $t$ is subject to large quantum fluctuations. This happens because there 
is only a single
length scale in two-dimensional quantum gravity, set by $1/\sqrt{\lambda}$, where $\lambda$ is the cosmological constant or,
for fixed two-volume, by $\sqrt{N_2}$. The expectation value of $N_1(t)$ and its standard deviation are therefore of the same order
of magnitude. A snapshot of the volume function $N_1(t)$, $t=0,1,\dots ,\tau\!-\! 1$ from a Monte Carlo simulation with $\tau\! =\! 130$ and $N_2\! =\! 200k$
is shown in Fig.\ \ref{ric2dcdt-fig:volume-profiles}, together with the average spatial volume $\bar{N}_1(t) \! =\! N_2/(2\tau)$ (keeping in mind that 
$N_2\! =\! 2\sum_t N_1(t)$). Potentially problematic for the curvature measurements are path integral configurations that contain spatial
slices of very small volume, on the order of $6\delta$ and below. However, 
the presence of fluctuations means that for fixed time $t$ there is no strict lower bound\footnote{at or above the kinematical 
minimum $N_1\! =\! 3$ required for a simplicial manifold} on the spatial volumes $N_1(t)$ that can occur. 
Moreover, even exact knowledge of
the distribution of $N_1$ in the ensemble would not be sufficient to determine how the sphere distance measurements are affected, 
because the double-spheres extend over many time slices, as will typical shortcuts between pairs of points. 

In practical terms, we will address this challenge by numerically establishing an upper bound $\delta_{\textrm max}$ on $\delta$ for given 
$N_2$ and $\tau$, such that ``topological shortcuts'' occur very rarely, i.e.\ at a prescribed low rate, and have a negligible
influence on the curvature profile. 
This analysis requires a prescription of how to define such shortcuts and find them on a given geometric configuration, an issue we turn to next.    

\subsection{Intersection number of geodesics}
\label{ric2dcdt-inter}

\begin{figure}[t]
	\centering
	\begin{subfigure}[t]{0.45\textwidth}
	\centering
		\includegraphics[height=0.8\textwidth]{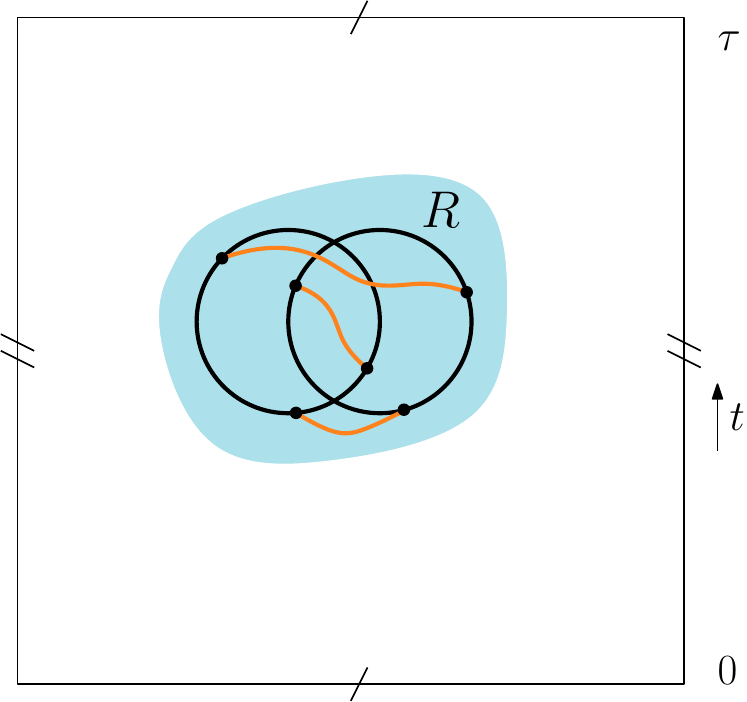}
		\caption{
		}
		\label{ric2dcdt-fig:torus-topo-safe}
	\end{subfigure}
	\hfill
	\begin{subfigure}[t]{0.45\textwidth}
	\centering
		\includegraphics[height=0.8\textwidth]{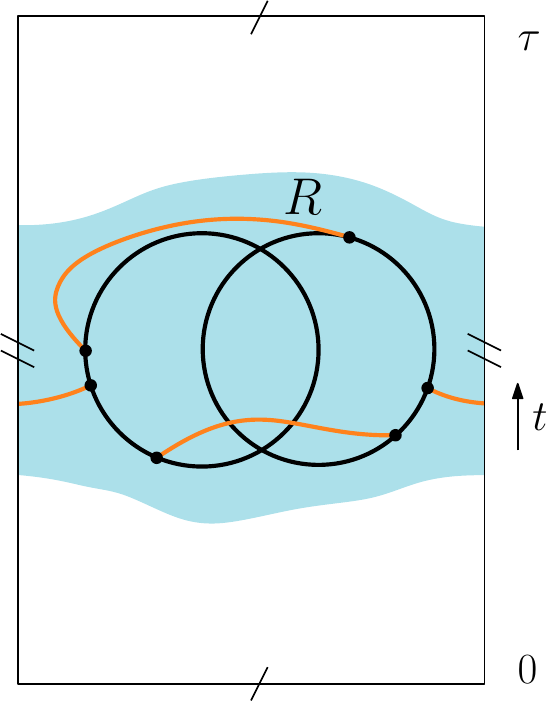}
		\caption{
		}
		\label{ric2dcdt-fig:torus-topo-unsafe}
	\end{subfigure}
	\caption{Schematic representation of a circle pair on a toroidal CDT geometry, including some shortest geodesics
	between pairs of points from the two circles. In the regular case (no effect of global topology) all shortest geodesics are contained 
	in a disc-shaped region $R$ (left). If the spatial extension becomes too short compared to the diameter of the double circle, not all
	shortest geodesics are contained in a contractible disc (right). This topological effect affects the computation of the average
	sphere distance.
	}
	\label{ric2dcdt-fig:torus-topo}
\end{figure}

Fig.\ \ref{ric2dcdt-fig:torus-topo} illustrates the geometric situation we want to analyze, consisting of a pair of spheres (circles) $S_p^\delta$, $S_{p'}^\delta$ 
of radius $\delta$ located somewhere on a curved CDT toroidal geometry of time extension $\tau$ (vertical direction) and with spatial slices   
of fluctuating size $N_1$ (fluctuations not shown), both periodically identified as indicated. We assume $\delta\! \leq\! \tau/6$, 
such that the time periodicity does not affect the
average sphere distance. If the size of all spatial slices is large enough, every geodesic between a point $q\!\in\! S_p^\delta$ and a point
$q'\!\in\! S_{p'}^\delta$ is completely contained in a simply connected region $R$ that includes the two circles and the union of their interiors (light blue region in Fig.\ \ref{ric2dcdt-fig:torus-topo-safe}). By contrast, in Fig.\ \ref{ric2dcdt-fig:torus-topo-unsafe} some of the spatial slices are too small, in the sense
that at least for some point pairs $(q,q')$ the shortest geodesic crosses the vertical line along which the torus has been cut open. In this case, a
region $R$ containing both the circles and all shortest geodesics between pairs $(q,q')$ cannot be disc-shaped, but instead must wind around the
spatial direction of the torus, forming a noncontractible annulus.   

Given a CDT geometry $T$ and a pair of circles $(S_p^\delta, S_{p'}^\delta)$
on it, the key idea for monitoring the occurrence of this latter phenomenon is to consider a suitable
closed curve $\gamma$ in $T$ that winds around the time direction once and lies \emph{outside} the union $C$ of the circles and their interiors, defined by 
$C\! :=\! \left\{q \,|\, d(q, p) \leq \delta \lor d(q,p') \leq \delta\right\}$. Then, if no shortest geodesic between any point pair 
$(q,q')\in S_p^\delta \times S_{p'}^\delta$ ever crosses $\gamma$, we are dealing with the ``regular'' case depicted in Fig.\ \ref{ric2dcdt-fig:torus-topo-safe},
which is unaffected by the periodic identification of the spatial slices. In fact, this criterion is slightly too restrictive; if $\gamma$ passes near
$C$, it can happen that a shortest geodesic crosses it twice (or an even number of times) in opposite directions, 
despite being part of a perfectly regular configuration. In this case, a continuous deformation of the curve $\gamma$ would eliminate the intersections.
This suggests that we should not simply count the number of intersections, but work with oriented curves  
and their corresponding (oriented) intersection number. 

In the Riemannian case, the intersection number $c_p(\alpha,\beta)$ of two parametrized curves $\alpha(s)$ and $\beta(t)$ at an isolated intersection point $p$
is 1 ($-1$) if the ordered pair of tangent vectors $(\dot{\alpha},\dot{\beta})$ forms a right-handed (left-handed) basis of the tangent space at $p$.
On a triangulation $T$, where geodesics between pairs of vertices by definition follow the links of $T$, it is convenient to choose the reference curve
$\gamma$ along dual links, to avoid any nontrivial overlaps of the curves along links (see Fig.\ \ref{app-in-fig:crossing-primal-dual} below). 
More details on the choice and construction of the curve $\gamma$ can be found in Appendix \ref{app-sec:intersection-numbers}. 

\begin{figure}[t]
	\centering
	\begin{subfigure}[t]{0.45\textwidth}
	\centering
		\includegraphics[height=0.8\textwidth]{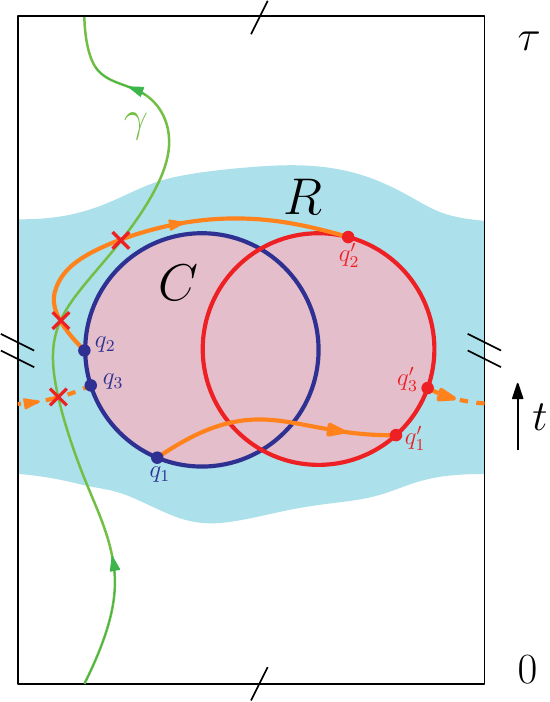}
		\caption{
		}
		\label{ric2dcdt-fig:torus-topo-cut-unsafe}
	\end{subfigure}
	\hfill
	\begin{subfigure}[t]{0.45\textwidth}
	\centering
		\includegraphics[height=0.8\textwidth]{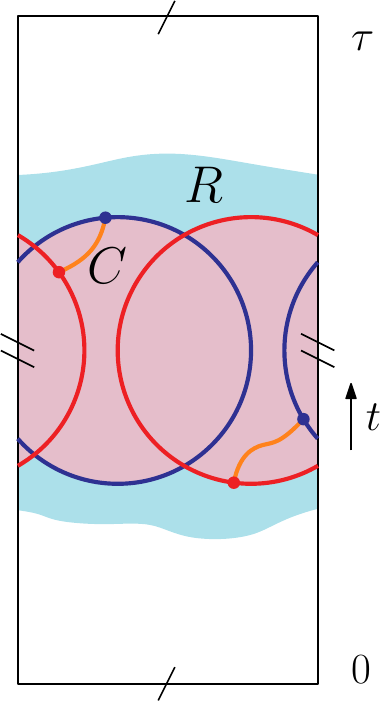}
		\caption{
		}
		\label{ric2dcdt-fig:torus-topo-wrap}
	\end{subfigure}
	\caption{Configuration of two spheres $S_p^\delta$ (blue) and $S_{p'}^\delta$ (red) on a CDT geometry $T$. Left: schematic illustration of a level-1 violation. 
	We can find a timelike curve $\gamma$ (green) outside the region $C$ (pink), but there exist shortest geodesics with nonzero intersection number 
	(like the dashed orange path), amounting to topological shortcuts. 
	Right: a level-2 violation, where the region $C$
	self-overlaps and no curve $\gamma$ with the required properties exists. The two shortest geodesics shown (orange) represent topological shortcuts. 
	}
	\label{ric2dcdt-fig:torus-topo-cut}
\end{figure}

Given a shortest geodesic $\phi (q,q')$ between two vertices $q$ and $q'$ on a CDT configuration $T$, oriented to run from $q$ to $q'$, say, 
and a (oriented) reference curve
$\gamma$ along dual links, we define their total intersection number $c(\phi,\gamma)$ as the sum over all intersection points $p$ of the intersection
number at $p$, $c(\phi,\gamma)\! =\! \sum_p c_p(\phi,\gamma)$. Following our reasoning above, if for all shortest geodesics $\phi$ between point pairs
$(q,q')$ from a pair of circles we have $c(\phi,\gamma)\! =\! 0$, we are in the regular situation where the computation of the average sphere distance 
is not affected by the global topology. By contrast, the occurrence of nonvanishing intersection numbers $c(\phi,\gamma)$ for a given pair of
spheres $(S_p^\delta, S_{p'}^\delta)$ indicates the presence of shortcuts. We will call this a ``level-1 violation'' of the regular case. An example is 
sketched in Fig.\ \ref{ric2dcdt-fig:torus-topo-cut-unsafe}, where the shortest geodesic between $q_3'$ to $q_3$ has a nonzero intersection number with the curve $\gamma$.

We still need to discuss a situation that occurs for even larger values of $\delta$, 
where no curve $\gamma$ can be found because the region $C$ wraps around the spatial
direction of the torus and creates a self-overlap, as illustrated by Fig.\ \ref{ric2dcdt-fig:torus-topo-wrap}. It is clear that in this case there will be many shortcuts,   
affecting the average sphere distance in a significant way. We will call this a ``level-2 violation'' of the regular case. 

\subsection{Intersection number measurements in CDT}
\label{ric2dcdt-measure}
Recall that our objective is to quantify the influence of the global topology, more specifically, of the spatial compactness of the CDT geometries,
on the expectation value of the curvature profile. We have seen that the effect is directly related to the presence of topological shortcuts, which
for increasing $\delta$ lead to ever smaller average sphere distances and therefore a smaller value of the curvature profile, compared with the
case of trivial topology. From the comparison with the classical case (Fig.\ \ref{ric2dcdt-fig:flattorus}), we expect the effect of the compactification to be large
overall, although the shape of the fall-off in $\langle \bar{d}_{\textrm av}(\delta)/\delta \rangle$ will presumably be modulated by the presence of the volume 
fluctuations of the spatial slices. Since we now have a computational tool at hand to check for the presence of shortcuts 
(see Appendix \ref{app-sec:intersection-numbers} for details on
how this is implemented on CDT geometries), we can address the following questions. Firstly, can the fall-offs for large $\delta$ we observe in the curvature 
profiles of Figs.\ \ref{ric2dcdt-fig:ric2dcdt} and \ref{ric2dcdt-fig:dric2dcdt} be attributed entirely to the compact topology? 
Secondly, from the parts of the curvature profiles where the influence of the
shortcuts is negligible, can we extrapolate to the curvature profile of an infinitely extended quantum geometry, representing a ``quantum plane''?    
As we will argue next, from the data we have collected, which of course are subject to numerical limitations in terms of system size and algorithm efficiency,
our qualified answer to both of these questions is in the affirmative.

Our numerical set-up is the same as for the pure average sphere distance measurements described in Sec.\ \ref{ric2dcdt-sec:numerical}. Namely, we generate
sequences $\{T_n\}$ of triangulations of fixed time extension $\tau$ and spacetime volume $N_2$, and on each geometry $T_n$ compute for each 
$\delta\!\in\! [1,\delta_{\textrm max}]$, with $\delmax \leq \tau/6$, the exact average sphere distance of a pair of spheres of radius $\delta$, located randomly on $T_n$. 
Here, we will perform measurements on 5$k$ different triangulations at given $N_2$ and $\tau$, now enhanced by a ``shortcut finder'', which we use as follows.
If for a given pair $(S_p^\delta, S_{p'}^\delta )$ of circles 
at least one point pair $(q,q') \in S_p^\delta \times S_{p'}^\delta$ is found to be connected by a shortest path of nonzero intersection number, 
the corresponding data point for $\bar{d}(S_p^{\delta},S_{p'}^{\delta})/\delta$ is marked as having a level-1 violation. 
If the region $C$ of the circles and their interiors is found to wrap around the spatial direction, with self-touchings or self-overlaps,
the corresponding data point for $\bar{d}(S_p^{\delta},S_{p'}^{\delta})/\delta$ is marked as having a level-2 violation.
After collecting all data for given $\tau$ and $N_2$, we determine the fractions $f_1$ and $f_2$ of data points that suffer from a level-1 or level-2 violation
respectively, as a function of $\delta$. (Note that a level-2 violation of a given pair of circles always implies the presence of a level-1 violation, meaning
that $f_2(\delta)\! \leq f_1\! (\delta),\,\forall\delta$.)

These fractions give us estimates of the frequency with which average sphere distance measurements are affected by topological shortcuts, but
they do not provide direct quantitative information about how much they lower the expectation value $\langle \bar{d}_{\textrm av}(\delta)/\delta \rangle$.
However, taking into account the behavior of the classical torus illustrated in Fig.\ \ref{ric2dcdt-fig:flattorus}, one would expect that shortly after the onset of level-1
violations, their number (and hence their effect on the average sphere distance) grows rapidly as a function of $\delta$. Since there is no sharp onset
for the presence of shortcuts in the quantum theory, as we have already discussed, we should redefine ``onset'' to mean ``larger than some small 
threshold fraction $\tilde{f}_1$''. In addition, we can use the level-2 violations as estimates for when the measurements will be strongly affected by shortcuts.
Classically, it is clear that when the spheres start to self-overlap, a large part of the distance measurements $d(q,q')$ will be shortcuts (a rough estimate
would be about one third). Their large impact is confirmed by Fig.\ \ref{ric2dcdt-fig:flattorus}, where this point corresponds to $\delta\! =\! 1/3$,
about halfway down the slope shown for the torus. 
In an attempt to translate this to a criterion for the quantum theory, one may expect that beyond a small, but nonvanishing
threshold value for the fraction $f_2$, one finds a large influence of topological shortcuts on the curvature profile.

\begin{figure}[t]
	\centering
	\begin{subfigure}[t]{0.48\textwidth}
	\centering
		\includegraphics[width=0.95\textwidth]{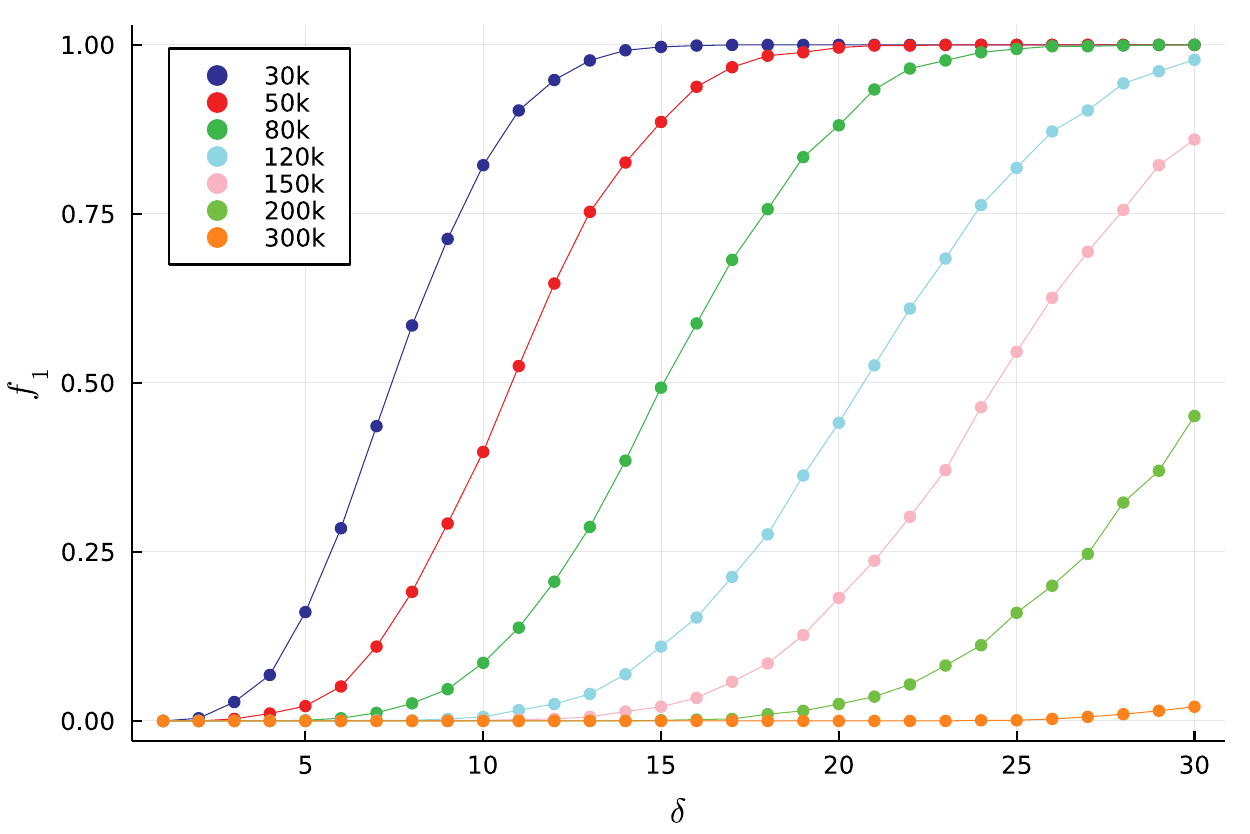}
		\label{ric2dcdt-fig:violations-primal-lvl-1}
	\end{subfigure}
	\hfill
	\begin{subfigure}[t]{0.48\textwidth}
	\centering
		\includegraphics[width=0.95\textwidth]{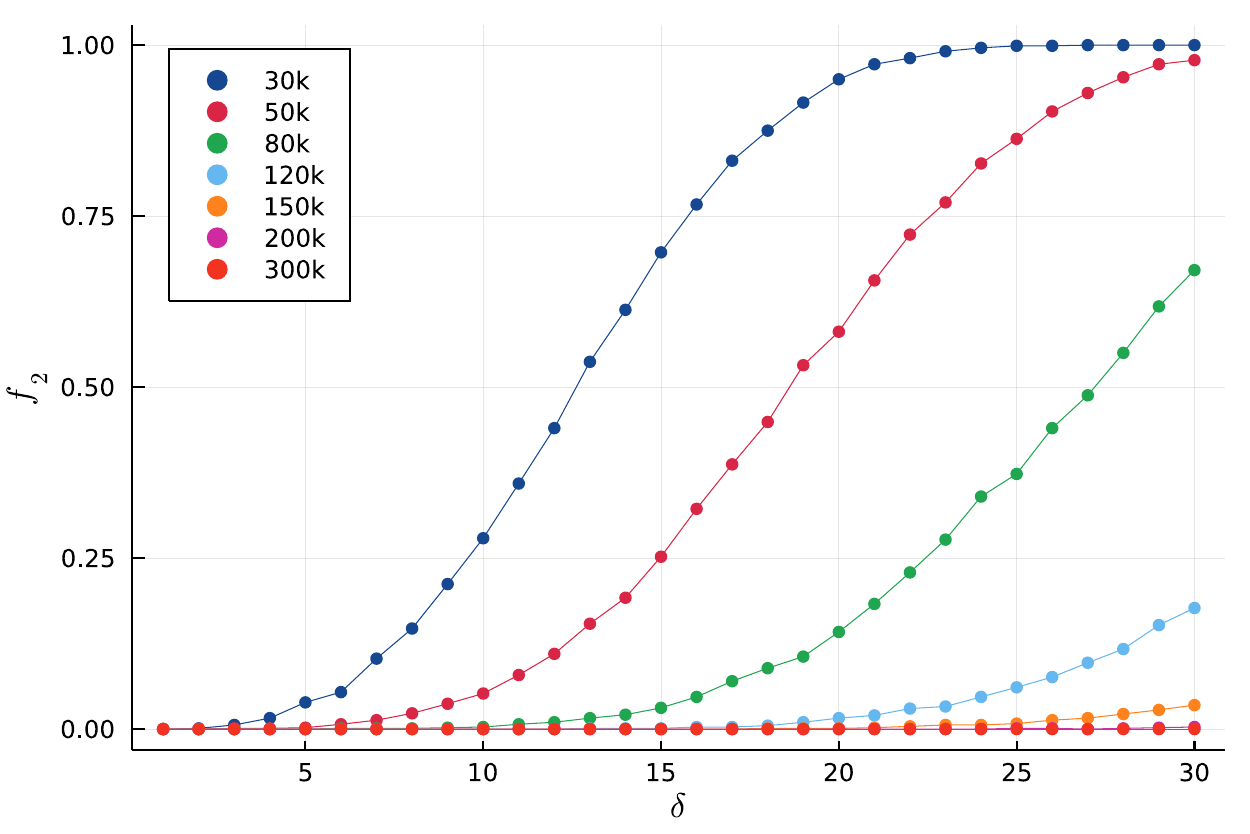}
		\label{ric2dcdt-fig:violations-primal-lvl-2}
	\end{subfigure}
	\caption{Fraction of level-1 violations $f_1(\delta)$ (left) and level-2 violations $f_2(\delta)$ (right) in CDT simulations with $\tau\! =\! 183$ 
	time slices and volumes $N_2\!\in\! [30k,300k]$.
	}
	\label{ric2dcdt-fig:violations-primal}
\end{figure}

The data we have found are consistent with these expectations. Our measurements of $f_1(\delta)$ and $f_2(\delta)$ at $\tau\! =\! 183$ and for
a range of volumes are shown in Fig.\ \ref{ric2dcdt-fig:violations-primal}. 
All curves for $f_1(\delta)$ start out close to zero, stay there for some range of $\delta$ that grows with the volume $N_2$, then undergo a rather rapid rise before
saturating to a value near 1. For the two smallest volumes, there is effectively no $\delta$-range that is not affected by shortcuts, especially considering
that short-distance lattice artifacts extend at least to $\delta\!\approx\! 5$, where anyway one should not trust the average sphere distance measurements. 
At the same time, for the largest system size $N_2\! =\! 300k$ the fraction $f_1$ does not exceed $\sim 0.02$, even at the largest measured value 
$\delta\! =\! 30$, which means that its influence on the curvature profile is very small. The curves for $f_2(\delta)$ have a similar shape, but appear
stretched along the $\delta$-axis by factors between 1.5 and 2.0. 

These findings motivate the introduction of a threshold value $\tilde{f}_1\! =\! 0.01$ for the fraction $f_1$ that characterises the onset of level-1 violations
in the quantum theory. In other words, our working assumption is that average sphere distance measurements whose fraction of shortcuts is below this threshold
are not significantly influenced by the topology. This seems a safe choice, also in view of the fact that at the onset of the shortcut phenomenon
(as $\delta$ increases), the difference in length between a shortcut $\phi(q,q')$ and the corresponding curve with vanishing intersection number between 
$q$ and $q'$ will be small. We have not performed a more detailed analysis comparing the two types of distance for given point pairs $(q,q')$, which is
somewhat involved technically and beyond the scope of this chapter. 
However, note that the threshold cannot be chosen much larger, because one then quickly reaches $\delta$-values
where level-2 violations become significant (c.f.\ Fig.\ \ref{ric2dcdt-fig:violations-primal}). For given $N_2$ and $\tau$, the threshold $\tilde{f}_1$ translates
into an upper bound on $\delta$, which we will again denote by $\delta_{\textrm max}$.\footnote{Whenever the bound $\delta_{\textrm max}$ coming from the time
identification is lower than the one coming from the threshold $\tilde{f}_1$, we implement the lower one of the two.} 
We have checked that for a fixed number of spatial slices
($\tau\! =\! 123$, 183 or 243), $\delta_{\textrm max}$ to good approximation scales linearly with the system size $N_2$, as one would expect.

\begin{figure}[t]
	\centering
	\includegraphics[width=0.7\textwidth]{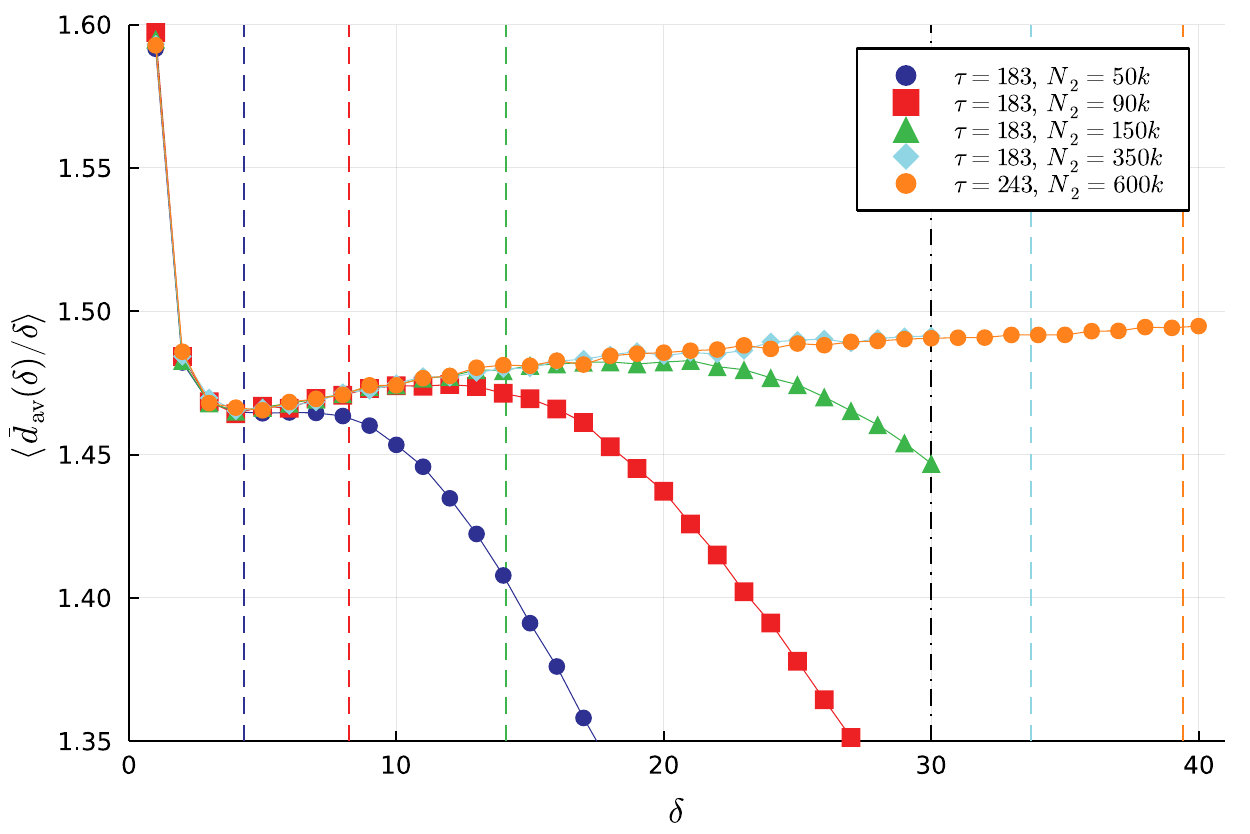}
	\caption{Curvature profiles $\langle \bar{d}_{\textrm av}(\delta)/\delta \rangle$ of Fig.\ \ref{ric2dcdt-fig:ric2dcdt}, supplemented by upper bounds on $\delta$
	(vertical dashed lines), 
	indicating the $\delta$-ranges where topological effects are negligible.}
	\label{ric2dcdt-fig:ric2dcdt-delmax}
\end{figure}

We will now re-examine the results for the curvature profiles presented Sec.\ \ref{ric2dcdt-sub:results} in the light of the preceding analysis. We have added to each 
of the curves in Fig.\ \ref{ric2dcdt-fig:ric2dcdt} the relevant upper bound $\delta_{\textrm max}(N_2,\tau)$ in the form of a vertical dashed line, as shown in 
Fig.\ \ref{ric2dcdt-fig:ric2dcdt-delmax}. 
An immediate observation is that the curvature profile for each of the three smaller volumes only starts sloping downward beyond its associated
bound $\delmax (N_2,\tau)$. This supports the hypothesis that these fall-offs are a direct consequence of the compact spatial topology of the torus, and would not
be present if instead we had studied CDT quantum gravity with spatial slices of the topology of an interval. In addition,
before a curvature profile reaches its bound $\delmax$, it is (within measurement errors) identical to all the other profiles. 
On the basis of the limited range of volumes we have been able to consider, this points to the existence of a common, universal curvature profile in the
limit of infinite volume. 
Interestingly, the resulting, enveloping curve is not a constant, but (beyond the initial dip) increases monotonically and at roughly the same rate 
over the range of $\delta$ we have been able to study. This appears to be a new type of quantum behavior with no obvious classical analogue,
which we will discuss further in Sec.\ \ref{ric2dcdt-sec:summary}.

Our current numerical limitations do not allow us to extend the measured profiles to larger $\delta$, to understand whether and at what rate 
the slope continues. On the one hand, 
the computation of the average sphere distance becomes quickly more expensive as $\delta$ increases, slowing down the data collection 
significantly. On the other hand, to accommodate larger values of $\delta_{\textrm max}$ the size of the torus would have to be scaled up in both directions,
further increasing simulation times. The combination of these factors makes it difficult to obtain statistics of sufficient quality for link distances much larger than 
the $\delmax\! =\! 40$ we have investigated here.

\subsection{Direction-dependent measurements}
\label{ric2dcdt-sec:anisotropy}

The quantum Ricci curvature $K_q(p,p')$ defined in Eq.\ \eqref{cp-2dexp} depends on a pair of points $(p,p')$, which can be thought of as a
generalized ``vector'', with a certain length (the distance between the points) and ``direction'' (along the shortest path between $p$ and $p'$). 
In order to turn it into a proper observable, we integrated $K_q(p,p')$ over all point pairs in the triangulation, subject to the condition 
$d(p,p')\! =\!\delta$, thereby losing any directional information. To nevertheless get an idea of the directional dependence of the quantum Ricci curvature, we 
can investigate separately two different types of ``direction'' by referring to the anisotropic, discrete structure of the underlying 
CDT geometries,\footnote{We refer the interested reader to \cite{curien2020geometric} for a detailed analytical investigation of the properties of two-dimensional causal random geometries.} following the treatment in four dimensions \cite{klitgaard2020how}. There is no claim that this construction leads to genuine observables, 
amongst other things, because these geometries do not possess a distinguished time direction. The results should therefore be interpreted with the 
appropriate caution.

\begin{figure}[t]
	\centering
	\includegraphics[width=0.7\textwidth]{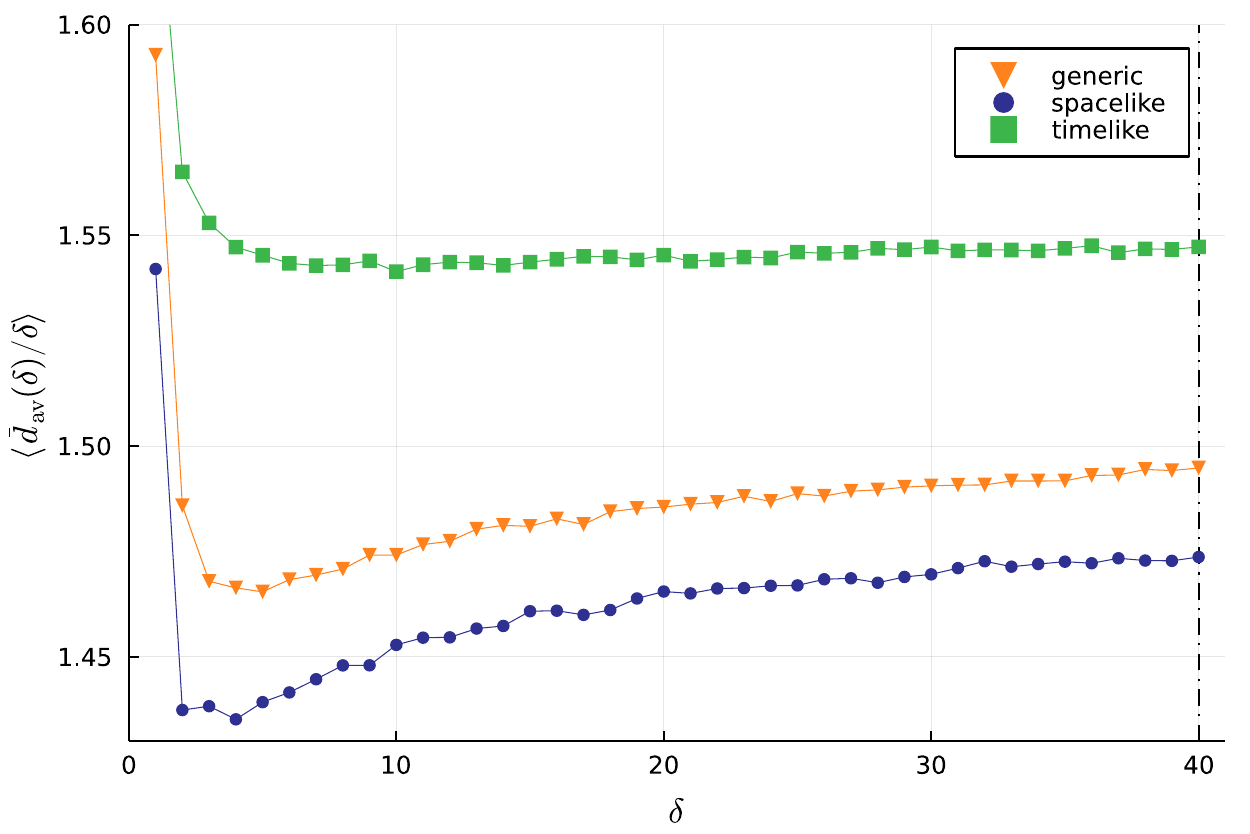}
	\caption{Direction-dependent curvature profiles $\langle \bar{d}_{\textrm av}(\delta)/\delta \rangle$ as a function of the link distance $\delta$, for timelike
	(green squares) and spacelike (blue dots) sphere separation, as described in the text, both for $N_2\! =\! 500k$. For comparison, we have added 
	the curvature profile for a generic orientation, measured at $N_2\! =\! 600k$, from Fig.\ \ref{ric2dcdt-fig:ric2dcdt} (orange triangles). 	
	(Error bars are on the order of the dot size.)}
	\label{ric2dcdt-fig:ric-anisotropic}
\end{figure}

We will consider the case where the circle centers $(p,p')$ are separated by $\delta$ spacelike edges contained in the same spatial slice of constant $t$,
and the case where they are separated by $\delta$ timelike edges, such that $\left|t_p\! -\! t_{p'}\right| \! = \! \delta$. The detailed geometry of
the corresponding double-circle configurations differs slightly, due to the anisotropy of the underlying lattice. For example, a timelike separated
circle pair always contains pairs of points with $d(q,q')\! =\! 3\delta$. This is not the case when a circle pair is spacelike separated,
because points $q$ and $q'$ in the same spatial slice can in general be linked by a geodesic which is not contained completely in the same slice. 
As noted in \cite{klitgaard2020how,klitgaard2022new}, discretization anisotropies are reflected in different values of the (non-universal)
constant $c_q$ of Eq.\ \eqref{cp-2dexp}, corresponding to different vertical offsets in the curve of the normalized average sphere distance. The
interesting question is then whether up to such a relative shift the quantum Ricci curvatures in the space- and timelike directions behave in the same
way, as was observed in CDT quantum gravity in four dimensions \cite{klitgaard2020how}.

We have performed $25k$ curvature profile measurements each for timelike and spacelike separations of the circle centers (Fig.\ \ref{ric2dcdt-fig:ric-anisotropic}).
Both were taken in the range $\delta\! \in\! [1, \delmax\! =\! 40]$ and at volume $N_2\! =\! 500k$, with $\tau\! =\! 243$ for the timelike and $\tau\! =\! 200$
for the spacelike case. For comparison, we have included the direction-averaged data from Fig.\ \ref{ric2dcdt-fig:ric2dcdt} for $N_2\! =\! 600k$ and $\tau\! =\! 243$.
We observe that the data for the two direction-dependent cases are not completely dissimilar, but are not parallel to each other either. The spacelike
curve resembles that of the direction-averaged data, with an initial dip, followed by a monotonic increase, whose slope however seems to decrease 
as a function of the link distance $\delta$.  
By contrast, the initial dip of the timelike curve is not very pronounced. It is followed by an almost flat regime, with a very slight upward slope. 
The quality of the data is not sufficient to establish whether or not this is compatible with a constant behavior. 
At this stage, this leaves open the possibility that two-dimensional CDT quantum gravity in the time 
direction -- if such a notion can be defined appropriately -- may be quantum Ricci-flat. 
It would be premature to draw any strong conclusions from our exploratory study, but the role of anisotropy in the curvature behavior of CDT in
two dimensions clearly deserves further attention.

\section{Summary and conclusion}
\label{ric2dcdt-sec:summary}

We set out to understand the curvature properties of the quantum spacetime generated dynamically in two-dimensional
CDT quantum gravity, working with standard boundary conditions, where both space and time are compactified
and the curved spacetimes summed over in the nonperturbative path integral therefore have the topology of a two-torus.
Using Monte Carlo simulations for discrete spacetime volumes of up to $N_2\! =\! 600k$, we measured average sphere distances
and the associated curvature profiles as functions of the link distance and the dual link distance. 
Our main objective was to distinguish between local and global features of the quantum Ricci curvature, and to isolate the former, to the
extent this is possible.
By ``global'' we mean effects that are present purely as a consequence of the compactification. We saw that topological effects in the time direction are easily
taken care of by putting a sharp upper bound on $\delta$, which sets the linear size of the local neighborhood that contributes to
a measurement of the average sphere distance, from which then the local quantum curvature is extracted.

A nontrivial part of the analysis dealt with establishing a similar bound to exclude the influence of the compactification in
the spatial direction. Here, large quantum fluctuations of the volume of the one-dimensional spatial slices prevent the existence of
a sharp bound. This motivated us to introduce a threshold criterion for the measured average sphere distances, which
involved the monitoring of what we called topological shortcuts. These are shortest geodesics between pairs of points on a double
circle, which wind around the torus in the ``wrong'' way, in the sense that the geodesic would not exist if the local configuration was 
part of an infinitely extended planar triangulation. We classified a data point for an average sphere distance measurement --
contributing to the ensemble average -- as admissible if the fraction of point pairs 
connected by a topological shortcut did not exceed 1\%. From the statistics of these ``level-1 violations'' we extracted
volume-dependent bounds $\delta_{\textrm max}$, below which the measurements can be regarded as 
effectively free of topological effects.

Considering only data points that do not exceed these $\delta$-bounds
and therefore can be regarded as describing pure, quasi-local geometry, a single universal curvature profile 
$\langle \bar{d}_{\textrm av}(\delta)/\delta \rangle$ is seen to emerge (orange dots in Fig.\ \ref{ric2dcdt-fig:ric2dcdt-delmax}). Disregarding
lattice artifacts for $\delta\! \lesssim \! 5$, this curve increases monotonically throughout the investigated range 
$5\!\leq\! \delta\!\leq\! 40$, ostensibly indicating a small negative curvature.  
Assuming that the underlying quantum geometry is approximately homogeneous, a natural point of comparison is
a classical space of constant negative curvature. However, we dismissed this interpretation, since it would entail the presence of
a dynamically generated scale (a curvature radius), for which an obvious source is lacking. 
We were thus led to conclude that the curvature behavior of the CDT quantum geometry falls into a new, nonclassical category, 
that of \emph{quantum flatness}. The fact that no rescaling in $\delta$ seems necessary to obtain a common curvature profile indicates 
a scale-independent, fractal behavior of the underlying quantum geometry, at least with respect to its (quasi-)local curvature properties. 
This does not contradict any
earlier findings on the geometric properties of two-dimensional CDT quantum gravity, which have largely concentrated on the dynamics of the spatial
volume, which is a \emph{global} geometric quantity.\footnote{Our use of ``local''
and ``global'' for geometric observables refers to the scale probed, in our case $\delta$. This is a meaningful distinction, even though in the
absence of other reference systems all observables include an averaging/integral over spacetime and therefore are always global (=nonlocal) 
in the standard classical sense.} 

Our result runs counter to the idea that the quantum-gravitating torus (on sufficiently coarse-grained scales and in the sense of expectation
values) should resemble a flat, classical torus in terms of its local curvature properties. 
Having finally a well-defined, nonperturbative notion of
renormalized curvature at our disposal \cite{klitgaard2018introducing,klitgaard2018implementing,klitgaard2020how}, we have been able to test this idea, but it appears to be incorrect.  
Instead, on the basis of our measurements using both the link distance and the dual link distance and on theoretical grounds, we have concluded 
that the quantum geometry is characterised by a universal, non-constant curvature profile.
This behavior, dubbed ``quantum flatness'', appears to be a genuine nonperturbative feature of two-dimensional CDT quantum gravity on the torus,
rather than a numerical fluke. It would be interesting to better understand its origin in theoretical, analytical terms. 
Further input for this could come from a more detailed analysis of the anisotropic
features of the quantum Ricci curvature, extending our preliminary investigation in Sec.\ \ref{ric2dcdt-sec:anisotropy}. 

A natural extension to studying the curvature profile of pure two-dimensional CDT is to add matter fields to the system, and to investigate to what extent the curvature profile is affected, if at all. It was shown in \cite{ambjorn2000crossing} that adding sufficiently many Ising spin degrees of freedom causes the CDT quantum torus to separate into two distinct regions: a `bulk' region consisting of time slices where most of the spacetime volume is concentrated, and a `stalk' region consisting of time slices with volumes near the cutoff scale. This `pinching' of the geometry may leave an imprint on the curvature profile of the system. A preliminary investigation of the quantum Ricci curvature on such ensembles was made in \cite{vanderfeltz2021matter}. An important obstacle in this setting is that the finite time extent (i.e. number of time slices with effectively nonzero spatial volume) limits the range of scales at which the average sphere distance can be computed. This makes it difficult to investigate the scaling behavior of the curvature profile of this system, and it would be interesting to address this issue in future work. 

Another interesting question is whether quantum flatness is a phenomenon particular to two-dimensional Lorentzian quantum 
gravity or perhaps a more general feature of strongly fluctuating quantum geometry. A natural testing ground would be
fully-fledged CDT quantum gravity on a four-torus, to which our set-up and methodology naturally generalize, and which
has been at the focus of recent research \cite{ambjorn2021cdt}. Lastly, the present study provides further evidence for the quantum Ricci 
curvature as a powerful new ingredient to help us understand the nature of quantum geometry and quantum gravity.

\chapter{Curvature in the presence of defects}\label{ch:defects}

\section{Introduction}
\label{defects-sec:intro}

In this chapter we examine a class of two-dimensional, compact model spaces of non-negative curvature, where 
the rotational $SO(3)$-symmetry of the constantly curved sphere is broken to a residual discrete subgroup associated
with the symmetries of a regular, convex polyhedron. A primary aim is to quantify how the \textit{distribution of curvature} is
reflected in the curvature profiles of the corresponding spaces, with the two-dimensional case chosen for simplicity. 
According to the Gauss-Bonnet theorem, the total
Gaussian curvature of any two-dimensional space of spherical topology is given by $4\pi$. On a round two-sphere,
this curvature is distributed completely evenly. The surface of a regular tetrahedron represents the
opposite extreme: it is flat almost everywhere, except for four isolated, singular points, each associated with a Gaussian
curvature (deficit angle) of $\pi$. There are two main questions which we will try to answer in what follows: can we tell the two spaces apart by comparing their volume [curvature?] profiles? And how
effective is the averaging in ``smearing out'' the effect of the isolated singularities?

We refer to Chapter \ref{cp-sec:qrc} for a detailed discussion on curvature profiles and the average sphere distance, including the key formulas used in their definition.
In Sec.\ \ref{defects-sec:cone}, we analyze the influence of an isolated conical singularity on the average sphere distance,
which is a crucial ingredient in the curvature measurements. It allows us to compute the curvature profiles 
at short distance scales $\delta$ of a class of regular polyhedral surfaces associated with Platonic solids in Sec.\ \ref{defects-sec:measure}.
In Sec.\ \ref{defects-sec:tetra} we discuss the construction of geodesics
on the surface of a regular tetrahedron, which is needed to determine the geodesic circles appearing in the
curvature construction. This turns out to be a nontrivial task, which can be tackled by unfolding the surface onto the
flat plane. It enables us to determine geo\-desic circles of arbitrary location and radius, which then serves as an input
for the numerical computation of the curvature profile. 
The final Sec.\ \ref{defects-sec:final} contains a summary and our conclusions.

\section{The influence of conical singularities}
\label{defects-sec:cone}

From the point of view of its intrinsic geometric properties, the surface of a regular tetrahedron is flat everywhere, apart from its
corners, where curvature is concentrated in the manner of a delta function. The curvature that appears along its edges (with the
exception of the four corners) in a
standard embedding picture of the tetrahedron in Euclidean $\R^3$ is only extrinsic and not of interest in our present context.
The same is true for the surfaces of other regular polyhedra. It implies that the intrinsic geometry of the 
neighborhood of any of their corners is identical to that of a cone with a conical curvature singularity at its ``tip''. 
To obtain the curvature profile of such a classical
two-dimensional space, we must understand how the presence of one or more conical singularities affects the average sphere
distance of a pair of circles on them. 

Our quantitative assessment starts by examining the influence of a single conical
singularity. Clearly, since the metric of a cone is flat everywhere except at the point where the singularity is located,
if the pair $(S_p^{\delta},S_{p'}^{\delta})$ of $\delta$-circles depicted in Fig.\ \ref{cp-fig:avg-sphere-dist} is sufficiently far away
from the singularity, the normalized average sphere distance is that of flat space, 
$\bar{d}(S_p^{\delta},S_{p'}^{\delta})/\delta\!\approx \! 1.5746$. On the other hand, if the conical singularity lies
somewhere inside the double circle, one would expect it to have a nontrivial effect. However, somewhat contrary to
na\"ive expectation, the curvature singularity has an influence on the average sphere distance even when it lies
strictly outside the two circles, as long as it is sufficiently close to them. As we will show in more detail below, this is due to the 
presence of ``geodesic shortcuts'' associated with the conical singularity.

\begin{figure}
\centering
\begin{subfigure}[t]{0.45\textwidth}
\centering
\includegraphics[height=0.6\linewidth]{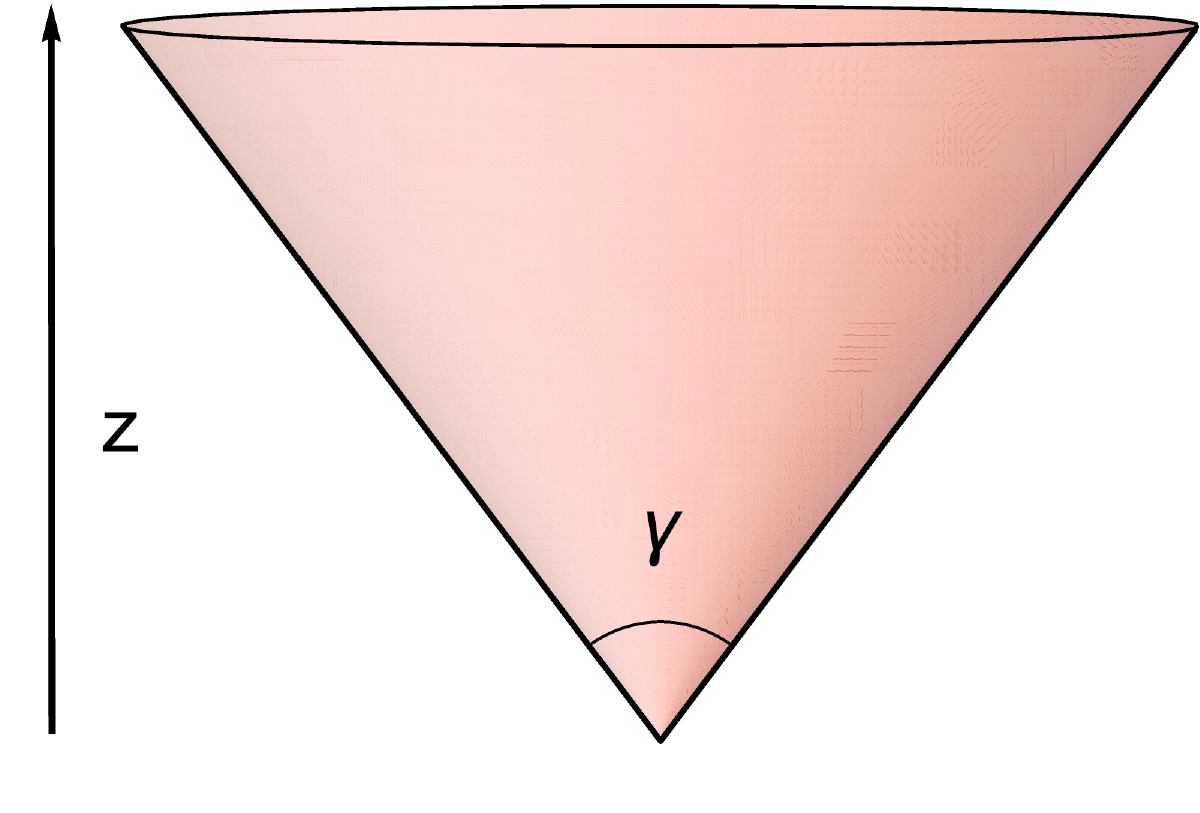}
\end{subfigure}
\hspace{0.05\textwidth}
\begin{subfigure}[t]{0.45\textwidth}
\centering
\includegraphics[height=0.6\linewidth]{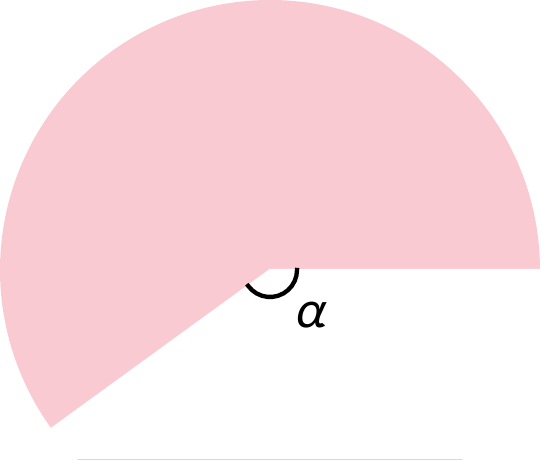}
\end{subfigure}
\caption{A geometry with a conical singularity can be represented as a
cone in Euclidean space with apex angle $\gamma$  (left) 
or a flat plane with an angle $\alpha$ removed (right).}
\label{defects-fig:cones}
\end{figure}

\subsection{Geometry of the cone}

One way of describing a cone with apex angle $\gamma$ is as the set of all points $(x,y,z)\in\R^3$ satisfying 
$z\! =\!\cot (\gamma/2) \sqrt{x^2+y^2}$, where $0\! <\!\gamma\!\leq\!\pi$ (Fig.\ \ref{defects-fig:cones}, left).
The apex or tip of the cone lies at the origin of the Cartesian
coordinate system, and $\gamma$ is the angle formed in the plane $x\! =\! 0$ by its intersection with the cone.
A point on the cone can be labelled by its Euclidean distance $r$ to the origin and a rotation angle 
$\theta =\arctan y/x \in [0,2\pi]$. The induced metric on the cone in terms of these coordinates is given by
\begin{equation}
ds^2 = dr^2 + \sin^2 ( \tfrac{\gamma}{2})\, r^2\,d\theta^2.
\label{defects-eq:metric}
\end{equation}
An alternative description of the cone is obtained by cutting out an infinite wedge with angle $\alpha$ 
from the Euclidean plane $\R^2$ (Fig.\ \ref{defects-fig:cones}, right) and identifying boundary points pairwise across the wedge. 
The result is a cone with a curvature singularity characterized by the deficit angle $\alpha$, which is
a direct measure of Gaussian curvature. Since we are only interested in singularities with positive 
curvature, the relevant angle range is $0\! \leq \! \alpha\!<\! 2\pi$. The relation with the apex angle $\gamma$
is $\alpha = 2\pi (1-\sin (\gamma/2))$. 

The useful feature of the ``planar'' representation of the cone in terms of a part of 
$\R^2$ is the fact that geodesics are simply given by straight lines. A knowledge of geodesics is
needed to determine the geodesic distances between points.\footnote{Here and in the remainder of the chapter,
a geodesic is defined as a locally shortest curve that does not contain any singularities, with the possible exception of
its endpoints.}
The only minor difficulty one has to take care of
is to continue a geodesic correctly across the wedge if necessary. Placing the singularity at the origin of $\R^2$ and
using standard spherical coordinates $(r,\varphi)$, the metric on the cone in this parametrization is
\begin{equation}
ds^2=dr^2+r^2 d\varphi^2,
\label{defects-eq:flat}
\end{equation}
where the range of the angle $\varphi$ is limited to $\varphi\in [0,2\pi-\alpha]$. Using the notation $r$ is justified, because
the radial geodesic distance here is identical to the one we used for the cone embedded in $\R^3$. Comparing the
two metrics \eqref{defects-eq:metric} and \eqref{defects-eq:flat}, we see that they are simply related by a constant rescaling of the 
angles $\theta$ and $\varphi$, namely, $\varphi = \sin(\gamma/2) \theta$. 
The distance between two points $(r_1,\varphi_1)$ and $(r_2,\varphi_2)$ is simply their Euclidean distance
\begin{equation}
d\left( (r_1,\varphi_1), (r_2,\varphi_2) \right) = \sqrt{r_1^2+r_2^2-2 r_1 r_2 \cos (\varphi_2 -\varphi_1) },
\label{defects-eudist}
\end{equation}
whenever the absolute angle difference $|\varphi_2-\varphi_1|$ does not exceed $\pi -\alpha/2$. If it does,
the argument of the cosine should be substituted by the angle $2\pi-\alpha-|\varphi_2 -\varphi_1|$. 
By rescaling the angles, we can immediately derive a distance function for points on the cone labelled by $r$ and $\theta$, namely,
\begin{equation}
 d\left( (r_1,\theta_1), (r_2,\theta_2) \right) = \sqrt{r_1^2+r_2^2-2 r_1 r_2 \cos (\sin(\tfrac{\gamma}{2}) \Delta (\theta_1,\theta_2)) },
\label{defects-conedist}
\end{equation} 
where
\begin{equation}
\Delta (\theta_1,\theta_2):= \min (|\theta_2-\theta_1|, 2\pi- |\theta_2-\theta_1| ).
\label{defects-Delta}
\end{equation}
In what follows, we will work with the cone parametrization in terms of coordinates $(r,\theta)$, which is more
convenient in the applications we will consider.

\subsection{Computing average sphere distances}
\label{defects-subsec:sd}

\begin{figure}
\centering
\begin{subfigure}[t]{0.47\textwidth}
\centering
\includegraphics[height=0.9\linewidth]{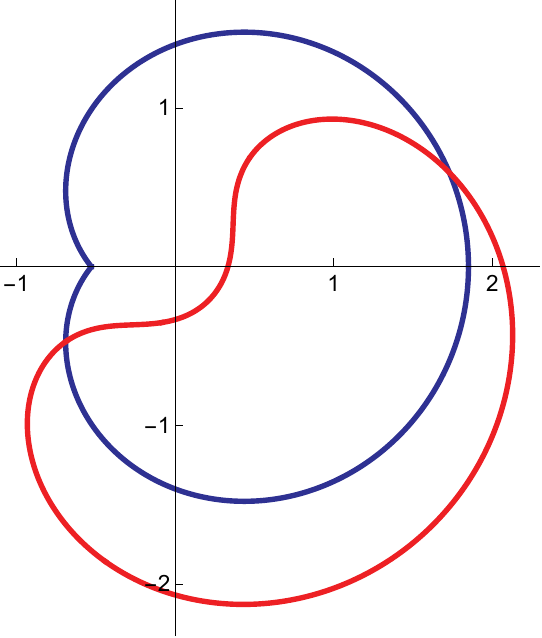}
\end{subfigure}
\hspace{0.04\textwidth}
\begin{subfigure}[t]{0.47\textwidth}
\centering
\includegraphics[height=0.9\linewidth]{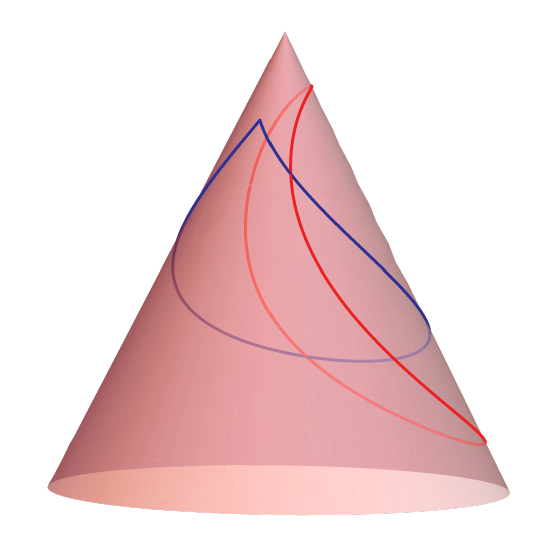}
\end{subfigure}
\caption{Two unit circles in the $(r, \theta)$-plane with a singularity at the origin, with the corresponding embedding in $\mathbb{R}$. The blue circle encloses the singularity and shows a discontinuous jump of its tangent vector at $\theta = \pi$.}
\label{defects-fig:tangent-vector-jump}
\end{figure}

To compute the average sphere distances \eqref{cp-sdist} on the cone $M$, we must first parametrize 
geodesic circles on $M$, where the geodesic circle of radius $\delta$ centered at the point $p$ consists of all points $q$ at distance 
$\delta$ from $p$, $S_p^\delta =\{q\in M |\, d(p,q)\! =\! \delta\}$. 
Let us first discuss the parametrization of a geodesic
circle which encloses the singularity at the origin. This case is slightly simpler, because the angle $\theta$ of $M$
can be used to label the points of $S_p^\delta$ uniquely. Using the rotational invariance of the set-up, we can without
loss of generality choose the center $p$ to have coordinates $p=(r_1,\theta_1\! =\! 0)$, where $r_1 <\delta$. 
Labelling a point $q$ on the circle
$S_p^\delta$ by $q=(r_2,\theta_2)$, it must by assumption satisfy $d(p,q)\! =\! \delta$, which is a quadratic equation
for $r_2$. Because of relation \eqref{defects-Delta}, we should distinguish between the cases $0\leq\theta_2\leq\pi$ and
$\pi\leq\theta_2\leq 2\pi$. The unique solutions $r_2(\theta_2)$ are
\begin{equation}
r_2 (\theta_2)\!  = \! \begin{cases}
r_1\cos\!\left( \theta_2 \sin (\frac{\gamma}{2})\right)\! +\! \sqrt{\delta^2\! -\! r_1^2 
\sin^2\!\left( \theta_2\sin (\frac{\gamma}{2})\right)}, \;\;\; & 0 \leq\theta_2\leq \pi , \\
r_1\cos\!\left( (2\pi\! -\!\theta_2) \sin (\frac{\gamma}{2})\right)\! +\! \sqrt{\delta^2\! -\! r_1^2 \sin^2\!
\left( (2\pi\! -\!\theta_2)\sin (\frac{\gamma}{2})\right)},
 & \pi\leq\theta_2\leq 2\pi .
\end{cases}
\label{defects-r2second}
\end{equation}
They can be used to uniquely parametrize the points along $S_p^\delta$ by the curve
\begin{equation}
c^\mu(\theta)=(r_2(\theta),\theta),\;\;\;\;\; 0\leq \theta \leq 2\pi ,
\label{defects-curve1}
\end{equation}
with curve parameter $\theta$, and where for ease of notation we have suppressed the dependence of the curve on $p$ and $\delta$.
Note that this curve is smooth everywhere apart from the point $\theta\! =\!\pi$,
where its tangent vector $dc^\mu/d\theta$ jumps in a discontinuous way, resulting in a kink in the curve $c(\theta)$ (Fig. \ref{defects-fig:tangent-vector-jump}).
If the center $p$ has coordinates $(r_1,\theta_1)$, for some nonvanishing angle $\theta_1$, the angle
$\theta_2$ on the right-hand sides of the relations \eqref{defects-r2second} should be substituted by $| \theta_2 -\theta_1|$.

The parametrization of a geodesic circle which does not enclose the curvature singularity can also be parametrized by
a $\theta$-angle, but not uniquely. As we will see below, a convenient choice in any integration along $S_p^\delta$ 
is to parametrize the circle along two separate segments and then add their contributions. 
Again we choose the point $p=(r_1,0)$ as the center of the circle, whose radial
coordinate now satisfies $r_1 \geq\delta$. When solving the equation $d(p,q)\! =\! \delta$ for $r_2$ as before, there are
two differences. First, the angle $\theta_2$ can only
vary in the interval $[-\theta_\textrm{m},\theta_\textrm{m}]$, where the maximal angle $\theta_\textrm{m}$ is defined by
$\sin (\sin{ (\gamma/2)}\, \theta_\textrm{m})\! =\! \delta/r_1$.
This is easy to see by examining the geometry of the situation in the flat-space coordinates $(r,\varphi)$.  
Second, there are two solutions $r_2(\theta_2)$ for every value $\theta_2$ in the interior of this interval, namely,
\begin{equation}
r^\pm_2(\theta_2)=
r_1\cos\!\left( \theta_2 \sin (\tfrac{\gamma}{2})\right)\! \pm \! \sqrt{\delta^2\! -\! r_1^2 
\sin^2\!\left( \theta_2\sin (\tfrac{\gamma}{2})\right)},
\;\; -\theta_\textrm{m} \leq\theta_2\leq \theta_\textrm{m}. 
\label{defects-r2both}
\end{equation}
One easily verifies that the argument of the square root cannot become negative in the angle range considered.
A straightforward way to parametrize the points $q$ along the circle is by splitting the circle into two segments $c_+(\theta)$ and $c_-(\theta)$,
corresponding to the two solutions \eqref{defects-r2both},
\begin{eqnarray}
&c_+^\mu(\theta) = \left(r_2^+(\theta), \theta \right), \;\;&  -\theta_\textrm{m} \leq \theta \leq  \theta_\textrm{m}, \\
&c_-^\mu(\theta) = \left(r_2^-(\theta), \theta \right), \;\; &  -\theta_\textrm{m} \leq \theta \leq \theta_\textrm{m}.
\label{defects-eq:circle-param-nenc}
\end{eqnarray}
If the angle $\theta_1$ of the point $p\! =\! (r_1,\theta_1)$ is nonvanishing, the angles $\theta_2$ in the above
expressions should again be substituted appropriately.
As a cross check of the circle parametrizations, one can use them to compute the lengths (one-dimensional volumes) of the circles,
which at any rate are needed in the computation of the average sphere distances \eqref{cp-sdist}. For the case that the circle encloses
the singularity, one finds
\begin{align}
\hspace{-1.2cm} vol (S^\delta_p) & =\int_0^{2 \pi}\!\!\! d\theta\, \sqrt{g_{\mu\nu} \tfrac{dc^\mu}{d\theta} \tfrac{dc^\nu}{d\theta}} \nonumber \\
&= \int_0^\pi \! d\theta\, \delta \sin(\tfrac{\gamma}{2}) (1\! +\! {\cal R}(\theta))+
\!\int_\pi^{2 \pi} \!\!\! d\theta\, \delta \sin(\tfrac{\gamma}{2}) (1\! +\! {\cal R}(2\pi -\theta))\\
&=2\pi\delta \sin (\tfrac{\gamma}{2})+2\delta \arcsin \left( \tfrac{r_1}{\delta}\sin \left( \pi \sin (\tfrac{\gamma}{2} )  \right)  \right),
\hspace{2.5cm} [r_1 <\delta ] \nonumber 
\label{defects-vol1}
\end{align}
where $g_{\mu\nu}$ refers to the metric \eqref{defects-eq:metric} and we have introduced the shorthand notation
\begin{equation}
{\cal R}(\theta)= \frac{r_1 \cos\left( \theta \sin (\tfrac{\gamma}{2}) \right)}{\sqrt{\delta^2-r_1^2 \sin^2 \left( \theta \sin (\tfrac{\gamma}{2}) \right)}}.
\label{defects-rshort}
\end{equation}
For the case that the circle does not enclose the singularity, the corresponding computation yields
\begin{align}
\hspace{-1.1cm} vol (S^\delta_p) & =\int_{-\theta_\textrm{m}}^{\theta_\textrm{m}}\!\!\! d\theta\, 
\sqrt{g_{\mu\nu} \tfrac{dc^\mu_+}{d\theta} \tfrac{dc^\nu_+}{d\theta}} 
+ \int_{-\theta_\textrm{m}}^{\theta_\textrm{m}}\!\!\! d\theta\, 
\sqrt{g_{\mu\nu} \tfrac{dc^\mu_-}{d\theta} \tfrac{dc^\nu_-}{d\theta}} \nonumber \\
& =\int_{-\theta_\textrm{m}}^{\theta_\textrm{m}}\! d\theta\, \delta \sin(\tfrac{\gamma}{2}) (1\! +\! {\cal R}(\theta))+
\!\int_{-\theta_\textrm{m}}^{\theta_\textrm{m}} \!\!\! d\theta\, \delta \sin(\tfrac{\gamma}{2}) (-1\! +\! {\cal R}(\theta))\\
&=2\pi\delta, \hspace{9cm} [r_1 \geq \delta ] \nonumber 
\label{defects-vol2}
\end{align}
which is the expected flat-space result.\footnote{Note that 
the tangent vectors $dc^\mu_\pm /d\theta$ diverge at the endpoints $\pm \theta_\textrm{m}$ of the two arches.To avoid problems in the numerical implementation of the sphere distance calculations we removed a tiny interval of width $\sim\ 10^{-8}$ from the integration range around such endpoints. This does not affect results at the accuracy considered.}

We now have all ingredients in hand to perform explicit computations of the average sphere distance \eqref{cp-sdist} for 
pairs $(p,p')$ of points at distance $\delta$. 

We perform the double-integral numerically in \textsc{Mathematica}, where the intermediate accuracy is kept at 8 significant digits. Apart from the strength of the singularity (captured by the deficit angle $\alpha$)
and the linear distance $\delta$, $\bar{d}(S_p^{\delta},S_{p'}^{\delta})$ depends on the distance
of $p$ from the singularity and on the orientation of the second circle relative to the first. 

Apart from the strength of the singularity (captured by the deficit angle $\alpha$)
and the linear distance $\delta$, $\bar{d}(S_p^{\delta},S_{p'}^{\delta})$ depends on the distance
of $p$ from the singularity and on the orientation of the second circle relative to the first. 
\begin{figure}
\centering{}
\begin{subfigure}[t]{0.49\textwidth}
\hspace{0.05\textwidth}\includegraphics[height=0.55\linewidth]{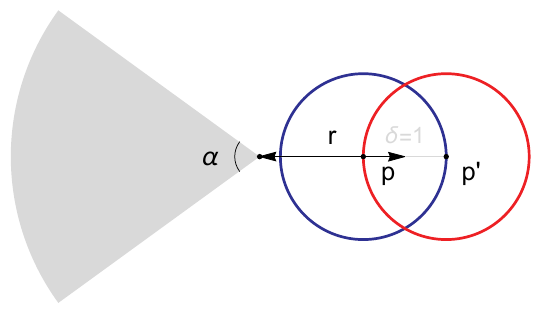} \\ \\
\includegraphics[width=1\linewidth]{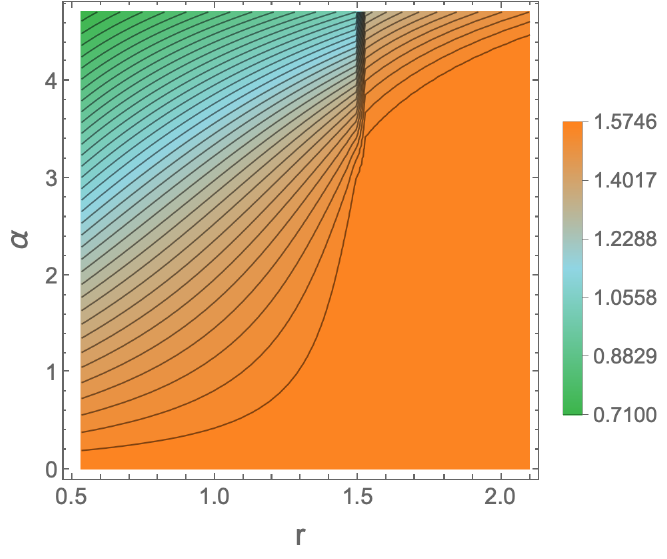}
\end{subfigure}
\begin{subfigure}[t]{0.49\textwidth}
\hspace{0.05\textwidth}\includegraphics[height=0.55\linewidth]{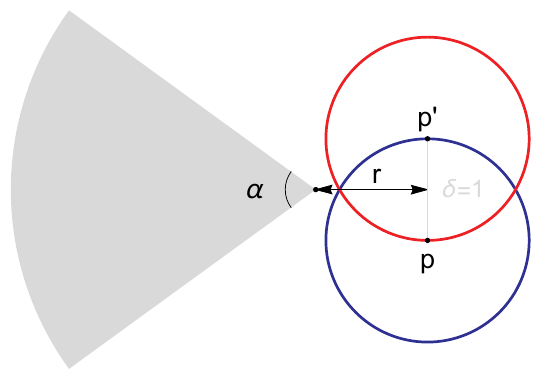} \\ \\
\includegraphics[width=1\linewidth]{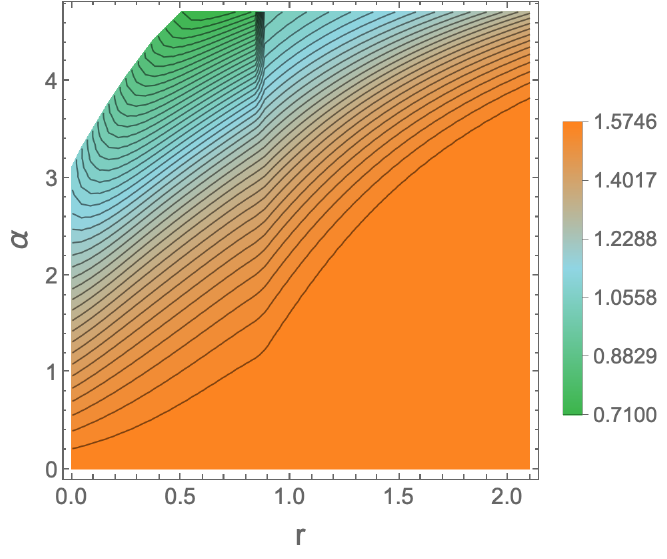}
\end{subfigure}
\caption{Contour plots of the average sphere distance $\bar{d} (S_{p}^\delta,S_{p'}^\delta)$ as a function of $\alpha$ and $r$, 
for two particular circle arrangements,
as described in the text. The distance $r$ is given in units of $\delta$ and the deficit angle $\alpha$ in radians.}
\label{defects-fig:contour-plots}
\end{figure}

To illustrate the behavior of $\bar{d}$,
Fig.\ \ref{defects-fig:contour-plots} shows contour plots for two particular orientations of the double circle relative to the singularity,
where the average sphere distance is given as a function of the deficit angle $\alpha$ and the distance $r$ of the midpoint between
$p$ and $p'$ to the singularity. To fix the overall scale, we have set $\delta\! =\! 1$. Recall that 
for $\alpha\! =\! 0$, there is no singularity, and that in flat space $\bar{d}\!\approx\! 1.5746$. 
As indicated by the darker shades in Fig.\ \ref{defects-fig:contour-plots}, the value of $\bar{d}$
decreases with increasing $\alpha$ and decreasing distance to the singularity, characteristic of a positive and growing
quantum Ricci curvature. The illustrations of the circle positions use the planar representation of the
cone with a wedge removed. In Fig.\ \ref{defects-fig:contour-plots}, left, the circle centers are chosen collinear with the singularity. For $r\! \geq\! 1.5$,
neither of the circles encloses the singularity and there is no or little deviation from the flat-space behavior, unless $\alpha$
increases beyond $\pi$. In this case the wedge will ``cut into'' the first circle, giving rise to a nontrivial effect. The region
$r\! < \! 0.5$ is not well defined, since it would imply that $p$ lies inside the wedge. In Fig.\ \ref{defects-fig:contour-plots}, right, the two circles are
arranged symmetrically with respect to the singularity. What is noteworthy here is the fact that even when the singularity
lies outside the circle configuration, i.e. $r\! >\! \sqrt{3}/2\approx 0.866$, and the size of the deficit angle is only moderate, there is a nontrivial
effect on $\bar{d}$. We will look closer at this phenomenon in the next subsection. The top left-hand corner of the contour
plot is not defined, since both $p$ and $p'$ would lie inside the wedge.

\subsection{Domain of influence}

So far, the developments of this section have been concerned with the geometry in the vicinity 
of an isolated curvature singularity on an infinitely extended cone.
However, we are interested in determining the curvature profiles of compact, flat spaces with
several such singularities, more precisely, the regular polyhedral surfaces of the so-called Platonic solids: the tetrahedron, octahedron,
cube, dodecahedron and icosahedron. The method developed for computing average sphere distances on a cone can be
used on these polyhedra too, but only if the double circles are sufficiently small to not be influenced by more than one of the
conical singularities at the corners of these polyhedra. For a surface with a given area, determined by the length of an edge,
this imposes an upper bound on the size $\delta$ of the circles. Since we need to average over all double circles of size $\delta$ to
obtain the curvature profile, we must determine the size $\delta_\textrm{max}$ below which the computation of
$\bar{d} (S_{p}^\delta,S_{p'}^\delta)$ will never be influenced by more than one of the corner points.

\begin{figure}
\centering
\begin{subfigure}[t]{0.45\textwidth}
\includegraphics[width=\linewidth]{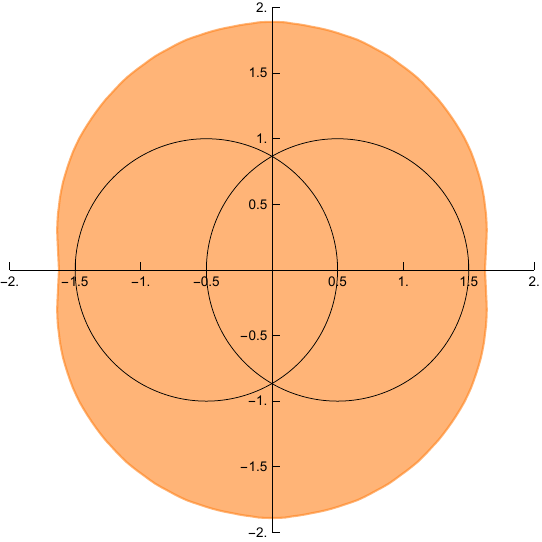}
\end{subfigure}
\hspace{0.02\textwidth}
\begin{subfigure}[t]{0.45\textwidth}
\includegraphics[width=\linewidth]{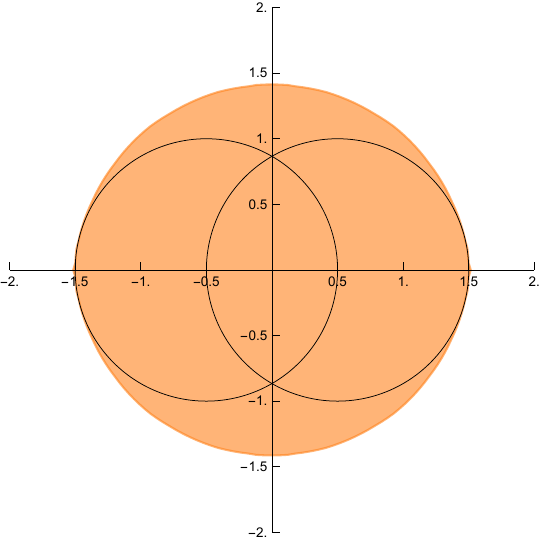}
\end{subfigure}
\caption{The domains of influence of a curvature singularity on the tetrahedron (left) and the octahedron (right). 
}
\label{defects-fig:domain}
\end{figure}

This leads to the notion of the \textit{domain of influence} of a curvature singularity with angle $\alpha$, defined as the
open two-dimensional region, including and surrounding a circle pair $(S_{p}^\delta,S_{p'}^\delta)$, which consists of all points
where a singularity of strength $\alpha$ would cause $\bar{d}/\delta$ to deviate from its flat-space value.
Fig.\ \ref{defects-fig:domain} illustrates the situation for $\alpha\! =\! \pi$ and $\alpha\! =\! 2\pi/3$, associated with the deficit angles
on the tetrahedral and octahedral surfaces respectively. The reason why a curvature singularity can
influence the average sphere distance, even when it lies outside either of the circles, is twofold. Firstly, a 
geodesic between a pair $(q,q')$ of points from the two circles can of course pass through a region that lies outside the circles
and be influenced by a singularity located there. Secondly, because of the presence of the singularity, geodesics between points
will not always be unique, but can occur in pairs that pass on either side of the singularity. This can lead to a geodesic shortcut between
given points $q$ and $q'$, compared with the situation in flat space.

The relevant quantity we need to determine for our purposes is the maximal diameter of the domain of influence for a
given angle $\alpha$. When measuring the curvature profile with the method described above, we must make sure that this diameter
does not exceed the edge length $L$ of the given polyhedron, which is the same as the distance between neighboring singularities.
If we call ${\cal D}(\alpha)$ the maximal diameter in units of $\delta$, it follows that the largest allowed circle size is 
$\delta_\textrm{max}\! =\! L/{\cal D}(\alpha)$. The maximal diameter for the tetrahedral surface 
can be determined from geometric considerations and is given by ${\cal D}(\pi)\! =\! \sqrt{9+4 \sqrt{2}}\! \approx\! 3.828$, corresponding
to the vertical axis in the left diagram of Fig.\ \ref{defects-fig:domain}. As the deficit angle $\alpha$ decreases towards $2\pi/3$, the domain of
influence shrinks along both axes until it reaches the left- and rightmost points of the double circle. The domain keeps shrinking along the
vertical axis to a value below 3 (Fig.\ \ref{defects-fig:domain}, right). It cannot shrink further along the horizontal direction since a singularity 
inside one (or both) of the circles always leads to a nonflat result for $\bar{d}/\delta$. We conclude that for the octahedron and
the remaining platonic solids with $\alpha\! =\! \pi/2$, $\pi/3$ and $\pi/5$, we have ${\cal D}(\alpha)\! =\! 3$.

\section{Measuring curvature profiles}
\label{defects-sec:measure}

The insights gained above will now be applied to evaluate curvature profiles for the surfaces of the five convex regular 
polyhedra depicted in Table \ref{defects-tab:platonic}. Each surface consists of identical polygons, whose edges have all the same 
length. Its Gaussian curvature is distributed equally over all vertices, which implies that each deficit angle has
size $\alpha\! =\! 4\pi/\#$vertices. Since we are interested in the effects of how the same amount of curvature is distributed
over a given spatial volume,
we compare the results for surfaces of the same area, which we take to be $4\pi$, the area of a two-sphere
of unit radius. The resulting values for the edge lengths $L$ and the associated maximal circle radii $\delta_\textrm{max}$
can be found in Table \ref{defects-tab:platonic}. From the limits on $\delta$ it is clear that we will only obtain partial curvature
profiles. It is not straightforward to extend the present method to include a second conical singularity, and
we have not attempted to do so. However, it turns out that for the special case of the tetrahedral surface there is an 
alternative way to extend the $\delta$-range significantly, as will be discussed in Sec.\ \ref{defects-sec:tetra} below. The reason is that one  
can attain a good and computationally explicit control over geodesics. 
\begin{center}
\newcommand\ph{0.09}
\begin{table}[ht]
\small
  \begin{tabular}{ | c | c | c | c | c | c | c | c |}
    \hline
 Platonic solid    &    surface &    \begin{tabular}[c]{@{}c@{}}elementary\\region\end{tabular} &\parbox{1.2cm}{\hspace{0.5cm}$\#$ \\vertices} & 
   \parbox{0.9cm}{\hspace{0.3cm}$\#$ \\edges}  &   \parbox{0.9cm}{\hspace{0.3cm}$\#$ \\faces}  
& $L$   &  $\delta_\textrm{max}$ \\ \hline
tetrahedron & \raisebox{-0.4 \height}{\includegraphics[height=\ph\textwidth]{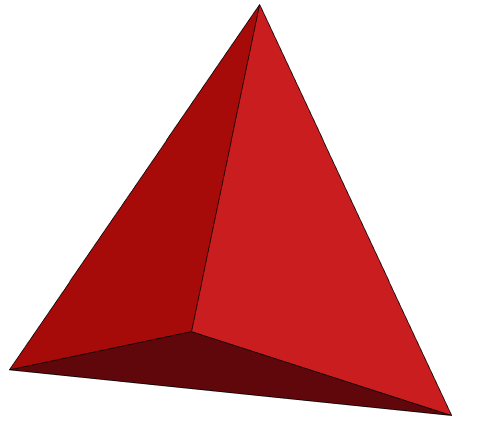}}&
\raisebox{-0.4 \height}{ \includegraphics[height=\ph\textwidth]{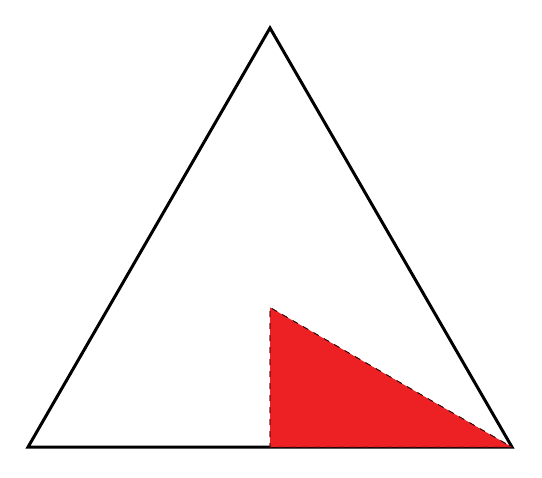}} &4&6&4& 2.694&0.7036 \\ \hline
octahedron & \raisebox{-0.4 \height}{ \includegraphics[height=\ph\textwidth]{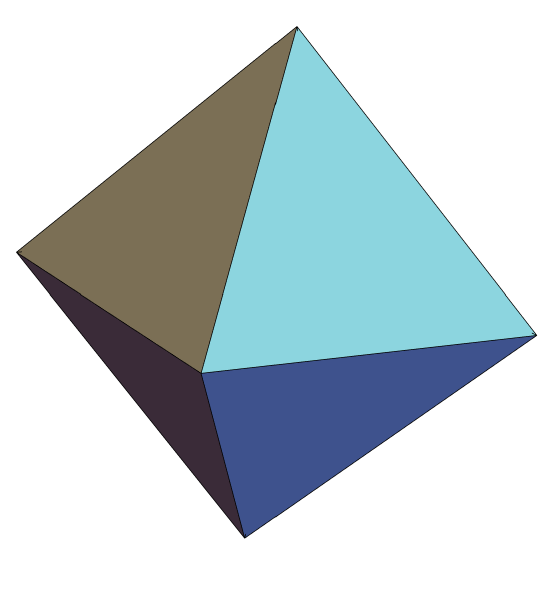} }& 
\raisebox{-0.4 \height}{ \includegraphics[height=\ph\textwidth]{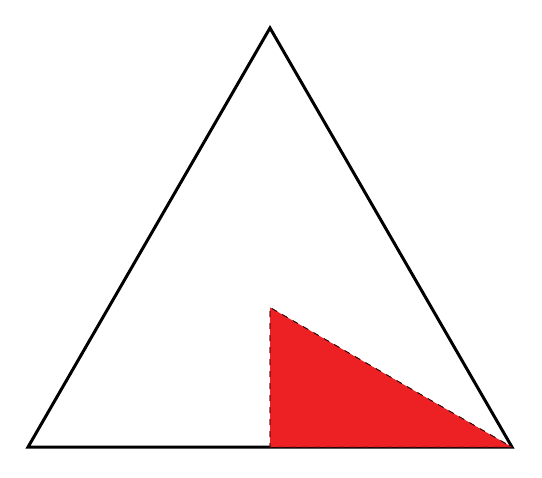}} &6&12&8& 1.905 &0.6349 \\ \hline
 cube &\raisebox{-0.4 \height}{ \includegraphics[height=\ph\textwidth]{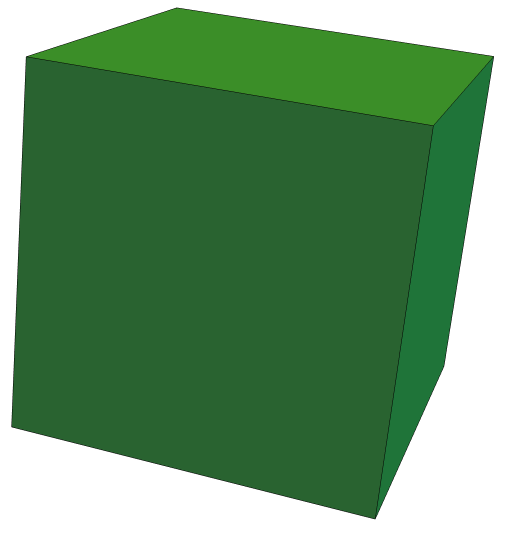} }& 
 \raisebox{-0.4 \height}{ \includegraphics[height=\ph\textwidth]{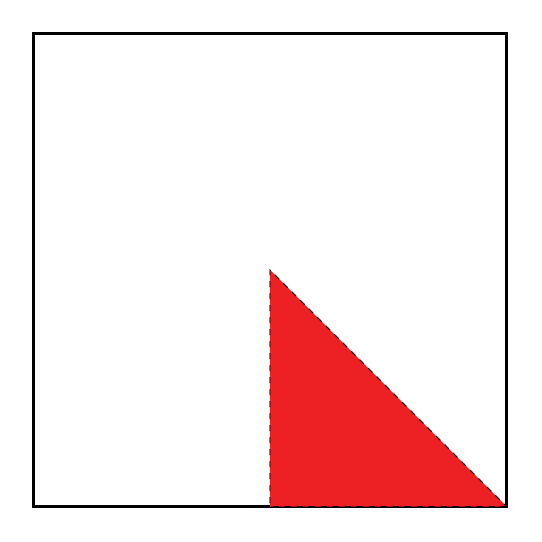} }   &8&12&6& 1.447&0.4824 \\ \hline
 icosahedron &   \raisebox{-0.4 \height}{ \includegraphics[height=\ph\textwidth]{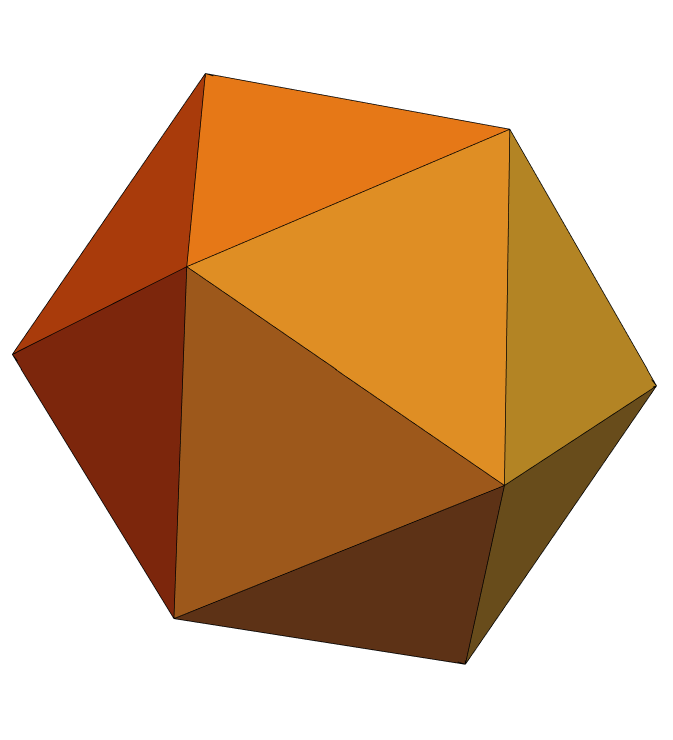} }  & 
 \raisebox{-0.4 \height}{ \includegraphics[height=\ph\textwidth]{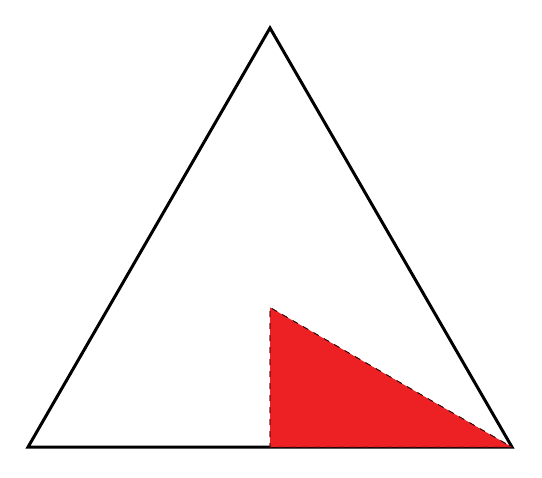} }   &12&30&20& 1.205&0.4015 \\ \hline
 dodecahedron  &  \raisebox{-0.4 \height}{  \includegraphics[height=\ph\textwidth]{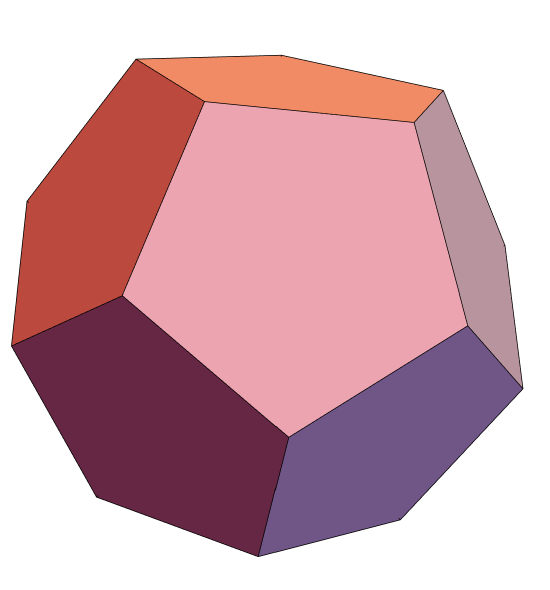} }& 
 \raisebox{-0.4 \height}{   \includegraphics[height=\ph\textwidth]{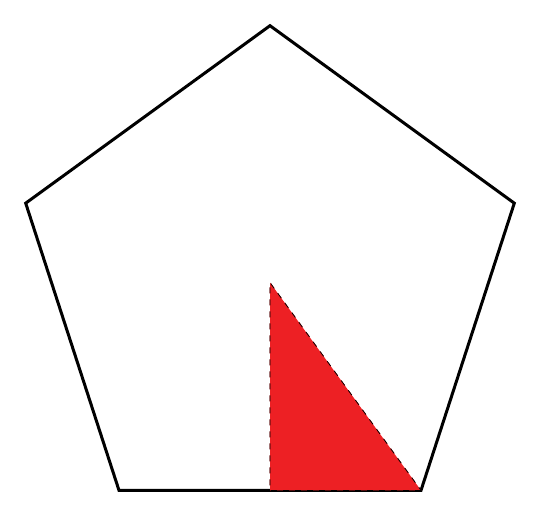} }&20  &30  &12  & 0.780 &0.2601 \\\hline
  \end{tabular}
\caption{Polyhedral surfaces of the Platonic solids, with some of their properties. }
  \label{defects-tab:platonic}
  \normalsize
  \end{table}
\end{center}
To compute the spatial average of the average sphere distance on a give polyhedral surface, we use a uniform sampling 
with respect to the flat metric (disregarding the singular points), somewhat similar to what has been done in a quantum context 
\cite{klitgaard2018implementing,klitgaard2020how}. Because of the discrete symmetries of the polyhedra, it suffices to pick the first circle center $p$ of a 
double-circle configuration from what we will call the elementary region.
This is a triangular subregion of a face, whose corners are given by the midpoint of an edge, one of the endpoints of
the same edge (a singular vertex), and the midpoint of the face, as indicated in Table \ref{defects-tab:platonic}. 
The data for a given value of $\delta\in\, ]0,\delta_\textrm{max}]$ are collected as follows:

\begin{enumerate}
\item Sprinkle $n$ points $p_{i}$, $i\! =\! 1,\dots,n$, randomly and uniformly into the elementary region.
\item Construct geodesic circles $S^\delta_{p_{i}}$, using the parametrizations found in Sec.\ \ref{defects-subsec:sd}.
\item For every $p_{i}$, pick a point $p'_{i}$ on $S^\delta_{p_{i}}$ uniformly at random, and construct the geodesic
circle $S^\delta_{p'_{i}}$ around it. 
\item Compute $\bar{d}(S_{p_{i}}^\delta,S_{p'_{i}}^\delta )$ for all pairs $(p_{i}, p'_{i})$. 
\item Average over the results to obtain $\bar{d}_\textrm{av}(\delta)$.
\end{enumerate}
\vspace{0.1cm}

\begin{figure}
\centering
\begin{subfigure}{0.48\textwidth}
\includegraphics[width=\linewidth]{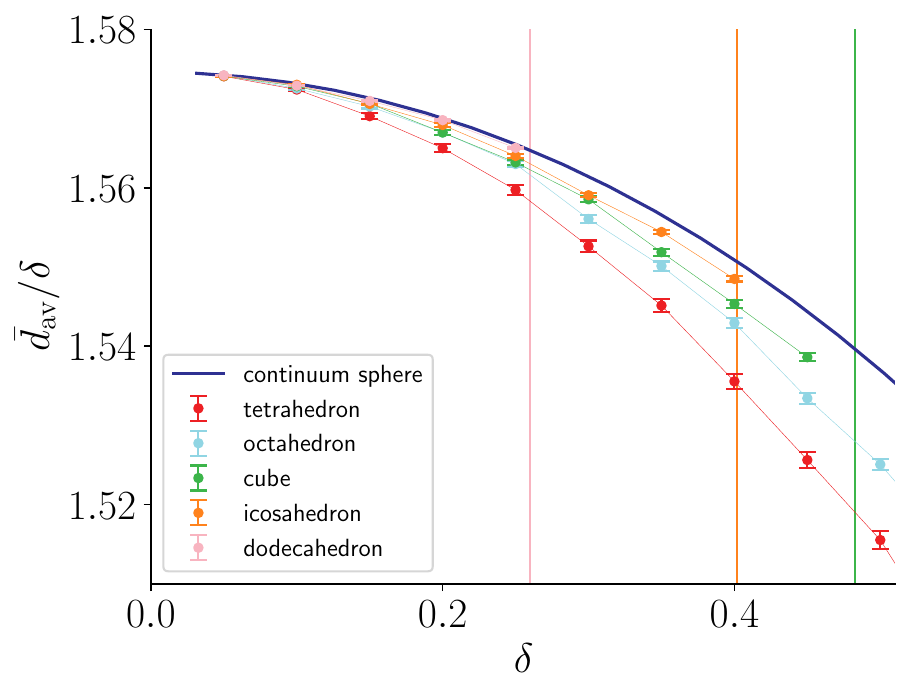}
\label{defects-fig:psr-a}
\end{subfigure}
\hspace{0.02\textwidth}
\begin{subfigure}{0.48\textwidth}
\includegraphics[width=\linewidth]{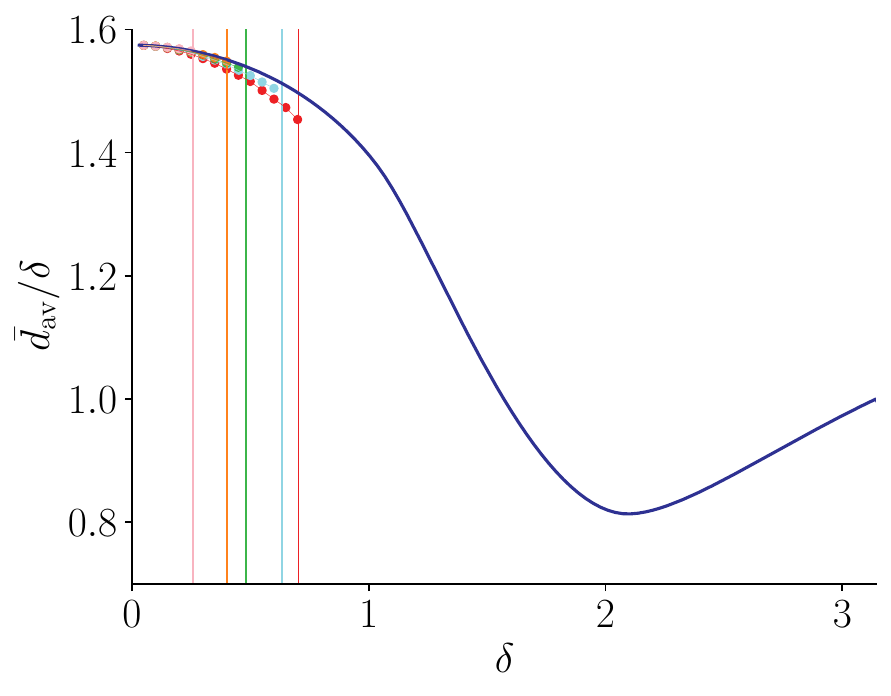}
\label{defects-fig:psr-b}
\end{subfigure}
\caption{Measurements of the curvature profiles $\bar{d}_\textrm{av}(\delta)/\delta$ 
of the surfaces of the five Platonic solids, compared to that of the continuum sphere of the same
area (top curve), plotted in the $\delta$-ranges $[0,0.5]$ (left) and $[0,\pi ]$ (right). The connecting lines serve as a guide to the eye.}
\label{defects-fig:platonic}
\end{figure}

\noindent We have collected 10.000 measurements for each data point, with a step size of $\delta\! =\! 0.05$. The results for the curvature profiles $\bar{d}_\textrm{av}/\delta$ 
for the five regular polyhedra are shown in Fig.\ \ref{defects-fig:platonic}, together with the curvature 
profile of the continuum two-sphere for comparison. On the left, we show the behavior at small distances $\delta\! \lesssim \! 0.5$.
It is clearly distinct for the five spaces, but qualitatively similar. 
(The vertical lines indicate the cutoff values $\delta_\textrm{max}$ for the various cases.) 
The data for the dodecahedron resemble those of the continuum sphere most closely, at least within the limited $\delta$-range
considered. Generally speaking, the larger the number of vertices over which the deficit angles are distributed, 
the closer is the match with the sphere curve. Note also that all curves seem to converge to the flat-space value as $\delta\rightarrow 0$,
as one would expect. 

In Fig.\ \ref{defects-fig:platonic}, right, we have zoomed out for a more global comparison with the curvature
profile of the sphere in the range $\delta\in [0,\pi]$. As mentioned above, we will in the following section introduce a different method
to determine geodesic circles on the tetrahedron, which will allow us extend its curvature profile over most of the range depicted here.
Based on the limited data presented here, we conclude that the averaged quantum Ricci curvature of all the investigated spaces is positive,
which is not surprising. 
It is largest for the tetrahedron, for which the decrease of the curve for $\bar{d}_\textrm{av}(\delta)/\delta$ is steepest. 

\section{Unfolding the tetrahedron}
\label{defects-sec:tetra}

The study of geodesics on regular convex polyhedra, and the classification of closed geodesics in particular, 
is an active area of mathematical research (for example, see \cite{fuchs2007closed,fuchs2014periodic}). 
Our primary interest are shortest geodesics between arbitrary pairs of points, a problem that is most straightforwardly tackled for the
tetrahedron. This is fortunate, since the tetrahedron is the most interesting case from our point of view. Of all the Platonic solids, its curvature is 
distributed least homogeneously, while the results on the curvature profiles of the previous section suggest that all other cases lie 
in between those of the tetrahedron and the smooth sphere.  
\begin{figure}[t]
\centering
\begin{subfigure}[t]{0.45\textwidth}
\centering
\includegraphics[height=0.7\linewidth]{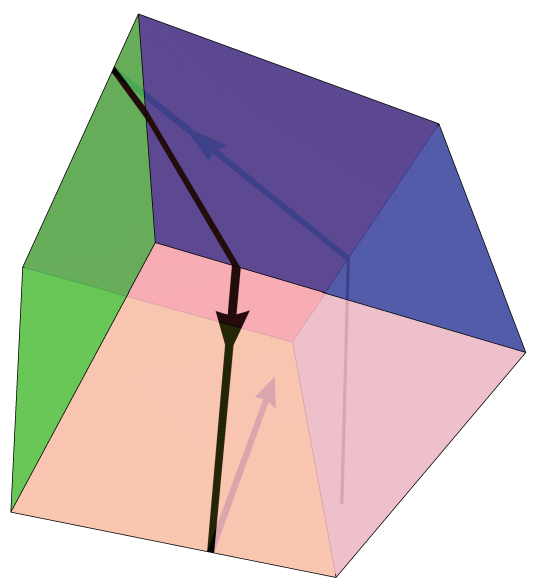}
\end{subfigure}
\hspace{0.02\textwidth}
\begin{subfigure}[t]{0.45\textwidth}
\centering
\includegraphics[height=0.7\linewidth]{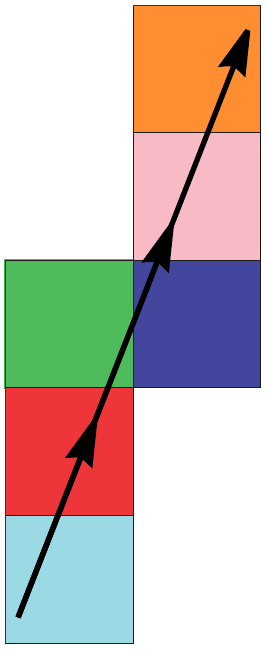}
\end{subfigure}
\caption{Geodesic on a cube (left) and its corresponding development in the plane (right).}
\label{defects-fig:develop-cube}
\end{figure}

A general technique one can use to study geodesics on regular polyhedra is by ``rolling''  or ``unfolding'' the polyhedron onto the flat two-dimensional
plane along a given geodesic. The so-called development of a polyhedron along a geodesic pro\-ceeds as follows \cite{fuchs2007closed}: 
starting from an arbitrary point $p$ contained in some face $F_0$ of the polyhedral surface and an initial direction, one follows the 
corresponding geodesic (straight line) in $F_0$ until it hits an edge to a neighboring face $F_1$. Since all faces are flat, the adjacent pair of 
$F_0$ and $F_1$ can in an isometric way be put down in the plane. Following the geodesic as it passes through $F_1$, it will meet another 
edge to some neighboring face $F_2$, which likewise can be folded out into the plane, and so forth.
The result of this development is a contiguous chain of faces in the plane, which is traversed by the geodesic, taking the form of a straight line.
Fig.\ \ref{defects-fig:develop-cube} illustrates the procedure for a geodesic running along the surface of a cube. The six sides of the cube have a 
different color coding and will in general appear repeatedly along a given geodesic.

\begin{figure}
\centering
\begin{subfigure}[t]{0.45\textwidth}
\includegraphics[height=0.8\linewidth]{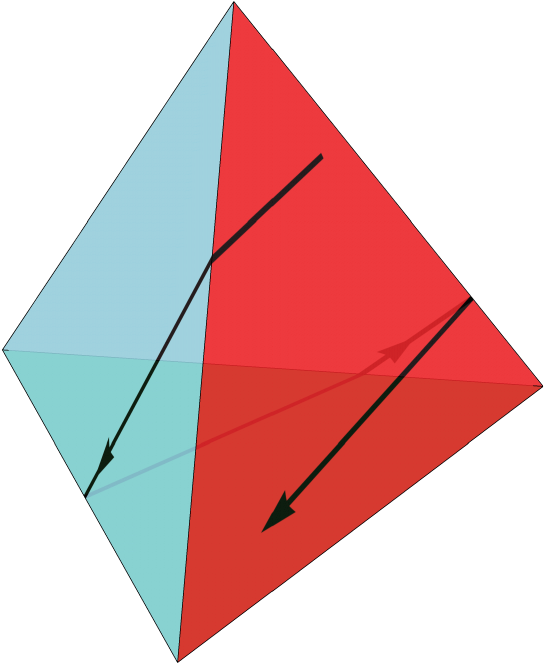}
\end{subfigure}
\hspace{0.02\textwidth}
\begin{subfigure}[t]{0.45\textwidth}
\includegraphics[height=0.8\linewidth]{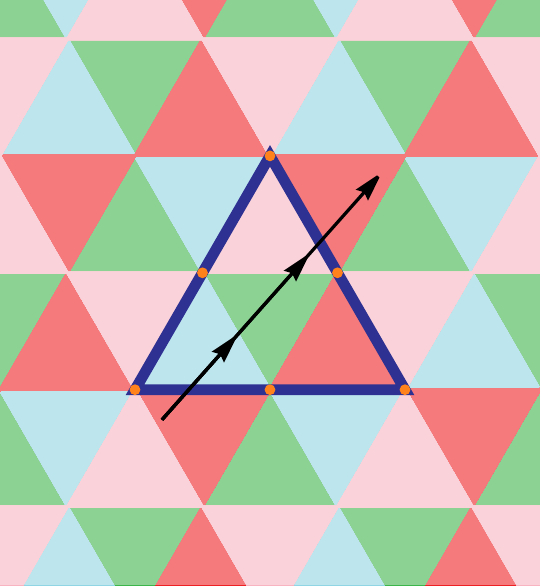}
\end{subfigure}
\caption{Geodesic on a tetrahedron (left) and its development in the plane, which forms part of a regular tiling in terms of the
four faces of the tetrahedron (right). The thick line encloses a fundamental domain $\cal F$.}
\label{defects-fig:develop-tetra}
\end{figure}

The tetrahedron is special among the regular polyhedra in the sense that it has an everywhere consistent development, 
which is independent of the geodesic chosen. The full development in all directions constitutes a regular tessellation of the flat plane by 
equilateral triangles, which come in four types or colors, corresponding to the four faces of the tetrahedron. Fig.\ \ref{defects-fig:develop-tetra}
shows an example of a geodesic on the tetrahedron and the corresponding straight line in the regular tiling. 
We have also indicated a triangle-shaped \emph{fundamental domain}  $\cal F$, consisting of four triangles that make up a single copy of the tetrahedron. 
The plane can also be thought of as a tessellation by copies of $\cal F$ with alternating orientation, either
with the orientation shown in the figure or an upside-down version rotated by 180 degrees. This planar set-up will allow us to 
construct geodesic circles and compute average sphere distances for radii $\delta$ of up to one tetrahedral edge length, which
is almost four times the range we could cover with the previous method.  

\subsection{Geodesic circles on the tetrahedron}
\label{defects-sec:geo}

\begin{figure}
\centering
\begin{subfigure}[t]{0.48\textwidth}
\centering
\includegraphics[height=5.5cm]{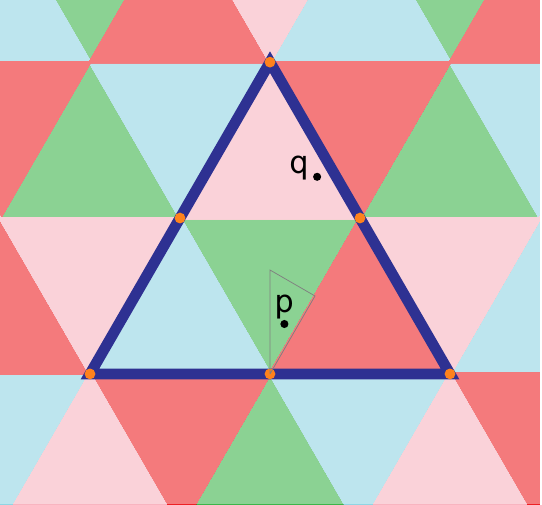}
\end{subfigure}
\hspace{0.02\textwidth}
\begin{subfigure}[t]{0.48\textwidth}
\centering
\includegraphics[height=5.5cm]{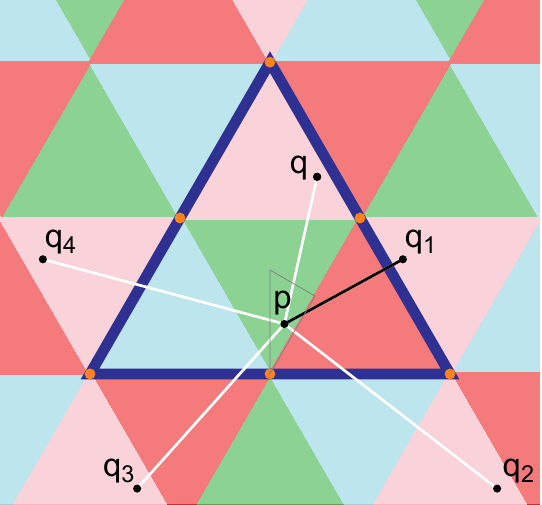}
\end{subfigure}
\caption{Fundamental domain $\cal F$ with a point $p$ lying in the elementary region of the center triangle of $\cal F$, and a second point $q$ (left). 
To determine the distance between $p$ and $q$, we must consider copies $q_i$ of the point $q$ in neighboring copies of $\cal F$ (right).
The shortest line is drawn in black. }
\label{defects-fig:tetra}
\end{figure}

The new representation in terms of unfolded tetrahedra allows us relatively easily to determine geodesic circles that enclose more than
one singularity. A more elementary step we need to understand first is how to determine the distance between two arbitrary points
$p$ and $q$ on the surface of the tetrahedron, which is given by the length of the shortest geodesic between them. 
Let us represent the surface by the fundamental domain $\cal F$ of Fig.\ \ref{defects-fig:develop-tetra}. 
Because of the symmetries of the tetrahedron, 
we can without loss of generality choose the point $p$ to lie in (or on the boundary of) the elementary triangular region in the central
triangle of $\cal F$. The second point $q$ can be located anywhere in $\cal F$ (Fig.\ \ref{defects-fig:tetra}, left). 
The key observation now is that the unique straight line 
one can draw from $p$ to $q$ in $\cal F$ clearly corresponds to a geodesic on the tetrahedron, but not necessarily the shortest one.
This can happen because the tetrahedral surface is obtained by appropriate pairwise identifications of the six boundary edges
of the domain $\cal F$, and the shortest geodesic to $q$ may cross one or more of these boundaries. 
One can easily find this geodesic by going back to the tessellation of the plane and considering a sufficiently large 
neighborhood of the fundamental domain $\cal F$, consisting of $\cal F$ itself and neighboring
copies ${\cal F}_i$, $i\! =\! 1,2,\dots$, of $\cal F$. On each of these copies, we mark the unique copy 
$q_i$ of the point $q$ (Fig.\ \ref{defects-fig:tetra}, right). 
The length of the shortest straight line between $p$ and either $q$ or any of the $q_i$ is the searched-for geodesic distance
between $p$ and $q$. 

This way of representing the tetrahedron and its geodesics is not unlike our previous representation of the
geometry of an isolated conical singularity in terms of a plane with a wedge removed: the nature of the geodesics is
maximally simple (straight lines), but one pays a price in the form of nontrivial identifications.

Building on the above insight on how geodesic distances are determined, we next discuss the nature of geodesic circles of radius $\delta$
based at a point $p$.
It is best illustrated by referring to the convex set ${\cal C}_p$ 
of all points $q$ \emph{in the tessellated plane}, for given $p$ in the elementary region, which are 
closer to $p$ than any of their copies (in $\cal F$ or any of the ${\cal F}_i$). To construct ${\cal C}_p$, one needs to recall the status of vertices
in the tessellated plane. On the original tetrahedron, they corresponded to singularities with a deficit angle $\alpha\! =\! \pi$. 
Cutting open the tetrahedron along three of its edges and putting it in the plane results in a copy of the fundamental domain $\cal F$, which
can be thought of as a region in the plane with three such wedges removed. Since all wedges have an angle $\pi$, one
fundamental domain can be glued into another's ``missing wedge''. Repeating this gluing throughout the plane leads to a tiling of the plane 
which preserves all neighborhood relations between pairs of tetrahedral faces sharing a common edge. 
What is not preserved is the number of faces meeting at a given vertex, which is three on the original tetrahedron and six
in the plane.

\begin{figure}
\centering

\includegraphics[height=6cm]{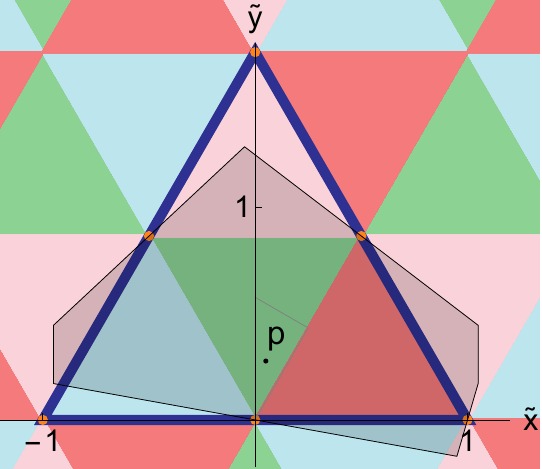}

\caption{The convex polygon ${\cal C}_p$ superimposed on the tessellated plane. The axes show the orthonormal coordinate system $(\tilde{x}, \tilde{y})$ used to construct the region.}
\label{defects-fig:tetra-cp}
\end{figure}

The region ${\cal C}_p$ in the tessellated plane can be constructed from symmetry considerations. 
For a general point $p$ inside the elementary
region, its boundary is a convex polygon
with six corner points $S_i$, $i\! =\! 1,\dots,6$. Its sides are formed by four straight line segments through the four vertices closest
to $p$, where each line segment is perpendicular to the line from $p$ to the vertex in question. 
In addition, there are two parallel lines (vertical lines in Fig.\ \ref{defects-fig:tetra-cp}) equidistant from $p$, which
get mapped onto each other by one of the translation symmetries of the tessellated plane. 
Note that by construction the area of ${\cal C}_p$ is the same as the surface area of the tetrahedron.

To arrive at a quantitative description, we introduce an orthonormal coordinate system $(\tilde{x},\tilde{y})$ in the plane,
whose origin $(0,0)$ is the midpoint between the two bottom edges of the fundamental domain $\cal F$ (Fig.\ \ref{defects-fig:tetra-cp}).
For definiteness, let us assume that all edges have unit length, $L\! =\! 1$.
Note that the elementary triangular region, which by assumption contains the point $p\! =\! (x,y)$, 
is spanned by the three corner points $(0,0)$, $(0,1/\sqrt{3})$ and $(1/4,\sqrt{3}/4)$, which means that the ranges of
$x$ and $y$ are limited to $x\!\in\! [0,1/4]$ and $y\!\in\! [0,1/\sqrt{3}]$.
In this coordinate system, 
the coordinates of the corner points of the region ${\cal C}_p$ are given by
\begin{equation}
S_1\! =\! \Big( x-1, \frac{x (1\! -\! x)}{y} \Big),\; 
S_2\! =\!\Big( x-1, \tfrac{1}{2} \big(\sqrt{3}-\frac{1-4 x^2}{\sqrt{3}\! -\! 2y} \big)\! \Big),\;
S_3\! =\!\Big( -x, \tfrac{1}{2} \big(\sqrt{3}+\frac{1-4 x^2}{\sqrt{3}\! -\! 2y} \big)\! \Big),\nonumber 
\end{equation}
\begin{equation}
S_4\! =\!\Big( x+1, \tfrac{1}{2} \big(\sqrt{3}-\frac{1-4 x^2}{\sqrt{3}\! -\! 2y} \big)\! \Big),\;
S_5\! =\! \Big( x+1, \frac{x (1\! -\! x)}{y} \Big),\; 
S_6\! =\! \Big( 1-x, -\frac{x (1\! -\! x)}{y} \Big),\; 
\label{defects-spoints}
\end{equation}
for $y\not= 0$, and by
\begin{equation}
S_1\! =\! (-1,0),\; S_2\! =\! \Big(\! -1,\frac{1}{\sqrt{3}}\Big),\; S_3\! =\! \Big( 0,\frac{2}{\sqrt{3}} \Big),\; 
S_4\! =\!\Big( 1,\frac{1}{\sqrt{3}}\Big),\; S_5\! =\! S_6\! =\! (1,0),
\label{defects-snot}
\end{equation}
for $y\! =\! 0$. In the latter case, corresponding to $p\! =\! (0,0)$, ${\cal C}_p$ becomes a five-cornered polygon that is equal to one half
of a regular hexagon.

We have now set the stage for analyzing geodesic circles $S_p^\delta$ centered at $p$. In what follows, we will exclude
the case $p\! =\! (0,0)$, where the circle center would coincide with a singularity of the tetrahedron. For added simplicity,
we will also assume that $p$ lies in the interior of the elementary region, which is the generic case. An analogous
treatment of points along the boundary is completely straightforward. 

Since in an open 
neighborhood of $p$ space is flat and Euclidean, for sufficiently small $\delta$ the set of points $S_p^\delta$ is
an ordinary circle of radius $\delta$ and circumference $2\pi\delta$. 
As we increase $\delta$ continuously, this continues to be the case until the circle
reaches the singularity closest to $p$, located at $(\tilde{x},\tilde{y})\! =\! (0,0)$, where it also meets the line segment that
forms part of the boundary of ${\cal C}_p$. The corresponding circle radius is $\delta\! =\! \sqrt{x^2+y^2}$. 
As we increase $\delta$ further, we need to take into account the nontrivial deficit angle associated with the point
$(0,0)$. It implies that the piece of line segment to one side of the vertex $(0,0)$ should be identified with the piece to the other side.
For the circle of radius $\delta$ we draw in the tessellated plane, it means that the circle segment beyond this line segment
is \emph{not} part of $S_p^\delta$. The ``circle'' $S_p^\delta$ with radius $\delta\! > \! \sqrt{x^2+y^2}$ therefore has a length
that is shorter than $2\pi\delta$. The story repeats itself when we increase $\delta$ further and $S_p^\delta$ meets the 
singularity at $(\tilde{x},\tilde{y})\! =\! (1/2,\sqrt{3}/2)$, which is second-closest to $p$. From this radius onwards, a second
circle segment will be removed from the ``na\"ive'' circle we can draw around $p$ in the plane. The same happens when $S_p^\delta$
reaches and passes the remaining two singularities, at $(-1/2,\sqrt{3}/2)$ and $(-1,0)$ respectively. 

Since the tetrahedron has a finite diameter, the size of $S_p^\delta$ must be zero above some maximal value of $\delta$,
given by the distance of the point(s) furthest away from the given point $p$, the antipode(s) of $p$.\footnote{The antipode is a
single point on the tetrahedron, but appears in multiple images on the boundary of ${\cal C}_p$.} In this context, it should be noted that
the three points $S_1$, $S_5$ and $S_6$ of Eq.\ \eqref{defects-spoints} are equidistant to $p$, with distance $d_1$, as are the three points 
$S_2$, $S_3$ and $S_4$, with distance $d_2$. Which of the two distances is larger and therefore sets the upper limit for the radius $\delta$ 
depends on the location of $p$ in the elementary region. It turns out that the extremal cases of the closest and the furthest
antipodes are associated with points $p$ on the boundary of the elementary region. The antipode is closest when $p$ lies
in the middle of an edge, $p\! =\! (1/4,\sqrt{3}/4)$, in which case the antipode has distance $d_1\! =\! d_2\! =\! 1$ and also
lies in the middle of an edge. By contrast, the point $p$ with the most distant antipode, with $d_1\! =\! d_2\! =\! 2/\sqrt{3} \approx \!
1.155$, lies at the center of a face, with $p\! =\! (0,1/\sqrt{3})$. 

\begin{figure}[t]
\includegraphics[height=0.41\linewidth]{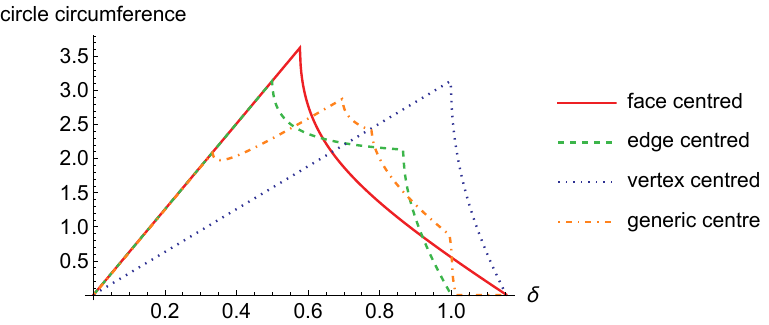}
\caption{The circumference of geodesic circles $S_p^\delta$ on the tetrahedron as a function of the radius $\delta$, both 
in units of edge length, for various locations of the center $p$.}
\label{defects-fig:tetra-circ}
\end{figure}

These results are illustrated in Fig.\ \ref{defects-fig:tetra-circ}, which shows the
circumference of $S_p^\delta$ as a function of the geodesic radius $\delta$ for various choices of $p$. 
The computation is done by identifying which arcs of the ``na\"ive'' circle at radius $\delta$, parametrized by an
ordinary rotation angle $\theta$, contribute to $S_p^\delta$
and by adding up the arc lengths. For all curves, we see
an initial linear rise, characteristic of a circle in the flat plane, with the exception of $p$ at a vertex, for which the linear
slope is flatter, corresponding to that of a cone with deficit angle $\pi$. For a generic location of the center $p$, 
there subsequently are four cusps, corresponding to the radii where the circle $S_p^\delta$ meets one of the singularities,
as discussed above. For non-generic $p$ some of the cusps can merge. When $p$ lies at the center of an edge, it is clear that 
the two closest singularities are reached simultaneously at $\delta\! =\! 0.5$, and the two remaining ones at $\delta\! =\! \sqrt{3}/2\! 
\approx \! 0.866$. In this case, the
antipode has the minimal distance 1 from $p$, which is why the curve in Fig.\ \ref{defects-fig:tetra-circ} stops at $\delta\! =\! 1$. 
When $p$ lies at the center of a face, the three closest singularities are reached
simultaneously at $\delta\! =\! 1/\sqrt{3}\!\approx \! 0.577$, giving rise to a single cusp. The antipode, which in this case is maximally far away from $p$, 
coincides with the remaining vertex, and the corresponding curve ends at $\delta\! \approx\! 1.155$. The same is true for the
dual case, where the construction starts at a vertex. For comparison, the analogous curve for a smooth two-sphere of the same area as
the tetrahedron would end at $\delta\! \approx\! 1.166$.

The above analysis and Fig.\ \ref{defects-fig:tetra-circ} underline the strong inhomogeneity and aniso\-tro\-py of the tetrahedral surface 
and at the same time fix an upper bound, $\delta\! =\! 1$, up to which the notion of a geodesic circle $S_p^\delta$ is
meaningful for \emph{arbitrary} locations of $p$. This is relevant when we start taking spatial averages of average distances between
circles of radius $\delta$ in order to obtain the curvature profile of the tetrahedron, which is the subject of the next section. 

\begin{figure}
\centering
\begin{subfigure}{0.45\textwidth}
\includegraphics[width=0.95\linewidth]{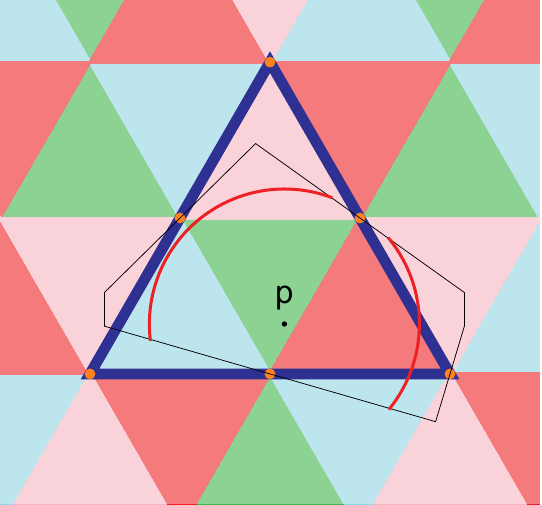}
\end{subfigure}
\hspace{0.02\textwidth}
\begin{subfigure}{0.45\textwidth}
\includegraphics[width=0.95\linewidth]{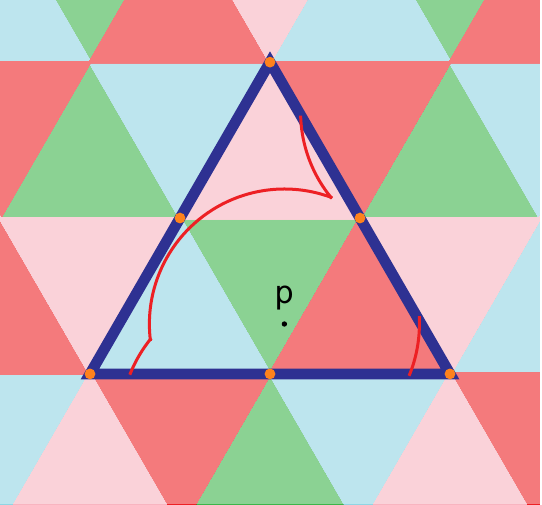}
\end{subfigure}
\caption{Constructing a geodesic circle on the tetrahedron. }
\label{defects-fig:wedge}
\end{figure}

\subsection{Computing the curvature profile}

Putting together the construction of geodesic circles $S_p^\delta$ and our earlier observation about the need to choose
the shortest geodesic between two points, we can now embark on computing average sphere distances for $\delta\!\in\! 
[0,1]$ in units of edge length. To obtain a pair of overlapping 
circles $S_p^\delta$, $S_{p'}^\delta$ for a given value of $\delta$, we start by picking a point $p$ 
randomly from the elementary region and constructing $S_p^\delta$, following the prescription of Sec.\ \ref{defects-sec:geo}. 
In general, this will be given by a set of arcs in ${\cal C}_p$, parametrized by corresponding angle intervals in terms of the 
rotation angle $\theta$.
To simplify subsequent computations, we map all arc sections that 
happen to lie in a neighboring domain ${\cal F}_i$ back to the fundamental domain $\cal F$ by appropriate
reflections and translations, which are symmetries of the tessellated plane (Fig.\ \ref{defects-fig:wedge}). 

Next, we randomly pick a point $p'$ on $S_p^\delta$, which will serve as the center of the second circle $S_{p'}^\delta$. We then again construct the arcs forming $S_{p'}^\delta$ using a region similar to ${\cal C}_p$, where the center point $p'$ is now generically not in the triangular elementary region. The analogous region ${\cal C}_{p'}$ can be found by an appropriate symmetry transformation. In principle it would be possible to then map all arc sections of $S_{p'}^\delta$ back into $\cal F$, but for the purpose of finding the minimum distance between $S_{p}^\delta$ and $S_{p'}^\delta$ it is simpler to perform two subsequent reflections of the arc sections through all six vertices on the boundary of $\cal F$. This puts 36 representatives of arcs of $S_{p'}^\delta$ in a neighborhood around $\cal F$, where some of these representatives can be discarded since several distinct pairs of subsequent reflections produce the same arc. The minimum distance from $S_p^\delta$ to any point $q' \in S_{p'}^\delta$ is guaranteed to be found among the remaining unique representatives.

Finding the average sphere distance $\bar{d}(S_p^{\delta},S_{p'}^{\delta})$ of Eq.\ \eqref{cp-sdist} can now be done by 
integrating the Euclidean distance between all pairs of points $(q,q')\!\in\! S_p^{\delta}\times S_{p'}^{\delta}$.

\begin{figure}
\centering
\includegraphics[width=0.4\textwidth]{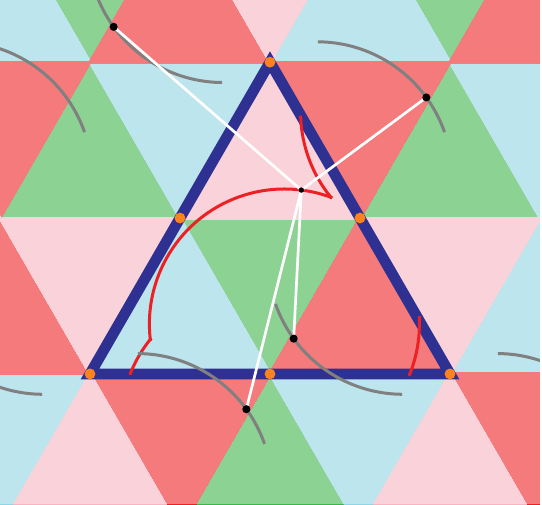}
\includegraphics[width=0.55\textwidth]{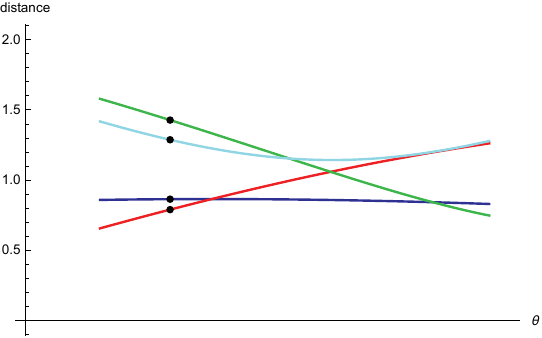}
\caption{Computing the distance in the plane of a point $q \in S_{p}^\delta$ to an arc $q'(\theta)$ 
of the circle $S_{p'}^\delta$ in $\cal F$, as well as to several
copies $q_i'(\theta)$ of the arc in a neighborhood of $\cal F$. Each colored curve segment of distance measurements
in the graph on the right corresponds to different reflection of the arc through one of the vertices. }
\label{defects-fig:tetra-point}
\end{figure}

We show first how to measure the distance between a fixed point $q\! \in\! S_p^{\delta}$ and the second circle 
$S_{p'}^{\delta}$. The latter consists of a set of smooth arcs $q'(\theta')$, not necessarily contained in $\cal F$.

For sufficiently small $\delta$, and when $S_{p'}^{\delta}$ does not
enclose any singularities, there may be just a single ``arc'', given by an entire smooth circle and parametrized by $\theta'\! \in\! [0,2\pi]$.
In general there will be several arcs, parametrized by smaller angle intervals. Whatever the case may be,
we determine the distance 
between $q$ and each such arc separately as follows. Consider a specific arc parametrized by $q'(\theta')\! \in\! [\theta'_{i},\theta'_{f}]$.
It is straightforward to
determine its distance from $q$ as a function of $\theta'$, using the Euclidean distance in the plane. However, recall from
our considerations at the beginning of Sec.\ \ref{defects-sec:geo} 
that the corresponding geodesic (straight line) between $q$ and any specific point $q'(\theta')$ along the arc 
need not be the shortest one between the corresponding points on the tetrahedron. 
To find the shortest geodesic in a systematic way, we repeat the distance measurements with
each copy (representative) $q'_i(\theta')$ of the arc in the neighborhood around $\cal F$.

Fig.\ \ref{defects-fig:tetra-point} shows an example where we have collected the distance measurements from four copies of an arc and
plotted the corresponding curve segments as a function of $\theta'\! \in\! [\theta'_{i},\theta'_{f}]$. 
The next step is to determine the continuous curve segment that to each $\theta'$ in this interval
assigns the minimum distance from $q$ to one of the four points $\{ q_1'(\theta'),\dots, q_4'(\theta') \}$.
This may be a single curve segment, which throughout the $\theta'$-interval lies below all 
other curve segments. Alternatively, it may consist of contributions from several mutually intersecting segments. 

The same construction must be applied to the remaining smooth arcs along $S_{p'}^{\delta}$. Putting the individual 
minimizing curve segments together results in a single continuous distance-minimizing curve. Integrating it over 
 $S_{p'}^{\delta}$, and dividing it by the length of $S_{p'}^{\delta}$ gives the average distance of $q$ to $S_{p'}^{\delta}$.  

To complete the computation of the average sphere distance \eqref{cp-sdist}, we still need to vary the point $q$ 
over the arcs of the first circle, $S_{p}^{\delta}$. The analysis mirrors the one we just performed for the second circle.
It produces a two-dimensional version of the distance-minimizing curve, namely, a distance-minimizing ``sheet''
parametrized by two angles $(\theta,\theta')$,
obtained by computing and comparing distances $d(q(\theta),q'(\theta'))$ between all pairs of points from each pair of arcs from
$S_{p}^{\delta}$ and $S_{p'}^{\delta}$ respectively, and their representatives in neighboring domains. These contributions must be combined, integrated and normalized to yield a single data point $\bar{d}(S_p^{\delta},S_{p'}^{\delta})$ 
at radius $\delta$. 

\begin{figure}[t]
\centering
\includegraphics[width=0.8\linewidth]{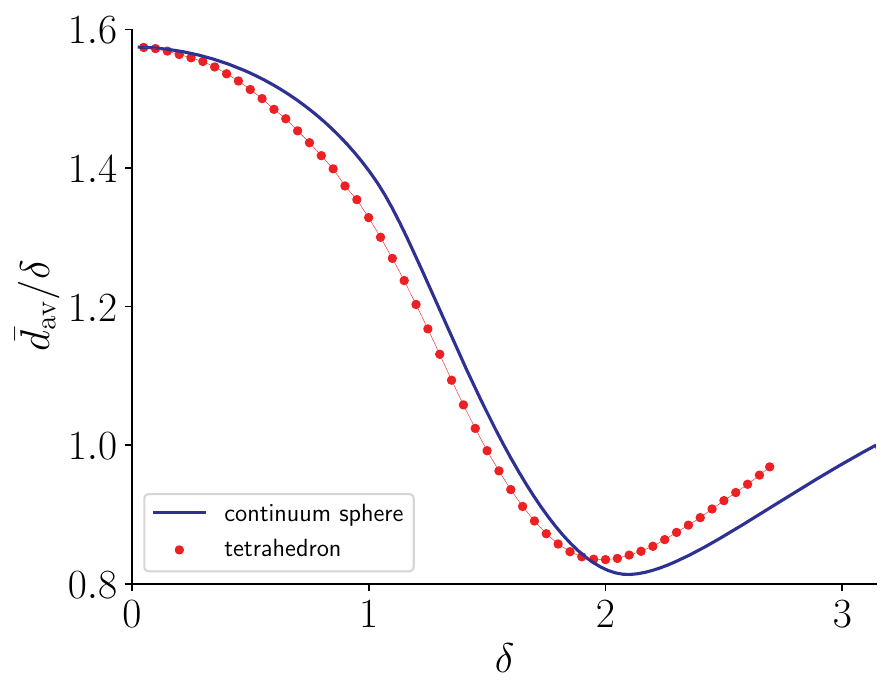}
\caption{Measurement of the curvature profile $\bar{d}_\textrm{av}(\delta)/\delta$ 
of the surface of a tetrahedron, compared to that of a sphere of the same
area. (Error bars too small to be shown.) }
\label{defects-fig:tetra-all}
\end{figure}

To collect the data for the curvature profile, we follow the same procedure as in Sec.\ \ref{defects-sec:measure} to
generate double-sphere configurations that are distributed uniformly at random over the tetrahedron. 
We took random samples of 10.000 average sphere distance measurements at each of 54 evenly spaced
values of $\delta$. For easier comparison with the results obtained for the Platonic solids with our earlier 
method (Fig.\ \ref{defects-fig:platonic}), we have reverted to ``volume-normalized'' units where the edge length of the tetrahedron
is given by $L\! =\! 2.694$ (cf.\ Table 1). In these units, the step size for measurements is $\delta\! =\! 0.05$. 
The final result for the curvature profile $\bar{d}_\textrm{av}(\delta)/\delta$ is shown in Fig.\ \ref{defects-fig:tetra-all}, with the curvature profile of
the smooth two-sphere for comparison. 
For small $\delta$, the data are compatible with those obtained with our earlier method. 

Overall, the curvature profile of the tetrahedron is qualitatively similar to that of the sphere, but clearly distinct from it,
in a way that cannot be absorbed by a simple linear rescaling of the $\delta$-axis.
The similarity demonstrates in explicit terms the robustness of this global observable, by which we mean its property of ``averaging out'' a
rather extreme curvature distribution, like that of the tetrahedron. It strengthens earlier observations of such a behavior on
large Delaunay triangulations of a sphere \cite{klitgaard2018introducing}. These are also examples of piecewise flat spaces, but with a curvature distribution
that is much closer to that of a continuum sphere, with small deficit angles everywhere. 

As the measurements of Sec.\ \ref{defects-sec:measure} already indicated, the curve for $\bar{d}_\textrm{av}/\delta$ falls more steeply for
small $\delta$,
which means that the (averaged) quasi-local quantum Ricci curvature of the tetrahedron is larger than that of the sphere.  
Features of the sphere curve for larger values $\delta\! \gtrsim \! 0.5$ account for large-scale geometric and topological properties
of the sphere and are reflected in a similar behavior of the tetrahedron, although the minimum of the curve is reached for 
smaller $\delta$ and its value is larger. 
When interpreting the large-scale behavior, one should keep in mind that on the two-sphere the circles $S_p^\delta$ have maximal size for
$\delta\! =\! \pi/2$, after which they start shrinking until for $\delta\! =\! \pi$ they degenerate into points. 
The latter is reflected in the value $\bar{d}_\textrm{av}(\pi)/\pi\! =\! 1$ on the sphere, the endpoint of the curve in Fig.\ \ref{defects-fig:tetra-all}. 
The fact that each point $p$ on the sphere has
an antipode at geodesic distance $\pi$ is of course a consequence of the highly symmetric character of its geometry. As we
have already seen, this is different on the tetrahedron, where the distance of the antipode depends on the point $p$ and there is
no analogue of the distinguished value $\delta\! =\! \pi$. 
The restriction on the $\delta$-range we set for the tetrahedron was precisely the distance of the closest antipode. 
We could in principle have chosen to go beyond this point, assigning the value zero to data points whose antipodal distance
is smaller than a given $\delta$. This would have extended the tetrahedron's curve to $\delta\! \approx \! 3.110$, but at the expense of
a somewhat unclear interpretation. 

\section{Summary and conclusion}
\label{defects-sec:final}

Motivated by the issue of observables in nonperturbative quantum gravity, 
we have advocated the study of a new, global geometric observable for curved metric spaces, the curvature profile.
It is obtained by integrating the quasi-local, scale-dependent quantum Ricci curvature introduced in earlier work, 
and has the interpretation of an averaged Ricci scalar depending on a scale $\delta$. On smooth classical
manifolds, the information contained in the curvature profile for infinitesimal and small $\delta$ is simply that of the 
averaged Ricci scalar, while for larger $\delta$ it captures the geometric and topological properties of the metric space 
in a coarse-grained manner. This scale dependence is very important for the corresponding observable in the quantum
theory, where it can help us to identify the transition from a pure quantum regime to a semiclassical one. In order
to be able to identify the latter, it is important to get a better understanding of the behavior of
the curvature profile on classical spaces.    

In the present chapter, we have specifically examined the influence of the distribution of the local curvature of the underlying
metric space on the classical curvature profile. The easiest set-up to compute and compare this effect explicitly is that of 
two-dimensional compact surfaces. We have investigated several regular polyhedral surfaces homeomorphic to the two-sphere, which 
contain a number of conical singularities. We were able to analyze the case of the tetrahedron completely, because we 
could set up a relatively simple method to compute the geodesic distance between two given points, making use
of a representation of the geodesics in the tessellated plane.\footnote{A related problem for the more difficult case of the
cube has been addressed in \cite{goldstone2021shortest}.} The curvature profile is distinct from that of the sphere, although its overall
features are similar. Perhaps the most remarkable feature is how well it resembles the profile of a smooth sphere. 
The property of averaging or coarse-graining well over regions where the curvature is singular is a
desirable feature from the point of view of the nonperturbative quantum theory, where the quantum geometry 
on Planckian scales tends to be extremely singular and ill-defined locally. Of course, the quantum Ricci curvature
underlying the construction of the curvature profile was introduced precisely to address and potentially mitigate this issue. 
For the other Platonic solids we could determine the curvature profile only for small $\delta$, but from the
limited evidence it appears that distributing the Gaussian curvature over more vertices leads to profiles that are even closer to
that of the sphere. 

In conclusion, we have for the first time computed a curvature profile for a curved classical space that is not maximally 
homogeneous and isotropic. It gives us a first quantitative gauge of how deviations from a maximally symmetric
situation are reflected in the curvature profile, which constitutes a kind of global fingerprint of a given geometry. 
It will be interesting to add the curvature profiles of other classical geometries to our reference catalogue, to help
us bridge the divide between results from nonperturbative quantum gravity and invariant properties of classical spacetimes.



\part{CDT in 2+1 dimensions}
\label{part:cdt3d}
\chapter{On the nature of spatial universes in 3D Lorentzian quantum gravity}\label{ch:slice3d}
\section{Introduction}
A complete theory of quantum gravity may offer insights into how the spacetime we observe and inhabit can emerge from first principles. 
The nonperturbative gravitational path integral is a promising route toward such a theory, formulated within a purely
quantum field-theoretic setting \cite{loll2022quantum}.
If one is interested in concrete Planckian or near-Planckian results in the full, four-dimensional theory, like information on 
the spectra of diffeomorphism-invariant observables, 
Causal Dynamical Triangulations or \emph{CDT quantum gravity} \cite{ambjorn2012nonperturbative,loll2019quantum} is arguably the path integral 
approach that is furthest developed. Recall that the continuum path integral for pure gravity is given by
\begin{equation}
Z=\int\limits_{{\cal G}(M)}\!\! {\cal D} [g]\, \textrm{e}^{\, i \seh [g]},\;\;\;\;\;\;
\seh [g]=\frac{1}{16\pi G_\textrm{N}}\, \int\limits_M d^4 x\, \sqrt{ -\det(g)}\, (R -2 \Lambda ),
\label{pathint}
\end{equation}
where ${\cal G}(M)$ denotes the space of diffeomorphism-equivalence classes $[g]$ of Lorentzian metrics $g_{\mu\nu}(x)$ on the manifold $M$,
and $\seh $ is the Einstein-Hilbert action. In the CDT set-up this formal expression 
is given a precise meaning, namely, as the continuum limit of a regularized version of (\ref{pathint}), 
with ${\cal G}(M)$ approximated by a 
space of piecewise flat Lorentzian spacetimes. Although the primary physical interest is in spacetime dimension $D\! =\! 4$, 
the CDT path integral has also been studied in two and three dimensions. 

Hallmarks of this strictly nonperturbative approach are (i) the presence of a well-defined analytic continuation or ``Wick rotation'', mapping 
the regularized 
path integral to a real partition function, which enables its analytical evaluation in $D\! =\! 2$ \cite{ambjorn1998nonperturbative} and numerical
evaluation in $D\! =\! 2$, 3 and 4 \cite{ambjorn2000nonperturbative,ambjorn2001dynamically}, (ii) its formulation on a space of \emph{geometries}, 
avoiding the need to gauge-fix the diffeomorphism symmetry and isolate its physical
degrees of freedom, (iii) following the logic of critical phenomena, a high degree of uniqueness and universality if a continuum limit can be
shown to exist, (iv) a nonperturbative cure of the conformal divergence, which by default renders Euclidean path integrals in $D\!\geq\! 3$ ill defined \cite{dasgupta2001propertime}, and (v) unitarity, in the form of reflection posi\-tivity of the regularized path integral, with respect to a notion of discrete proper time \cite{ambjorn2001dynamically,ambjorn2012nonperturbative}. 

In terms of results in $D\! =\! 4$, in addition to the presence of second-order phase transitions \cite{ambjorn2011secondorder,ambjorn2012second,coumbe2016exploring}, 
necessary for the existence of a continuum limit,
an important finding of CDT is the emergence of an extended four-dimensional universe \cite{ambjorn2004emergence,ambjorn2021cdt}. 
With the standard choice $M\! =\! S^1\!\times\! S^3$ for the topology, in terms of the quantum observables measured so far 
(spectral and Hausdorff dimensions \cite{ambjorn2005spectral,ambjorn2005reconstructing}, shape of the universe, including quantum 
fluctuations \cite{ambjorn2008planckian,ambjorn2008nonperturbative}, average Ricci curvature \cite{klitgaard2020how}), 
its behavior on sufficiently coarse-grained scales is compatible with that of a de Sitter universe.
This is remarkable because it represents nontrivial evidence of a classical limit,
one of the high hurdles to clear for any nonperturbative and manifestly background-independent approach to quantum gravity.

After Wick rotation, the gravitational path integrals of CDT become partition functions of  
statistical systems, whose elementary geometric building blocks (flat $D$-dimensional simplices, see Sec.\ \ref{sec:phases} for further details) 
are assembled into piecewise flat manifolds $T$ -- the triangulations --
each one contributing with a Boltzmann weight $\exp (-\seh [T])$.\footnote{Note that $\seh [T]$ is the so-called bare action of the
regularized theory, depending on bare coupling constants, which in the continuum limit will typically undergo renormalization.} 
There are a couple of reasons why such seemingly simple ingredients can give rise to interesting continuum theories of quantum gravity and quantum
geometry. On the one hand, there is the highly nontrivial combinatorics of how the simplicial building blocks can be glued together to yield 
distinct curved spacetimes $T$. Especially in dimension $D\!\geq\! 3$, this reflects the complexities of local geometry and curvature, already familiar 
from the classical theory. On the other hand, there is
a complicated interplay between ``energy'' (the bare action) and ``entropy'' (the number of distinct triangulations for a given value of the bare action),
which depends on the values of the bare coupling constants, i.e.\ the point in phase space at which the path integral is evaluated. 

An enormous amount has been learned about such nonperturbative systems of geometry over the last 35 years, beginning with the Euclidean
analogue and precursor of the Lorentzian CDT theory, based on Euclidean dynamical triangulations or dynamical triangulations (DT) for short 
\cite{david1985planar,kazakov1985critical,ambjorn1997quantumb}. A crucial role in the exploration of these systems 
has been played by Monte Carlo methods, which are employed to numerically evaluate the path integral and expectation values of
observables by importance sampling \cite{binder1997applications,newman1999monte}. This is also true in dimension $D\! =\! 2$,
where in addition a variety of nonperturbative analytical solution techniques are available, e.g.\ combinatorial, matrix model and transfer matrix methods \cite{difrancesco19952d,ambjorn1998nonperturbative,ambjorn1999euclidean}, leading to compatible results.

Monte Carlo simulations should be seen as numerical experiments, providing tests and feedback for the construction of the 
theory. For full quantum gravity, the quantitative information on the nonperturbative sector obtained from numerical analysis 
is extremely valuable, since it cannot currently be 
substituted by anything else. Although we do not know in detail what a theory of quantum gravity will eventually look like, it seems unlikely that
it will come in closed analytic form. A potential scenario would be akin to QCD, where we manage to
extract nonperturbative information about the theory's spectrum (of suitable quantum-geometric observables) with ever greater accuracy,
using a background-independent analogue of lattice gauge theory such as (C)DT. 
Despite its conventional, quantum field-theoretic setting and the absence of any exotic ingredients, this type of lattice gravity has already uncovered 
unexpected features of strongly quantum-fluctuating geometry, like the dynamical dimensional reduction of 
spacetime near the Planck scale \cite{ambjorn2005spectral}, which is conjectured to be universal \cite{carlip2017dimension}. 

The focus of the present chapter will be the Lorentzian CDT path integral in \emph{three} spacetime dimensions \cite{ambjorn2001nonperturbative}.
More specifically, as a stepping stone towards a more detailed geometric understanding of this quantum gravity model, we will
investigate the geometry of its two-dimensional spatial hypersurfaces. A key question is whether in
a continuum limit the behavior of these surfaces falls into one of the known universality classes \cite{goldenfeld2019lectures} 
of nonperturbative quantum gravity in two dimensions, or whether there is evidence for a different type of quantum dynamics.  
The two universality classes in question are that of (the scaling limit of) two-dimensional DT \cite{david1985planar,ambjorn1997quantumb}, 
which also contains Liouville quantum gravity, and that of two-dimensional CDT quantum gravity \cite{ambjorn1998nonperturbative,ambjorn2013universality}.    

Our study will be numerical in nature, but -- depending on the outcome -- may well provide input for further analytical work, which could be 
technically feasible because of the effective two-dimensional character of the spatial slices. Note
that we do not claim that there is a direct physical interpretation of the properties of these spatial geometries from a three-dimensional point of view
(inasmuch as a lower-dimensional toy model of quantum gravity can be called ``physical'' at all).
Although in our set-up a spatial slice at constant proper time is an invariantly defined concept\footnote{It is defined as the set of all points at a given proper-time 
distance to a given initial spatial surface or an initial singularity, in the spirit of similar constructions in the continuum \cite{andersson1998cosmological}. 
Note also that the proper-time slicing is not related to any gauge-fixing, since the CDT set-up is manifestly diffeomorphism-invariant 
(see e.g.\ \cite{loll2019quantum} for a detailed discussion).}, 
it is not clear to what extent its properties can be thought of as ``observable'', because of the highly nonlocal construction of the hypersurfaces and because 
of their singular nature (``moments in time'') from the point of view of the quantum theory. 
To obtain true quantum observables in a three-dimensional, spacetime sense would presumably require some smearing in the time direction. 
 
Nevertheless, our measurements within the slices of constant time are perfectly well defined operationally and 
give us a quantitative handle on the influence of the three-dimensional quantum geometry in which the spatial slices are embedded, as we will 
demonstrate. We will start by investigating the distribution of the vertex order in the slices, which counts the number of spatial edges meeting
at a vertex. This quantity is not per se related to a continuum observable, but can be compared with known exact results for the 
ensembles of two-dimensional DT and CDT geometries.
The core of this chapter consists of measuring and analyzing the following quantum observables: (i) the entropy exponent $\gamma$, also known as the string susceptibility, which determines the subexponential growth
of the partition function at fixed two-volume $A$, as a function of $A$; (ii) the Hausdorff dimension $d_H$, obtained by
comparing volumes with their linear extension, where we distinguish between a local and
a global variant; (iii) the so-called curvature profile $\mathcal{R}(\delta)$ of the spatial slices, measuring the average 
quantum Ricci curvature \cite{klitgaard2018introducing} of the surfaces as a function of a linear coarse-graining scale $\delta$.
We find convincing evidence that the effective dynamics of the spatial slices in the so-called degenerate phase of three-dimensional CDT
quantum gravity is described by two-dimensional DT quantum gravity. However, we do not find a match with any known two-dimensional
system of quantum geometry in the so-called de Sitter phase, where the dynamics of the hypersurfaces is much richer due to the nontrivial
influence of the embedding three-geometry. Further research is needed to determine the continuum nature of the effective spatial dynamics in
this phase.

The remainder of the chapter is structured as follows. 
In the next section, we recall the main ingredients of CDT quantum gravity in $D\! =\! 3$ and review previous research on the subject,
and what it has revealed about its phase structure and physical characteristics. 
In Sec.\ \ref{sec:impl}, we discuss the numerical implementation of the three-dimensional CDT path integral in terms of Markov chain Monte Carlo methods. 
Sec.\ \ref{sec:2d-qg} contains a detailed description of the properties of the spatial slices that we have studied numerically. 
We present the results of our measurements, and describe the overall picture that emerges from them. 
In Sec. \ref{sec:disc} we summarize and discuss our findings.
A couple of technical discussions have been relegated to appendices, to improve the readability of the main part of the chapter.

\section{Three-dimensional CDT quantum gravity}
\label{sec:phases}

Quantum gravity in three spacetime dimensions \cite{carlip1998quantum} provides an interesting test case for the full gravitational path integral. 
Although the pure gravity theory does not have any local propagating degrees of freedom, the path integral
has the same functional form in terms of the three-dimensional metric as its four-dimensional counterpart (\ref{pathint}), and therefore looks equally 
ill-behaved with regard to its behavior under renormalization.
How to reconcile the difficulties of solving this metric path integral with the ``topological'' nature of three-dimensional gravity\footnote{More 
precisely, the physical degrees of freedom of three-dimensional gravity are global modes of the metric, described by Teichm\"uller parameters, 
which are present when the genus of the spatial slices is larger than or equal to 1. The present chapter uses spherical slices, without such parameters.}, 
which leads to considerable simplifications in a first-order, Chern-Simons formulation, without any quantum field-theoretic divergences, 
is only partially understood (see \cite{carlip2005quantum} for a discussion). 

The CDT formulation has thrown some light on the nonperturbative aspects of this question, uncovering both similarities and 
differences between the three-dimensional and the physical, 
four-dimensional theory \cite{ambjorn2001nonperturbative,ambjorn2001computer,ambjorn20023d}. 
For a better understanding of the issues involved and to set the stage for the main part of the chapter, let us briefly recall the set-up in three dimensions.
After applying the Wick rotation mentioned in the previous section, the regularized CDT path integral in $D\! =\! 3$ takes the form of a partition function
\begin{equation}
Z = \sum\limits_{\text{triang.}\, T}\frac{1}{C_T}\, \textrm{e}^{-\seh [T]}, \;\;\;\;\;\;\;
\seh [T]=-k_0 N_0(T)+k_3 N_3 (T),
\label{cdtpi}
\end{equation}
where $\seh [T]$ denotes the Regge form of the Einstein-Hilbert action on the piecewise flat triangulation $T$, $N_0(T)$ and $N_3(T)$
are the numbers of vertices (``zero-simplices'') and tetrahedra (``three-simplices''), and $C_T$ is the order of the automorphism group of $T$. 
The coupling $k_0$ is proportional to the inverse bare Newton constant and $k_3$ depends linearly on the bare cosmological constant (see
\cite{ambjorn2001nonperturbative} for details). The sum is taken over simplicial manifolds
of a given, fixed topology, which in our case will be $S^1\!\times\! S^2$, a periodically identified time interval times a two-sphere.

\begin{figure}[t]
	\centering
	\includegraphics[width=0.6\textwidth]{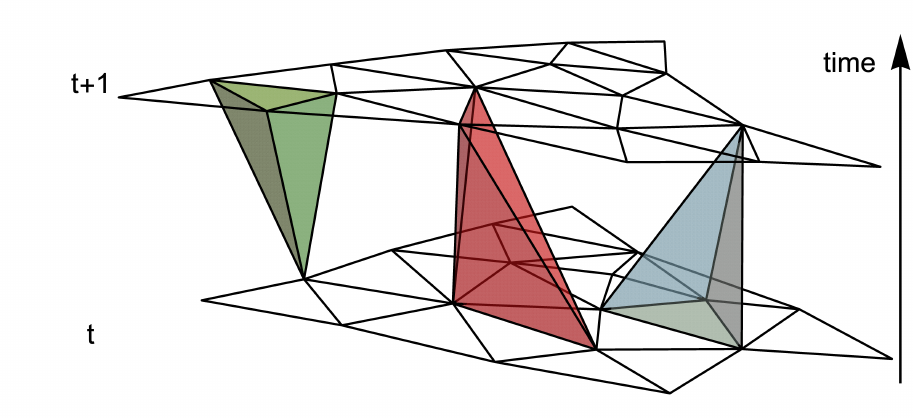}
	\caption{Three types of tetrahedral building blocks of three-dimensional CDT: type (1,3)  (left), type (2,2) (center), 
	and type (3,1) (right). Note that the two-dimensional triangulations at times $t$ and $t\! +\! 1$ are not drawn isometrically; 
	they are in general curved surfaces.}
	\label{fig:simplices}
\end{figure}

The triangulated configurations $T$ of the Lorentzian CDT path integral have a discrete product structure, representing a simplicial
version of global hyperbolicity, and are assembled from flat, Minkowskian tetrahedra \cite{ambjorn2012nonperturbative,loll2019quantum}. 
A given spacetime geometry can be thought of as a sequence 
of two-dimensional curved, spacelike triangulations, labeled by an integer proper time $t=1,2,3,\dots, \ttot$, and made of equilateral triangles. 
The spacetime volume between each pair of adjacent constant-time slices is completely filled in with tetrahedra,
resulting in a ``sandwich'' of simplicial three-dimensional spacetime with topology $[0,1] \times S^2$. The tetrahedral edges linking 
neighboring spatial slices are \emph{timelike} (and all of equal length), while the edges lying within a spatial slice are of 
course \emph{spacelike} (and also of equal length).
We can therefore classify the building blocks according to their constituent vertices. A tetrahedron of type $(p,q)$ is defined as having
$p$ vertices in slice $t$ and $q$ vertices in slice $t\! +\! 1$,
giving rise to the types (1,3), (2,2) and (3,1), as illustrated by Fig.\ \ref{fig:simplices}. Note that up to time reversal a (1,3)- and a (3,1)-tetrahedron are
geometrically identical.
Since the analytic continuation only affects the edge length assignments and not the topology of the triangulation, this 
characterization of the tetrahedra continues to be meaningful after the Wick rotation.

The key findings of the original, mostly numerical investigation of three-dimensional CDT quantum gravity on $S^1\!\times\! S^2$ inside 
the range $k_0\!\in\! [3,7]$ were as follows \cite{ambjorn2001nonperturbative,ambjorn2001computer,ambjorn20023d}. 
After fine-tuning the bare ``cosmological'' constant $k_3$ to its critical value from inside the region of convergence\footnote{This region
exists because the number of three-dimensional CDT configurations is exponentially bounded as a function of the discrete 
volume $N_3$ \cite{durhuus2015exponential}.} of the partition function Z, 
a two-phase structure was found, consisting of what we shall call a \emph{degenerate phase} for $k_0\!\geq k_0^\textrm{c}$ 
and a \emph{de Sitter phase} for $k_0\!\leq k_0^\textrm{c}$. These two phases are very reminiscent of corresponding phases in
four-dimensional CDT quantum gravity \cite{ambjorn2012nonperturbative} with regard to their volume profiles, i.e.\ the behavior of
their spatial volume $V_2$ as a function of the proper time $t$. In the degenerate phase, $V_2(t)$ oscillates wildly, indicating
that spacetime disintegrates into a sequence of uncorrelated two-dimensional geometries (see also Fig.\ \ref{fig:volprofs} below). 
By contrast, in the de Sitter phase a nontrivial 
``blob'' forms when the time extension is chosen sufficiently large, whose shape matches that of a Euclidean de Sitter space (with
$\langle V_2(t)\rangle \propto \cos^2 (\mathit{c}\, t)$), analogous to what has been observed in 
$D\! =\! 4$.\footnote{Volume profiles for nonperiodic boundary conditions in time, with random, fixed two-spheres of various
sizes as spatial boundaries
have been considered in \cite{cooperman2014first,cooperman2017second}.} 
This constitutes nontrivial evidence that a well-defined and macroscopically three-dimensional ground state of geometry\footnote{in the
sense of minimizing the effective Euclidean action governing the nonperturbative quantum dynamics} exists nonperturbatively.
However, the transition at the critical point $k_0^c$ appears to be a first- and not a second-order phase transition,
and no fine-tuning of the inverse gravitational coupling $k_0$ is needed to obtain a continuum limit. This is 
in line with the expectation that no higher-order transitions are present, due to the absence of propagating degrees of freedom. 

Following these results, three-dimensional CDT quantum gravity has been studied from various perspectives. 
Considerable effort has been focused on the transfer matrix associated with a single time step $\Delta t\! =\! 1$, which 
captures the amplitude of going from one spatial two-geometry to an adjacent one. More precisely, one usually considers a simpler, reduced
transfer matrix, whose in- and out-states are labelled by the spatial two-volume (and possibly Teich\-m\"uller parameters). Given our knowledge about
the physical degrees of freedom of three-dimensional gravity, these are the parameters
that are \emph{expected} to be the only relevant ones in the continuum limit. 
The sandwich geometries contributing to the transfer matrix are closer to
two-dimensional quantities and therefore potentially more amenable to an analytic treatment.
In this spirit, a variant of the model was introduced in \cite{ambjorn2001lorentzian}, in which the (1,3)- and (3,1)-building blocks are 
substituted by (1,4)- and (4,1)-pyramids, something that is not expected to affect the universal properties of the model.  
The motivation for considering this variant is that taking a midsection of a sandwich geometry at half-integer time yields a quadrangulation, whose dual
graph is a configuration described by a Hermitian two-matrix model with $ABAB$-interaction, for which analytical results are available. 
However, the bicolored graph configurations generated by the matrix model form a much larger class than those coming from CDT sandwich geometries,
and correspond to geometries that in general violate the simplicial manifold conditions of the two-dimensional slices and of the 
interpolating three-dimensional piecewise flat geometries in specific ways. Three of these four conditions (following the enumeration in 
\cite{ambjorn2001lorentzian}) are considered mild, in the sense that 
violating them is conjectured not to affect the universality class of the CDT model. As corroborating evidence, \cite{ambjorn2001lorentzian} cites
new numerical simulations of the CDT model in $D\! =\! 3$, where these conditions are relaxed, but which
nevertheless reproduce
the results found in the de Sitter phase of the earlier work that used strict simplicial manifolds \cite{ambjorn2001nonperturbative}. 
Interestingly, they note that the degenerate phase completely disappears in the simulations of this generalized variant of CDT quantum gravity,
and conjecture that the presence of this phase constitutes a discretization artifact. Although we do not directly address
these various conjectures (and work with simplicial manifolds only) in this chapter, 
our results suggest that there may be more scope for different universality classes in three dimensions than has been considered up to now, and that it
may be fruitful to re-examine the influence of regularity conditions on continuum results in greater detail. (Of course, not all universality classes
may be associated with interesting models of quantum gravity.) A similar sentiment was expressed in
\cite{durhuus2020structure}, which gives a precise characterization of the bicolored two-dimensional cell complexes associated with midsections
of CDT geometries with spherical and disc-like spatial slices.

The configurations described by the $ABAB$-matrix model violate also a fourth regularity condition \cite{ambjorn2001lorentzian}, which is associated with a
considerable enlargement of the space of three-geometries. It allows for the appearance of spatial wormholes and a new phase, not present
in standard CDT quantum gravity, where these wormholes are abundant. Whether this phase is interesting from a physics point of view remains
to be understood. The association of CDT quantum gravity with the $ABAB$-matrix model was also used to
analyze the behavior of the bare coupling constants of the former under renormalization \cite{ambjorn2004renormalization}.
An asymmetric version of the matrix model was studied in
\cite{ambjorn20033d}, motivated by the search for a Hamiltonian associated with the reduced transfer matrix.  
Without invoking matrix models, a continuum Hamiltonian of this kind was derived for the first time in \cite{benedetti2007dimensional}, for spatial slices of cylinder topology,
albeit for a sub-ensemble of CDT configurations with certain ordering restrictions. 
The effective action for the two-volumes of spatial slices with toroidal topology was investigated in \cite{budd2013exploring}, and the dynamics of its Teichm\"uller parameters in \cite{budd2012thesis,budd2012effective}. The phase structure of a one-dimensional balls-in-boxes model, meant to
capture the effective dynamics of the two-volume of CDT quantum gravity, was analyzed in \cite{benedetti2017capturing}, and shown to reproduce
certain features of CDT as well as Ho\v rava-Lifshitz-inspired gravity models (see also \cite{benedetti2015spacetime}).
The spectral dimension of the CDT model in the de Sitter phase was measured and found to be compatible with the classical value of 3 on
large scales and to exhibit a dynamical dimensional reduction to a value compatible with 2 on short scales, similar to what happens in CDT for 
$D\! =\! 4$ \cite{benedetti2009spectral}. Lastly, a generalized model of CDT quantum gravity with causally well-behaved configurations, but without
a preferred proper-time slicing was defined and investigated numerically, and found to reproduce the volume profile of a de 
Sitter space \cite{jordan2013causal,jordan2013sitter}.

\section{Implementation}
\label{sec:impl}

\begin{figure}[t]
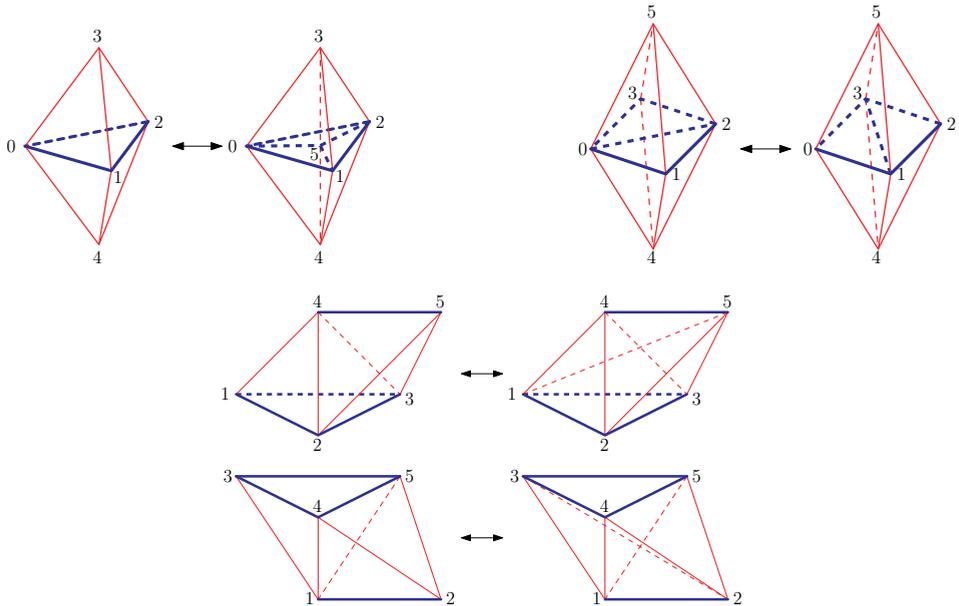

    \centering
    \includegraphics[width=0.40\textwidth]{3d-cdt-add-del.pdf} \hfill
    \includegraphics[width=0.40\textwidth]{3d-cdt-flip.pdf}
    \\ \vspace{1em}
    \includegraphics[width=0.55\textwidth]{3d-cdt-shift-ishift.pdf}
    \caption{Three basic local moves in 3D CDT quantum gravity together with their inverses, and their effect on spatial slices. Space\-like edges are 
    drawn in blue and timelike ones in red. Top left: subdivision of a spatial triangle into three; top right: flip of a spatial edge. The move at the
    bottom does not affect the spatial triangulations (time inverse not shown).}
    \label{fig:moves-3d}
\end{figure}

Expectation values of geometric observables $\mathcal{O}$ in CDT quantum gravity are computed as 
\begin{equation}
	\langle \mathcal{O} \rangle = \frac{1}{Z}\, \sum_{T} \frac{1}{C_T}\,\mathcal{O}[T]\,  \textrm{e}^{-S^\textrm{EH} [T]},
	\label{eq:q-exp-o}
\end{equation}
where $Z$ is the partition function defined in Eq.\ \eqref{cdtpi}. As already mentioned, the focus of this chapter is a set of observables pertaining to
the two-dimensional spatial triangulations of constant integer time $t$ of the three-dimensional CDT configurations, which will be the subject of  
Sec.~\ref{sec:2d-qg}. Since it is not known how to compute $Z$ analytically in three dimensions, 
we will compute statistical estimates of the expectation value of an observable 
by sampling the CDT ensemble through Monte Carlo simulations.\footnote{Our implementation code can be found at \cite{brunekreef2022jorenb}.} In these simulations, we construct a random walk in the ensemble of CDT geometries by performing local updates (``moves'') on a triangulation. The basic set of 
moves we used is shown in Fig.\ \ref{fig:moves-3d}, see also \cite{ambjorn2001nonperturbative}. If we impose so-called detailed balance \cite{metropolis1953equation} on the updating procedure, by accepting or rejecting such moves with an appropriate probability, this random walk corresponds to a sample of the ensemble where geometries appear with a relative rate according to their Boltzmann weight. Since
subsequent geometries in a random walk are almost identical, we must perform a large number of local moves on a given geometry to obtain a new and
sufficiently independent one. This procedure is iterated to obtain a sequence of independent triangulations, and an estimate of the expectation value of 
an observable is computed as the weighted average \eqref{eq:q-exp-o} over this sequence. 
To study the continuum properties of observables, the number of building blocks should be taken to infinity. Since this is impossible in practice, due to the 
finiteness of our computational resources, we use finite-size scaling methods \cite{newman1999monte} to estimate the behavior of the system in the continuum limit. For more details on computer simulations of three-dimensional CDT
we refer the interested reader to \cite{ambjorn2001computer}.

For a given value of the gravitational coupling $k_0$, the cosmological coupling $k_3$ is always tuned to its $k_0$-dependent pseudocritical 
value\footnote{which in the limit $N_3\!\rightarrow\!\infty$ would become the critical value $k_3^c(k_0)$}, which means that we are investigating a
one-dimensional phase space parametrized by $k_0$. The location of the critical point $k_0^c$ along this line, associated with the first-order
transition mentioned in Sec.\ \ref{sec:phases}, is not a universal quantity. For example, it depends on the regularity conditions imposed on the ensemble,
and the time extension $\ttot$ of the geometries \cite{ambjorn2001nonperturbative}. 
In our analysis of the spatial slices we will use standard CDT simplicial manifolds\footnote{The effects of relaxing the local manifold constraints have been investigated further in \cite{brunekreef2022phase}, where it was found that the order of the phase transition is likely unchanged, although the location of the critical point shifts to a smaller value as the restrictions are loosened.} and $\ttot\! =\! 3$ with periodic boundary conditions in time, for which we 
have found $k_0^c\! \approx\! 6.24$.

\begin{figure}[t]
	\centering
	\includegraphics[width=0.5\textwidth]{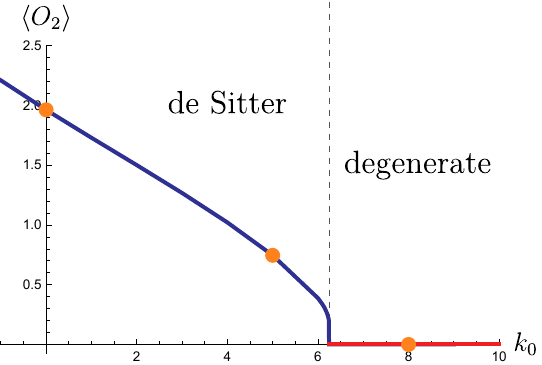}
	\caption{Expectation value of the order parameter $\optwo\! =\! N_{22}/N_{31}$ as a function of the bare coupling $k_0$, exhibiting a first-order phase transition 
	at $k_0^c\! \approx\! 6.24$. To the left of the transition is the de Sitter phase, and to its right the degenerate phase. Our measurements on
	spatial slices are taken for the $k_0$-values $0.0$, $5.0$ and $8.0$, as indicated.
	}
	\label{fig:phase-diagram}
\end{figure}

\begin{figure}[t]
	\centering
	\includegraphics[width=0.45\textwidth]{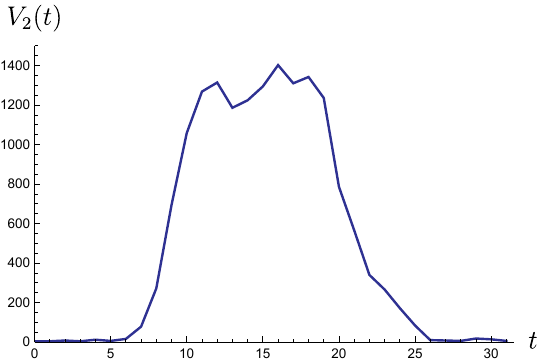}
	\hfill
	\includegraphics[width=0.45\textwidth]{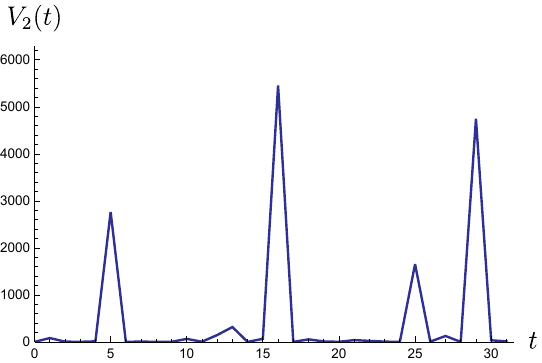}
	\caption{Volume profiles $V_2(t)$ of typical configurations appearing in the de Sitter phase (left) and the degenerate phase (right) of the
	three-dimensional CDT model on $S^1\!\times\! S^2$.}
	\label{fig:volprofs}
\end{figure}

A convenient order parameter to locate the phase transition is the ratio
\begin{equation}
	\optwo (T) = \frac{N_{22}(T)}{N_{31}(T)},
\end{equation}
where $N_{22}(T)$ and $N_{31}(T)$ denote the numbers of (2,2)- and (3,1)-simplices of the triangulation $T$ respectively. 
Its expectation value $\left\langle \optwo \right\rangle$ is
nonvanishing for small $k_0$ and drops to zero rapidly as the transition point $k_0^c$ is approached, beyond which it remains zero for all values 
of $k_0\! >\! k_0^c$ we have investigated. The measured values of $\left\langle \optwo \right\rangle$ as a function of $k_0$ are shown in 
Fig. \ref{fig:phase-diagram}, obtained in a system with $N_3\! =\! 64.000$ and $\ttot = 3$.
The regions to the left and right of the transition correspond to the de Sitter and degenerate phases introduced earlier. Snapshots of typical
volume profiles $V_2(t)$, counting the number of triangles in the spatial slice at time $t$, are depicted in Fig.\ \ref{fig:volprofs} for a system with
$N_{31}\! =\! 16.000$ and $\ttot\! =\! 32$.  
The volumes of neighboring slices in the degenerate phase are largely uncorrelated, while they tend to align in the de Sitter phase.
When taking an ensemble average of the latter with the ``centers of volume'' aligned, the expectation value $\langle V_2(t)\rangle$ matches that of
a three-dimensional Euclidean de Sitter universe in a proper-time parametrization \cite{ambjorn2001nonperturbative}. 

We will perform measurements of the spatial slices for three different $k_0$-values, two in the de Sitter phase at $k_0\! =\! 0.0$ and $5.0$
and one in the degenerate phase at $k_0\! =\! 8.0$, cf.\ Fig.\ \ref{fig:phase-diagram}. The volumes $N_{31}$ of the systems we investigate take values in the range $[1.000, 96.000]$.
In choosing these particular values of $k_0$, we are staying away from the direct vicinity of the first-order transition at $k_0^c\! \approx\! 6.24$, to avoid
that the system jumps between the two phases during the simulation. The two points chosen in the de Sitter phase are spaced well apart, while taking into
account that the Monte Carlo algorithm becomes increasingly inefficient as $k_0$ is lowered. 
It may be worth pointing out that the expression for the discretized, bare Einstein-Hilbert action in Eq.\ \eqref{cdtpi} is highly non-unique, and that
the value $k_0\! =\! 0.0$ is therefore in no way physically distinguished.

The observed decoupling of neighboring slices provides a strong argument for an effective slice
behavior that is characteristic for the universality class of Euclidean dynamical triangulations, 
something our data in Sec.\ \ref{sec:2d-qg} below will confirm. 
As a cross check that the system's behavior stays the same throughout the degenerate phase, we have performed a short series of 
measurements for all observables (except the curvature profile) at the much higher value of 
$k_0\! =\! 15.0$, which found no differences compared with $k_0\! =\! 8.0$.
Little is known about the slice behavior in the de Sitter phase;
a preliminary investigation of the Hausdorff dimension for small slice volumes $V_2(t)\sim 1.000$
was made in \cite{ambjorn2001nonperturbative}, resulting in the estimate $d_H\! =\! 3.4\pm 0.4$. Furthermore,
a measure of homogeneity for the spatial slices was formulated and implemented in \cite{cooperman2014scaledependent}, but with inconclusive results. 

A final technical issue to be discussed before turning to a description of the measurement results is that of volume-fixing. 
As usual, we perform all measurements at fixed spacetime volume; more precisely, since the Monte Carlo moves are not volume-preserving, 
the simulations are run in the vicinity of a target volume. The latter can be stated in terms of the total volume $N_3$ or 
in terms of $N_{31}$, which is what we will do below. Both prescriptions are essentially equivalent, since in the case of
periodic boundary conditions in time we have the identity $N_3\! =\! 2 N_{31} + N_{22}$, and since for fixed $k_0$ the ratio between
$N_{22}$ and $N_{31}$ is approximately constant, and can be read off the graph in Fig.\ \ref{fig:phase-diagram}. Note also that
$N_{31}$ is equal to the total number of \emph{spatial} triangles in the triangulation, i.e.\ the sum over all $t$ of $V_2(t)$.
The approximate volume-fixing is implemented by adding a term
\begin{equation}
	S_\textrm{fix}[T] = \epsilon\, \big(N_{31}(T)-\tilde{N}_{31}\big)^2
\label{vcon}	
\end{equation}
to the bare action, where $\tilde{N}_{31}$ denotes the desired target volume and the value of the small, positive parameter $\epsilon$ determines the
typical size of the fluctuations of $N_{31}$ around $\tilde{N}_{31}$. In the simulations performed for this chapter, we generally set $\epsilon$ to values on the order of $10^{-5}$.

In addition, since we are interested in the intrinsic properties of the spatial slices and in extracting their continuum behavior from finite-size scaling,
we must collect measurement data at different, fixed \emph{slice} volumes.
Instead of adding further volume-fixing terms for the individual slices to the action, which would run the risk of introducing an unwanted bias, 
we let the individual slices fluctuate freely, subject only to the total volume constraint (\ref{vcon}), but take data only when a slice hits
a precise desired value $\tilde{V}_2$.
More concretely, in between two measurements we first perform a fixed number of attempted moves\footnote{These attempted moves will be accepted or rejected according to the detailed balance condition mentioned earlier.}, collectively referred to as a \emph{sweep}. Different observables can have different autocorrelation times (measured in Monte Carlo steps) and therefore require different sweep sizes. A typical sweep size is taken to be on the order of 1.000 times the target 
volume $\tilde{N}_{31}$. After completion of a sweep, we continue performing local updates until one of the slices hits the target two-volume $\tilde{V}_2$. 
We then perform a measurement of the observable under consideration on this slice, and subsequently start a new sweep.

The choice of the target three-volume $\tilde{N}_{31}$ that maximizes the probability of encountering slices with target two-volume $\tilde{V}_2$
depends on the phase. In the de Sitter phase and for the small time extension $\ttot$ we have used, the total volume spreads roughly evenly 
over the available slices and an appropriate choice is $\tilde{N}_{31} \! = \! \ttot \cdot \tilde{V}_2$. In the degenerate phase, the volume tends to
concentrate on one of the slices, and a good choice is $\tilde{N}_{31}\! =\! \tilde{N}_2 + m$, where for $\ttot\! =\! 3$ and $\tilde{V}_2\! >\! 1.000$
we found that $m\! =\! 100$ is a convenient choice.

To maximize the volume of the spatial slices, we work with $\ttot\! =\! 3$, 
the minimal number allowed by our simulation code. 
Both this choice and our choice of periodic boundary in time can in principle have an influence on the behavior of observables, even if our
measurements are confined to individual slices. Investigations of the transfer matrix in four-dimensional CDT quantum gravity 
have indicated that such a set-up can be appropriate, at least for selected observables \cite{ambjorn2013transfer}.
As an extra check, we have performed a few measurements on systems with larger $\ttot\!\lesssim\! 32$, and found 
the same behavior for the slice geometries. This is reassuring, but not a substitute for a more systematic investigation of the influence of
these global choices, which goes beyond the scope of this chapter. This proviso should be kept in mind when interpreting the outcomes 
of our research, which will be presented next.

\section{Geometric observables on spatial hypersurfaces}
\label{sec:2d-qg}

The main objective of this chapter is a detailed measurement of the geometric properties of the two-dimensional spatial hypersurfaces
of constant integer proper time $t$ in three-dimensional CDT quantum gravity, both in the de Sitter and the degenerate phase. 
We will compare our findings with known results for nonperturbative
models of two-dimensional quantum gravity. There are two pure-gravity systems in $D\! =\! 2$ available for reference, Euclidean DT and Lorentzian CDT
quantum gravity. However, especially in the de Sitter phase there are no stringent reasons why the slice geometries should match these known
systems, since they are part of a larger, three-dimensional geometry. For example, the ambient geometry may induce extrinsic curvature terms 
on the spatial slices, which are not present in intrinsically two-dimensional situations. 
The following subsections will deal in turn with the four quantities we have studied on the spatial slices: the vertex order, the entropy exponent, 
the Hausdorff dimension and the curvature profile.

\subsection{Vertex order}
\label{subsec:coord}

We first examined the distribution of the vertex order, which counts the number $q(v)$ of spacelike edges meeting at a given vertex $v$.
For the simplicial manifolds we use in the simulations, the possible vertex orders are $q(v)\! =\! 3,4,5,\dots $.
As already mentioned earlier, the distribution of the $q(v)$ is not strictly speaking an observable. It is a non-universal property of the discrete lattice, which  
depends on the details of the lattice discretization and does not have an obvious continuum counterpart. For example, using  
quadrangulations instead of triangulations leads to the same continuum theory of two-dimensional Euclidean DT quantum gravity \cite{miermont2013brownian}, 
but the two models have different distributions of vertex orders. We have studied this quantity nevertheless, since it is known analytically for both the DT and CDT ensembles and gives us a first idea of whether and how our system changes as a function of the bare coupling $k_0$. 

\begin{figure}[t]
	\centering
	\includegraphics[width=0.6\textwidth]{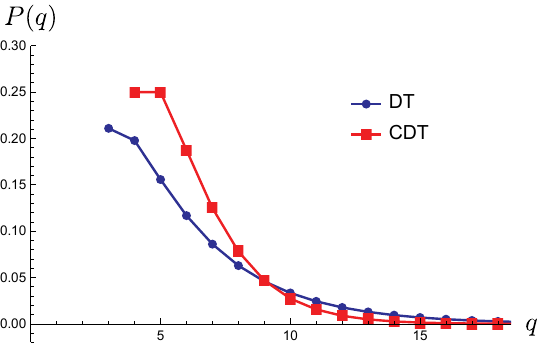}
	\caption{Probability distribution of the vertex order $q$ for the ensembles of two-dimensional Euclidean DT and Lorentzian CDT quantum gravity, in the
	limit of infinitely many triangles.}
	\label{fig:coord-theory}
\end{figure}

The normalized probability distribution $P(q)$ for the vertex order in two-dimensional DT quantum gravity in the thermodynamic limit
has been determined analytically as \cite{boulatov1986analytical}
\begin{equation}
	P_\textrm{DT}(q) = 16 \left(\frac{3}{16}\right)^q  \frac{(q-2)(2q-2)!}{q!(q-1)!}, \quad \quad q \geq 2,
	\label{eq:dt-coord}
\end{equation}
with a large-$q$ behavior $\sim\! \left(\frac{3}{4}\right)^q$. The corresponding probability distribution for two-dimen\-sio\-nal CDT quantum gravity is given by \cite{ambjorn1999new}
\begin{equation}
	P_\textrm{CDT}(q) = \frac{q-3}{2^{q-2}}, \quad \quad q \geq 4,
    \label{eq:cdt-coord}
\end{equation}
with a fall-off behavior $\sim\! \left(\frac{1}{2}\right)^q$ for large $q$. Both distributions are shown in Fig.\ \ref{fig:coord-theory}.

\begin{figure}[t]
	\centering
	\includegraphics[width=0.6\textwidth]{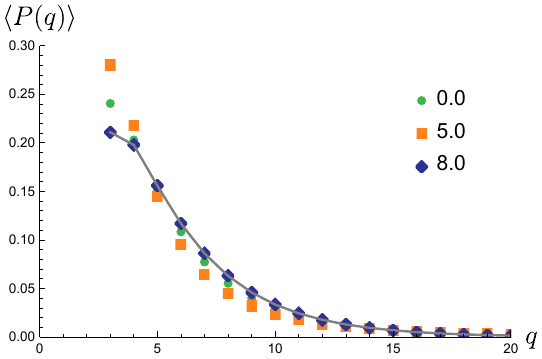}
	\caption{Measured probability distribution of the vertex order $q$ on spatial slices of volume $V_2\! =\! 4.000$ in three-dimensional CDT, for 
	$k_0\! =\! 0.0$, $5.0$ and $8.0$
	and $q\! \leq\! 20$. For comparison, the line connecting the exact values for DT from Fig.\ \ref{fig:coord-theory} are included.}
	\label{fig:cdt-coord-results-nonlog}
\end{figure}

\begin{figure}[t]
	\centering
	\includegraphics[width=0.6\textwidth]{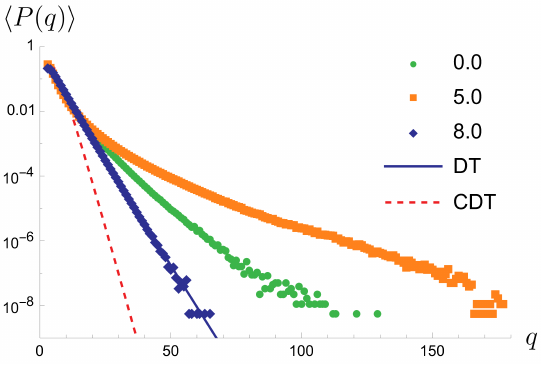}
	\caption{Measured probability distribution of the vertex order $q$ (logarithmic scale) on spatial slices of volume $V_2\! =\! 4.000$ in three-dimensional CDT, for $k_0\! =\! 0.0$, $5.0$ and $8.0$. The unbroken straight line is the analytical result \eqref{eq:dt-coord} for two-dimensional DT, and the dashed straight line the analytical result \eqref{eq:cdt-coord} for two-dimensional CDT.}
	\label{fig:cdt-coord-results}
\end{figure}

We took measurements at the three chosen phase space points $k_0 = 0.0, 5.0$ and $8.0$, with a target volume $\tilde{V}_2\! =\! 4.000$ for the spatial slices,
corresponding to a target three-volume $\targn\! =\! 12.000$ in the de Sitter phase, and $\targn\! =\! 4.400$ in the degenerate phase. 
The sweep size was set to $100 \cdot \targn$ in each case. For each value of $k_0$, we collected measurements of $P(q)$ for 100k different slices, 
by recording for each slice the full set of vertex orders $q(v)$ for all vertices and normalizing the resulting histogram. We then approximated the
expectation value $\langle P(q)\rangle$ by taking the ensemble average over this data set according to the prescription (\ref{eq:q-exp-o}). 
The results of the measurements for $q\!\in\! [3,20]$ are shown in Fig.\ \ref{fig:cdt-coord-results-nonlog}, and are clearly distinct for the three
$k_0$-values. While the distribution for $k_0\! =\! 8.0$ in the degenerate phase is a very good match for the analytical result, 
the distributions for $k_0\! =\! 0.0$ and $5.0$ in the de Sitter phase are not good matches and are also different from each other.
This is confirmed when plotting the measurements on a logarithmic scale, taking into account a much larger range of $q\!\leq \! 180$,
as depicted in Fig.\ \ref{fig:cdt-coord-results}, which also includes the known 
distributions for Euclidean DT and CDT as thin straight lines.
We see that within measurement accuracy, the distribution of vertex orders in the degenerate phase of the model is indistinguishable from Eq.\ 
\eqref{eq:dt-coord}, a result that is also compatible with the numerical simulations of pure Euclidean DT quantum gravity 
performed in \cite{boulatov1986analytical}. By contrast, the distributions obtained in the de Sitter phase exhibit a very different behavior, at least for large $q$.
High-order vertices are relatively speaking more probable, and the data cannot be fitted to a single exponential over the $q$-range we have explored. 
Moreover, unlike what happens in the degenerate phase, the distributions within the de Sitter phase depend on the value of $k_0$.

To obtain a more detailed picture of the dependence on $k_0$, 
we performed a series of shorter simulation runs at additional points in the de Sitter phase. 
When entering the de Sitter phase from the degenerate phase by crossing the phase transition at $k_0^c$, the vertex order distribution 
jumps discontinuously to a shape similar to that for $k_0\! =\! 5.0$. 
As $k_0$ is decreased further inside this phase, the distribution changes shape in a continuous way; the distributions we measured for points 
in the interval $0.0\! < \! k_0\! <\! 5.0$ interpolate in a straightforward manner between the ones for $k_0 = 0.0$ and $k_0 = 5.0$ 
shown in Figs.\ \ref{fig:cdt-coord-results-nonlog} and \ref{fig:cdt-coord-results}. 

To summarize, the measurements of the vertex order distribution in the degenerate phase produce an excellent match with the known one for
DT quantum gravity. By contrast, the distributions found in the de Sitter phase do not match those of the standard DT or CDT ensembles in $D\! =\! 2$.
As mentioned earlier, this does not necessarily mean that the slice geometries in the de Sitter phase do not lie in either of the associated universality classes, 
but it is a first indication that they may not. By looking at genuine observables next, we will be able to make more definite statements about the
universal geometric properties of the slice geometries.

\subsection{Entropy exponent}
\label{entresults}

An important parameter characterizing two-dimensional systems of random geometry is the entropy exponent $\gamma$,
which contains information about the behavior of the partition function at fixed two-volume $N_2$ (number of triangles), in the limit as $N_2$ 
becomes large.\footnote{We use $N_2$ to denote two-volumes in two-dimensional models of quantum gravity and $V_2$ to denote
the two-volume of a spatial slice in three-dimensional quantum gravity.}
Recall that the path integral of DT quantum gravity in $D\! =\! 2$ with bare cosmological constant $\lambda$ can be written as the infinite sum
\begin{equation}
	Z(\lambda ) = \sum_{N_2} Z(N_2)\, \textrm{e}^{-\lambda N_2},
\label{z2part}
\end{equation}
which is the (discrete) Laplace transform of the partition function $Z(N_2)$ for fixed volume. For large $N_2$, $Z(N_2)$ behaves like
\begin{equation}
	Z(N_2) \sim \textrm{e}^{\lambda^c N_2} N_2^{\gamma-3} \left(1+{\mathcal O}(1/N_2)\right),
	\label{eq:fixed-n-partition}
\end{equation}
whose leading exponential growth is governed by a (non-universal) critical cosmological constant $\lambda^c\! >\! 0$ and whose
subleading power-law behavior defines the universal entropy exponent $\gamma$, which for DT quantum gravity is given by $\gamma\! =\! -1/2$.
The asymptotic functional form (\ref{eq:fixed-n-partition}) continues to hold when conformal matter of central charge $c\! <\! 1$ is added to the Euclidean
quantum gravity model, giving rise to the entropy exponents \cite{knizhnik1988fractal}
\begin{equation}
	\gamma = \frac{1}{12} \left(c-1 - \sqrt{(25-c)(1-c)}\right), 
	\label{eq:gamma_s}
\end{equation}
with $c\! =\! 0$ corresponding to the pure-gravity case. Two-dimensional CDT quantum gravity, which is not described by formula (\ref{eq:gamma_s}), 
is characterized by $\gamma\! = \! 1/2$
\cite{ambjorn1998nonperturbative}. 
Computer simulations have demonstrated that adding matter with $c\! =\! 4$ to the CDT system induces a phase transition in the 
geometry \cite{ambjorn2000crossing}, but the corresponding entropy exponent is not known.

According to \cite{jain1992worldsheet}, the distribution of so-called baby universes 
in two-dimensional Euclidean quantum gravity -- parts of a geometry that are connected to the larger bulk geometry via a thin neck -- 
depends in a simple way on the 
entropy exponent $\gamma$. This insight was used subsequently to formulate a prescription of how to extract $\gamma$ by measuring the
distribution of minimal-neck baby universes (``minbus'') in the DT ensemble with the help of Monte Carlo simulations \cite{ambjorn1993baby}. 
A minbu is a simply connected subset of disk topology
of a two-dimensional triangulation $T$, which is connected to the rest of $T$ along a loop consisting of three edges, 
which is the minimal circumference of  a neck allowed by the simplicial manifold conditions.  

As shown in \cite{jain1992worldsheet,ambjorn1993baby}, it follows from relation (\ref{eq:fixed-n-partition}) that the average number 
$\bar{n}_{N_2}(B)$ of minbus of volume $B$ (counting the number of triangles in the minbu) in a spherical triangulation of 
volume $N_2$ for sufficiently large $B$, $N_2$ behaves like 
\begin{equation}
	\label{eq:minbu-dist}
	\bar{n}_{N_2} (B) \sim (N_2-B)^{\gamma-2} B^{\gamma-2}.
\end{equation}
By measuring the distribution of minbus across a range of volumes $B$ for fixed $N_2$ in a DT ensemble and fitting the results to the function
(\ref{eq:minbu-dist}), the expected results $\gamma\! =\! -1/2$ for pure gravity and $\gamma\! =\! -1/3$ for gravity coupled to Ising spins ($c\! =\! 1/2$)
were reproduced within measuring accuracy \cite{ambjorn1993baby}. 

We have carried out a similar analysis on the spatial slices of triangulations generated by Monte Carlo simulations of 
three-dimensional CDT quantum gravity. There is no obvious reason why (\ref{eq:fixed-n-partition}) should hold for some ``effective'' fixed-volume
partition function for the two-volume $N_2\! =\! V_2$ of a single spatial slice in this three-dimensional system. However, if the number of (2,2)-simplices
drops essentially to zero \emph{and} neighboring slices decouple, as is the case in the degenerate phase, the three-dimensional partition function 
at fixed volume will depend only on $N_{31}$ (equal to the total two-volume), which makes it plausible that (\ref{eq:fixed-n-partition}) holds on 
individual spatial slices, with $N_2\! =\! V_2$. In the de Sitter phase, if the spatial geometries can be described in terms of two-dimensional DT 
quantum gravity, the minbu method will presumably also lead to $\gamma\! =\! -1/2$.

\begin{figure}[t]
	\centering
	\includegraphics[width=0.45\textwidth]{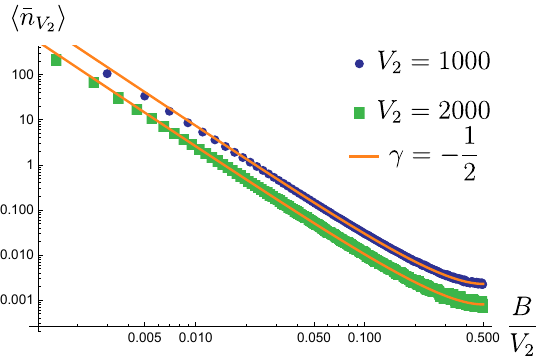}
	\\ \vspace{1cm}
	\includegraphics[width=0.45\textwidth]{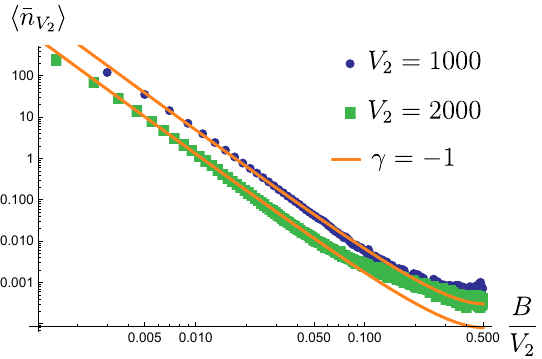}
	\hfill
	\includegraphics[width=0.45\textwidth]{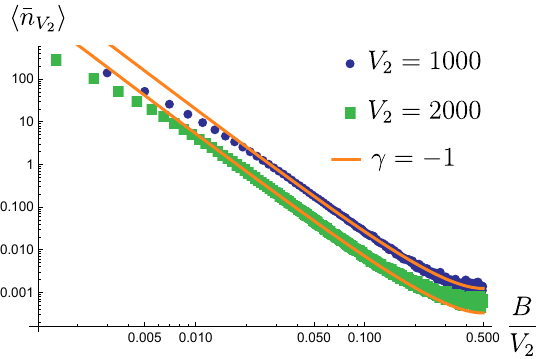}
	\caption{Expectation values of the distribution $\bar{n}_{V_2}$ of minbu sizes for spatial slices of volume $V_2\! =\! 1.000$ and $2.000$
	in three-dimensional CDT configurations, in the degenerate phase ($k_0\! =\! 8.0$, top) and the de Sitter phase ($k_0\! =\! 0.0$, bottom left,
	and $k_0\! =\! 5.0$, bottom right), on a log-log scale. The continuous lines are best fits of the form \eqref{eq:minbu-dist} for specific values of 
	$\gamma$.
	Error bars are smaller than the dot size.}
	\label{fig:minbu-dist}
\end{figure}

We measured the distribution $ \bar{n}_{V_2}(B)$ of minbu sizes $B$ for target slice volumes $\tilde{V}_2\! =\! 1.000$ and $2.000$ 
at the three phase space points $k_0 = 0.0, 5.0$ and $8.0$.
The sweep size was set to $10^4\! \cdot\! \tilde{V}_2$ for measurements in the degenerate phase, and $10^5\! \cdot\! \tilde{V}_2$ in the de Sitter phase. 
We used longer sweeps in the de Sitter phase because the observed autocorrelations were much larger, especially at $k_0\! =\! 0.0$, 
where the algorithm is much less efficient. We collected on the order of $5\! \cdot\! 10^4$ minbu size histograms for 
$\tilde{V}_2\! =\! 1.000$ and $1.5\! \cdot\! 10^5$ histograms for $\tilde{V}_2\! =\! 2.000$ in the de Sitter phase, and
$4 \cdot 10^5$ histograms for $\tilde{V}_2\! =\! 1.000$ and $2 \cdot 10^5$ histograms for $\tilde{V}_2\! =\! 2.000$ in the degenerate phase.
The resulting expectation values $\langle \bar{n}_{V_2}\rangle$ of the minbu size distribution 
as a function of the normalized ratio $B/V_2\in [0,1/2]$ are shown in Fig.\ \ref{fig:minbu-dist}, together with
best fits of the form \eqref{eq:minbu-dist} for specific values of $\gamma$. The best fits were determined following the procedure used 
in \cite{ambjorn1993baby}, and involved a subleading correction term to the power law $B^{\gamma -2}$, as is described in more detail in
Appendix \ref{app:entropy}.

In the degenerate phase (Fig.\ \ref{fig:minbu-dist}, top), the choice $\gamma\! =\! -1/2$ fits the data extremely well throughout the entire range
of $B/V_2$, with the exception of the smallest minbu sizes. Our results coincide within error bars with those reported in Table 1 
of \cite{ambjorn1993baby}. This confirms that the spatial slices in the degenerate phase exhibit behavior consistent with DT quantum gravity in two
dimensions. 

In the de Sitter phase, there is no $\gamma$-value that leads to a good fit over the full range of $B/V_2$,
even when we disregard the region of small minbu size $B$, where Eq.\ \eqref{eq:minbu-dist} is known to be inaccurate. 
The plots for $k\! =\! 0.0$ and $k_0\! =\! 5.0$ (Fig.\ \ref{fig:minbu-dist}, bottom) illustrate the optimum of what can be achieved, namely,
a fit that works reasonably well for an intermediate range of $B/V_2$, in this case, a fit corresponding to $\gamma\! =\! -1$.
However, this clearly does not fit the data for large minbus near $B/V_2\!=\! 1/2$, especially not for the smaller value of $k_0$, 
and the discrepancy seems to get worse with increasing volume $V_2$. 

We conclude that the minbu distribution in the de Sitter phase does not follow the functional form of the right-hand side of relation \eqref{eq:minbu-dist},
at least not for the slice volumes we have considered (and which seem to be sufficient in the degenerate phase). 
A possible explanation is that the ``effective'' partition function $Z(V_2)$, which is obtained from the three-dimensional CDT partition function $Z(N_3)$ for
fixed three-volume by integrating out all degrees of freedom except for a single slice volume $V_2$, does not have the asymptotic form (\ref{eq:fixed-n-partition}).
This would imply that it is not in the universality class of a DT model with central charge $c\! <\! 1$ or of CDT quantum gravity. 
A more subtle scenario would be that
$Z(V_2)$ does behave according to Eq.\ \eqref{eq:fixed-n-partition} (and perhaps does correspond to a known gravity model in $D\! =\! 2$), but that the 
derivation of the minbu distribution \eqref{eq:minbu-dist} is invalidated by the presence of correlations between the bulk and a baby universe 
that exist because of the embedding of the spatial slice in a three-dimensional simplicial geometry. Such correlations are not present in the
degenerate phase because of the absence of (2,2)-simplices. In the following two subsections, we will look at observables which also characterize the
intrinsic geometry of the spatial slices, but whose determination is less subtle than that of the entropy exponent.

\subsection{Hausdorff dimension}

The Hausdorff dimension is a notion of fractal dimension that can be used to characterize a quantum geometry in an invariant manner. Broadly speaking,
it is extracted by comparing volumes with their characteristic linear size, measured in terms of a geodesic distance. There have been many studies of
the Hausdorff dimension in the context of DT and CDT quantum gravity, including extensive investigations 
in two-dimensional Euclidean DT models with and without matter
(see, for example, \cite{ambjorn1995scaling,ambjorn1995fractal,ambjorn1997quantumb,barkley2019precision} and references therein). Following \cite{ambjorn1995scaling}, we will investigate a local and a global (``cosmological'') variant of this observable on the spatial slices. 
From analytical considerations, it is known that in two-dimensional Euclidean DT quantum gravity without matter 
both types of Hausdorff dimension are equal to four
(i.e.\ different from the topological dimension of the triangular building blocks),
while for two-dimensional CDT quantum gravity they are both equal to two \cite{ambjorn1998nonperturbative,durhuus2010spectral}.

When measuring the Hausdorff dimension numerically, one can use either the link distance or the dual link distance as a discrete implementation
of the geodesic distance. In known systems of pure gravity in $D\! =\! 2$, they lead to equivalent notions of geodesic distance in the continuum limit, but
for finite lattice sizes, one particular choice may be more convenient. This is true for our investigation below, where we will use the dual link distance, which is defined
between dual vertices (equivalently, centers of triangles) and given by the length of the shortest path along dual links between the vertices. 

When comparing our results with previous measurements of the Hausdorff dimension in the context of two-dimensional Euclidean DT \cite{catterall1995scaling,ambjorn1995fractal}, one must take into account that the latter employed a larger ensemble of geometries.
This generalized ensemble allows for local gluings of the equilateral triangles that violate the strict simplicial manifold conditions\footnote{two triangles
cannot share more than one edge, any two vertices cannot be connected by more than one edge}. When characterizing the triangulations in terms of their dual, trivalent graphs, 
the generalization consists in allowing for tadpole and self-energy insertions. It has been demonstrated to reduce finite-size effects \cite{ambjorn1995new},
and is justified by the fact that the model on the enlarged ensemble can be shown to lie in the same universality class (see e.g.\ \cite{schneider1999universality}
and references therein). Since no analogous result is available in three-dimensional CDT quantum gravity, it is prudent to use only simplicial manifolds,
which implies that the spatial slices are simplicial manifolds too. This may affect the quality of our results, compared with the earlier,
purely two-dimensional investigations.
Note also that nonlocal minbu surgery moves were used in \cite{ambjorn1995fractal} to complement the standard local Monte Carlo moves and reduce
autocorrelation times, something we cannot easily implement on two-dimensional embedded triangulations.

\subsubsection{Local Hausdorff dimension}
\label{sssec:mhd}

A key quantity in determining the local Hausdorff dimension $d_h$ of a two-dimensional triangulation $T$ is the shell volume $S(r)$,
which in our implementation counts the number of dual vertices (equivalently, triangles) at dual link distance $r$ from a given dual vertex.   
The corresponding observable is the quantity $\bar{S}(r)$, obtained by averaging $S(r)$ 
over all dual vertices of $T$. 
The reason for using the dual link distance is that shells with respect to the link distance quickly cover a large fraction of the geometry as $r$ grows.
This implies that the average shell volumes $\bar{S}(r)$ cover only a small range of radii $r$ before dropping to zero,
yielding too few data points to make reliable estimates of either Hausdorff dimension.
The local Hausdorff dimension is extracted from
the expectation value $\bar{S}(r)$ in the ensemble at fixed two-volume $N_2$ according to 
\begin{equation}
\langle \bar{S}(r)\rangle_{N_2} \sim r^{d_h -1},
\label{eq:haus-micro}
\end{equation} 
for small $r$. In other words, $d_h$ captures the initial, volume-independent power-law growth of small geodesic spherical shells, 
where $r$ must be sufficiently large
to avoid dominance by discretization artifacts and sufficiently small to avoid significant corrections to the simple power law behavior (\ref{eq:haus-micro}),
if such a behavior is indeed present.

We have measured the expectation values of average shell volumes as a function of the dual link distance $r$, 
at slice volumes $V_2\! =\! 16k$ and $32k$ and
at the three chosen phase space points $k_0$, which is all straightforward. The local Hausdorff dimension $d_h$ was extracted
by fitting the measured data to the functional form
\begin{equation}
\langle \bar{S}(r)\rangle_{V_2}	 = c \cdot (r+a)^{d_h-1},
	\label{eq:haus-micro-fit}
\end{equation}
where -- following \cite{ambjorn1995fractal} -- we have introduced an offset $a$ to account for short-distance discretization artifacts,  and
$c$ is a multiplicative parameter.
The dependence of the Hausdorff dimension on the chosen fitting range $r\!\in\! [r_{\textrm{min}},r_{\textrm{max}}]$ will be analyzed in more detail below.

\begin{figure}[t]
	\centering
	\includegraphics[width=0.45\textwidth]{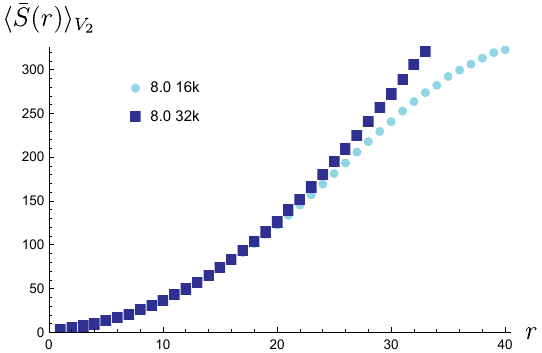}
	\hfill
	\includegraphics[width=0.45\textwidth]{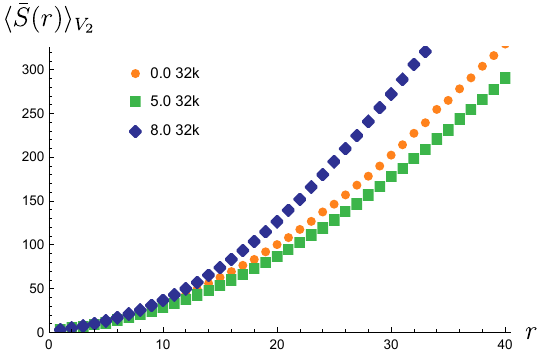}
	\caption{Expectation values $\langle \bar{S}(r)\rangle_{V_2}$ of average shell volumes in the degenerate phase $(k_0\! =\! 8.0)$ for slice volumes $V_2\! = \!16k, 32k$ (left), and for all three phase space points $k_0\! =\! 0.0$, $5.0$ and $8.0$ for slice volume $V_2\! =\! 32k$ (right). Error bars are smaller than
	dot size.}
	\label{fig:haus-micro-all}
\end{figure}

The measured expectation values of the average shell volume are shown in Fig.\ \ref{fig:haus-micro-all} for $0\! \leq\! r\! \leq\! 40$. 
The plot on the left, describing the behavior of the system in the degenerate phase, illustrates the difference between the data for slice volumes
$16k$ and $32k$. The data points for the smaller volume start deviating from the common behavior around $r\!\approx\! 20$, 
indicating that a power-law fit becomes
inadequate beyond this point. The plot on the right illustrates the dependence of the initial slope on the value of $k_0$. Note that although the curve for
$k_0\! =\! 0.0$ lies in between the two other curves, the corresponding Hausdorff dimension obtained from fitting to the functional form
\eqref{eq:haus-micro-fit} comes out lower than that for $k_0\! =\! 5.0$, see below. 

Since the criteria for fixing the fitting range $r\!\in\! [r_{\textrm{min}},r_{\textrm{max}}]$ for Eq.\ \eqref{eq:haus-micro-fit} are only approximate, it is important to understand which choice is most appropriate and how stable the results for $d_h$ are when the range is varied. 
Some earlier work used the range $r\!\in\! [5,15]$ in terms of the dual link distance, and for a system of volume $N_2\! =\! 64k$
\cite{ambjorn1998quantum}.

\begin{figure}[t]
	\centering
	\includegraphics[width=0.45\textwidth]{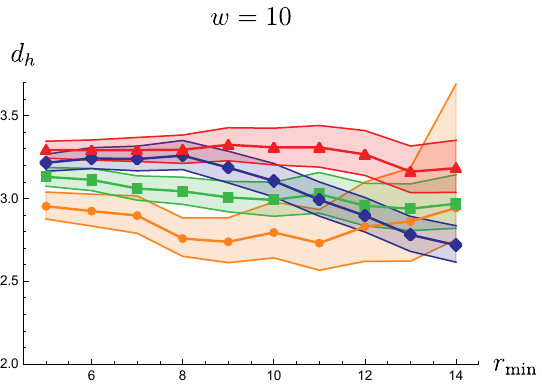}
	\hfill
	\includegraphics[width=0.45\textwidth]{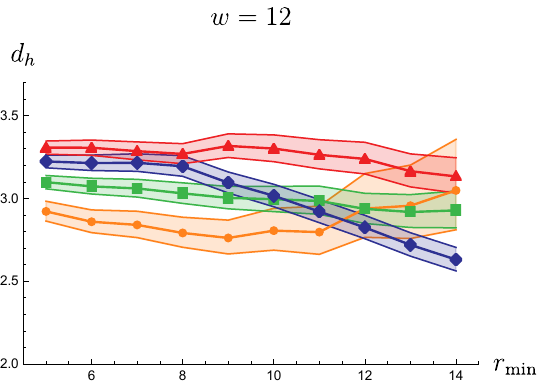}
	\\ \vspace{1em}
	\includegraphics[width=0.4\textwidth]{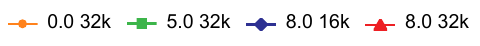}
	\caption{Best fit values for $d_h$ from fitting Eq.\ \eqref{eq:haus-micro-fit} to the measured expectation values $\langle \bar{S}(r)\rangle_{V_2}$ in the range $r\!\in\! [r_\textrm{min}, r_\textrm{min}\! +\! w]$ for spatial slices in the degenerate phase ($k_0\! =\! 8.0$, slice volumes $V_2\! =\! 16k,32k$) and in the de Sitter phase ($k_0\! =\! 0.0,5.0$, slice volume $V_2\! =\! 32k$). The error bars correspond to the 95\% confidence intervals of a $\chi^2$-test. }
	\label{fig:haus-micro-width}
\end{figure}

To investigate the influence of the fitting range in a systematic way, 
we have performed fits for a set of ranges $r\!\in\! [r_\textrm{min}, r_\textrm{min}\! +\! w]$ of varying width $w$, and with $r_\textrm{min}\!\in\! [5,14]$. 
The resulting best fit values for the local Hausdorff dimension $d_h$ as a function of $r_\textrm{min}$ are shown in Fig.\ \ref{fig:haus-micro-width},
for two different widths $w\! =\! 10$ and $12$. 

\begin{figure}[t]
	\centering
	\includegraphics[width=0.45\textwidth]{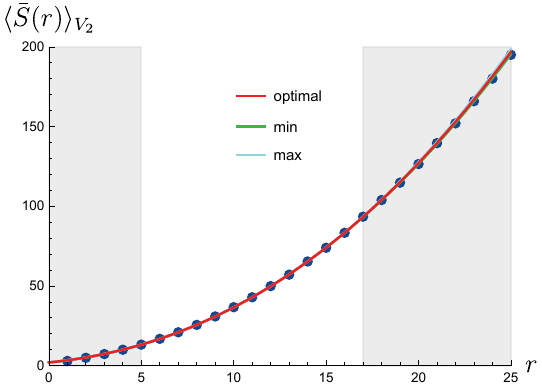}
	\caption{Expectation values $\langle \bar{S}(r)\rangle_{V_2}$ of average shell volumes in the degenerate phase ($k_0\! =\! 8.0$) and slice volume $V_2\! =\! 32k$. The curves are plots of the fit function \eqref{eq:haus-micro-fit} for three different sets of fit parameters: the optimal fit which minimizes $\chi^2$, and the two fits at the boundaries of the 95\% confidence interval. The fit is performed to the data points in the range $5 \leq r \leq 17$, as indicated by the unshaded region.}
	\label{fig:haus-micro-fit-extremes}
\end{figure}

Starting our analysis in the degenerate phase, we observe a good stability of the value of $d_h$ when $r_\textrm{min}$ is increased away from its
lowest, ``canonical'' cutoff value of 5, which indicates that the region $r_\textrm{min}\! \gtrsim\! 5$ is not affected by short-distance artifacts and 
that data in the corresponding interval $r\!\in\! [r_\textrm{min}, r_\textrm{min}\! +\! w]$ are well approximated by a pure power law. Shifting the fitting
range to start beyond $r_\textrm{min}\! =\! 7$ changes the extracted best $d_h$, mildly for the larger volume $V_2\! =\! 32k$ and more strongly for $V_2\! =\! 16k$.
Taking into account Fig.\ \ref{fig:haus-micro-all}, left, this indicates
that one is leaving the region where the functional form (\ref{eq:haus-micro-fit}) is an appropriate fit. Lastly, setting $w\! =\! 12$ instead of 10 reduces
the error bars without appreciably changing $d_h$, and is therefore preferable. To conclude, the optimal choice of range among the possibilities 
we have investigated at $k_0\! =\! 8.0$ appears to be $r\!\in\! [5,17]$. The associated local Hausdorff dimension for $V_2\! =\! 32k$ is given by
\begin{equation}
	d_h = 3.31(4), \quad \quad \textrm{degenerate phase } (k_0=8.0),
\label{dhvalue}
\end{equation}
obtained from fitting to (\ref{eq:haus-micro-fit}), with fit parameters $a\! =\! 4.0(2)$ and $c\! =\! 0.08(1)$.
To illustrate the excellent quality of the fit, Fig.\ \ref{fig:haus-micro-fit-extremes} shows the measured data together with the best-fit curve and the
fits at the edges of the 95\% confidence interval, which basically fall on top of each other 
(see Appendix \ref{app:ci} for details on how to compute such confidence intervals). The fits for the data at $k_0\! =\! 5.0$ and $k_0\! =\! 0.0$
are of similar quality. 
We postpone a discussion of the compatibility of this result with the analytical value $d_h\! =\! 4$ for DT quantum gravity to later, after having
investigated the global Hausdorff dimension.

Turning next to a discussion of the de Sitter phase, Fig.\ \ref{fig:haus-micro-width} shows the best fit values for $d_h$ we extracted from the shell volume data.
They are consistently smaller than those 
in the degenerate phase, and within the de Sitter phase decrease further with decreasing $k_0$. 
We observe only a shorter range of values $r_\textrm{min}\! \gtrsim\! 5$ where the Hausdorff dimension is reasonably stable. 
Examining the corresponding curves in Fig.\ \ref{fig:haus-micro-all}, right, 
there does not seem to be a very extended initial-growth regime, before the curves straighten out to become approximately linear.
It is possible that finite-size effects affect the data points at the upper end of the chosen ranges $[r_{\textrm{min}},r_{\textrm{max}}]$ 
and cause the observed deviations from a simple power-law behavior. Alternatively, this behavior may be a bona-fide feature of the embedded slices.
The error bars for larger $r_{\textrm{min}}$ are significantly larger
than in the degenerate phase, especially for $k_0\! =\! 0.0$. At least in part, this appears to be due to  
a statistical uncertainty of the measured shell volumes for increasing $r$.
As before, the error bars are smaller for $w=12$ than for $w=10$. 
For $k_0\! =\! 5.0$, the local Hausdorff dimension seems to decrease slightly with growing $r_{\textrm{min}}$, but in view of the width of the
95\% confidence interval this may not have any significance. 

From a best fit in the interval $r\!\in\! [5,17]$, we find for the local Hausdorff dimension
\begin{align}
	d_h &= 2.91(5), \quad \quad \textrm{de Sitter phase } (k_0=0.0), \\
	d_h &= 3.10(4), \quad \quad \textrm{de Sitter phase } (k_0=5.0),
\end{align}
for the fit parameters $a\! =\! 2.5(3)$, $c\! =\! 0.26(5)$ and $a\! =\! 3.9(3)$, $c\! =\! 0.11(2)$ respectively.
For the time being, we take note of these results and refer to Sec.\ \ref{HDresults} below for a summary and attempted interpretation of
all Hausdorff dimension measurements.

\subsubsection{Global Hausdorff dimension}

The global Hausdorff dimension of a two-dimensional triangulation describes its behavior as a whole and can again be characterized by the average volume 
$\bar{S}_{N_2}(r)$ of spherical shells of radius $r$, where we have added an explicit subscript $N_2$ to emphasis that the total volume of the triangulation will
now play an important role. Given an ensemble of geometries of volume $N_2$, a global Hausdorff dimension $d_H$ can be extracted if in the limit of large $N_2$
the eigenvalue of the distribution of shell volumes over the entire $r$-range can be described by the functional form
\begin{equation}
\langle \bar{S}_{N_2}(r)\rangle =N_2^{1-1/d_H} \mathcal{F}(x),  \quad \quad x = \frac{r}{N_2^{1/d_H}},
\label{eq:haus-global}
\end{equation}
where $\mathcal F$ is a universal function that depends on the rescaled geodesic distance $x$. 
This is known to be the case for DT quantum gravity in two dimensions, where $\mathcal F$ has been computed explicitly \cite{ambjorn1995fractal},
but there is no guarantee that the scaling law (\ref{eq:haus-global}) holds for general systems of geometries in $D\! =\! 2$. 
Even when a global Hausdorff dimension $d_H$ can
be assigned in this manner, it need not be equal to the local Hausdorff dimension $d_h$ \cite{ambjorn1995fractal,ambjorn1990summing}.

We will attempt to extract a global Hausdorff dimension for the spatial slices in three-dimensional CDT 
by performing a finite-size scaling analysis where we collect data for $\langle \bar{S}_{V_2}(r)\rangle$ for the full range of radii $r$ 
and several slice sizes $V_2$, and then try to rescale the resulting distributions according to Eq.\ \eqref{eq:haus-global}. 
If we can find a single value $d_H$ such that the rescaled distributions fall on top of each other for all volumes $V_2$, we define this $d_H$ 
to be the global Hausdorff dimension of the system.

Following \cite{ambjorn1995fractal}, we will work with the normalized shell volume distributions $n_{V_2}(r)\! :=\! \langle \bar{S}_{V_2}(r)\rangle /V_2$,
for which the scaling law \eqref{eq:haus-global} assumes the form
\begin{equation}
n_{V_2}(r) = V_2^{-1/d_H} \mathcal{F}(x),  \quad \quad x = \frac{r}{V_2^{1/d_H}}.
\label{normal}
\end{equation}
Note that the measured distributions $n_{V_2}(r)$ are functions of a discrete variable $r \in \mathbb{N}_0$. To perform a smooth rescaling, we first construct continuous functions that interpolate between these discrete values, which by slight abuse of notation we continue to call $n_{V_2}(r)$. 
Following the methodology of \cite{barkley2019precision}, we then rescale, for each system volume $V_2$ separately,
the corresponding distribution $n_{V_2}(r)$ such that it maximally overlaps with the normalized distribution $n_{V_\textrm{max}} (r)$ for the largest slice volume $V_\textrm{max}$ in the simulation, which we are using as a reference distribution.
We denote these rescaled distance profiles by $\tilde{n}_{V_2}(\tilde{r})$, where $\tilde{r}$ is a rescaled length variable. They take the form
\begin{equation}
	\tilde{n}_{V_2}(\tilde{r}_{V_2}) =  \left(\frac{V_2}{V_\textrm{max}}\right)^{1/d} n_{V_2}(\tilde{r}), \quad \quad 
	\tilde{r}_{V_2} = \left(\frac{V_2}{V_\textrm{max}}\right)^{1/d}(r+a)-a,
\label{averaging}
\end{equation}
where the two fit parameters are a rescaling dimension $d$ and a phenomenological shift $a$ \cite{ambjorn1995fractal,barkley2019precision} 
that corrects for discretization effects at small $r$, similar to the prescription (\ref{eq:haus-micro-fit}) we used for the local Hausdorff dimension.
\begin{figure}[t]
	\centering
	\includegraphics[width=0.45\textwidth]{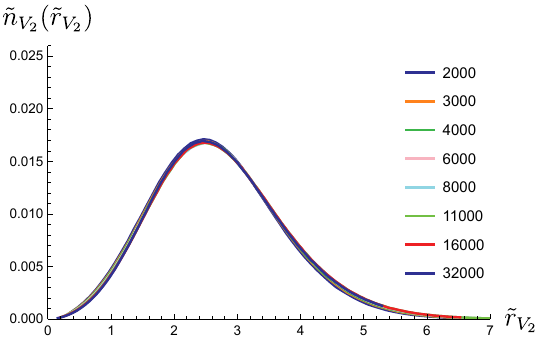} 
	\\
	\includegraphics[width=0.45\textwidth]{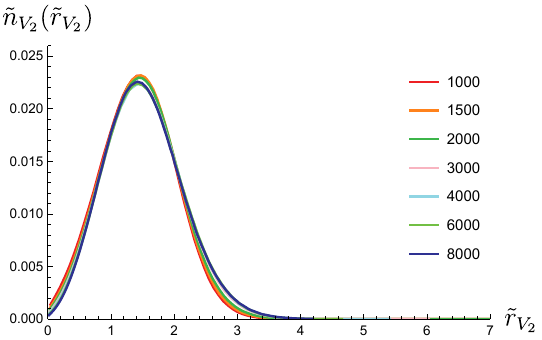}
	\hfill
	\includegraphics[width=0.45\textwidth]{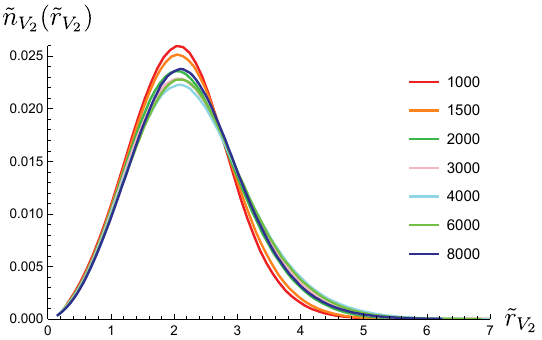}
	\caption{Distributions $\tilde{n}_{V_2}(\tilde{r}_{V_2})$ of shell volumes rescaled with the averaged parameters $(\bar{d},\bar{a})$ 
	in the degenerate phase ($k_0\! =\! 8.0$, top) and the de Sitter phase 
	($k_0\!=\! 5.0$, bottom left, and $k_0\!=\! 0.0$, bottom right).}
	\label{fig:haus-global-collapse}
\end{figure}
Note that $\tilde{r}_{V_\textrm{max}}$ is equal to the original discrete length parameter $r$, so we can use them interchangeably, 
and that $\tilde{n}_{V_\textrm{max}}\! =\! n_{V_\textrm{max}}$.  For each system size $V_2\! <\! V_\textrm{max}$ we determine the corresponding fit parameters 
$a$, $d$ from the condition that the sum of the squared differences $\left(\tilde{n}_{V_2}(\tilde{r}_{V_2})-\tilde{n}_{V_\textrm{max}}(r) \right)^2$ 
should be minimized, where $\tilde{r}_{V_2}$ depends implicitly on the discrete parameter $r$. 
We perform the fit in the range of integers $r$ where $\tilde{n}_{V_\textrm{max}}(r) > \frac{1}{5}\, \textrm{max}_{r}\,\tilde{n}_{V_\textrm{max}}(r)$, 
which means we disregard contributions from the tails of the distribution, where discretization effects are more likely to be present.
If over a large range of $V_2$ the rescaled distributions overlap to a common curve or are reasonably close to doing so, 
we take the mean $\bar{d}$ of all the rescaling dimensions $d$ and define this to be the global Hausdorff dimension, $d_H\! :=\! \bar{d}$.
We also average the shift parameters to obtain one optimal shift $\bar{a}$ for the system. 
If we do not find sufficient overlap, the method fails and we cannot assign a global Hausdorff dimension. 

We measured the expectation value of the shell volume distribution for eight different slice volumes $V_2\! \in\! [1.000,32.000]$ in the degenerate phase, and for seven slice volumes $V_2\!\in\! [1.000,8.000]$ in the de Sitter phase. Again, the autocorrelation times are much larger in the de Sitter phase, and the resulting uncertainties especially large near the peaks of the distributions. Since the location and height of these peaks are important features for finding the appropriate rescaling dimension required for a collapse, large uncertainties in this region imply large error bars for the best fit parameters. 
This led to our choice for the smaller volume range in the de Sitter phase.

\begin{figure}[t]
	\centering
	\includegraphics[width=0.45\textwidth]{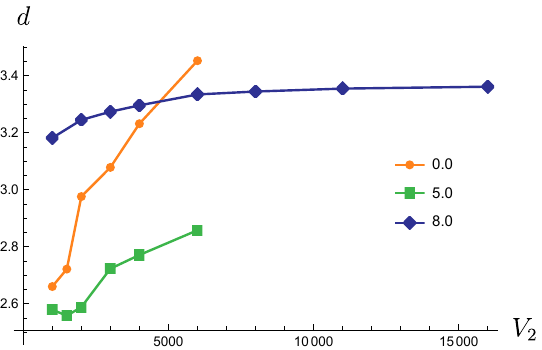}
	\caption{Values for $d$ extracted by comparing measured distributions of shell volumes at a given volume $V_2$ with those at the top volume,
	for $k_0\! =\! 0.0$, $5.0$ and $8.0$, as explained in the text.}
	\label{fig:dcomparison}
\end{figure}

For each of the three phase space points, Fig.\ \ref{fig:haus-global-collapse} shows a collection of curves for a range of volumes $V_2$, which have
been rescaled according to (\ref{averaging}), using the averaged pair $(\bar{d},\bar{a})$ for all of them.  
The topmost graph, for $k_0\! =\! 8.0$, makes a convincing case for the presence of finite-size scaling in the
degenerate phase. Using $\bar{d}\! =\! 3.30$ and $\bar{a}\! =\! 2.64$ as the joint rescaling parameters leads to a curve collapse of good, although not
perfect quality for slice volumes up to $32k$. 
This is not the case for the rescaled curves in the de Sitter phase (Fig.\ \ref{fig:haus-global-collapse}, bottom), which were obtained for
the averaged fit parameters $\bar{d}\! =\! 2.68$ and $\bar{a}\!=\! -0.46$ at $k_0\! =\! 5.0$ and $\bar{d}\! =\! 3.02$ and $\bar{a}\! =\! 2.08$ at $k_0\! =\! 0.0$. 
Although the slice volumes span a significantly smaller range than in the degenerate phase, our rescaled data do not support the presence of finite-size scaling
in this phase. In addition to the mismatch among the curves, we also note that the distributions in the two phases look different as a function of
the rescaled radius $x\! =\! r/V_2^{1/\bar{d}}$. While in the degenerate phase its range extends to around 6.7, and the peak is located near 2.4, 
the slice geometries in the de Sitter phase have a smaller linear extension, with $x$ reaching on the order of 4 (5.7) and the peak located near 1.4 (2.0) 
for $k_0\! =\! 5.0$ (0.0).
One could be tempted to disregard the lack of overlap between the curves for different volumes and simply \emph{define} the global Hausdorff 
dimension to be equal to the average $\bar{d}\! =\! 2.68$ for $k_0\! =\! 5.0$ and $\bar{d}\! =\! 3.02$ for $k_0\! =\! 0.0$. 
However, Fig.\ \ref{fig:dcomparison} shows that this would be misguided, since the values for $d$ 
exhibit a strong dependence on the volume in the range we have investigated. Especially the curve for $k_0\! =\! 0.0$ shows a steep rise, with
little indication of asymptoting to a constant value, very different from the curve for the degenerate phase, which we have included for comparison. 
Note also that the $d$-values extracted from measurements in the de Sitter phase are not in any obvious way related to the values we found for the
local Hausdorff dimension, namely, $d_h\! =\! 3.10(4)$ for $k_0\! =\! 5.0$ and $d_h\! =\! 2.91(5)$ for $k_0\! =\! 0.0$.          
This appears to be yet another indication that the behavior of the shell distributions is not governed by a single scale and that the scaling hypothesis
(\ref{eq:haus-global}) is simply not valid in the de Sitter phase. 

Based on these observations, we are unable to associate a global Hausdorff dimension with the phase space points in the de Sitter phase.
By contrast, finite-size scaling is observed in the degenerate phase, and the associated global Hausdorff dimension is given by  
\begin{equation}
	d_H = 3.30(2), \quad \quad \textrm{degenerate phase } (k_0 = 8.0),
\end{equation}
which within error margins coincides with the local Hausdorff dimension $d_h\! =\! 3.31(4)$ 
of Eq.\ \eqref{dhvalue} we determined in the previous subsection.  

\subsubsection{Discussion of results}
\label{HDresults}

Let us consider first the results we obtained in the degenerate phase. We found clear evidence for the presence of finite-size scaling when analyzing
the global Hausdorff dimension, and mutually consistent values for both Hausdorff dimensions, with $d_h\! =\! 3.31(4)$ for the local and $d_H\! =\! 3.30(2)$ 
for the global variant. Despite the large discrepancy with the analytical value $d_h\! =\! d_H \! =\! 4$ for two-dimensional DT quantum gravity, we
nevertheless believe that the observed Hausdorff dimension of the spatial slices is compatible with this model of quantum gravity. 
This interpretation is supported by known difficulties in numerically extracting the Hausdorff dimension in two-dimensional systems 
of random geometry (see \cite{loll2015locally} and references therein), with
a tendency of the Hausdorff dimension measurements to underestimate its true value.\footnote{More precisely, this is true for DT models; 
numerical measurements of the Hausdorff dimension for two-dimensional CDT found $d_H\!\approx\! 2$ \cite{ambjorn1999new} and $d_H\! =\! 2.2(2)$ \cite{loll2015locally}.} 
In previous works, these difficulties have motivated the use of a
more general geometric ensemble, the introduction of additional minbu moves and of a phenomenological shift parameter, as well as refined fitting
techniques \cite{ambjorn1995fractal,catterall1995scaling,barkley2019precision}. 
As mentioned earlier, several of these improvements are unfortunately not directly applicable in our case, because our two-dimensional geometries 
are parts of larger, three-dimensional triangulations. 

Earlier numerical results for pure DT quantum gravity whose derivation most closely
resembles our treatment are the finite-size scalings obtained for $N_2\!\leq\! 32k$ from collapsing curves for the shell volume distributions in terms of the 
dual link distance in \cite{catterall1995scaling}. Depending on the fitting 
method, they yielded the values $d_H\! =\! 3.150(31)$ and $d_H\! =\! 3.411(89)$. Unlike ours, this work used a generalized ensemble, but the
final results for the Hausdorff dimension are broadly in line with our findings. Another aspect well illustrated by \cite{catterall1995scaling} is the
increase of the measured Hausdorff dimension with the system volume, something we also observed in the degenerate phase (Fig.\ \ref{fig:haus-micro-width}).

Our results in the de Sitter phase are less clear-cut, but point to a system that is in a different universality class from DT quantum gravity. 
The values we determined for the local Hausdorff dimensions, $d_h\! =\! 3.10(4)$ for $k_0\! =\! 5.0$ and  $d_h\! =\! 2.91(5)$ for $k_0\! =\! 0.0$,
are even further removed from 4, but this might conceivably still be due to some even more serious underestimate than in the degenerate phase.
More significant is the absence of finite-size scaling and the ensuing impossibility to associate a consistent global Hausdorff dimension to the
system. The most likely explanation is the absence of a single scale governing the dynamics, which would imply a different universality class from
that of the degenerate phase. 
Discarding an interpretation of the de Sitter phase in terms of DT quantum gravity leaves us without any obvious alternative candidate theory to explain these results. 
There are a number of two-dimensional CDT models with matter coupling that have a local and/or global Hausdorff dimension of or near 3,
including CDT with eight Ising spins \cite{ambjorn2000crossing}, several massless scalar fields \cite{ambjorn2012pseudotopological},
or restricted hard dimers \cite{ambjorn2014restricted}. Also two pure-gravity models of so-called locally causal dynamical triangulations,
generalizing the strict slicing of two-dimensional CDT, were found to have Hausdorff dimensions near 3 \cite{loll2015locally}. 
There is nothing obvious that would link one of these models to the Euclidean slices
we are dealing with here, but we cannot exclude this possibility either without a further investigation, which however would take us beyond the scope of the
present chapter.

\subsection{Curvature profile}

\begin{figure}[t]
	\centering
	\includegraphics[width=0.65\textwidth]{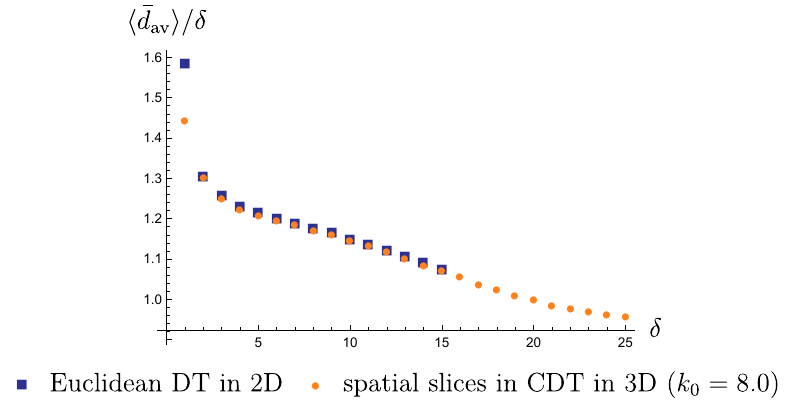}
	\caption{Expectation value $\langle \dbav / \delta \rangle$ of the normalized average sphere distance, measured in two-dimensional 
	DT quantum gravity (blue squares) \cite{klitgaard2018implementing}, 
	and for the spatial slices of three-dimensional CDT quantum gravity in the degenerate phase (orange dots), both for volume $V_2\! =\! 60k$. 
	(Error bars are smaller than dot size.)}
	\label{fig:ricci-60k}
\end{figure}

The last quantity we studied on the spatial hypersurfaces of three-dimensional CDT is the curvature profile, which we defined in Chapter \ref{ch:curv-profs} of this thesis. We measured the expectation value
$\langle \dbav / \delta \rangle$ of the curvature profile of the spatial slices at the phase space points $k_0\! =\! 0.0$, 5.0 and 8.0. 
We used slice volumes in the range $V_2\!\in\! [4k, 60k]$ in the degenerate phase and $V_2\!\in\! [4k,20k]$ in the de Sitter phase.
For the volumes $V_2\! =\! 20, 30, 40$ and $60k$, we compared the curvature profiles in the degenerate phase at $k_0\! =\! 8.0$ to those of
two-dimensional DT quantum gravity \cite{klitgaard2018implementing}, and in each case found agreement within statistical error bars \footnote{We thank N.\ Klitgaard for making the original data available to us.}. 
For illustration, the results for $V_2\! =\! 60k$ are shown in Fig.\ \ref{fig:ricci-60k}.
There is an excellent match of our present data, taken for $\delta\! \in\! [1, 25]$, with those of DT quantum gravity in the range where they overlap,
except at $\delta\! =\! 1$. The fact that this agreement extends even into most of the region of discretization artifacts at small $\delta$ 
provides additional evidence that the spatial hypersurfaces in the degenerate phase behave like two-dimensional DT geometries. 

The monotonically decreasing curvature profiles we found in both the degenerate and the de Sitter phase clearly indicate the presence of 
positive curvature, as can be seen from Eq.\ \eqref{cp-profile}. 
Motivated by the fact that curvature profiles of two-dimensional DT quantum gravity can best be fitted to those of a five-dimensional continuum sphere
with some effective curvature radius $\rho_\textrm{eff}$ \cite{klitgaard2018implementing}, we tried to do the same for our data. 
As expected from the match of the curvature profiles, our results for the effective curvature radii in the degenerate phase are close 
to the values listed in Table 1 of \cite{klitgaard2018implementing}. Note that a rough estimate for the onset of 
finite-size effects in measuring $\langle \dbav / \delta \rangle$ on a sphere of curvature radius $\rho$ is $\delta\!\approx\rho$, where 
the extension $3\delta$ of the double circle of Fig.\ \ref{cp-fig:avg-sphere-dist} is 
approximately equal to $\pi\rho$, half of the circumference of the sphere. This is in good agreement with the findings 
in \cite{klitgaard2018implementing}.

\begin{figure}[t]
	\centering
	\includegraphics[width=0.45\textwidth]{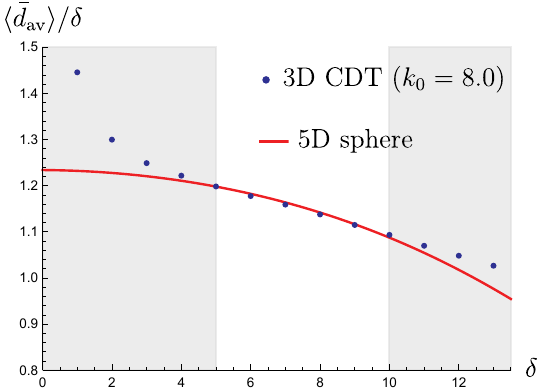}
    \hfill
    \includegraphics[width=0.45\textwidth]{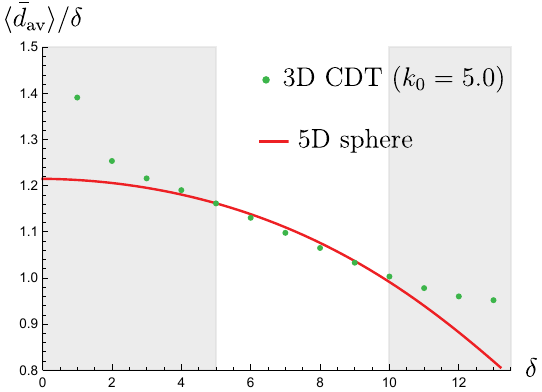}
	\caption{Comparing measured curvature profiles in the range $\delta\! \in\! [5, 10]$ with those 
	of five-dimensional continuum spheres, for slice volume $V_2\! =\! 20k$. Left: fit to a sphere with $\rho_\textrm{eff}\! =\! 13.5$ in the degenerate phase 
	($k_0\! =\! 8.0$). Right: fit to a sphere with $\rho_\textrm{eff}\! =\!11.1$ in the de Sitter phase ($k_0\! =\! 5.0$).}
	\label{fig:curvfits}
\end{figure}

Consistent with this argument, we found that for $V_2\! =\! 60k$, a fitting range $\delta\!\in\! [5,15]$ is appropriate,
while for the smaller slice volume $V_2\! =\! 20k$, which is the maximal size available for measurements in the de Sitter phase, 
the smaller range $\delta\!\in\! [5,10]$ should be used. 
The measured curvature profile for $V_2\! =\! 20k$ at $k_0\! =\! 5.0$ in the de Sitter phase is shown in Fig.\ \ref{fig:curvfits}, 
where for comparison we present it alongside 
the result for the same slice volume at $k_0\! =\! 8.0$ in the degenerate phase. 
The continuous lines are best fits to a 5D continuum sphere. Following \cite{klitgaard2018implementing}, an additive shift was used such that the 
data point at $\delta\!=\!5$ always lies on the continuum curve. 
Because of the small fitting range we cannot and do not claim that the data taken at this (or even smaller) volume represent convincing evidence for
the curvature behavior of a sphere, and the effective curvature radii extracted ($\rho_\textrm{eff}\! =\!13.5$ for $k_0\! =\! 8.0$, 
$\rho_\textrm{eff}\! =\!11.1$ for $k_0\! =\! 5.0$) should be taken with a large grain of salt.\footnote{The data at $k_0\! =\! 0.0$ are somewhat similar
to those at $k_0\! =\! 5.0$, but their quality is even worse, and we do not show them here.}   
In the degenerate phase we can say a bit more, as we have seen, since we can go up to volume $V_2\! =\! 60k$ and essentially reproduce the
results of DT quantum gravity. With regard to the de Sitter phase, we can at this stage only conclude that the curvature profiles are not in contradiction 
with those of a 5D continuum sphere, but it is clear that the quality of our results is not sufficient to definitely say that it is a sphere, let alone determine
its dimension and curvature radius reliably. This would require us to probe much larger systems, which especially in the de Sitter phase was not feasible
in our set-up.

\section{Summary and conclusion}
\label{sec:disc}

We set out to gain a more detailed understanding of the properties of three-dimensional CDT quantum gravity by studying the intrinsic geometric properties of its
spatial slices at integer proper time. We worked with the ``classic'' ensemble of three-dimensional simplicial manifolds, 
which implies that the spatial hypermanifolds under
consideration also satisfy manifold conditions, and can be characterized by dual trivalent graphs without tadpoles or self-energy insertions.  
The original work on this quantum gravity model found two distinct phases on either side of a first-order transition as a function of the bare inverse
gravitational coupling $k_0$ \cite{ambjorn2001nonperturbative}: a de Sitter phase of extended geometry for $k_0\! <\! k_0^c$ and a degenerate phase
for $k_0\! >\! k_0^c$, characterized by a strongly fluctuating volume profile, the almost complete absence of (2,2)-tetrahedra and an approximate
decoupling of nearby spatial slices. 

We investigated the intrinsic geometric properties of the spatial slices in both phases, well away from the critical coupling $k_0^c\!\approx\! 6.24$,
at $k_0\! =\! 0.0$ and $5.0$ in the de Sitter and at $k_0\! =\! 8.0$ in the degenerate phase. The quantities considered were the expectation values
of the average coordination number of vertices in the slices, of the entropy exponent extracted from the distribution of minimal-neck baby universes, of the local 
and global Hausdorff dimension, and of the curvature profile, obtained by averaging the quantum Ricci curvature. They are for the most part well studied in
two-dimensional DT and CDT quantum gravity, the primary systems of reference for our results on the spatial slices. 
One aim of our investigation, motivated by 
the observed decoupling behavior of the spatial slices, was to verify that the behavior of the slices in the degenerate phase lies in the same
universality class as DT quantum gravity in $D\! =\! 2$. What happens in the de Sitter phase, and to what extent the embedding three-dimensional 
geometry influences the effective dynamics of the hypersurfaces in this phase is much less clear a priori.

Summarizing our results, we found convincing evidence that the behavior of the spatial slices in the degenerate phase is indeed compatible with that of
two-dimensional DT quantum gravity. The measured distribution of the vertex order follows the analytical prediction almost perfectly 
(Fig.\ \ref{fig:cdt-coord-results-nonlog}). The same is true for the distribution of (sufficiently large) minbu sizes (Fig.\ \ref{fig:minbu-dist}, top), yielding
an entropy exponent $\gamma\! =\! -1/2$, the known value for pure Euclidean gravity in $D\! =\! 2$. Measurement of the local and global Hausdorff dimension yielded mutually compatible results, with $d_h\! =\! 3.31(4)$ and $d_H\! =\! 3.30(2)$, exhibiting finite-size scaling for the latter. 
We argued that the discrepancy between the measured values and the analytically known value $d_h\!\equiv\! d_H\! =\! 4$ is in line with expectations for
a simplicial manifold ensemble and the relatively small volumes under consideration here. Finally, the measured curvature profile matched very well that of a
previous investigation of DT quantum gravity in $D\! =\! 2$ (Fig.\ \ref{fig:ricci-60k}), at least up to the slice volume $V_2\! =\! 60k$ we could investigate.

By contrast, in the de Sitter phase we could not establish a corresponding overall match of the behavior of the measured observables with that of any 
known quantum gravity 
model in two dimensions. In particular, we did not find any evidence that the spatial slices behave according to DT quantum gravity in $D\! =\! 2$,
which one may argue is the most natural hypothesis, given the absence in the slices of a preferred direction or a time-space asymmetry, 
which characterizes two-dimensional
CDT configurations. We also found that the behavior within the de Sitter phase depends on the value of the bare coupling constant $k_0$. 
Since it was tangential to our main focus, we did not study the nature of this $k_0$-dependence more closely, which may be a worthwhile project in itself. 
Within the limited range of couplings $k_0\!\in\! [3.0,6.0]$, earlier work found some evidence that 
results inside the de Sitter phase can be mapped onto each other by a $k_0$-dependent rescaling of the time- and space-like length units \cite{ambjorn2001nonperturbative}. It would be interesting to understand whether this also extends to the value $k_0\! =\! 0$ we have been using,
or even to negative $k_0$.

Returning to the specifics of our results, the vertex order in the de Sitter phase was found to obey a very different distribution from that in the degenerate phase, with large coordination numbers being more prevalent (Fig.\ \ref{fig:cdt-coord-results}). We also saw that the distribution for $k_0\! =\! 5.0$ is even
further removed from that for $k_0\! =\! 8.0$ than the distribution for $k_0\! =\! 0.0$.
The method of determining the entropy exponent $\gamma$ from the distribution of minbu sizes does not appear to be applicable in the de Sitter phase,
which we conjectured to be due to the presence of correlations in the three-dimensional embedding triangulations. Although this does not 
necessarily disprove a DT-like behavior of the spatial slices, it does not present any evidence in favour of it either. The measured local Hausdorff
dimensions, given by $d_h\! =\! 3.10(4)$ for $k_0\! =\! 5.0$ and $d_h\! =\! 2.91(5)$ for $k_0\! =\! 0.0$ are significantly smaller than the value found in the
degenerate phase, and point to a different continuum limit than that of two-dimensional DT quantum gravity. 
Of course, it could be the case that the de Sitter phase is subject to much larger discretization artifacts, because of the genuinely three-dimensional 
nature of the underlying geometries, and that one needs to go to larger slice volumes to get a better approximation of continuum behavior. 
However, even taking this possibility into account, a yet stronger indication that
the slices do not exhibit DT-like behavior comes from the absence of finite-size scaling at fixed $k_0$ to extract a global Hausdorff dimension, from
which we deduced that the slice dynamics is likely governed by more than just one scale. 
Finally, the measurement of the curvature profiles showed the presence of a positive average quantum Ricci scalar. Matching with
a continuum sphere was in principle possible, but should at this stage be regarded as inconclusive, since it was based on only a handful of measurement 
points, which could be compatible with other curvature profiles also. It therefore cannot serve as evidence that the system is equivalent to
two-dimensional DT quantum gravity. 

Having largely dismissed an interpretation of the spatial slices in the de Sitter phase in terms of two-dimensional DT or CDT quantum gravity does not leave 
any obvious alternatives to associate them with (the universality classes of) other known systems of random geometry. 
DT gravity coupled to matter with a conformal charge $c\! <\! 1$ is disfavoured,
because the apparent absence of finite-size scaling for the global Hausdorff dimension of the spatial slices contradicts the scale-invariance of these systems.   
On the one hand, the quality of our data is far removed from the precision measurements of the Hausdorff dimension of such 
systems \cite{barkley2019precision}, 
and we cannot entirely exclude that finite-size scaling will appear at much larger volumes than the ones we could probe here.
On the other hand, we would urge caution when comparing to ensembles with less stringent regularity conditions, like those used in \cite{barkley2019precision}: even if the relaxation of simplicial
manifold conditions does not change the universality class in specific two-dimensional models, this may not hold in general 
in three-dimensional quantum gravity models, and may depend on the details of the regularity conditions.
This is part of a more general question, namely, are there natural larger ensembles of three-dimensional triangulations which contain the 
simplicial manifold ensemble of CDT, but lie in the same universality class? Conversely, are there strictly smaller ensembles contained in that
of standard CDT quantum gravity, which still belong to the same universality class? Larger ensembles may facilitate
numerical simulations and lead to faster convergence, while smaller ensembles may be easier to enumerate and handle analytically (see \cite{budd2022family} for a recent example in three-dimensional DT quantum gravity). 
Of course, there may be more than one physically interesting universality class associated with three-dimensional Lorentzian random geometries, 
like the wormhole phase described by the ABAB-matrix model \cite{ambjorn2001lorentzian} already mentioned in the introduction.

Returning to ensembles of simplicial manifolds, our results in the de Sitter phase may indicate the existence of another, new model of
two-dimensional quantum geometry, where the embedding three-dimensional geometry induces some effective dynamics on the spatial slices,
presumably through $k_0$-dependent extrinsic curvature contributions. Further research is needed to understand whether such an induced
model exists and whether it can in turn be interpreted as a two-dimensional quantum field theory with properties like locality and unitarity, as is
the case in the degenerate phase.
Contrasting our relatively straightforward verification of the DT nature of the spatial dynamics in the degenerate phase with
the difficulties we encountered when investigating the de Sitter phase highlights 
the fact that quantum geometry in three dimensions -- here by leaving its imprint on the hypersurfaces -- is significantly more complex and 
complicated than quantum geometry in two dimensions. Much remains to be done to illuminate its mathematical and physical properties. 


\part{Techniques for random geometry}
\label{part:mat}
\chapter{One-matrix differential reformulation of two-matrix models}\label{ch:diff-mat}

\section{Introduction}
Two-matrix models have been widely studied in the literature since their introduction by Itzykson and Zuber \cite{itzykson1980planar}, see for instance \cite{ Mehta:1981jm, eynard1997eigenvalue, eynard2003large,Bertola:2003cn,Bertola:2007is, Duits:2012vg}. Their Feynman expansions count ribbon graphs with two kinds of vertices.
They match, for different choices of potentials, the partition functions of some statistical models such as the Ising model \cite{KAZAKOV-Ising}, colored triangulations \cite{DiFrancesco:1998bp}, hard particles \cite{Bouttier_2002} etc. on 
random surfaces (thereby relating to 2D quantum gravity). 

In quantum field theory, differential formulations such as
\be
\label{dm-eq:central-identity}
\begin{split}
\frac{\cZ}{\cZ_0} &= \frac{1}{\cZ_0} \int \D\phi \, \e^{- \int \d^d x \, \d^d y\,\frac 1 2 \phi(x)  K(x,y)  \phi(y) - \int \d^d x \, V\left(\phi(x)\right)}\\ &= \left[\e^{- \int \d^d w \, V\left(\frac{\pd}{\pd {J(w)}}\right) } \e^{\int \d^d x\, \d^d y\,  J(x) K^{-1}(x,y)  J(y)}\right]_{J=0}\,,
\end{split}
\ee
are widely used (Zee coins it the ``central identity of quantum field theory'' in his textbook \cite{zee2010quantum}).
Here the $\phi$ are fields on a $D$-dimensional spacetime, $\int\D\phi$ is a functional integration, the interaction potential $V$ is a polynomial in the fields, $K$ is the propagator, and $\cZ_0$ is a normalization.   
Such formulations allow expressing Gaussian expectation values and Wick's formula, the key ingredient of perturbative expansions, as well-defined {algebraic} expressions that do not suffer the potential divergences of the combinatorially equivalent integrals (see for instance \cite{salmhofer1999renormalization, bauerschmidt2019introduction, gurau2015analyticity,gurau2014renormalization}). They moreover provide a practical tool for computing Feynman diagrams. 

In particular, the differential formulation provides well-defined expressions for the Gaussian expectation values of \emph{two-matrix models}, while the integrals are not well-defined and only understood formally through Wick's theorem.

Remarkably, as we will show in this chapter, two-matrix models can be expressed in terms of a single matrix in a differential formulation:
the two matrices necessary to control the interaction between the two kinds of vertices (or equivalently impose bipartiteness of the ribbon graphs) 
are no longer needed in the differential formulations. Since the necessity of using two matrices complicates the resolution of two-matrix models, the existence of a one-matrix differential formulation raises the question whether this allows for a simpler resolution.

With the help of the one-matrix differential reformulation, it is possible to diagonalize the two-matrix models directly in the differential formulation.
The steps leading to the expression of the partition function as a Slater determinant and thereby to the resolution by biorthogonal polynomials involve elegant and interesting equations whose equivalence is not always manifest. 
We detail their relations and take the opportunity to review and clarify some aspects of the differential formulation that can be found in the matrix-model literature, as well as the formula needed to diagonalize the differential operators with respect to matrix elements for Hermitian matrices, implicitly shown by Itzykson and Zuber in \cite{itzykson1980planar} and stated by Zuber in~\cite{zuber2008large}. These computations do not seem to indicate any simplification in the resolution of two-matrix models in differential formulation as initially hoped: 
the role played by the two matrices is now played by the matrix and the derivatives with respect to the matrix, leading to the same resolution by biorthogonal polynomials.

On the other hand, the analysis of the equivalence between the different formulas for one-matrix models does provide new insights on two-matrix models. Indeed, the proof of equivalence between two of the diagonalized differential formulations of the one-matrix models involves transforming certain derivatives to variables in the Slater determinant formulation. The computation is not trivial and is detailed in the last section \ref{dm-sec:slater}. The same computation can 
be applied with little modifications to two-matrix models, and leads to a new determinant formulation of their partition functions. 
Building on this, we comment on a potential new resolution using orthogonal polynomials instead of biorthogonal polynomials.

In order to show that the different expressions of the partition functions indeed coincide, we verify that all the prefactors and normalizations match. These constants are often omitted in the literature as they are sometimes tedious to compute and do not contribute to the critical behaviors or the combinatorial interpretations.
An advantage of the computations in differential formulations is that they most often do not need an overall normalization, as seen for instance in \eqref{dm-eq:central-identity}.


\

\begin{figure}[t!]
\centering
\begin{align}
\resizebox{\textwidth}{!}{$
    \xymatrix@C-1.5pc{ 
    &&& 
	{\begin{array}{c} \frac1{\vd}\e^{\pd^2}\vd\e^{V} \\ \eqref{dm-eq:diff-diagonal-new-herm} \end{array}}
    \ar@{-}[rr]^(.44){\phi}\ar@{-}[dd]|\hole^(.36){\kappa} && 
	{\diagblock{\frac{1}{\vd}\det \e^{\d^2} x_i^{j}\e^{V}}{dm-eq:Slater-first-expr}}  \ar@{-}[rr]^{\sigma} 
	\ar@{-}[dd]|\hole^(.36){\kappa} &&
	{\diagblock{\det\d^{i} \e^{\d^{2}} x^{j} \e^{V}}{dm-eq:Slater-diff-one-mat}} \ar@{-}[dd]^{\kappa}
	\\
	{\diagblock{\e^{\pd_M^2} \e^{V^M}}{dm-eq:diff-one-mat-2}} \ar@{-}[dd]^{\alpha} \ar@{->}[rrru]^{\eta} &&
	{\diagblock{\e^{\pd^2}\vd^2\e^{V}}{dm-eq:diff-diagonal-usual-herm}} \ar@{-}[rrrr]^{\phi}\ar@{-}[dd]^{\alpha}  &&&& 
	{\diagblock{\det\e^{\d^2}x^{i+j}\e^{V}}{dm-eq:Slater-diff-one-mat-bis}}\ar@{-}[ur] ^{\lambda} \ar@{-}[dd]^(.34){\alpha}
	\\
    &&& 
    {\diagblock{\frac1{\vd} \int K \vd \e^V}{dm-eq:diff-diagonal-new-herm-heat}} \ar@{-}[rr]^(.44){\phi} && {\diagblock{\frac1{\vd}\det\int K_i y^j \e^V} {dm-eq:Slater-first-expr-heat}} \ar@{-}[rr]|\hole^(.39){\sigma}
    && {\diagblock{\det\int \partial^i_x K y^{j} \e^{V^y}}{dm-eq:C-second-onematrix}}
    \\
	{\diagblock{\int_M\e^{-S}}{dm-eq:one-mat}} \ar@{->}[rr]^{\delta} &&
	{\diagblock{\int_\x\vd^2\e^{-S}}{dm-eq:diagonal-hermitian}} 
	\ar@{-}[rrrr]^{\phi} &&&& 
	{\diagblock{\det\!\int  \e^{y^2} y^{i+j}\e^V}{dm-eq:slater-1-mat-usual}}  \ar@{-}[ur]
	} \nonumber
	$}
\end{align}
\normalsize
\caption{Diagrammatic overview of the various expressions for one-matrix models.}
\label{dm-fig:diagram-one-matrix}
\end{figure}
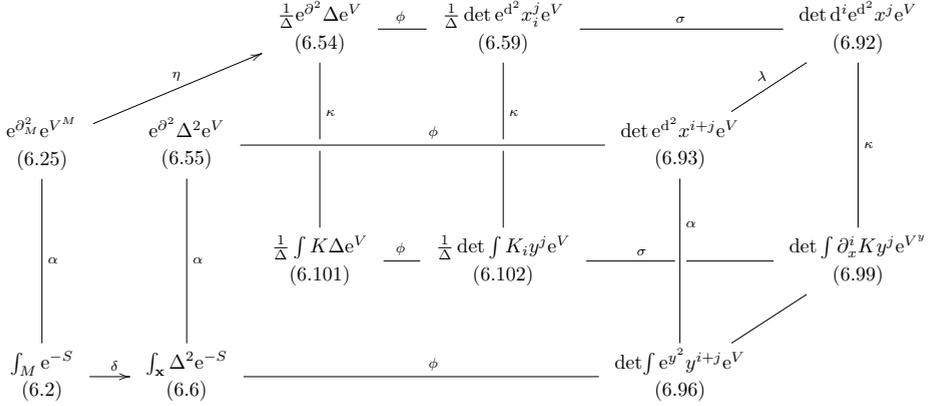

\begin{figure}[t!]
\centering
\begin{align}
\resizebox{\textwidth}{!}{$
    \xymatrix@C-1.5pc{
    &&&&&& {\diagblock{\det \e^{V_1^d} x^r \d^s x^j \e^{V_2}}{dm-eq:general-new-expression-p-s-diff}} \ar@{-}[d]^{\lambda}
    \\
	& {\diagblock{\e^{V_1^{\partial_M}} \e^{V_2^M}}{dm-eq:diff-1-2-model}} \ar@{->}[r]^{\eta}
	& {\diagblock{\frac{1}{\vd} \e^{V_1^\partial} \vd \e^{V_2}}{dm-eq:diff-diagonal-new}}
	\ar@{-}[rr]^(.46){\phi}\ar@{-}[dd]|\hole^(.36){\kappa} 
	&& {\diagblock{\frac{1}{\vd} \det \e^{V_1^\d} x_i^{j} \e^{V_2}}{dm-eq:Slater-first-expr}} \ar@{-}[rr]^{\sigma} \ar@{-}[dd]|\hole^(.36){\kappa} 
	&& {\diagblock{\det \d^i \e^{V_1^\d} x^j \e^{V_2}}{dm-eq:Slater-diff}}  \ar@{-}[dd]^{\kappa}
	\\
	{\diagblock{\e^{\partial_A \partial_B} \e^{V_{1,2}^{A,B}}}{dm-eq:Two-mat-diff-form}} \ar@{-}[dd]^{\alpha}\ar@{-}[ur]^{\beta}
	& {\diagblock{\e^{\partial_\ba \partial_\bb} \vd_\ba \vd_\bb \e^{V_{1,2}^{\ba, \bb}}}{dm-eq:diff-of-diag-two-mat}} \ar@{-}[rrrr]^{\phi}\ar@{-}[dd]^{\alpha} 
	&&&& {\diagblock{\det \e^{\partial_a \partial_b} a^i b^j \e^{V_{1,2}} }{dm-eq:diff-of-diag-two-mat-dev}}\ar@{-}[dd]^(.36){\alpha} \ar@{-}[ur]^{\beta}
	\\
	&& {\diagblock{\frac{1}{\vd} \int K_{V_1} \vd \e^{V_2}}{dm-eq:app-c1}} \ar@{-}[rr]^(.46){\phi} 
	&& {\diagblock{\frac{1}{\vd} \det \int K_{V_1}^i\,  y^j \e^{V_2}}{dm-eq:app-c2}}	\ar@{-}[rr]^(.39){\sigma}|\hole 
	&& {\diagblock{\det \int \partial_x^i K_{V_1^x} y^j \e^{V_2^y}}{dm-eq:app-c3}}
	\\
	{\diagblock{\int_{A,B}\e^{-S}}{dm-eq:2-matrix_model}} \ar@{->}[r]^(.46){\delta}
	& {\diagblock{\int\vd_\ba \vd_\bb \e^{-S}}{dm-eq:diag:2-mat}} \ar@{-}[rrrr]^{\phi} 
	&&&& 
	{\diagblock{\det \int \e^{xy} y^i e^{V_1^y} x^j \e^{V_{2}^{x}}}{dm-eq:slater-2-mat-usual}} \ar@{-}[ur] \ar@{-}[d]^{\tilde \lambda}
	\\
	&&&&& {\diagblock{\det \int \e^{xy}\e^{V_1^y}x^r \d^s_x x^j \e^{V_2^x}}{dm-eq:general-new-expression-p-s-int}}
	} \nonumber
	$}
\end{align}
\caption{Diagrammatic overview of the various expressions for two-matrix models.}
\label{dm-fig:diagram-two-matrix}
\end{figure}
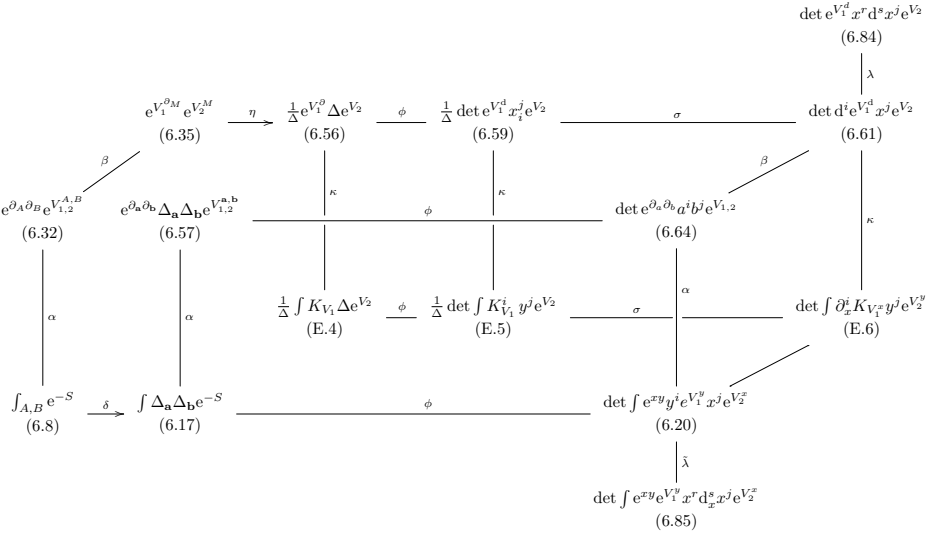

For the benefit of the reader, we first present in Fig.\ \ref{dm-fig:diagram-one-matrix} and Fig.\ \ref{dm-fig:diagram-two-matrix} a diagrammatic overview of all the different representations of the one- and two-matrix models that we consider in this chapter, with lines connecting two representations when we have a direct computation for translating between them. The expressions are shown in a schematic way, omitting normalization terms, traces, factors of $N$, and the like. The full form of each expression can be found in the main text of the chapter. Function arguments and integration variables are omitted where possible. If confusion may arise, the arguments of the potential functions $V$ (for one-matrix models) and $V_1, V_2$ (for two-matrix models) are indicated by superscripts. The action is understood to be $S=N\, \tr(\frac{1}{2}M^2-V(M))$ for one-matrix models, and $S=N\, \tr(AB-V_1(A)-V_2(B))$ for two-matrix models. Elements of the matrices appearing inside the determinants are labeled  $i,j$. In the external expressions \eqref{dm-eq:general-new-expression-p-s-diff} and \eqref{dm-eq:general-new-expression-p-s-int}, the $r, s, p$ are related to the $i$ of \eqref{dm-eq:Slater-diff} and \eqref{dm-eq:slater-2-mat-usual} as $i=(p-1)r+s$ for a polynomial potential $V_1(x) = \sum_k^p \frac{\alpha_k}{k} x^k$ of degree $p$.

The structure is the same in both diagrams. In the bottom layer, we have all the integral representations of the matrix model under consideration. In the top layer, we show the corresponding expressions in differential form. $\vd$ denotes the Vandermonde determinant. The front layers gather the expressions obtained after diagonalizing in integral form, while the back layers gather the expressions obtained after diagonalizing in differential form. 

Several lines connecting parts within a diagram share common features, indicated by a common greek letter. The lines $\alpha$ symbolize the direct  equivalence between integral and differential formulations. Lines $\beta$ symbolize the equivalence between the one- and two-matrix differential formulations. Lines  $\delta$ involve a diagonalization procedure in integral formulation, leading to the appearance of Vandermonde determinants. Similarly, lines $\eta$ refer to a diagonalization in the differential formulation. When moving from left to right along lines marked $\phi$, the common strategy is to expand the Vandermonde determinants, thereby moving the integration or differentiation inside the matrix elements of the determinant.
The appearance of $\kappa$ indicates that a heat kernel method was used to transform between an integral and a differential expression. The steps marked $\sigma$ absorb the $1/\vd$ prefactor by extracting the symmetric part of the determinant.
Finally, for vertical lines labeled $\lambda$, it is shown how to transform some of the derivatives to variables and vice-versa, while $\tilde \lambda$ corresponds to the equivalent computation in integral formulation. These are the computations leading to our newly found determinantal expressions for two-matrix models, that could potentially lead to a new resolution in terms of orthogonal polynomials.

\renewcommand{\eqref}[1]{Eq.\ (\ref{#1})}

\section{Brief presentation of the models}

We start with a reminder on the two models presenting all relevant details, in particular their perturbative expansions and diagonalizations as well as the reformulation of two-matrix models in terms of biorthogonal polynomials.

\subsection{One-matrix models}
\label{dm-sub:oneMatDef}

\subsubsection{The model} 

The partition function of the Hermitian one-matrix model with potential $V$ is
\be
\label{dm-eq:one-mat}
Z_V = \int_{\bH_N} \frac{\d M}{a_N} \e^{-N \tr (\frac 1 2 M^2- V(M))},
\ee
where $\mathbb{H}_N$ is the set of Hermitian matrices of size $N\times N$, the
potential is $V(x)=\sum_{k\ge 1} \frac {\lambda_k} k x^k$, the measure is $\d M=\prod_{i=1}^N \d M_{ii} \prod_{i<j} \d \Re M_{ij}\d \Im M_{ij}$, and the normalization  $a_N= {2^{N/2}} (\pi/N)^{{N^2}/2}$ is such that $Z_{V=0}=1$ (by explicit computation of the  Gaussian integral).
The free energy of the model is $F_V= \log Z_V$.

\subsubsection{Diagrammatic expansion}  
\label{dm-subsub:FeynmanOne}

By expanding the exponential of the potential $V$, exchanging the summation and the integration, and using Wick's theorem, one obtains a Feynman expansion of the one-matrix model over ribbon graphs, which can be seen as graphs with an additional cyclic ordering of the edges around each vertex. Given a ribbon graph, a cyclic sequence of edges such that every edge in the sequence also precedes the edge that follows in the ordering around a vertex of the graph is called a \emph{face} (for more details see \cite{ZvonkinMapEnumeration, BouttierMapEnumeration, eynard2018random, eynard2016counting, thurigen2021renormalization}). In the expansion of the partition function, the vertices of valency $k$ (that is, with $k$ edges attached) are counted with a weight $N \frac{\lambda_k} k$ and the edges with a weight $1/N$, as they correspond to propagators
\be
\label{dm-eq:propa-one-usual}
\int_{\bH_N} \frac{\d M}{a_N} \e^{-\frac N 2 \tr ( M^2)} M_{ij} M_{kl} = \frac 1 N \delta _{i,l}\delta_{j,k}.
\ee
To each face corresponds a trace of these Kronecker symbols, resulting in an additional factor $N$ per face. The Feynman expansion of the free energy involves only connected graphs, which is not true for the partition function. 
Denoting by $V_k(\Gamma)$ the number of vertices of valency $k$ of a ribbon graph $\Gamma$, the weight (or amplitude) of a graph in the perturbative expansion is therefore 
\[
\amp(\Gamma) = \frac{N^{\chi(\Gamma)}}{|\mathrm{Aut}(\Gamma)|} \prod_{k\ge 1} \lambda_k^{V_k(\Gamma)},
\]
where $\chi(\Gamma)$ is the Euler characteristic
\[
\chi = V - E + F = 2K - 2g,
\]
in which $V=\sum_k V_k$ is the total number of vertices, $E$ is the total number of edges, $F$ the total number of faces, and $K$ the number of connected components of the ribbon graph ($K=1$ for the graph expansion of the free energy), and $g$, the genus of the ribbon graph, is a non-negative integer. The combinatorial factor $|\mathrm{Aut}(\Gamma)|$ is the number of automorphisms of the ribbon graph $\Gamma$ (see \cite{ZvonkinMapEnumeration, BouttierMapEnumeration, eynard2018random}). 

\subsubsection{Clarification regarding matrix integrals}
\label{dm-subsub:formal}

All matrix integrals in this chapter are understood as \emph{formal} integrals: They are perturbative expansions labelled by Feynman diagrams, that is, formal generating functions for ribbon graphs (see \cite{eynard2016counting, orantin2015chain, eynard2018random} and references therein). 
Every equality between partition functions is understood as equality between their perturbative expansions, term by term, regardless of the analytic properties of these series. The questions of whether the matrix integrals are well-defined, whether the series are summable when the coupling constants are in some domain and in certain limits (e.g.~large $N$), and whether an equality between these different quantities holds non-perturbatively are not addressed in this chapter.

\subsubsection{Diagonalization} 

One of the standard ways of solving this model is to first  diagonalize the Hermitian matrices as $M = U^{\dagger} X U$, where $U\in \U(N)$ and $X=\textrm{diag}(\x)$, $\x=(x_1, \ldots, x_N)\in \bR^N$. Due to the unitary invariance, the integration over unitary matrices factors out the volume of the unitary group, and the partition function is expressed as an integral over the $N$ real eigenvalues,
\be
\label{dm-eq:diagonal-hermitian}
Z_V =  \int_{\bR^N} \frac{\d\x}{b_N}\, \vd^{2}(\x)\, \e^{-N(\frac 1 2 \x^2 - V(\x))}, 
\ee
\sloppy where the measure is $\d\x=\prod_{i=1}^{N} \d x_{i}$, the normalization is $b_N={(2\pi)^{N/2}\prod_{j=1}^{N}j!}\,{N^{-N^2/2}}$ (this can be computed from Mehta's integral (3.3.10) in \cite{mehta2004random}), 
$\Delta(\x)=\prod_{i<j} (x_j - x_i)$ is the Vandermonde determinant,  $\x^2=\sum_{i=1}^{N}x_{i}^{2}$, and we use the notation
\[V(\x)\equiv\sum_{i=1}^N V(x_i).
\]
To solve the model one may apply for example the saddle point method, loop or Schwinger-Dyson equations,  or
the method of orthogonal polynomials \cite{eynard2018random}.
Here we focus on the steps leading to the resolution by orthogonal polynomials relying on the diagonalization.

\subsection{Two-matrix models}

\subsubsection{Partition function}  
The models we are interested in have the following form:
\be
\label{dm-eq:2-matrix_model}
Z_{V_1, V_2} = \int_{\bH_N\times \bH_N}\frac{\d A\, \d B}{c_N} \e^{-N \tr(AB- V_1(A) - V_2(B))}
\ee
where $V_1(x)=\sum_{k\ge 1} \frac{\alpha_k} k x^k$ and $V_2(x)=\sum_{k\ge 1} \frac{\beta_k} k  x^k$, and the normalization $c_N$ is chosen such that for $V_1=V_2=0$, $Z_{0,0}=1$. The Gaussian part is formal and should be understood as follows. 
For two real variables, we formally have for $\alpha\in \mathbb{R}$:
\be
\label{dm-eq:Gaussian-mixt-0}
\int_{\bR^2}\d x\, \d y\, \e^{- \alpha  N  xy } = \frac{2\pi} {\imath \alpha N},
\ee
and more generally for $n,m\ge 0$:
\begin{align}
\label{dm-eq:Gaussian-mixt-00}
\int_{\bR^2}\d x\, \d y\, \e^{- \alpha N  xy }\, x^n y^m &= \frac {\delta_{n,m}  }{(- N)^n} \frac{\partial^{n}}{\partial \alpha^{n}} \int_{\bR^2}\d x\, \d y\, \e^{-  \alpha N  xy } \\
&=  \frac {\delta_{n,m}  } {(- N)^n} \frac{\partial^{n}}{\partial \alpha^{n}} \frac{2\pi}{\imath \alpha N} 
,
\end{align}
so that
\be
\label{dm-eq:Gaussian-mixt}
\int_{\bR^2}\d x\, \d y\, \e^{- \c N  xy }\, x^n y^m = \delta_{n,m}\,  \frac{2\pi} \imath \,  \left(\frac 1 {\c N}\right)^{n+1}n!.
\ee
This leads to the normalization:
\be
\label{dm-eq:normal-two-mat-mod}
c_N= \int_{\bH_N\times \bH_N}\d A\, \d B\, \e^{-N \tr(AB)}= 2^N  \left( \frac \pi {\imath N}\right)^{N^2},
\ee
and to the propagator
\be
\label{dm-eq:propagator-two-mat-usual}
\int_{\bH_N\times \bH_N} \frac{\d A\, \d B}{c_N}\,  \e^{-N \tr(AB)}\, A_{ab}B_{cd} = \frac 1 N \delta_{a,d}\delta_{b,c}.
\ee

The integrals in \eqref{dm-eq:Gaussian-mixt-0}, \eqref{dm-eq:Gaussian-mixt}, \eqref{dm-eq:normal-two-mat-mod} and \eqref{dm-eq:propagator-two-mat-usual} are not well-defined. On the other hand, the values we set formally here are the ones needed in order to recover the correct combinatorial expansions with the correct overall normalizations of the partition functions.\footnote{The formal propagator \eqref{dm-eq:Gaussian-mixt} can be found in [\cite{DiFrancesco:1998bp}, Eq.~(4.2)], but we have added the factor ${2\pi} /{\i}$ to agree with the value of \eqref{dm-eq:2-matrix_model} when $V_1(x)=V_2(x)= - \epsilon x^2/2$, in the limit where $\epsilon \rightarrow 0$ see e.g.~\cite{eynard20062matrix}, App. A.} In the differential formulation, the equivalent expressions are well-defined and automatically normalized, as will be detailed in the rest of the text.

\subsubsection{Diagrammatic expansion}  

The Feynman graph expansion of a two-matrix model involves a summation over ribbon graphs that have two kinds of vertices, respectively associated to the traces of the matrices $A$ and $B$, so that the edges only link vertices of different kinds (the ribbon graphs are \emph{bipartite}
\footnote{As a convention, if there are quadratic terms in $V_1$ and $V_2$, we interpret them as interactions corresponding to bi-valent vertices in the graphs.}). 
Denoting by $V^A_k(\Gamma)$ and $V^B_k(\Gamma)$ the number of vertices of valency $k$ of each type of vertices of a ribbon graph $\Gamma$, the weight of a Feynman graph is given by 
\be
\label{dm-eq:amp-two-mat-graph}
\amp(\Gamma) = \frac{N^{\chi(\Gamma)}}{|\mathrm{Aut}(\Gamma)|} \prod_{k\ge 1} \alpha_k^{V^A_k(\Gamma)}\beta_k^{V^B_k(\Gamma)},
\ee
the graphs contributing to the free-energy being connected.

\subsubsection{Diagonalization} 

The first step in the resolution is to  diagonalize $A$ and then use the Harish-Chandra--Itzykson--Zuber formula \cite{HarishChandra,itzykson1980planar} to diagonalize $B$:
\be 
\int_{\U(N)} \d U\,  \e^{\c\tr(AUBU^\dagger)} 
= {\c^{-\frac{N(N-1)}2}}\prod_{k=1}^{N-1} k!\  \frac{\det \{\e^{\c\, a_i b_j}\}_{1\le i,j\le N}}{\Delta(\ba)\Delta(\bb)},
\label{dm-eq:itz-zub}
\ee
where $\d U$ is the normalized Haar measure on the unitary group $\U(N)$ and $\c$ is a complex coefficient.
As shown in \cite{Mehta:1981jm}, for $\c=N$ this leads to 
\be
\label{dm-eq:diag:2-mat}
Z_{V_1,V_2}=  \int_{\bR^{2N}} \frac{\d\ba\,\d\bb}{N!\,d_N} \, \Delta(\ba)\Delta(\bb) \, 
\e^{-N (\ba\cdot \bb -V_1(\ba) - V_2(\bb))},
\ee
where $d_N=(2\pi / \imath)^N \prod_{j=1}^{N-1} j!\,   N^{-N(N+1)/2}$.

\subsubsection{Determinant form} 
\label{dm-subsub:det-form-usual}

Let us detail the steps leading to the usual resolution using biorthogonal polynomials \cite{Mehta:1981jm}. The Leibniz determinant formula  
\[\label{dm-eq:Leibnizformula}
\det(\{f_{i,j}\}_{1\le i,j \le N}) = \sum_{\sigma\in S_N} (-1)^\sigma \prod_{i=1}^N\,f_{i,\sigma(i)}
\]
where $(-1)^\sigma$ is the sign of the permutation $\sigma$,  is then used to develop the Vandermonde determinants $\vd(\x)=\det \{x_i^{j-1}\}_{1\le i,j \le N}$, so that \eqref{dm-eq:diag:2-mat} reads:
\begin{align}
Z_{V_1,V_2}
&= \frac {1} {N!\,d_N} \sum_{\sigma, \sigma'\in S_N }(-1)^{{\sigma+ \sigma'}} \prod_{i=1}^N\int_{\bR} \d x \int_{\bR} \d y\, x^{\sigma(i)-1}y^{\sigma'(i)-1} \, \e^{-N (xy-V_1(x) - V_2(y))}\nonumber\\
&=\frac {1}{d_N}  \sum_{\sigma \in S_N }(-1)^{{\sigma}} \prod_{i=1}^N\int \d x\, \d y\,  x^{\sigma(i)-1}y^{i-1} \, \e^{-N (xy-V_1(x) - V_2(y))},
\end{align}
leading to the following expression of the partition function as a Hankel determinant (again by Eq.~\ref{dm-eq:Leibnizformula}):
\begin{equation}
\label{dm-eq:slater-2-mat-usual}
Z_{V_1,V_2}= \frac {1}{d_N} \det\left\{ \int \d x\, \d y\, \e^{-N xy} x^{i}  \e^{N V_1( x)} y^{j} \e^{N V_2(y)}  \right\}_{0\le i,j\le N-1}.
\end{equation}

\subsubsection{Biorthogonal polynomials}
\label{dm-subsub:ortho-pol-usual}

Using the properties of the determinant, one can subtract from each row or column a linear combination of rows or columns of lower indices without affecting the result, so that one can replace $x^i$ and $y^{j} $ by any polynomials $P_i(x)$, $Q_j(y)$ of degrees $i$ and $j$ respectively with unit leading coefficients (such polynomials are said to be \emph{monic}):
\begin{align}
\label{dm-eq:bil-form-int}
Z_{V_1,V_2} &= \frac {1}{d_N} \det\left(\left\{ \langle P_i \vert Q_j\rangle  \right\}_{0\le i,j\le N-1}\right), \\
\langle P_i \vert Q_j\rangle &= \int \d x\, \d y\, \e^{-N xy} P_i(x)  \e^{N V_1( x)} Q_j(y) \e^{N V_2(y)}.
\end{align}
 In particular, the monic polynomials $P_i, Q_j$ can be chosen to be orthogonal for the 
formal symmetric bilinear form on the right-hand side of \eqref{dm-eq:bil-form-int}, in which case the determinant \eqref{dm-eq:slater-2-mat-usual} is the product of the diagonal terms: 
\[
\langle P_i \vert Q_j\rangle=h_i \delta_{i,j}, \qquad\qquad  Z_{V_1,V_2}= \frac {1}{d_N}\prod_{j=1}^N h_i. 
\]
Two sequences of polynomials satisfying this orthogonality relation are said to be biorthogonal. They are determined recursively for specific choices of the potentials $V_1$ and $V_2$ \cite{Mehta:1981jm, eynard1997eigenvalue, DiFrancesco:1998bp, Bouttier_2002, Bertola:2007is}.

\section{Differential reformulation of matrix integrals}
\label{dm-sec:diff-reformulation}

In this section we show how to  reformulate two-matrix models in terms of a single matrix and its derivatives. We first review the usual differential formulations of one- and two-matrix models. Then we present the new one-matrix differential formulation of the latter. We finally show how to diagonalize the models in this formulation.

\subsection{The differential formulation}
\label{dm-sub:diff-formulation}

It is known (see e.g.~\cite{gurau2015analyticity}) that a Hermitian one-matrix model admits the following reformulation:
\be
\label{dm-eq:diff-one-mat}
Z_V = \int_{\bH_N}\frac{\d M}{a_N} \e^{-N \tr (\frac 1 2 M^{2}-V(M))} = \left[ \e^{+\frac1 {2N} \tr (\frac{\pd}{\pd M})^2} \e^{N \tr\, V(M)} \right]_{M=0},
\ee
where $\tr \bigl(\frac{\pd}{\pd M}\bigr)^2=\sum_{a,b=1}^N \frac{\pd}{\pd M_{ab}} \frac{\pd}{\pd M_{ba}}$. The role of the matrices and derivatives can be exchanged:
\be
\label{dm-eq:diff-one-mat-2}
Z_V =  \left[ \e^{N  \tr V (\frac{\pd}{\pd M})} \e^{\frac 1 {2N} \tr\, (M^2)} \right]_{M=0},
\ee
which is the formulation analogous to \eqref{dm-eq:central-identity} for a one-matrix model.
Note that the differential expressions are clearly well-normalized, as for $V=0$, the only contributing term in the series-expansion of the exponential is 1. 

As mentioned in Sec.~\ref{dm-subsub:formal}, the sign ``$=$'' means that the equality holds at the perturbative level: the Feynman expansions coincide term by term. While the perturbative expansion of the matrix model is obtained by expanding the exponential of $V$ in series and exchanging  the summation and the integration (Sec.~\ref{dm-subsub:FeynmanOne}), in the differential formulation, the perturbative expansion is obtained by exchanging the summation and the evaluation $M=0$.

More precisely, recalling that $V(x)=\sum_{k\ge 1} \frac {\lambda_k} k x^k$, the perturbative expansion of the right-hand side of \eqref{dm-eq:diff-one-mat} reads:
\be
\label{dm-eq:dvt-proof}
\begin{split}
\left[ \e^{\frac1 {2N} \tr (\frac{\pd}{\pd M})^2}\right. &\left.\e^{N \tr\, V(M)} \right]_{M=0} \\
&=\sum_{\{n_k\ge 0\}}\prod_k \frac{(N\lambda_k/k)^{n_k}}{n_k!} \left[ \e^{\frac1 {2N} \tr (\frac{\pd}{\pd M})^2} \prod_k  \tr(M^k)^{n_k} \right]_{M=0}  \\
&= \sum_{\{n_k\ge 0\}}\prod_k \frac{(N\lambda_k/k)^{n_k}}{n_k!} \\
& \quad \quad \cdot \sum_{i\ge 0}\frac 1 {i!}\left[ \left(\frac1 {2N} \tr (\frac{\pd}{\pd M})^2\right)^i \prod_k  \tr(M^k)^{n_k} \right]_{M=0}  \\
&= \sum_{\{n_k\}} \prod_k \frac{(N\lambda_k/k)^{n_k}}{n_k!}\frac 1 {({\ell(\bn)}/2)!} \\
& \quad \quad \cdot \Biggl[\left(\frac1 {2N} \tr (\frac{\pd}{\pd M})^2\right)^{ \frac {\ell(\bn)}2} \prod_k\tr(M^k)^{n_k}\Biggr],
\end{split}
\ee
where $\ell(\bn) = \sum_k k n_k$. Developing the term originating from the propagator:
\[
\left(\tr\frac{\pd}{\pd M}\frac{\pd}{\pd M}\right)^{\frac {\ell(\bn)}2}= \sum_{p=1}^{ {\ell(\bn)}/2}\sum_{i_p, j_p=1}^N \prod_{p=1}^{ {\ell(\bn)}/2} \frac{\pd}{\pd M_{i_p j_p}}\frac{\pd}{\pd M_{j_p i_p}}, 
\]
we see that the expression between brackets in the last line of \eqref{dm-eq:dvt-proof} is a sum over all possible ways to pair the derivatives $\frac{\pd}{\pd M_{ij}}$  and the matrix elements $M_{kl}$ originating from the interaction potential, with:
\[
\frac 1 {2N} \tr\frac{\pd}{\pd M_{i_p j_p}}\frac{\pd}{\pd M_{j_p i_p}}M_{ab}M_{cd} = \frac 1 {2N}(\delta_{i_p,a}\delta_{j_p,b}\delta_{j_p,c}\delta_{i_p,d} + \delta_{i_p,c}\delta_{j_p,d}\delta_{j_p,a}\delta_{i_p,b}).
\]
Since each $i_p$ and $j_p$ appear only in one term of the sort, the sums over $i_p,j_p$ can be carried out with the following  simplifications:
\[
\sum_{i_p,j_p} \frac 1 {2N} \tr\frac{\pd}{\pd M_{i_p j_p}}\frac{\pd}{\pd M_{j_p i_p}}M_{ab}M_{cd} = \frac 1 {N}\delta_{a,d}\delta_{b,c},
\]
where we recognize the propagator \eqref{dm-eq:propa-one-usual}. The combinatorial factor $\frac 1 {(  {\ell(\bn)}/2)!}$ compensates for the number of permutations of the traces of pairs of derivatives (the index $p$), as all permutations give the same result. In other words, the term in square brackets in the last line of \eqref{dm-eq:dvt-proof} coincides with a Gaussian expectation through Wick's theorem:
\begin{equation}
\label{dm-eq:Wick-One}
\begin{split}
\frac 1 {({\ell(\bn)}/2)!} \left(\frac1 {2N} \tr\frac{\pd}{\pd M}\frac{\pd}{\pd M}\right)^{ {\ell(\bn)}/2} \prod_k\tr(M^k)^{n_k} \\= \int_{\bH_N} \frac{\d M}{a_N} \e^{-\frac N 2 \tr(M^2)} \prod_k\tr(M^k)^{n_k},
\end{split}
\end{equation}
which is expressed as a sum over ribbon graphs with $n_k$ vertices of valency $k$ and ${\ell(\bn)}/2 = \frac1 2{\sum_k k n_k}$ edges. We thus recover the usual perturbative expansion of \eqref{dm-eq:one-mat}, which proves \eqref{dm-eq:diff-one-mat} at the perturbative level. \eqref{dm-eq:diff-one-mat-2} is shown the same way, with:
\begin{equation}
\begin{split}
\frac 1 {({\ell(\bn)}/2)!} \prod_k\left(\tr \left(\frac{\pd}{\pd M}\right)^k\right)^{n_k }\left(\frac1 {2N} \tr (M^2 )\right)^{ \frac{\ell(\bn)}2} \\= \int_{\bH_N} \frac{\d M}{a_N} \e^{-\frac N 2 \tr(M^2)} \prod_k\tr(M^k)^{n_k}.
\end{split}
\end{equation}

\subsection{Differential formulation of two-matrix models}
\label{dm-sub:diff-two-matrix}

For a two-matrix model the standard differential formulation is simply
\begin{equation}
\label{dm-eq:Two-mat-diff-form}
\begin{split}
Z_{V_1,V_2}= \int_{\bH_N^2}\frac{\d A\, \d B}{c_N}\, \e^{-N \tr(AB-V_1(A) - V_2(B))} 
 \\= \left[ \e^{\frac 1 N \tr\, (\frac{\pd}{\pd A} \frac{\pd}{\pd B})} \e^{N \tr\left(V_1(A)+V_2(B)\right)} \right]_{A,B=0}.
 \end{split}
\end{equation}
Again, the differential expression is clearly well-normalized. \eqref{dm-eq:Two-mat-diff-form} is shown perturbatively using the fact that now the propagator  corresponds to:
\[
\sum_{i_p,j_p} \frac 1 {N} \tr\frac{\pd}{\pd A_{i_p j_p}}\frac{\pd}{\pd B_{j_p i_p}}A_{ab}B_{cd} = \frac 1 {N}\delta_{a,d}\delta_{b,c},
\]
so that through Wick's theorem, denoting by $\ell(\ba + \bb) = \sum_k k (a_k+b_k)$, we have
\begin{equation}
\begin{split}
\label{dm-eq:Wick-Two}
\frac 1 {\frac{\ell(\ba + \bb)} 2!}\left(\frac1 {N} \tr\frac{\pd}{\pd A}\frac{\pd}{\pd B}\right)^{\frac{\ell(\ba + \bb)} 2} \prod_k\tr(A^k)^{a_k}\tr(B^k)^{b_k} \\
= \int \frac{\d A\d B}{c_N} \e^{ - N  \tr(AB)} \prod_k\tr(A^k)^{a_k}\tr(B^k)^{b_k},
\end{split}
\end{equation}
the other steps being the same as for one-matrix models. This last quantity is expressed as a sum over bipartite ribbon graphs with $a_k$ and $b_k$ vertices of valency $k$ associated to the matrix $A$ respectively $B$ and $\sum_k k a_k= \sum_k k b_k$ edges.

While the right-hand side of \eqref{dm-eq:Wick-Two} is formal and understood \emph{via} Wick's theorem and the formal propagator \eqref{dm-eq:propagator-two-mat-usual}, the left-hand side is well-defined.

\

In the differential formulation however, it is no longer necessary to use two matrices: this is important in the integral formulation to impose 
bipartiteness of the Feynman ribbon graphs. This  bipartiteness can also be implemented in a differential formulation using the fact that derivatives only act on matrices. The resulting differential formulation thereby only involves a \emph{single matrix}: 
\be
\label{dm-eq:diff-1-2-model}
Z_{V_1,V_2} = \left[ \e^{N \tr\, V_1(\frac 1 {\sqrt N} \frac{\pd}{\pd M})} \e^{N \tr\, V_2(\frac 1 {\sqrt{N}} M)} \right]_{M=0}.
\ee
This can be proven by showing that this differential formula generates the same ribbon graphs, together with the same combinatorial weights \eqref{dm-eq:amp-two-mat-graph}. It can also be proven in the differential formulation starting from the right-hand side of \eqref{dm-eq:Two-mat-diff-form}:
\begin{align}
\begin{split}
Z_{V_1,V_2} &= \left[ \e^{\frac 1 N \tr\frac{\pd}{\pd A}\frac{\pd}{\pd B}}\,\e^{N \tr\, V_1(A)}\, \e^{N \tr\, V_2(B)} \right]_{A=B=0} \\
&= \biggl[ \left[\e^{\frac 1 N \tr\frac{\pd}{\pd A}\frac{\pd}{\pd B}}\,\e^{N \tr\, V_1(A)}\right]_{A=0} \e^{N \tr\, V_2(B)} \biggr]_{B=0},
\end{split}
\end{align}
where, recalling that $\ell(\ba)=\sum_k k a_k$:
\begin{align}
 &\left[\e^{\frac 1 N \tr\frac{\pd}{\pd A}\frac{\pd}{\pd B}}\,\e^{N \tr\, V_1(A)}\right]_{A=0} \nonumber \\
 &\quad \quad= \sum_{\{a_k\ge 0 \}}\sum_{i\ge 0 } \prod_k \frac{(N \alpha_k/k)^{a_k}}{a_k!}\frac{1}{i!} \left[ \Bigl(\frac 1 N \tr\frac{\pd}{\pd A}\frac{\pd}{\pd B}\Bigr)^{i}\prod_k (\tr A^k )^{a_k}  \right]_{A=0} \\
 &\quad \quad= \sum_{\{a_k\ge 0 \}} \prod_k \frac{(N \alpha_k/k)^{a_k}}{a_k!}\frac{1}{\ell(\ba)!} \Bigl(\frac 1 N \tr\frac{\pd}{\pd A}\frac{\pd}{\pd B}\Bigr)^{\ell(\ba)}\prod_k (\tr A^k )^{a_k} .
\end{align}
Assuming that we have proven that: 
\[
\label{dm-eq:central-point-proof}
\Bigl(\tr\frac{\pd}{\pd A}\frac{\pd}{\pd B}\Bigr)^{\ell(\ba)} \prod_k \Bigl(\tr \left(A^k\right) \Bigr)^{a_k}   = \ell(\ba)! \prod_k \left(\tr \left(\frac{\pd}{\pd B}\right)^k\right)^{a_k}, 
\]
we obtain 
\be
\label{dm-eq:proof3C}
\left[\e^{\frac 1 N \tr\frac{\pd}{\pd A}\frac{\pd}{\pd B}}\,\e^{N \tr\, V_1(A)}\right]_{A=0}  = \e^{N V_1(\frac 1 N \frac{\pd}{\pd B})}, 
\ee
and thereby 
\[
Z_{V_1,V_2} = \left[ \e^{N \tr\, V_1(\frac 1 N \frac{\pd}{\pd B})} \e^{N \tr\, V_2(B)} \right]_{B=0}. 
\]
The sought formula \eqref{dm-eq:diff-1-2-model} is obtained by change of variable $M=\sqrt N B$.\footnote{Note that in this differential formulation, changes of variables are made without ``Jacobian'', since both sides are clearly normalized.}

Let us now prove \eqref{dm-eq:central-point-proof}.
We write $$\left(\tr\frac{\pd}{\pd A}\frac{\pd}{\pd B}\right)^{\ell(\ba)}= \sum_{p=1}^{\ell(\ba)}\sum_{i_p, j_p=1}^N \prod_{p=1}^{\ell(\ba)}\, \frac{\pd}{\pd A_{i_p j_p}}\frac{\pd}{\pd B_{j_p i_p}}. $$
For every value taken by $\{i_p, j_p\}_{p}$, the quantity 
$ \prod_{p=1}^{\ell(\ba)} \frac{\pd}{\pd A_{i_p j_p}}\frac{\pd}{\pd B_{j_p i_p}} \prod_k \left(\tr\left(A^k\right) \right)^{a_k}$ is the sum over all possible ways of assigning all the derivatives $\frac{\pd}{\pd B_{j_p i_p}}\frac{\pd}{\pd A_{i_p j_p}}$ to the matrix elements $A_{kl}$ in the product of traces. Every time a term of the form $\frac{\pd}{\pd B_{j_p i_p}}\frac{\pd}{\pd A_{i_p j_p}}$  acts on a matrix element $A_{kl}$, the latter is replaced in the product of traces by 
$\frac{\pd}{\pd B_{l k}}\delta_{i_p, k}\delta_{j_p, l}.$ All the indices $i_p, j_p$ belong to a single Kronecker symbol, so that the sum  $\sum_{p=1}^{\ell(\ba)}\sum_{i_p, j_p=1}^N$ simply eliminates all Kronecker symbols. We are left with 
\begin{align}
\Bigl(\tr\frac{\pd}{\pd A}\frac{\pd}{\pd B}\Bigr)^{\ell(\ba)} \prod_k \tr (A^k) ^{a_k} &= \hspace{-0.2cm}\sum_{\substack{{\textrm{possible}}
\\
{\textrm{ assignments}}}} \prod_k \left(\tr \Bigl(\bigl(\frac{\pd}{\pd B} \bigr)^T\Bigr)^k \right)^{a_k} \\
&= \ell(\ba)!\, \prod_k \left(\tr \bigl(\frac{\pd}{\pd B} \bigr)^k \right)^{a_k},
\end{align}
where the last equality holds because there are $(\sum_k k a_k)!$ ways of assigning the derivatives $\frac{\pd}{\pd B_{j_p i_p}}\frac{\pd}{\pd A_{i_p j_p}}$ to the matrix elements $A_{kl}$ in the product of traces, and the result for each assignment is independent of the specific assignment chosen. This concludes the proof.

\subsection{Diagonalization in the differential formulation}
\label{dm-sub:diff-formulation-diagonalization}

It is known \cite{itzykson1980planar,zuber2008large} that for Hermitian matrices $A$, 
the differential operator 
$D_k(\frac{\pd}{\pd A}) = \tr (\frac{\pd}{\pd A})^k$ 
acting on  $U(N)$-invariant functions
diagonalizes as
\be
\label{dm-eq:diagonaldiffoperator}
D_k\Bigl(\frac{\pd}{\pd a_i}\Bigr) = \frac 1 {\vd(\ba)} \sum_{i=1}^{N}\frac{\pd^{k}}{\pd a_{i}^{k}} \vd(\ba). 
\ee
This implies in particular that the differential operator in \eqref{dm-eq:diff-1-2-model} has the diagonal form
\be 
\label{dm-eq:diag-pot-diff}
\tr\, V_1\left( \frac1{\sqrt{N}} \frac{\pd}{\pd M} \right) 
= \sum_{k\ge1}\frac{\alpha_k}{k \sqrt{N}^k} D_k\Bigl(\frac{\pd}{\pd M}\Bigr) 
= \frac 1 {\vd(\bfm)} \sum_{i=1}^N V_1\left(\frac1{\sqrt{N}} \frac{\pd}{\pd m_i} \right) \vd(\bfm),
\ee
since it acts on $\exp(V_2( M))$, which depends on traces of $M$ and is thus $U(N)$-invariant.

Since this is only briefly mentioned in the final remarks of \cite{zuber2008large}, let us make the argument more precise here.
One defines the operator $\widehat{D}_k$ for some positive integer $k$ as the operator which acts on the exponential trace of the product of two Hermitian matrices $A$ and $B$ as
\[\label{dm-eq:defdiffoperator}
\widehat{D}_k  \e^{\c\tr A B} \equiv \c^k \tr(B^k) \e^{\c\tr A B}
\]
and $\c$ is a complex coefficient. 
From this definition it follows that this operator has the explicit representation as a functional of matrix derivatives
\be
\label{dm-eq:def-Dk-1}
D_k\Bigl(\frac{\pd}{\pd A}\Bigr) = \tr \Bigl(\frac{\pd}{\pd A}\Bigr)^k 
= \sum_{i_1,...,i_k} \frac{\pd}{\pd A_{i_1 i_2}}\frac{\pd}{\pd A_{i_2 i_3}}\dots\frac{\pd}{\pd A_{i_k i_1}} \, 
\ee
on functions of Hermitian matrices $f(A)$ whose function space is spanned by the basis $\{\e^{\c\tr A B}\}_{B\in\mathbb{H}_N}$. For $\c=\i$, any Hermitian matrix-valued function belongs to that space by Fourier transformation on Hermitian matrices (see e.g.~\cite{zhang2021harmonic})
\be
\label{dm-eq:Fourier-f-invariant}
f(A) =\frac {1}{2^N \pi^{N^2}} \int_{\mathbb{H}_N} \d B\, \tilde{f}(B) \,\e^{\i\tr A B}\,.
\ee

Consider now the representation of the operator $\hat{D}_k$ on the subspace of  $\U(N)$-invariant functions $f(A)=f(UAU^{-1})$, or equivalently 
\be 
\label{dm-eq:After-Fourier-HCIZ}
f(A)=\int_{\U(N)}\d U\,f(UAU^{-1}) 
= \frac {1}{2^N \pi^{N^2}}\int_{\mathbb{H}_N} \d B \tilde{f}(B) \int_{\U(N)} \d U\,\e^{\c\tr (U A U^{-1} B)}\, .
\ee
The domain of such invariant functions is $N$-dimensional, and the functions can be represented on diagonal matrices. 
The defining equation \eqref{dm-eq:defdiffoperator} on the invariant subspace thus has the form
\be
\widehat{D}_k \int_{\U(N)} \d U \e^{\c\tr (U A U^{-1} B)} 
\equiv \c^k \tr(B^k) \int \d U \e^{\c\tr (U A U^{-1} B)}
\]
Using the Harish-Chandra--Itzykson--Zuber integral formula \eqref{dm-eq:itz-zub},
it follows directly from a Laplace expansion of the determinant,
\begin{align}\label{dm-eq:Laplaceexpansion}
\sum_l\Bigl(\frac{\pd}{\pd a_l}\Bigr)^k \det(\e^{\c a_m b_n})_{m,n} 
&= \sum_l \Bigl(\frac{\pd}{\pd a_l}\Bigr)^k \sum_{j}(-1)^{l+j} \e^{\c a_l b_j} \det(\e^{\c a_m b_n})_{\substack{m\ne l,\\n\ne j}} \\
&= \sum_{j} (\c b_j)^k \sum_l 
(-1)^{j+l} \e^{\c a_l b_j} \det(\e^{\c a_m b_n})_{\substack{m\ne l,\\n\ne j}} \\
&= \c^k\tr(B^k) \det(\e^{\c a_m b_n})_{m,n} \nonumber\, ,
\end{align}
that the functional of eigenvalue derivatives $D_k(\frac{\pd}{\pd a_i})$ \eqref{dm-eq:diagonaldiffoperator} is the representation of $\widehat{D}_k$ on $\U(N)$-invariant functions in diagonalized variables, that is on eigenvalues.

\

Another equivalent way to prove this formula is to see the function $f$  as a function of the matrix elements $\{A_{i,j}\}$ in the canonical basis and consider the eigenvalues $\{a_1,\ldots,a_N\}$ of $A\in\bH_N$,  together with $N(N-1)$ variables  that determine uniquely $U\in \U(N)$ such that $A =U \textrm{Diag}(\{a_i\})U^\dagger$,  as new variables after a change of variables (see the detailed discussion in Chapter 3 of \cite{mehta2004random}). Differentiating $\sum_{l,k} U^{\dagger}_{il} A_{lk}U_{kj}$ with respect to the $a_p$, we find
$
\sum_{l,k} U^{\dagger}_{il} \frac{\partial A_{lk}}{\partial a_p}U_{kj} = \delta_{ij} \delta_{jp},
$
which is inverted as
$
 \frac{\partial A_{ij}}{\partial a_p} = U_{ip}U^{\dagger}_{pj}. 
$
From this, the derivative of any function of $A$,
$ \frac{\partial}{\partial a_p} f(\{A_{ij}\})$ is well-defined, and we may verify explicitly that both the right-hand side $D^{(1)}$ of \eqref{dm-eq:def-Dk-1} acting on a unitary invariant function  expressed as \eqref{dm-eq:Fourier-f-invariant}, and the right-hand side $D^{(2)}$ of \eqref{dm-eq:diagonaldiffoperator} acting on \eqref{dm-eq:After-Fourier-HCIZ} expressed in eigenvalue-variables using \eqref{dm-eq:itz-zub} give the same expression: 
\[
D^{(1)} f(A) 
=  
D^{(2)} f (A) 
= 
\frac {\i^k}{2^N \pi^{N^2}} \int_{\bH_N} \d B \tilde f (B) \tr(B^k)\e^{\imath \tr(AB)}.
\]
This requires using the fact that the Fourier transform $\tilde f (B)$ is also unitary invariant.

\

We use this formula to diagonalize the differential formulations.
For a one-matrix model, applying \eqref{dm-eq:diagonaldiffoperator} to \eqref{dm-eq:diff-one-mat}, we obtain
\be
\label{dm-eq:diff-diagonal-new-herm}
Z_V=\left[ \e^{\frac{1}{2N} \tr \frac{\pd^{2}}{\pd M^{2}}} \e^{N V(M)} \right]_{M=0} 
= \left[ \frac 1 {\vd(\x)} 
\e^{\frac 1 {2N}(\frac{\pd}{\pd \x})^2}
\vd(\x) \,\e^{N V(\x)} \right]_{\x=0}
\ee
where $(\frac{\pd}{\pd \x})^2 = \sum_i \frac{\pd^{2}}{\pd x_{i}^{2}}$.
The formula is clearly well-normalized, since $(\frac{\pd}{\pd \x})^2 \vd(\x) = 0$.
Note that the differential reformulation of the usual integral over eigenvalues \eqref{dm-eq:diagonal-hermitian} reads
\be
\label{dm-eq:diff-diagonal-usual-herm}
Z_V = \int_{\bR^N} \frac{\d \x}{b_N} \e^{- \frac {N} {2} \x^2} \vd^{2}(\x)\, \e^{NV(\x)} = e_N\left[ \e^{\frac 1 {2N}(\frac{\pd}{\pd \x})^2}\vd^{2}(\x)\, \e^{N V(\x)} \right]_{\x=0} ,
\ee
where the normalization $
e_N = N^{N(N-1)/2} / \prod_{j=1}^N j!
$  is computed in Appendix \ref{dm-app:normalizations}.

\

For the diagonalization of two-matrix models it is crucial that we have a differential formulation  using only one matrix such as \eqref{dm-eq:diff-1-2-model}.
It is not possible to simultaneously diagonalize differential operators involving two matrices such as $\tr\, (\frac{\pd}{\pd A} \frac{\pd}{\pd B})$ in \eqref{dm-eq:Two-mat-diff-form}, at least not by the arguments used here.
However, the differential formulation in terms of a single matrix allows us to diagonalize  two-matrix models.

Applying \eqref{dm-eq:diag-pot-diff} to \eqref{dm-eq:diff-1-2-model} yields
\be
\label{dm-eq:diff-diagonal-new}
Z_{V_1,V_2}=  \left[ \frac 1 {\vd(\x)} \e^{N V_1( \frac 1 {\sqrt N} \frac{\pd}{\pd \x})} \vd(\x) \,\e^{NV_2(\frac 1 {\sqrt N}  \x)} \right]_{\x=0}
\, .\ee
For comparison,  the differential formulation of the usual integral over eigenvalues \eqref{dm-eq:diag:2-mat} is
\be
\label{dm-eq:diff-of-diag-two-mat}
Z_{V_1,V_2}=  e_N\left[ \e^{\frac 1 N \sum_{i=1 }^N\frac{\pd}{\pd a_i}\frac{\pd}{\pd b_i}}\,\Delta(\ba)\Delta(\bb)\e^{N V_1(\ba)}\, \e^{N V_2(\bb)} \right]_{\ba=\bb=0}
\ee
where the normalization is the same as for the one-matrix model (see Appendix \ref{dm-app:normalizations}).

The equivalence between \eqref{dm-eq:diff-diagonal-new-herm} and \eqref{dm-eq:diff-diagonal-usual-herm} on one hand, and between \eqref{dm-eq:diff-diagonal-new}  and \eqref{dm-eq:diff-of-diag-two-mat} on the other hand is not manifest in this form, however it becomes clear when formulated in terms of Slater determinants, as detailed later in  Sec.~\ref{dm-subsub:Equivalence-diag-Twomat} for the two-matrix case and in Sec.~\ref{dm-subsub:invert-diff-diag-gaussian} for the one-matrix model.

\section{Expressions as Slater determinants and applications}
\label{dm-sec:slater}

\subsection{Equivalent formulations of a two-matrix model}

In this first subsection, we detail the steps leading to the Slater determinant formulation of the partition functions of two-matrix models, in differential formulation.
{The same holds for the one-matrix models,  as the special case $V_1(x)=\frac{x^2}{2}$ (cf.~Sec.~\ref{dm-subsub:invert-diff-diag-gaussian}).}

\subsubsection{Determinant form} 
\label{dm-subsub:slater-diff-two-mat}
Starting from  \eqref{dm-eq:diff-diagonal-new}:
\[
Z_{V_1,V_2}=   \left[ \frac 1 {\vd(\x)} \left(\prod_{i=1}^N \e^{N V_1( \frac 1 {\sqrt N} \frac{\pd}{\pd x_i})}\right) \vd(\x) \left( \prod_{i=1}^N \e^{NV_2(\frac 1 {\sqrt N} x_i)}\right) \right]_{\x=0},
\]
 we incorporate the products on the left and right of the Vandermonde determinant into a Slater determinant: 
\be
\label{dm-eq:Slater-first-expr}
Z_{V_1,V_2}=   \left[ \frac 1 {\vd(\x)} \det \left\{ \e^{N V_1( \frac 1 {\sqrt N} \frac{\d}{\d x_i})} x_i^{j-1} \e^{N V_2(\frac 1 {\sqrt N} x_i)} \right\}_{1\le i,j\le N}  \right]_{\x=0}.
\ee

We can extract the symmetric part of the determinant by using the following identity (\cite{hua1963harmonic}, Thm.~1.2.4 p.~24):
\be
\label{dm-eq:identity-diff-det}
\left[\frac{1}{\vd(\x)} \det_{i,j}\bigl(f_j(x_i)\bigr)\right]_{\forall i,\, x_i=x} = f_N\,\det_{i,j}\left (f_j^{(i-1)}(x)\right),
\ee
with $f_N= 1/\prod_{j=1}^{N-1} j !$.\footnote{The factor $(-1)^{\frac {N(N-1)}{2}}$ in the reference is due to a different convention in the definition of the Vandermonde determinant.} This leads to the following Slater determinant:

\begin{equation}
\label{dm-eq:Slater-diff}
Z_{V_1,V_2}= f_N \det\left\{ \left[ \dxi{i}  \e^{N V_1(\frac 1 {\sqrt N} \dx)} x^{j} \e^{N V_2(\frac 1 {\sqrt N} x)} \right]_{x=0} \right\}_{0\le i,j \le N-1}.
\end{equation}
This is again well-normalized, since for $V_1=V_2=0$ the determinant is $\prod_{i=1}^{N-1} i!$.

\subsubsection{Biorthogonal polynomials}  

As in Sec.~\ref{dm-subsub:ortho-pol-usual}, $\dxi{i} $ and $x^{j} $ can be replaced in \eqref{dm-eq:Slater-diff} by any monic polynomials $P_i(\dx)$, $Q_j(x)$ respectively of degrees $i$ and $j$:
\begin{equation}
\label{dm-eq:Slater-diff-poly}
Z_{V_1,V_2}= f_N \det\left\{ \left[ P_i\Bigl(\dx\Bigr) \e^{N V_1( \frac 1 {\sqrt N} \frac{d}{d x})} Q_j(x) \e^{N V_2(\frac 1 {\sqrt N} x)} \right]_{x=0} \right\}_{0\le i,j \le N-1},
\end{equation}
the differential formulation of the bilinear form of \eqref{dm-eq:bil-form-int} being:
\begin{equation}
\langle f \vert g \rangle =  c_1 \left[ f \biggl(\frac 1 {\sqrt N}\dx\biggr) \e^{N V_1( \frac 1 {\sqrt N}\dx)} g\biggl(\frac x {\sqrt N}\biggr) \e^{N V_2(\frac 1 {\sqrt N} x)} \right]_{x=0}.
\end{equation}
From this, one can apply the method of biorthogonal polynomials. 

\subsubsection{Equivalence between the diagonalized formulations for two-matrix models}
\label{dm-subsub:Equivalence-diag-Twomat}

Expanding the Vandermonde determinants in \eqref{dm-eq:diff-of-diag-two-mat} as in Sec.~\ref{dm-subsub:det-form-usual}, we obtain:
\be
\label{dm-eq:diff-of-diag-two-mat-dev}
Z_{V_1,V_2}=  e_N N!  \det\left\{ \left[ \e^{\frac 1 N \frac{\partial}{\partial a} \frac{\partial}{\partial b}}a^i  \e^{N V_1(a)} b ^{j} \e^{N V_2(b)} \right]_{a=b=0}\right\}_{0\le i,j \le N-1}.
\ee
This is seen to be equivalent to \eqref{dm-eq:Slater-diff},  since, from \eqref{dm-eq:proof3C} which also holds when $A,B$ are real variables (and seeing $N$ as a real coefficient):
\[
\left[ \dxi{i}  \e^{N V_1(
\frac 1 {\sqrt N} 
\frac{d}{d x})} x 
^{j} \e^{N V_2(\frac 1 {\sqrt N} x)} \right]_{x=0} = N^{i/2} \left[ \e^{\frac 1 {\sqrt N} \frac{\partial}{\partial a} \frac{\partial}{\partial b}}a^i  \e^{N V_1(a)} b ^{j} \e^{N V_2(\frac 1 {\sqrt N}b)} \right]_{a=b=0} ,
\]
which in turn by the change of variable $b = \sqrt N b'$ is equal to 
$
 N^{\frac{i+j}2} [ \e^{\frac 1  N \frac{\partial}{\partial a} \frac{\partial}{\partial b}}a^i  \e^{N V_1(a)} b ^{j} \e^{N V_2(b)} ]_{a=b=0}.
$
The normalizations do agree since $f_N = e_N N! / N^{N(N-1)/2}$.

\eqref{dm-eq:diff-of-diag-two-mat-dev}  is in turn seen to be equivalent to the Hankel determinant \eqref{dm-eq:slater-2-mat-usual}, as from \eqref{dm-eq:Two-mat-diff-form} for $N=1$:
\be \left[ \e^{\frac 1  N \frac{\partial}{\partial a} \frac{\partial}{\partial b}}a^i  \e^{N V_1(a)} b ^{j} \e^{N V_2(b)} \right]_{a=b=0} = \int_{\bR^2} \frac{\d x\, \d y}{c_1}\, \e^{-N xy} x^{i}  \e^{N V_1( x)} y^{j} \e^{N V_2(y)},
\ee
where we recall that $c_1=\frac{2\pi}{i N}$, so that $e_N N! / (c_1)^N = N^{\frac{N(N-1)} 2}f_N/(c_1)^N=1/d_N$.

\subsubsection{Expansion over Schur functions}
Performing steps similar to Sec.~\ref{dm-subsub:det-form-usual} but in the differential formulation and the opposite direction, we may re-express \eqref{dm-eq:Slater-diff} as:
\be
\label{dm-eq:Orlov}
Z_{V_1,V_2}= \frac{f_N}{N!} \left[{\Delta\biggl(\frac{\pd}{\pd \x}\biggr)} \e^{NV_1( \frac 1 {\sqrt N}\frac{\pd}{\pd \x})} \vd(\x) \,\e^{NV_2(\frac 1 {\sqrt N} \x)} \right]_{\x=0}.
\ee
As explained in (\cite{orlov2002tau}, Eq~(2.2.4)), the expression \eqref{dm-eq:Orlov} is easily seen --- by its action on Schur functions --- to be a scalar product on symmetric functions, with 
\[Z_{V_1,V_2}= f_N\left \langle \exp\left(N V_1\left( \frac 1 {\sqrt N}\,\bigcdot\,\right)\right), \exp\left(N V_2\left( \frac 1 {\sqrt N}\,\bigcdot\,\right)\right) \right\rangle . \] Using the orthogonality of Schur functions for this scalar product and the Cauchy-Littlewood formula, one obtains directly an explicit double-series expansion over Schur functions  (Eq.~(2.2.13) in \cite{orlov2002tau}), which guarantees the fact that $Z_{V_1,V_2}$ is a tau function of the Toda hierarchy.

\subsection{Integration over one variable in the determinant form of two-matrix models}
\label{dm-sub:integration}

In order to verify the equivalence --- in  differential formulation --- between \eqref{dm-eq:diff-diagonal-new-herm} and \eqref{dm-eq:diff-diagonal-usual-herm}, it is required to go to Slater determinant form, and then transform certain derivatives to variables.
We do this for general $V_1$, as the computation is the same in differential formulation for general $V_1$ or for $V_1(x)=x^2/2$ (corresponding to the one-matrix models). 
We also detail the computation in the integral formulation for general $V_1$, which is naturally more involved than for $V_1(x)=x^2/2$, in which case it is a simple Gaussian integration. This leads to a new determinant formulation of the partition functions of two-matrix models, and we comment on its potential use for resolutions involving orthogonal polynomials instead of biorthogonal polynomials. 

\subsubsection{General computation}
\label{dm-subsubsec:gen-comp}
We start from \eqref{dm-eq:Slater-diff}:
\begin{align}
\label{dm-eq:ex-p}
\begin{split}
Z &= f_N \det_{i,j}\left\{ \left[ \dxi{i}  \e^{ NV_1(\frac1 {\sqrt N}\dx)} x^{j} \e^{N V_2( \frac x{\sqrt N})} \right]_{x=0} \right\} \\
&=  f_N \det_{i,j}\left\{ \left[ \dxi{i} \e^{ N V_1(\frac 1 N \dx)} x^{j} \e^{N V_2( x)} \right]_{x=0} \right\},
\end{split}
\end{align}
where $0\le i,j\le N-1$, by change of variable.\footnote{The factors  $\sqrt N^{j-i}$ from the change of variable compensate when factorized out of the determinant.} In this subsection, we assume $V_1$ to be a polynomial:
\[
V_1(x)=\sum_{k =1}^p \frac{\alpha_k} k x^k, \qquad p\ge 1.
\]
We write, as in \eqref{dm-eq:proof3C} but for real variables:
\be
\label{dm-eq:second-expr-p}
\left[ \dxi{i}  \e^{N V_1(\frac 1 N\dx)} x^{j} \e^{N V_2( x)} \right]_{x=0} =\left[ \left[\e^{\dx\dy}y^i  \e^{N V_1(\frac 1 N y)}\right]_{y=0} x^{j} \e^{N V_2( x)} \right]_{x=0}.
\ee
We first notice that:
\be 
\dyi{r}   \e^{N V_1(\frac 1 N y)} = (N^{1-p}\alpha_p)^r S_{(p-1)r}(y)\,   \e^{NV_1(y)},
\ee
where $S_{(p-1)r}(y)$ is a monic polynomial of degree $(p-1)r$. \\

Let us consider the row $i=(p-1)r+s$, where $s\in\{0,1,\ldots, p-1\}$. We can replace in the determinant the right-hand side of \eqref{dm-eq:second-expr-p}  by 
\be
\Bigl[\Bigl[\e^{\dx\dy} y^s S_{(p-1)r}(y)   \e^{NV_1(\frac 1 Ny)}\Bigr]_{y=0} x^{j} \e^{N V_2( x)} \Bigr]_{x=0}
\ee
 and therefore by
\be
\label{dm-eq:interm-gen-comp}
Z=  f_N \det_{i,j}\left\{ (N^{1-p}\alpha_p)^{-r(i)}\left[\left[\e^{\dx\dy}  y^{s(i)} \dyi{r(i)}   \e^{NV_1(\frac 1 Ny)}\right]_{y=0} x^{j} \e^{N V_2( x)} \right]_{x=0} \right\},
\ee
where $0\leq i,j < N$. We now show that:
\be 
\label{dm-eq:point-to-prove-p-s}
\resizebox{0.95\textwidth}{!}{$
\left[ \left[\e^{\dx\dy} y^s \dyi{r}  \e^{ NV_1(\frac 1 N y)}\right]_{y=0} x^{j} \e^{N V_2( x)} \right]_{x=0}
= \left[ \e^{NV_1(\frac 1 N\dx)} x^r \dxi{s} \Bigl[x^{j} \e^{N V_2( x)} \Bigr]\right]_{x=0},
$}
\ee
that is, just as for \eqref{dm-eq:proof3C} or  \eqref{dm-eq:second-expr-p}, the result of the inner bracket $[...]_{y=0}$ on the left-hand side is obtained by ``replacing'' the $y$'s by $\dx$'s, but also the $\dy$'s by $y$'s, while reversing the left-to-right ordering.

Developing first the right hand side, we see that:
\begin{align}
\begin{split}
&\left[ \e^{NV_1(\frac 1 N\dx)} x^r \dxi{s} \left[x^{j} \e^{N V_2( x)} \right]\right]_{x=0}  \\ 
& \quad \quad= \sum_{\{n_k\ge 0\}_k} \prod_{k=1}^p \frac {(N^{1-k}\alpha_k/k)^{n_k}} {n_k!}  \left[ \dxi{\ell(\bn)} x^r \dxi{s} \left[x^{j} \e^{N V_2( x)} \right] \right]_{x=0},
\end{split}
\end{align}
where $\ell(\bn) = \sum_{k=1}^p k n_k$. The bracket on the right hand side of this equation is non-vanishing only if $\ell(\bn) = \sum_{k=1}^p k n_k\ge r$ and $r$ of the derivatives act on $x^r$, in which case: 
\[
\left[\dxi{\ell(\bn)} x^r \dxi{s} \left[x^{j} \e^{N V_2( x)} \right] \right]_{x=0} = r! \binom{\ell(\bn)}{r}\left[  \dxi{\ell(\bn)-r+s} x^{j} \e^{N V_2( x)} \right]_{x=0}. 
\]
The right hand side of \eqref{dm-eq:point-to-prove-p-s} is therefore equal to:
\be
\label{dm-eq:expansion-proof-gen-comp}
 \sum_{\bigl\{n_k \ge 0\; \mid\; \ell(\bn) \ge\, r\bigr\}}\prod_{k=1}^p \frac {(N^{1-k}\alpha_k/k)^{n_k}} {n_k!} \frac {\ell(\bn) !}{ (\ell(\bn)-r)! }  \left[  \dxi{\ell(\bn)-r+s}x^{j} \e^{N V_2( x)} \right]_{x=0}.
\ee

We now focus on the left-hand side of \eqref{dm-eq:point-to-prove-p-s}:
\begin{align}
\label{dm-eq:LHS-gen-comp}
\left[\e^{\dx\dy} y^s \dyi{r}  \e^{NV_1(\frac 1 N y)}\right] &= \sum_{q\ge 0} \frac 1 {q!}  \left[\dyi{q} y^s \dyi{r} \e^{NV_1(\frac 1 N y)}\right] \dxi{q} \\
&= \sum_{q\ge s} \frac {s!} {q!}  \binom{q}{s}\hspace{-0.1cm}\left[  \dyi{q+r-s}   \e^{NV_1(\frac 1 N y)}\right] \dxi{q},
\end{align}
in which the three brackets are evaluated in $y=0$. We develop:
\begin{align}
\label{dm-eq:LHS-gen-comp-21}
\left[  \dyi{q+r-s}   \e^{NV_1(\frac 1 N y)}\right]_{y=0} & = \sum_{\{n_k\ge 0\}_k} \prod_{k=1}^p \frac {(N^{1-k}\alpha_k/k)^{n_k}} {n_k!} \left[  \dyi{q+r-s}   y^{\ell(\bn)}\right]_{y=0}\\  &= \sum_{\substack{{\{n_k\ge 0\}\textrm{ s.t.}}\\[+0.5ex]{ \ell(\bn) = q + r - s}}} \prod_{k=1}^p \frac {(N^{1-k}\alpha_k/k)^{n_k}} {n_k!} \ell(\bn) !.
\label{dm-eq:LHS-gen-comp-22}
\end{align}
Replacing this in \eqref{dm-eq:LHS-gen-comp} and  then \eqref{dm-eq:LHS-gen-comp} in the left-hand side of \eqref{dm-eq:point-to-prove-p-s} leads to the same expansion \eqref{dm-eq:expansion-proof-gen-comp}. 
We have shown so far that: 
\[
Z_{V_1, V_2} = f_N\det\left\{ M_{i,j} \right\}_{0\le i,j \le N-1},
\]
where:
\begin{align}
\label{dm-eq:general-new-expression-p-s-diff}
M_{(p-1)r + s\; ,\; j} &= ({N^{p-1}}/ {\alpha_p})^{r} \left[ \e^{NV_1(\frac 1 N \dx)} x^r \dxi{s} \Bigl[x^{j} \e^{N V_2( x)} \Bigr]\right]_{x=0}  \\[+1ex]&= ({N^{p-1}}/ {\alpha_p})^{r} \int_{\bR^2}\frac{\d x\, \d y}{2\pi / \imath}\, \e^{-xy} \e^{NV_1(\frac 1 N y)} x^r \dxi{s} \Bigl[x^{j} \e^{N V_2( x)} \Bigr].
\label{dm-eq:general-new-expression-p-s-int}
\end{align}

The equivalence between \eqref{dm-eq:general-new-expression-p-s-diff} and \eqref{dm-eq:general-new-expression-p-s-int} is the usual one-matrix differential formulation of two-matrix models of Sec.~\ref{dm-sub:diff-two-matrix} but for $A,B$ two real variables, and seeing $x^r \dxi{s} [x^{j} \e^{N V_2( x)}]$ as a function of $x$.

\subsubsection{Computation in the integral formulation}
 Starting from \eqref{dm-eq:slater-2-mat-usual}, we change variable as $y' = N y $ and include the normalization by $2\pi/\imath$:
 \begin{equation}
Z_{V_1,V_2}= f_N \det\left\{ \int_{\bR^2} \frac{\d x\, \d y}{2\pi / \imath}\, \e^{- xy} x^{i}  \e^{N V_1( \frac 1 N x)} y^{j} \e^{N V_2(y)}  \right\}_{0\le i,j\le N-1},
\end{equation}
where we have used the fact that $f_N = (2\pi/\imath)^N d_N^{-1} N^{-\frac{N(N+1)}2}$.
The steps up to equation \eqref{dm-eq:interm-gen-comp} are precisely the same in integral form, so we do not detail them again. While a perturbative proof of \eqref{dm-eq:point-to-prove-p-s} is certainly possible, it does not seem to us that the proof above can directly be translated in integral form  (in particular the steps \eqref{dm-eq:LHS-gen-comp} to \eqref{dm-eq:LHS-gen-comp-22}), so we propose a different proof. In integral form,  \eqref{dm-eq:interm-gen-comp} reads
\be 
Z_{V_1,V_2}= f_N \det_{i,j} \left\{(N^{1-p}\alpha_p)^{-r(i)} \int_{\bR^2} \frac{\d x\, \d y}{2\pi / \imath}\, \e^{- xy} y^{s(i)} \left[ \dyi{r(i)}   \e^{ NV_1(y)}\right] x^{j} \e^{N V_2( x)} \right\},
\ee
again with $0\le i,j <  N$. By performing $r$ integrations by part:
\begin{align}
\label{dm-eq:point-proof-int-form}
\begin{split}
\int \d x\, \d y\, \e^{- xy} y^s &\left[\dyi{r}   \e^{ NV_1(y)}\right] x^{j} \e^{N V_2( x)}  \\
&= (-1)^r \int \d x\, \d y\, \left[\dyi{r} \e^{-xy} y^s\right] \e^{ NV_1(y)}    x^{j} \e^{N V_2( x)} ,
\end{split}
\end{align}
where the boundary terms are assumed to vanish every time due to the exponential terms. We use the following identity:
\be 
\dyi{r} \bigl[(ay)^s \e^{a xy}\bigr] = \dxi{s} \bigl[(ax)^r \e^{a xy}\bigr].
\ee 
Indeed, using Leibniz formula:
\begin{align}
\begin{split}
\dyi{r} \left[(ay)^s \e^{a xy}\right]  &= \sum_{k=0}^r \binom{r} {k} a^s\dyi{k}\left[ y^s \right]\dyi{r-k}\left[ \e^{a xy} \right]
  \\ 
  &= \e^{a xy} \sum_{k=0}^{\min(r,s)} \frac 1 {k!} \frac {s!}{(s-k)!}\frac {r!}{(r-k)!} \frac{(ay)^{s} (ax)^{r} }{(axy)^k}.
\end{split}
\end{align}
The same expression is obtained developing $ \dxi{s} \left[(ax)^r \e^{a xy}\right]$. Using this identity,  \eqref{dm-eq:point-proof-int-form} is equal to:
\begin{align}
\begin{split}
(-1)^{s} \int \d x\, \d y\, \e^{ NV_1(y)} &\left[\dxi{s} \e^{- xy} x^r\right]   x^{j} \e^{N V_2( x)}\\
& =\int \d x\, \d y\, \e^{- xy} y^s    \e^{ NV_1(y)} \left[\dxi{s}x^{j} \e^{N V_2( x)}\right] ,
\end{split}
\end{align}
where the equality is obtained by doing again  $s$ integrations by parts. 
We thus recover \eqref{dm-eq:general-new-expression-p-s-int}.

\subsection{Potential application to new orthogonal polynomial method}
\subsubsection{Equivalence between the diagonalized differential formulations for one-matrix models}
\label{dm-subsub:invert-diff-diag-gaussian}
 Performing the steps of Sec.~\ref{dm-subsub:slater-diff-two-mat}, but starting from the differential formulation \eqref{dm-eq:diff-diagonal-new-herm} of one-matrix models leads to:
\begin{equation}
\label{dm-eq:Slater-diff-one-mat}
Z_{V}= f_N\, \det\left\{ \left[ \dxi{i} \e^{\frac 1 {2N} \dxi{2}} x^{j} \e^{N V( x)} \right]_{x=0} \right\}_{0\le i,j \le N-1}.
\end{equation}
This is also the differential formulation of the Slater determinant expression \eqref{dm-eq:Slater-diff} for a two-matrix model with $V_1(x)=\frac{x^2}{2}$.

On the other hand, performing steps similar to Sec.~\ref{dm-subsub:det-form-usual} to express \eqref{dm-eq:diff-diagonal-usual-herm} in determinant form, we obtain: 
\begin{equation}
\label{dm-eq:Slater-diff-one-mat-bis}
Z_{V}= N! e_N\, \det\left\{ \left[ \e^{\frac 1 {2N} \dxi{2}} x^{i+j} \e^{N V( x)} \right]_{x=0} \right\}_{0\le i,j \le N-1}.
\end{equation}

From the simpler formulation \eqref{dm-eq:Slater-diff-one-mat-bis}, one can use \emph{orthogonal} polynomials satisfying
\be 
\label{dm-eq:orthogonal-differential}
\left[  \e^{\frac 1 {2N} \dxi{2}} P_i(x)P_j(x) \e^{N V( x)}\right]_{x=0} = \delta_{ij} p_i
\ee
instead of \emph{biorthogonal} polynomials \eqref{dm-eq:bil-form-int} satisfying 
\be
\label{dm-eq:biorthogonal-differential}
\left[ P_i\Bigl(\dx\Bigr) \e^{\frac 1 {2N} \dxi{2}} Q_j(x) \e^{N V( x)} \right]_{x=0} = \delta_{ij} p_i,
\ee
\emph{a priori} leading to a simpler resolution.

\

The computations of Sec.~\ref{dm-sub:integration} in the differential formulation, in the case where $V_1(x)=\frac{x^2}{2}$ ($p=2$, $\alpha_2=1$), prove the equivalence between \eqref{dm-eq:Slater-diff-one-mat} and \eqref{dm-eq:Slater-diff-one-mat-bis}. Indeed for $p=2$, $r(i)=i$ so that the coefficients  $({N^{p-1}}/ {\alpha_p})^{r(i)}=N^i$ in the determinant just provide an overall factor $N^{N(N-1)/2}$, and   $f_NN^{\frac {N(N-1)} 2}= N! e_N$.

\

In integral form, the proof is simpler, as the Gaussian integration for $V_1(x)=\frac{x^2}{2}$ can be carried out explicitly after changing variables for  $x'=N x / \imath$.\footnote{There is an important subtlety in changing variables from a real variable to a purely imaginary variable: in addition to the Jacobian of the change of variables, an additional factor $-1$ has to be taken into account. This is detailed in Appendix~\ref{dm-appendixB}, see \eqref{dm-eq:change-variables-pure-complex}. The problem does not appear for the one-matrix model in the computations in differential formulation of Sec.~\ref{dm-sub:integration}, or using the heat kernel \eqref{dm-eq:heat-kernel}.} Equivalently, one may recover directly the Slater determinant form of the partition function of one-matrix models in the integral formulation, that is  
\begin{equation}
\label{dm-eq:slater-1-mat-usual}
Z_{V_2}= \frac {N!}{b_N} \det\left\{ \int \d y\,  \e^{-N\frac {y^2}2}\e^{NV_2(y)}y^{i+j}  \right\}_{0\le i,j\le N-1},
\end{equation} 
from the differential formulation of the Slater determinant \eqref{dm-eq:Slater-diff} (after the change of variables \eqref{dm-eq:ex-p}),
\begin{equation}
\label{dm-eq:Slater-diff-onemat}
Z_{{x^2}/2,V_2}= f_N \det\left\{ \left[ \dxi{i}  \e^{\frac 1 {2N} \dxi{2}} x^{j} \e^{N V_2( x)} \right]_{x=0} \right\}_{0\le i,j \le N-1},
\end{equation}
by using the \emph{heat kernel} formulation:
\be
\label{dm-eq:heat-kernel}
\e^{t \dxi{2}} F(x) = \int \d y\, K_t(x-y) F(y), \qquad K_t(x-y) = \frac 1 {\sqrt{4\pi t}}e^{-\frac{(x-y)^2}{4t}}, 
\ee
for $t=\frac 1 {2N}$ and $F(x)= x^{j} \e^{N V_2(x)}$. We get: 
\be
\label{dm-eq:C-second-onematrix}
Z_{{x^2}/2,V_2}= f_N \det\left\{ \sqrt{\frac N {2\pi}}  \int \d y \left[ \frac{\partial^{i}}{\partial x^{i}}   \e^{-\frac{N (x-y)^2}{2}}\right]_{x=0}   y^{j} \e^{N V_2( y)} \right\}_{0\le i,j \le N-1},
\ee
where the bracket evaluating $x$ at 0 has been moved inside the integral, as only the derivative of the kernel depends on $x$. One can verify that 
\[
\left[ \frac{\partial^{i}}{\partial x^{i}}   \e^{-\frac{N (x-y)^2}{2}}\right]_{x=0}  =N^{i} \tilde Q_i(y) \e^{-\frac{N}2 y^2},
\]
where $\tilde Q_i(y)$ is a monic polynomial of degree $i$, so we do recover  \eqref{dm-eq:slater-1-mat-usual} as $f_N (\frac N {2\pi})^{\frac N 2} N^{\frac {N(N-1)}2} = N!/b_N$.

One can also use the heat kernel to find the integral representation of \eqref{dm-eq:diff-diagonal-new-herm} and the one-matrix version of \eqref{dm-eq:Slater-first-expr}, again obtained by setting $V_1(x) = \frac{x^2}{2}$. Applying the $N$-dimensional version of \eqref{dm-eq:heat-kernel} to \eqref{dm-eq:diff-diagonal-new-herm} for each $x_i$ we find
\be
Z_V = \left(\frac{N}{2\pi}\right)^{\frac{N}{2}} \left[ \frac{1}{\vd(\x)}  \int \d\y \, \e^{-\frac{N(\x-\y)^2}{2}} \vd(\y)\, e^{N V(\y)} \right]_{\x=0}.
\label{dm-eq:diff-diagonal-new-herm-heat}
\ee
Similarly, in the one-matrix version of \eqref{dm-eq:Slater-first-expr} we can use the heat kernel to find
\be
Z_{x^2/2,V_2} = \left[ \frac{1}{\vd(\x)} \det  \left\{ \frac{1}{\sqrt{2\pi}}\int \d y\, \e^{-\frac{(x_i-y)^2}{2}} y^{j-1} \e^{N V_2\left(\frac{1}{\sqrt{N}} y\right)} \right\}_{1\le i,j\le N} \right]_{\x=0}.
\label{dm-eq:Slater-first-expr-heat}
\ee

\subsubsection{Application to new orthogonal polynomial methods}
\label{dm-subsubsec:app-new-pol}

As we have seen at the beginning of the present section, the computations of Sec.~\ref{dm-sub:integration} in the case where $V_1(x)=\frac{x^2}{2}$, prove the equivalence in the differential formulation between \eqref{dm-eq:Slater-diff-one-mat} and \eqref{dm-eq:Slater-diff-one-mat-bis}, and the latter expression allows using orthogonal polynomials \eqref{dm-eq:orthogonal-differential} instead of biorthogonal polynomials \eqref{dm-eq:biorthogonal-differential}. One may naturally wonder if for potentials $V_1$ of degree higher than two, our new expressions \eqref{dm-eq:general-new-expression-p-s-diff}, \eqref{dm-eq:general-new-expression-p-s-int} could also allow the use of orthogonal polynomials instead of biorthogonal polynomials for two-matrix models.

\ 

To this aim, we reorganize the lines of the matrix $M$ in \eqref{dm-eq:general-new-expression-p-s-diff} according to their remainder modulo $p-1$, that is:
\begin{equation}
\label{dm-eq:new-determinant}
\begin{split}
Z_{V_1, V_2} = h_N \det\left\{ \tilde M_{i,j} \right\}_{0\le i,j \le N-1}, \quad \\ \mathrm{where}\quad 
\tilde M = \left(
\begin{array}{c}
 T_0 \\ \hline
 \vdots
 \\
 \hline
T_{p-1} 
\end{array}
\right) \quad \mathrm{and} \quad (T_s)_{r,j} = M_{(p-1)r + s\; ,\; j}. 
\end{split}
\end{equation}
where $h_N = f_N\, \mathrm{sgn} (N, p)$, $\mathrm{sgn} (N, p)$ being the parity of the permutation of rows, and we recall (\eqref{dm-eq:general-new-expression-p-s-diff}) that in differential formulation,
\be
\label{dm-eq:general-new-expression-p-s-recall}
M_{(p-1)r + s\; ,\; j} = ({N^{p-1}}/ {\alpha_p})^{r} \left[ \e^{NV_1(\dx)} x^r \dxi{s} \Bigl[x^{j} \e^{N V_2( x)} \Bigr]\right]_{x=0}.
\ee 
If $s_0$  and $r_0$ are respectively the remainder and the quotient of the Euclidean division of $N-1$ by $p-1$, then for $0\le s\le s_0$, the matrix $T_s$ has $r_0+1$ lines, while for $s_0< s < p-1$, $T_s$ has $r_0$ lines.  

\ 

By reorganizing the columns of $\tilde M$, and  the lines of $T_s$ independently for each $s$, we may replace $({N^{p-1}}/ {\alpha_p})^{-r}(T_s)_{r,j} = ({N^{p-1}}/ {\alpha_p})^{-r}M_{(p-1)r + s\; ,\; j}$ in the determinant by 
\begin{align}
\label{dm-eq:new-determinant-ortho}
\begin{split}
 \left[ \e^{NV_1(\frac 1 N \dx)} P_r^{(s)}(x) \dxi{s} \Bigl[Q_j(x) \e^{N V_2( x)} \Bigr]\right]_{x=0}  \\ =  \int\frac{\d x\, \d y}{2\pi / \imath}\, \e^{-xy} \e^{NV_1(\frac 1 N y)} P_r^{(s)}(x) \dxi{s} \Bigl[Q_j(x) \e^{N V_2( x)} \Bigr],
 \end{split}
\end{align}
where for $0\le s < p-1$, the $P_r^{(s)}(x)$ are monic polynomials of degree $r$, and $Q_j$ is a monic polynomial of degree $j$.

\ 

One may for instance choose $P_r^{(s)} = Q_r$ for all $0\le s < p-1$. We use the following notation for the determinant obtained this way:
\begin{equation}
\label{dm-eq:new-determinant-ortho-fin}
Z_{V_1, V_2} = h_N \det\left\{ W_{i,j} \right\}_{0\le i,j \le N-1}, \quad \mathrm{where}\quad 
W = \left(
\begin{array}{c}
 J_0 \\ \hline
 \vdots
 \\
 \hline
J_{p-1} 
\end{array}
\right). 
\end{equation}
For $s=0$, the elements $(J_0)_{r,j}$ of the resulting matrix $J_0$ define the symmetric bilinear form 
\begin{align}
\begin{split}
\langle Q_r \mid\mid Q_j \rangle &= \left[ \e^{NV_1(\frac 1 N \dx)} \e^{N V_2( x)} Q_r(x) Q_j(x)  \right]_{x=0} \\
&= \frac{\imath}{2\pi} \int\d x\,  Q_r(x) Q_j(x) \e^{N V_2( x)} \int \d y\, \e^{-xy} \e^{NV_1(\frac 1 N  y)},
\end{split}
\end{align}
and requiring  $\langle Q_r \mid\mid Q_j \rangle = h_r \delta_{r,j}$ defines a family of orthogonal polynomials for the following weight: 
\be 
\label{dm-eq:ortho-weight}
\e^{NV_1(\frac 1 N  \dx)}\e^{N V_2( x)} \quad \leftrightarrow\quad  \e^{N V_2( x)} \int \d y\, \e^{-xy} \e^{NV_1( \frac 1 N  y)}.
\ee
In addition to the usual three-terms recurrence $Q_{n+1}(x) = (x-\beta_n) Q_n - \frac{h_n}{h_{n-1}} Q_{n-1}$ satisfied by any family of orthogonal polynomials,  to obtain families of recurrence relations on the coefficients $\beta_n$ and $h_n$, one may for instance use the fact that 
\begin{align}
    \left[ \e^{NV_1(\frac 1 N \dx)} V_1'\Bigl(\frac 1 N\dx\Bigr)F(x) \right]_{x=0} & =     \left[ \left[ \e^{\dx \dy} V_1'\Bigl(\frac 1 Ny\Bigr)\e^{NV_1(\frac 1 N y)} \right]_{y=0}F(x)\right]_{x=0}\\&= \left[ \left[ \e^{\dx \dy} \dy\e^{NV_1(\frac 1 N y)} \right]_{y=0}F(x) \right]_{x=0} ,
\end{align} 
so that 
\be 
\label{dm-eq:equation-E-ortho-pol}
\left[ \e^{NV_1(\frac 1 N\dx)} V_1'\Bigl(\frac 1 N\dx\Bigr)F(x) \right]_{x=0} = \left[  \e^{NV_1(\frac 1 N\dx)}  xF(x) \right]_{x=0},
\ee 
applied to $F(x) = Q_r(x) Q_j(x) \e^{N V_2( x)}$ for any $r,j$. For instance for the case $V_1(x) = x^3 / 3$, for which the weight for the orthogonal polynomials includes the Airy function as a factor, this equation 
leads to the family of equations 
\begin{align}
\begin{split}
\frac 2 N  \langle V_2' Q_r' \mid\mid Q_j \rangle &+ \frac 2 N  \langle V_2' Q_r \mid\mid Q_j' \rangle +  \langle R\, Q_r \mid\mid Q_j\rangle\\
&+ \frac 1 {N^2}\Bigl(\langle Q_r'' \mid\mid Q_j \rangle + \langle Q_r \mid\mid Q_j'' \rangle + 2\langle  Q_r' \mid\mid Q_j' \rangle\Bigr)   = 0,
\end{split}
\end{align}
with $R(x) = \frac 1 N  V_2''(x) +  ( V_2'(x))^2 -  x$. 

\

The rest of the determinant \eqref{dm-eq:new-determinant-ortho-fin} for $s>0$, obtained from \eqref{dm-eq:new-determinant-ortho}, also involves sums of terms of the form $\langle g\, Q_r \mid\mid Q_j^{(k)} \rangle $ for some polynomials $g(x)$. Computing the determinant therefore requires knowledge about the derivatives of the families of orthogonal polynomials with weight \eqref{dm-eq:ortho-weight}. It is not clear at this point whether the family of orthogonal polynomials can indeed be constructed and whether the rest of the determinant can be computed this way.

On the other hand, everything seems to simplify drastically at large $N$: for the choice $P_r^{(s)} = Q_r$ for all $0\le s < p-1$, the leading contributions in $N$ to the matrix elements \eqref{dm-eq:new-determinant-ortho} for the  part of the determinant corresponding to $s>0$ are obtained when all $s$ derivatives act on $e^{NV_2(x)}$, raising a factor  $N^s(V_2'(x))^s$, so that the elements of the determinant simplify to
\be
\label{dm-eq:new-determinant-ortho-lead}
(J_s)_{r,j} \sim_{N\rightarrow \infty} \frac{N^{(p-1)r + s}}{\alpha_p^r}\left \langle (V_2')^s Q_r \mid\mid Q_j \right\rangle .
\ee 
In the same way, the leading contribution to the family of relations \eqref{dm-eq:equation-E-ortho-pol} seems to be given by 
\begin{align}
\label{dm-eq:equation-E-ortho-pol-largeN}
\left[ \e^{NV_1(\frac 1 N\dx)} e^{NV_2(x)}V_1'\bigl(V_2'(x)\bigr) Q_r(x)Q_j(x) \right]_{x=0} &= \left[  \e^{NV_1(\frac 1 N\dx)}  xF(x) \right]_{x=0}, \\
 \Leftrightarrow \quad \langle U Q_r \mid\mid Q_j\rangle &=0,
\end{align} 
 with $U(x) = V_1'\bigl(V_2'(x)\bigr) - x$.
However, this already goes beyond the original scope of the chapter, and we leave this for future work.

\section{Summary and conclusion}
We have studied a reformulation of one- and two-matrix models in a language of differential operators. This is especially interesting in the context of two-matrix models, where an integral over two matrix variables can be recast into an expression containing only one matrix variable and its derivatives. For such two-matrix models, we explicitly showed how the Harish-Chandra--Itzykson--Zuber formula can be used to diagonalize the matrix variables in both the differential and integral settings. Furthermore, we demonstrated that the partition functions for both one- and two-matrix models can be written in terms of Slater determinants. This Slater determinant expression leads to the decomposition of the partition function into inner products of (bi-)orthogonal polynomials.

We organize the many different (but equivalent) expressions of the one-matrix model partition function in Fig.\ \ref{dm-fig:diagram-one-matrix}, and those of the two-matrix model partition function in Fig.\ \ref{dm-fig:diagram-two-matrix}. This highlights the overall structure of our work, and points out the relations that exist between the various partition function expressions encountered in this chapter. 

The decomposition of matrix model partition functions in terms of (bi-)orthogonal polynomials can lead to exact solutions of a model's correlation functions, provided that one can explicitly find the appropriate family (or families) of polynomials. This can, for instance, be achieved by solving recurrence relations obtained from inner products between expressions involving the polynomials. This is generally simpler for one-matrix models, where one can make use of families of orthogonal polynomials, as opposed to families of biorthogonal polynomials for the two-matrix case. Our reformulation of the two-matrix model partition function in terms of a single matrix variable and its derivatives suggests that it may also be possible to solve such models using families of orthogonal instead of biorthogonal polynomials. Finally, we propose a possible starting point for such a procedure in Sec.\ \ref{dm-subsubsec:app-new-pol}.

\renewcommand{\eqref}[1]{(\ref{#1})} 


\chapter{Quantum gravity on the computer}\label{ch:monte-carlo}
The CDT partition function after Wick rotation takes the form of a sum over configurations with a real Boltzmann weight dictated by the Regge action. We can therefore view CDT quantum gravity as a \emph{statistical} theory of random geometry. As discussed in Sec.\ \ref{intro-sec:cdt} of the Introduction, we can compute quantum gravitational expectation values $\langle \mo \rangle$ of observables by determining their ensemble average:
\begin{equation}
	\langle \mo \rangle= \frac{1}{Z} \sum_{T \in \mathcal{T}} \frac{1}{C_T}\, \mo[T] \, e^{-S_E[T]},
\end{equation}
where the normalization factor $Z$ is the partition function \eqref{intro-eq:cdt-sum} of the system and $S_E[T]$ is the CDT Regge action \eqref{intro-eq:regge-cdt}. However, analytically computing expectation values of observables is often intractable. In these cases, we can make use of computer simulations to sample the geometric ensemble, allowing us to compute an estimator of an observable's quantum expectation value. The technical underpinnings of such simulations are comprehensively described in \cite{ambjorn2001dynamically}. This chapter aims to provide a hands-on introduction to the basic principles involved in CDT computer simulations, and to explain in detail how these principles can be put into practice. We start by discussing the general set-up of Markov chain Monte Carlo (MCMC) methods for random geometry. Next, we describe basic ingredients of CDT simulations in two and three dimensions. We subsequently explain some of the difficulties one encounters when requiring these simulations to run efficiently, and how a particular method for storing simplices in memory can be used to address these issues. Finally, we discuss the three stages of a simulation run: tuning of bare coupling constants to their pseudocritical value, evolution of the initial geometry to a well-thermalized state, and the measurement of observables on the geometry.

The latter part of this chapter is heavily based on our own implementation of CDT simulations in two and three dimensions. The associated C++ implementation code \cite{brunekreef2021jorenb,brunekreef2022jorenb} has been open-sourced and is freely available for use by quantum gravity enthousiasts.

\section{Monte Carlo methods for random geometry}
\label{mc-sec:mcmrg}
Our aim is to generate a sample $X$ of an ensemble $\mathcal{T}$ of triangulations, with the following probability distribution $P(T)$ for drawing a triangulation $T$ from the sample:
\begin{equation}
	P(T) = \frac{1}{Z} e^{-S_E(T)}.
	\label{mc-eq:prob-sample}
\end{equation}
By computing the sample mean of an observable $\mo$ over $X$ we obtain an estimator $\widehat{\langle \mo \rangle}$ of the ensemble average $\langle \mo \rangle$. In order to generate the sample $X$ we can make use of Markov chain Monte Carlo (MCMC) methods. The idea behind this approach is to construct a random walk (in the form of a Markov chain) through the space of triangulations, where we perform local updates to a geometry at each step of the walk. By enforcing certain conditions on this random walk, we can ensure that the probability of encountering a specific triangulation $T$ in the random walk approaches \eqref{mc-eq:prob-sample}. If we sample from this random walk with a sufficiently large number of local updates in between subsequent draws, we obtain a set of independent triangulations with the desired probability distribution. This set forms the sample $X$. It is useful to point out here that we work with \emph{labeled} triangulations in computer simulations, and that triangulations related by a relabeling of building blocks should be counted only once. This requires us to introduce a combinatorial factor to correct for this overcounting, so that for a labeled triangulation $T$ we have
\begin{equation}
	P_l(T) = \frac{1}{Z} \frac{1}{N_0(T)!} e^{-S_E(T)},
	\label{mc-eq:prob-labeled}
\end{equation}
where $N_0(T)$ is the number of vertices present in the geometry. We label the vertices instead of the higher-dimensional simplices, since the higher-dimensional simplices can be described by lists of vertex labels.

In practice, it is convenient to start the random walk from a simple triangulation that we construct ``by hand''. This is a special choice of starting point, and in order to avoid that any bias affects the measurement results one should minimize dependence on this choice of initial geometry. We therefore let the simulation run for a long time in order to let the system \emph{thermalize}. After this thermalization phase is complete, we have obtained an independent triangulation that can be used as the initial configuration for a series of measurements. In between subsequent measurements, we perform a large number of local updates, which we group together in batches, also called \emph{sweeps}. The size of a sweep should be taken large enough to ensure that significant autocorrelations with the previously measured geometry have disappeared. However, larger sweeps require longer simulation times, and we typically have to strike a balance between reducing autocorrelation influences and increasing execution time. Later in this chapter we discuss how to determine appropriate lengths for thermalization phases and sweeps.

We refer the interested reader to \cite{newman1999monte} for a more detailed explanation of the MCMC approach and associated practical issues like autocorrelation effects, and now proceed to briefly discuss the main elements relevant for our purposes.

\subsection{Random walks and detailed balance}
The first condition we should enforce on the random walk is \emph{ergodicity}, meaning that all triangulations in the geometric ensemble $\mathcal{T}$ can be reached in a finite number of steps. Without ergodicity, we would be restricted to a subspace of the ensemble, potentially making the estimators of quantum expectation values unreliable. The second condition relates to the transition probabilities of a single step in the random walk. Let us write the probability for transitioning from a triangulation $T$ to a new triangulation $T'$ in a single step by $p(T \to T')$. In order to let the probability distribution of the random walk converge to \eqref{mc-eq:prob-sample}, it is sufficient to demand that the transition probabilities satisfy the condition of \emph{detailed balance}:
\begin{equation}
	P_l(T) p(T \to T') = P_l(T') p(T' \to T).
	\label{mc-eq:detailed-balance}
\end{equation}
For the ensembles considered in this thesis, it is convenient to restrict the transitions to a basic set of small local updates to the triangulations, and as a consequence most transition probabilities are set to zero. For the allowed transitions a solution to the detailed balance equation \eqref{mc-eq:detailed-balance} is
\begin{align}
	\frac{P_l(T)}{P_l(T')} = \frac{p(T' \to T)}{p(T \to T')}.
	\label{mc-eq:transition-probs}
\end{align}
We furthermore allow for trivial transitions $T \to T$ from a triangulation to itself, i.e.\ $p(T \to T) \neq 0$. Labeling every step of the walk with an integer ``time'' parameter $n$, the random walk consists of a sequence of triangulations $T_n$. The updating procedure then works as follows. At step $n$ of the random walk, we have that $T_n = T$. We select one of the allowed local updates $T \to T'$ with a \emph{selection probability} $g(T \to T')$. This proposed update, also called a \emph{move}, is subsequently accepted with probability $A(T \to T')$, and rejected otherwise. The probability $A(T \to T')$ for a move to be accepted is called the \emph{acceptance ratio}. In case it is accepted, we set the next triangulation in the sequence to be $T_{n+1} = T'$, and if the move is rejected the triangulation is unchanged and we set $T_{n+1} = T$. 

The transition probability for the move $T \to T'$ can be decomposed as
\begin{equation}
	p(T \to T') = g(T \to T') A(T \to T').
\end{equation}
The selection probability $g(T \to T')$ depends on the set of basic moves and the details of the triangulation $T$ at hand. Subsequently, the acceptance ratio $A(T \to T')$ should be chosen in such a way to ensure detailed balance. A simple choice is the one made in the \emph{Metropolis-Hastings algorithm} \cite{metropolis1953equation,hastings1970monte}, and it amounts to setting
\begin{align}
	A(T \to T') &= \textrm{min}\left(1, \frac{g(T' \to T) P_l(T')}{g(T \to T') P_l(T)} \right), \\
	A(T' \to T) &= \textrm{min}\left(1, \frac{g(T \to T') P_l(T)}{g(T' \to T) P_l(T')} \right).
	\label{mc-eq:acceptance-ratios}
\end{align}
It is instructive to see how this works in the two example scenarios of two- and three-dimensional CDT, and how the choice of the move implementation affects the selection probabilities and associated acceptance ratios. In what follows, we denote by an $(n,m)$-move a geometric update on a neighborhood consisting of a number $n$ of $d$-simplices before the move and a number $m$ of $d$-simplices after the move.

\subsection{Metropolis algorithm for 2D CDT}
\label{mc-sec:metropolis-2d}
\begin{figure}[ht!]
	\centering
	\includegraphics[]{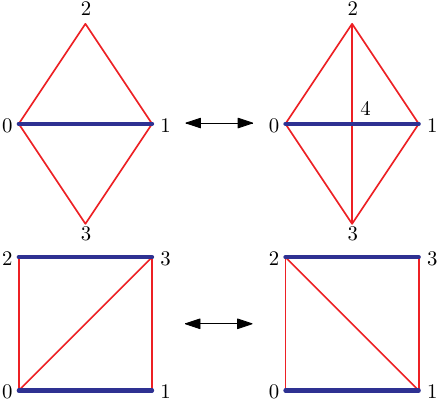}
	\caption{The two basic move sets used in 2D CDT: \code{add}/\code{delete} (top) and \code{flip} (bottom). The simplices consist of spacelike (blue) and timelike (red) links. The labels of the vertices are for identification purposes.}
	\label{mc-fig:2d-cdt-moves}
\end{figure}

We start by computing the probability $P_l(T)$ of encountering a certain labeled triangulation $T$ in the ensemble of 2D CDT with toroidal topology. Recalling the Regge action \eqref{intro-eq:cdt-regge-2d} for CDT in two dimensions and substituting this into Eq.\ \eqref{mc-eq:prob-labeled}, we find 
\begin{equation}
	P_l(T) = \frac{1}{Z} \frac{1}{N_0(T)!} e^{-\lambda N_2(T)}.
\end{equation}
This expression will be used to compute the left-hand side of Eq.\ \eqref{mc-eq:transition-probs} for a certain move and its inverse. Only two simple moves (and their inverses) suffice to construct an ergodic random walk through the space of two-dimensional CDT geometries with a fixed number of time slices and an either even or odd number of triangles. We illustrate the moves in Fig. \ref{mc-fig:2d-cdt-moves}, where we show spacelike links in blue and timelike links in red. Furthermore, time labels of the slices increase in the upwards direction, so that a 21-simplex is a triangle pointing up and a 12-simplex is a triangle pointing down.\footnote{Here we follow the shorthand simplex type naming convention introduced in Sec.\ \ref{intro-sec:cdt}, so that a $pq$-simplex contains $p$ vertices in the slice labeled $t$ and $q$ vertices in the slice labeled $(t+1)$.} The first move, denoted the $(2,4)$-move, takes two adjacent triangles sharing a spacelike link, and replaces it by four triangles. We also call this the \code{add} move. Its inverse, the $(4,2)$-move or \code{delete} move, takes four triangles sharing a single vertex of order four (the order of a vertex is defined as the number of triangles that contain this vertex), and replaces them by two triangles. The second move, called the $(2,2)$-move or \code{flip} move, takes two adjacent triangles sharing a timelike link, and flips this link between them. This move is only allowed when one of the two triangles is of 12-type and the other is of 21-type, otherwise we would introduce timelike links connecting vertices on the same time slice. The $(2,2)$-move is its own inverse. 

\subsubsection{The \code{add} and \code{delete} moves}
We start by considering the \code{add} move and its inverse, the \code{delete} move. Performing the \code{add} move on a triangulation $T$ consisting of $N_2$ triangles results in a triangulation $T'$ consisting of $(N_2+2)$ triangles. A simple method for implementing the \code{add} move given a starting triangulation $T$ is as follows. We randomly select a single triangle \code{t} in the triangulation, with uniform probability $1/N_2$ per triangle (where $N_2$ is the total number of triangles in the system). We then find the unique triangle \code{tv} that shares a timelike link with \code{t}, i.e.\ its \code{v}ertical neighbor. These triangles can also be described by their constituent vertices as marked in Fig. \ref{mc-fig:2d-cdt-moves}, so that we have \code{t} $=(012)$ and \code{tv} $=(031)$ (vertices are listed in counterclockwise order by convention). Note that \code{t} is of type 21 in this case, and the move should be understood to proceed analogously if the input triangle \code{t} is of type 12.

The proposed move is to insert a new vertex, labeled 4 in the Figure, and two triangles \code{tn} $= (412)$ and \code{tvn} $=(431)$ on the right of \code{t} and \code{tv}, respectively. The original triangles get their vertices reassigned as \code{t} $=(042)$ and \code{tv} $=(034)$. We denote the resulting triangulation $T'$, with $(N_2+2)$ triangles. Note that as a part of this update, one vertex and three links were added to the geometry. The new vertex requires a label in the full triangulation (these labels are generally different from the ones shown in the Figure), and we can choose to insert it anywhere in the ``list'' of existing vertices. This implies that there are $(N_0+1)$ ways of assigning this label, while arriving at the same abstract triangulation $T'$. Furthermore, the resulting triangulation is the same regardless of whether we selected \code{t} or its vertical neighbor \code{tv} initially, so we should multiply the probability of going from $T$ to $T'$ by two. The overall \emph{selection probability} for this move is therefore $g(T \to T') = \frac{2}{N_2 \cdot (N_0+1)} = \frac{1}{N_0 \cdot (N_0+1)}$, where we used that $N_2 = 2 N_0$ for 2D CDT with toroidal topology.

We do not need to perform any checks before proposing this move, since it can be carried out for any input triangle \code{t} in the geometry. This is not the case for its inverse, the (4,2)-move, since it requires that the vertex shared by the four initial triangles is of order four. We now present two distinct approaches for implementing the \code{delete} move, where we will see that this influences how we should choose the acceptance ratios. Since we used an initial triangulation $T$ on which we perform the \code{add} move, we take the resulting triangulation $T'$ as a starting point for the \code{delete} move in order to compute the terms in Eq.\ \eqref{mc-eq:acceptance-ratios}.

\paragraph{Method I: blind guessing.}
The simplest approach amounts to randomly selecting a vertex \code{v} in the triangulation $T'$ with uniform probability $\frac{1}{N_0(T')} = \frac{1}{N_0(T)+1}$. If \code{v} is not of order four, do not perform any update to the geometry, and let the next triangulation in the random walk again be $T'$. On the other hand, if \code{v} is of order four, find the four triangles that contain \code{v} = $(4)$, as shown in Fig. \ref{mc-fig:2d-cdt-moves}. Now delete the triangles $(412), (431)$ from the system, and reassign the vertices of the other two as $(042) \to (012), (034) \to (031)$. This move only brings us back to the triangulation $T$ that we started out with if the selected vertex \code{v} is the one that was added to the system by the \code{add} move discussed earlier, so that the selection probability for the move $T' \to T$ equals $g(T' \to T) = \frac{1}{N_0(T)+1}$. Plugging this into Eq.\ \eqref{mc-eq:acceptance-ratios}, we find the acceptance ratios
\begin{align}
	A(T \to T') &= \textrm{min}\left(1, \frac{N_0(T)}{N_0(T)+1} e^{-2\lambda} \right), \\
	A(T' \to T) &= \textrm{min}\left(1, \frac{N_0(T)+1}{N_0(T)} e^{2 \lambda} \right).
\end{align}
The exponential terms are due to the change in the action, and these would appear regardless of the choice of selection process. On the other hand, the ratio $N_0/(N_0+1)$ arises on combinatorial grounds, and is tied to the specific implementation of the moves.

\paragraph{Method II: bookkeeping.}
The previous method is simple to implement, but one may expect that it will often result in no change to the triangulation. After all, the move is aborted if the selected vertex is not of order four. In a thermalized configuration, approximately $1/4$ of the vertices are of order four \cite{ambjorn1999new}, so that on average the configuration is updated at most a quarter of the times that a \code{delete} move is proposed. We may expect that we can do better by keeping track of the order-four vertices throughout the simulation, and to select these sites as input for the \code{delete} move. The move is then always legal, and it remains to compute the selection probabilities and the appropriate acceptance ratios. The selection probability for the \code{add} move is unchanged, but the selection probability for the \code{delete} move now reads $g(T' \to T) = \frac{1}{N_\textit{vf}\,(T) + 1}$, where $N_\textit{vf}\,(T)$ is the number of vertices of order four in the original triangulation $T$. Computing the corresponding acceptance ratios, we find
\begin{align}
	A(T \to T') &= \textrm{min}\left(1, \frac{N_0(T)}{N_\textit{vf}\,(T)+1} e^{-2\lambda} \right), \\
	A(T' \to T) &= \textrm{min}\left(1, \frac{N_\textit{vf}\,(T)+1}{N_0(T)} e^{2 \lambda} \right).
\end{align}
Since $N_\textit{vf} \leq N_0$ by definition, we see that the acceptance ratio for the \code{delete} move has in fact gone down, while it has gone up for the \code{add} move. However, we should take into account that the \code{delete} move using method I was rejected in case the selected vertex was not of order four. In two-dimensional CDT, the critical value of the cosmological constant is $\lambda_c = \ln 2$, and we typically perform simulations at this fixed value for $\lambda$. For thermalized configurations, the probability for a random vertex in the configuration to have order four is equal to $1/4$, so that $N_0 \approx 4N_\textit{vf}$. If the system is large enough, we therefore conclude that the transition probabilities for the move $T' \to T$ are approximately the same for methods I and II. The acceptance ratio for the \code{add} move in method I is close to 1, and the fact that the acceptance ratios are bounded from above by 1 implies that $A(T \to T') = 1$ almost always in method II.

We see that in this model, the ``blind'' and ``bookkeeping'' methods are likely to be equally useful in the case of thermalized configurations. However, it is reasonable to expect that method II is more efficient in the thermalization phase, where order-four vertices are possibly rare. Furthermore, the arguments that showed that transition rates are approximately equal hinged upon the fact that the fraction of order-four vertices is approximately constant in this model. Similar relations are not necessarily present in higher-dimensional models, and it may be the case that bookkeeping-type approaches are more efficient for certain ranges of the coupling parameters. We end the discussion on the two-dimensional model by pointing out that attaining high acceptance rates for the moves does not guarantee optimal execution times in practice. The reason is that a simple check like the one occurring in method I is computationally cheap, and it may be the case that the overhead incurred by bookkeeping-type methods slows down the simulation to an extent that it offsets the potential benefits. In many situations, it may therefore \emph{a priori} be unclear which of the approaches provides the best performance, and one will need to carry out profiling runs in order to determine the optimal strategy for the purpose at hand.

\subsubsection{The \code{flip} move}
We now discuss the implementation of the $(2,2)$-move, also called the \code{flip} move. Again, we can opt for blind- or bookkeeping-type approaches. We restrict our attention to a specific method of the second kind here, since this is the approach used in the codebase \cite{brunekreef2021jorenb} that we will discuss later in this Chapter. The \code{flip} move is its own inverse, so we do not need to distinguish two directions of the move like for \code{add} and \code{delete}.

Instead of keeping track of all vertices of order four, we now maintain a list of all triangles that have a neighbor of opposite type on its \emph{right}. By ``type'' we here mean the orientation of the triangles with respect to the time ordering, i.e.\ a 12-type or 21-type simplex. The timelike link shared by two such triangles is a candidate for being flipped without destroying the causal structure present in the geometry. Since no triangles are created or destroyed in the \code{flip} move overall, we see that both the action and the combinatorial factor in Eq.\ \eqref{mc-eq:prob-labeled} are left unchanged in this process, so that the left-hand side of Eq.\ \eqref{mc-eq:transition-probs} equals 1. Given a triangulation $T$, suppose there are $N_\textit{tf}$ triangles with triangles of opposite type on their right. We select one such triangle \code{tl} $=(032)$ with uniform probability $1/N_\textit{tf}$, together with its right neighboring triangle \code{tr} $=(013)$ (although the triangle \code{tl} is of type 12 in this case, the extension to the type-21 case should be obvious). We then propose to flip the timelike link $(03)$ shared by \code{tl} and \code{tr} as shown in Fig. \ref{mc-fig:2d-cdt-moves}, resulting in a new triangulation $T'$ where $(03)$ is replaced by the link $(12)$. The selection probability for a move taking $T$ to a certain $T'$ is then equal to $g(T \to T') = 1/N_\textit{tf}$. Now we should pay attention to the fact that a \code{flip} move may change $N_\textit{tf}$. If the triangle \code{tll} on the left of \code{tl} was initially of the same type as \code{tl}, it is of opposite type after the move and $N_\textit{tf}$ increases by 1. Equivalently, $N_\textit{tf}$ decreases by 1 if \code{tll} and \code{tll} were initially of opposite type. By subsequently checking the type of the right neighbor \code{trr} of \code{tr} in a similar manner we can compute the updated value $N_\textit{tf}'$ of the triangulation $T'$. The probability of moving back from $T'$ to $T$ with a subsequent \code{flip} move is then equal to $g(T' \to T) = 1/N_\textit{tf}'$.

As a result, we find the following acceptance ratio for the \code{flip} move:
\begin{align}
	A(T \to T') &= \textrm{min}\left(1, \frac{N_\textit{tf}'}{N_\textit{tf}} \right),
\end{align}
and its reciprocal for the inverse move $T' \to T$.

\subsection{Metropolis algorithm for 3D CDT}
\label{mc-sec:3d-cdt-moves}
The probability for encountering a certain labeled triangulation $T$ in the 3D CDT ensemble is given by
\begin{equation}
	P_l(T) = \frac{1}{Z} \frac{1}{N_0(T)!} e^{k_0 N_0(T) - k_3 N_3(T)},
\end{equation}
where we see the appearance of the Regge action \eqref{intro-eq:cdt-regge-3d} for this model in the exponential. There are now three classes of moves, which we describe in more detail below when computing the appropriate acceptance ratios. 

\begin{figure}[ht!]
	\centering
	\includegraphics[width=0.6\textwidth]{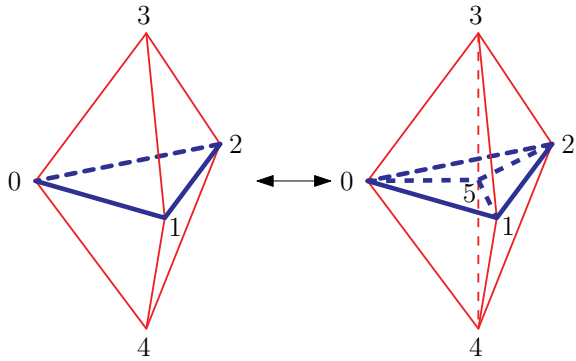}
	\caption{The \code{add} and \code{delete} moves in 3D CDT.}
	\label{mc-fig:3d-cdt-add-del}
\end{figure}

\begin{figure}[ht!]
	\centering
	\includegraphics[width=0.6\textwidth]{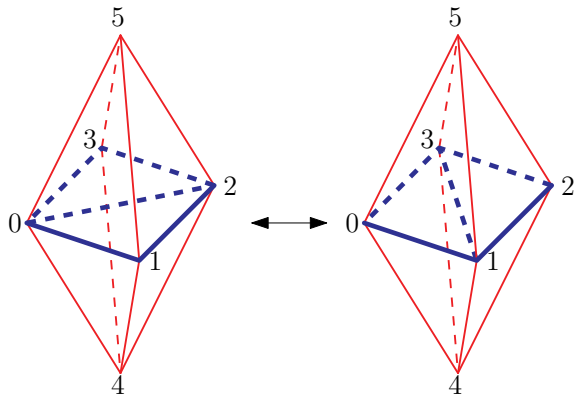}
	\caption{The \code{flip} move in 3D CDT.}
	\label{mc-fig:3d-cdt-flip}
\end{figure}

\begin{figure}[ht!]
	\centering
	\includegraphics[width=0.7\textwidth]{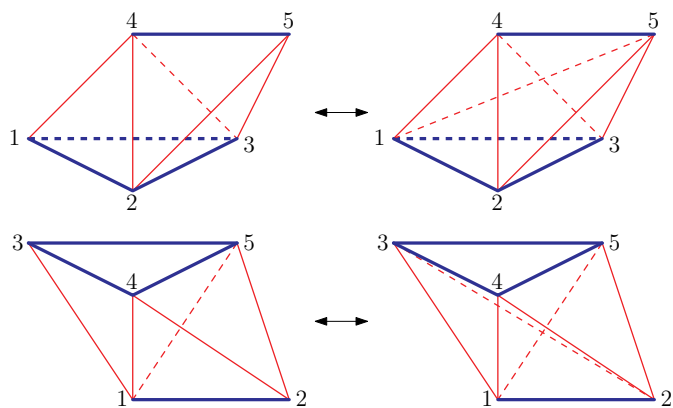}
	\caption{The \code{shift} and \code{ishift} moves in 3D CDT.}
	\label{mc-fig:3d-cdt-shift-ishift}
\end{figure}

The $(2,6)$- and $(6,2)$-moves --- forming the first class of moves --- are each other's inverse, and we also refer to them as the \code{add} and \code{delete} moves, since they are analogous to their 2D counterparts. We show a schematic representation of the \code{add} and \code{delete} moves in Fig. \ref{mc-fig:3d-cdt-add-del}. The second class consists of the $(4,4)$-move, which is its own inverse. This move flips a spatial link in the system, and we also refer to it as the \code{flip} move, shown in Fig. \ref{mc-fig:3d-cdt-flip}. The third class of moves is formed by the $(2,3)$- and $(3,2)$-moves, and we refer to these as the \code{shift} and \code{inverse shift} (or \code{ishift}) moves, respectively. These moves add (delete) a 22-simplex in a region consisting of one (two) 22-simplex paired with either a 31- or a 13-simplex, as presented in Fig. \ref{mc-fig:3d-cdt-shift-ishift}.

While developing the codebase \cite{brunekreef2022jorenb} for simulating 3D CDT, the choice was made to implement all moves using blind-type approaches for simplicity. It is possible to include bookkeeping-type methods, in which one maintains lists of special simplices where a specific move can be performed. However, the checks that need to be performed in order to keep these lists updated are nontrivial, and one needs to make sure that the lists are complete in order to fulfill the detailed balance condition. It is difficult to implement such checks properly in a computationally efficient way, and we therefore opted to make use of blind-type approaches for the time being. However, it may be the case that appropriate bookkeeping-type methods deliver substantial improvements to the execution time in some regions of the phase space, and a further investigation of this strategy is an important item on the development roadmap of the codebase.

\subsubsection{The \code{add} and \code{delete} moves}
We implement the \code{add} move by randomly selecting a 31-simplex \code{t} from a triangulation $T$ with uniform probability $1/N_{31}$. We denote its adjacent 13-simplex \code{tv}. If the move is carried out, a vertex is then added in the center of the shared spatial triangle of these two simplices, which effectively adds two 31-simplices and two 13-simplices to the system. The order of this newly created vertex is six, in the sense that the number of tetrahedra containing \code{v} equals six. We also note that this move adds two triangles to the \emph{spatial} slice at the base of \code{t}. We refer to Fig. \ref{mc-fig:3d-cdt-add-del} for a visual representation of this move. The resulting triangulation $T'$ contains $(N_0(T)+1)$ vertices and $(N_3(T) + 4)$ tetrahedra, and as a result we have that
\begin{equation}
	\frac{P_l(T')}{P_l(T)} = \frac{1}{N_0(T)+1} e^{k_0 - 4k_3}.
\end{equation}
The selection probability for the \code{add} move reads $g(T \to T') = \frac{1}{N_{31} \cdot (N_0 + 1)}$, where the factor $1/(N_0+1)$ again arises because we can choose what label to assign to the new vertex.

For the \code{delete} move we pick a random vertex \code{v} with uniform probability $1/(N_0(T)+1)$ from the triangulation $T'$. Vertices of order six can only exist in the center (labeled 5) of a configuration like the one shown in Fig. \ref{mc-fig:3d-cdt-add-del}, and any such vertex is therefore a candidate for removal with a \code{delete} move. Like in the two-dimensional implementation, we therefore abort the move if this vertex is not of order six. If \code{v} is indeed of order six, the proposed move is to delete four of the six surrounding tetrahedra as shown in Fig. \ref{mc-fig:3d-cdt-add-del}. The probability that this brings us back to the triangulation $T$ is therefore $g(T' \to T) = 1/(N_0(T) + 1)$, since this only occurs when the selected vertex is the one that was added when performing the previously discussed \code{add} move.

Using the above results, we can compute the appropriate acceptance ratios for the \code{add} and \code{delete} moves:
\begin{align}
	A(T \to T') &= \textrm{min}\left(1, \frac{N_{31}(T)}{N_0(T)+1} e^{k_0 - 4 k_3} \right), \\
	A(T' \to T) &= \textrm{min}\left(1, \frac{N_0(T)+1}{N_{31}(T)} e^{-k_0 + 4 k_3} \right).
\end{align}
A major difference with the two-dimensional case is that the couplings $k_0$ and $k_3$ are not fixed to a specific critical value. As discussed in Sec.\ \ref{intro-subsec:lat-to-cont}, $k_0$ can be set to some arbitrary value, and $k_3$ is subsequently tuned to the corresponding (pseudo-)critical value $\kcc(k_0)$ in order to approach the limit of infinite building blocks. The acceptance ratios are affected by this choice of $k_0$, resulting in a different distribution on the triangulations in the ensemble. This can have major effects on ensemble averages of observables, as we have seen repeatedly in Chapter \ref{ch:slice3d} of this thesis.

It is useful to point out here that certain moves in 3D CDT, including the \code{delete} move, can introduce degeneracies in the triangulation. For example, it is possible that two of the 31-simplices have a common neighbor tetrahedron \code{tn}. When the three 31-simplices are ``merged'' into one 31-simplex \code{t}, the simplex \code{tn} neighbors \code{t} on two of its faces. In the resulting geometry, there are two distinct spatial links connecting the same two vertices, which violates the simplicial manifold conditions. We typically exclude such configurations from consideration, and certain checks should be put into place in order to avoid these degeneracies. It is possible to consider extended ensembles where degenerate triangulations are allowed, but this may affect the behavior of the system in the continuum limit. Such degenerate 3D CDT ensembles were investigated in \cite{brunekreef2022phase}. In this thesis, we have only considered ensembles of proper simplicial manifolds, so all moves that potentially introduce degeneracies in the geometry are subject to the aforementioned checks. If a proposed move fails any of those checks, the move is rejected.

\subsection{The \code{flip} move}
As an input for the \code{flip} move, we pick a 31-simplex \code{t} from the triangulation $T$ with uniform probability $1/N_{31}$. We subsequently select with uniform probability $1/3$ one of its three neighboring tetrahedra \emph{in the same slab geometry}, i.e.\ we do not consider the 13-simplex that shares a spatial triangle with \code{t}. We denote the selected tetrahedron by \code{tn}. If \code{tn} is not of type 31 (this implies it is a 22-simplex), we abort the move. Conversely, if \code{tn} is a 31-simplex, we flip the spatial link shared by \code{t} and \code{tn}. This also induces a flip in the spatial link shared by the adjacent 13-simplices, which we see in Fig. \ref{mc-fig:3d-cdt-flip}. We again call the resulting triangulation $T'$. Note that the \code{flip} move induces a link flip in the two-dimensional spatial geometry at the base of \code{t}.

There is no change in the number of simplices during this move, so $P_l(T) = P_l(T')$. Furthermore, the selection probabilities for a \code{flip} move and its inverse are the same using this strategy of blindly selecting a site where the flip is performed. We therefore find for the acceptance ratios that $A(T \to T') = A(T' \to T) = 1$, and the move is always accepted if the aforementioned conditions are satisfied --- except, that is, if the selected flip move would introduce a degeneracy like the ones mentioned above, in which case the move is rejected.

\subsection{The \code{shift} and \code{ishift} moves}
The input for the \code{shift} move is either a 31-simplex or a 13-simplex. Both cases should be implemented in order to guarantee ergodicity, but their description is analogous so we restrict our attention to the case of a 31-simplex without loss of generality. We select this 31-simplex \code{t} from the triangulation $T$ with uniform probability $1/N_{31}$. Subsequently we select with uniform probability $1/3$ one of its three neighboring tetrahedra in the same slab geometry. Denote this neighboring tetrahedron \code{ta}. If \code{ta} is not of type 22 (i.e.\ it is of type 31), abort the move. If, on the other hand, \code{tn} is a 22-simplex, we propose the following change to the geometry. Using the labels from the top left configuration in Fig. \ref{mc-fig:3d-cdt-shift-ishift}, we can denote the 31-simplex \code{t} $=(1234)$ and the 22-simplex \code{ta} $=(2345)$. These two simplices get replaced by a 31-simplex \code{tn0} $=(1235)$ and two 22-simplices \code{tn1} $=(1245)$ and \code{tn2} $=(1345)$. This adds a 22-simplex and a timelike link to the system, as shown in Fig. \ref{mc-fig:3d-cdt-shift-ishift}.

No vertices are added during a \code{shift} move, so the combinatorial factor in Eq.\ \eqref{mc-eq:prob-labeled} is unaffected. However, the addition of a 22-simplex means that $N_3(T') = N_3(T) + 1$, so the action changes in this transition. The selection probability for this move is $g(T \to T') = \frac{1}{3 N_{31}(T)}$.

The inverse move, which we call \code{ishift}, also takes as input a uniformly selected 31-simplex \code{t}, now from the triangulation $T'$. We then pick two of its neighbors, denoted \code{tn0} and \code{tn1}. The order in which they are picked is irrelevant, so this choice can be made in three distinct ways. The move is aborted unless \code{tn0} and \code{tn1} are neighbors, and both of type 22. Otherwise, one of the 22-simplices is removed from the system, in a manner inverse to the update made during the \code{shift} move. We see that the selection probability for this move is again $g(T' \to T) = \frac{1}{3 N_{31}(T)}$. Taking into account the change in the action we then find from Eq.\ \eqref{mc-eq:acceptance-ratios} that 
\begin{align}
	A(T \to T') &= \textrm{min}\left(1, e^{-k_3} \right), \\
	A(T' \to T) &= \textrm{min}\left(1, e^{k_3} \right).
\end{align}

Note that the geometry of the two-dimensional spatial slices is not affected by the \code{shift} and \code{ishift} moves, and they rather change the connectivity between the slices. However, this has an effect on the entropy of the slice geometries, a fact that turned out to be important in Chapter \ref{ch:slice3d} of this thesis.

\subsection{Volume fixing}
\label{mc-sec:vol-fix}
It is often the case that we want to collect measurements at a certain fixed target volume $\tilde{N}$, especially in the context of a finite-size scaling analysis. However, there is no volume-preserving set of moves that is \emph{ergodic} in the space of CDT geometries of fixed size, so we must allow the volume to fluctuate around $\tilde{N}$. It is difficult to tune the cosmological coupling parameter to its pseudocritical value associated with a certain target volume, and even then the fluctuations around the target volume are large, so that we only rarely encounter systems of the right size. A standard solution is to include a \emph{volume-fixing} term $S_\textit{fix}$ in the action, which penalizes configurations that stray too far from the target volume. A useful choice is a quadratic volume-fixing term of the form
\begin{equation}
	S_\textit{fix} = \epsilon \left(N-\tilde{N}\right)^2,
\end{equation}
where $N$ is the current system size and $\epsilon$ sets the strength of the fixing. When this term is taken into account when computing acceptance ratios and the coupling is tuned to the pseudocritical value, the volume tends to fluctuate around the target $\tilde{N}$ with a typical fluctuation size determined by $\epsilon$ (with larger $\epsilon$ leading to smaller fluctuations, and vice versa).

\section{Implementation}
In the first part of this chapter we have shown how to implement Monte Carlo simulations of CDT in principle, with the detailed balance equations fully worked out for several choices of move implementations for the two- and three-dimensional versions of the model. However, several technical issues arise when putting these ideas into practice. In what follows, we discuss some of the challenges one encounters in this process. We furthermore explain the strategies that were used to address these challenges while implementing the simulation codebases \cite{brunekreef2021jorenb,brunekreef2022jorenb}. We do not aim to provide a full description of the C++ code itself, but rather focus on the generalities of our approach. 

The central classes from the user's point of view are \code{Universe} and \code{Simulation}. The class \code{Universe} represents the current state of the triangulation, and stores properties of the geometry in a convenient manner. It furthermore provides methods that carry out changes on the geometry. The \code{Simulation} class is responsible for all procedures related to the actual Monte Carlo simulation. It proposes moves and computes the detailed balance conditions, and if it decides a move should be accepted, it calls on the \code{Universe} to carry out the move at a given location. It furthermore triggers the measurement of observables when the time is right. The user can define custom observables as children (in the sense of object-oriented programming) of the standard parent class \code{Observable}. This class provides some of the basic functionality that one may require when implementing certain observables, like a breadth-first search algorithm for measuring distances or constructing metric spheres on the (dual) lattice.

Two classes that are mostly important ``under the hood'' are the \code{Pool} and \code{Bag}. These perform the memory management for all the simplices present in the triangulation, and the user can generally consider these as black boxes. It is useful, however, to understand the capabilities and limitations of these structures, so we explain them in further detail below.

We now briefly summarize what follows in the rest of this chapter. The first topic we discuss is that of storing and accessing a triangulation and its constituents in the computer memory. The approach we use is largely independent of the dimension of the triangulation. Dimension-specific considerations are required, however, when implementing the Monte Carlo moves that generate a random walk through the space of triangulations. We discuss these considerations in the second part of this section, both for 2D and 3D CDT. In the third part, we go into detail on the distinct phases of the simulation process: tuning of the coupling constants, thermalizing the system, and collection of measurements. We finally discuss the inclusion of observables in the fourth part, where we also briefly mention the breadth-first search (BFS) algorithm used for measuring distances on the lattice.

\subsection{Storage and access}
Our goal is to store a $d$-dimensional triangulation $T$ in computer memory. The triangulation consists of a collection of $d$-simplices $\sigma^{(d)} \in T$, in which every such $\sigma^{(d)}$ has $(d+1)$ neighboring $d$-simplices.\footnote{Note that we only  consider triangulations without boundary here.} We can also decompose the triangulation as a collection of $n$-simplices, where $n = 0, 1, \cdots, d$. An $n$-simplex can be described as a list of $(n+1)$ vertices, or 0-simplices. In a proper simplicial manifold, an $n$-simplex is unique in the sense that no two distinct simplices consisting of the same vertices can exist.

Consider a triangulation $T$ containing $N$ vertices. A straightforward choice is to assign all vertices $v \in T$ a distinct label from the set $\left\{0,1, \cdots, N-1\right\}$. The $n$-simplices $\sigma^{(n)} \in T$ can then be denoted by lists of the vertex labels comprising them. As an example, we take the simplest possible triangulation of a 2-sphere: the tetrahedron. This is a two-dimensional triangulation containing four vertices with labels $\{0,1,2,3\}$. It consists of four 2-simplices, or triangles, each with the other three triangles as its neighbors. The triangulation has six 1-simplices, or links. We show this triangulation and the lists of its 0-, 1-, and 2-simplices in Table \ref{mc-fig:tetrahedron}.
\begin{table}[ht!]
	\begin{minipage}{0.48\textwidth}
	\centering
	\includegraphics[width=0.5\textwidth]{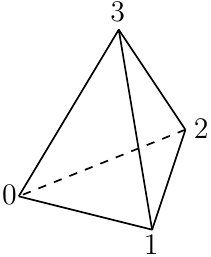}
	\end{minipage}
	\begin{minipage}{0.48\textwidth}
	\begin{tabular}{p{0.2\textwidth}|p{0.7\textwidth}}
	vertices & \{(0),(1),(2),(3)\} \\
	\hline
	links & \{(01), (02), (03), (12), (13), (23)\} \\
	\hline
	triangles & \{(012), (013), (023), (123)\}
	\end{tabular}
	\end{minipage}
	\caption{The surface of a tetrahedron, seen as the minimal triangulation of a 2-sphere.}
	\label{mc-fig:tetrahedron}
\end{table}
In principle, the list of triangles is sufficient to reconstruct the entire triangulation. For example, if we want to find the neighbors of a given triangle \code{t}, we search for all triangles that have two vertices in common with \code{t}. We can also make changes to the triangulation by updating the list of triangles, and potentially adding new vertex labels. For example, we can split the triangle $(013)$ in three by adding a new vertex, labeled 4, in its center. We subsequently remove $(013)$ from the list of triangles, and add three new triangles $(014), (034), (134)$ to it.

While this strategy is sound in principle, it is not hard to see that it is inefficient in practice. Finding the neighbors of a triangle \code{t} requires us to search through the list of triangles until we have found three objects that match the criterion of having two vertices in common with \code{t}, and the average time duration of such a search scales linearly with the number of triangles present in the system. In the language of complexity theory, we say that this operation runs in $O(N_2)$ time, where $N_2$ is the number of triangles in $T$. Our aim is to construct large triangulations in order to approach a continuum limit, and to perform enormous numbers of subsequent local updates on these geometries. When performing a single local update, we therefore clearly want to avoid operations that scale with some (positive) power of the system size whenever possible. This can be elevated to a general guiding principle for the simulation code:

\begin{mdframed}[align=center,userdefinedwidth=0.7\textwidth,skipabove=0.5cm]
\begin{center}
	\textbf{Operations required for performing Monte Carlo moves on the geometry should run in $O(1)$ time.}
\end{center}
\end{mdframed}
\vspace{0.5cm}
Here we again understand the big-O notation as it is used in complexity theory: we say that an operation on a $d$-dimensional CDT geometry runs in $O(1)$ time --- also referred to as \emph{constant time} --- if its execution time does not depend on the total number of $d$-simplices in the system. The requirement of constant execution time per basic operation precludes the use of searches through the full lists of simplices. One may expect that a straightforward object-oriented approach can solve this issue. For example, in the two-dimensional case we could formulate three classes of simplex objects: \code{Vertex}, \code{Link}, and \code{Triangle}. A \code{Triangle} object could then store lists with references to the three \code{Vertex} objects and the three \code{Link} objects that comprise it, and a list of references to the three \code{Triangle} objects adjacent to it. This is a step in the right direction, but not yet a solution to all our problems. The reason is that the analogous lists of references stored by \code{Vertex} and \code{Link} objects are not all of constant size. A single vertex can be a member of an arbitrary number of triangles, so that the list of triangles that contain a certain vertex is unbounded in length. One situation in which this is problematic is, for example, when executing a \code{delete} move in 2D CDT. Two triangles are removed from the system, and three vertices (the ones labeled $1,2,3$ in the top half of Fig. \ref{mc-fig:2d-cdt-moves}) should be updated to reflect this change. This again requires a search through the list of references to \code{Triangle} objects stored in the \code{Vertex} object, slowing down the procedure.\footnote{An additional motivation for avoiding the use of variable-length lists in the simplex objects is that they can incur an overhead in allocation time compared to fixed-length lists. However, advances in computer architecture and compiler design have brought down this overhead significantly, and it is likely that this is only a minor point of concern.}

The upshot is that we should find a strategy for storing a certain minimal amount of information required to perform all desired operations on lists of simplex objects in $O(1)$ time. It is convenient to maintain separate lists of simplices for each simplex dimensionality, and in what follows we consider a simplex to be of a fixed (but arbitrary) dimension $d$.

There are four basic types of operations that are required by Monte Carlo simulations of CDT:
\begin{enumerate}
	\item \textbf{Object creation:} add a $d$-simplex to a triangulation and assign it a unique label, thereby increasing the total list size $N_d$.
	\item \textbf{Object deletion:} remove a $d$-simplex given its label from a triangulation, thereby decreasing the total list size $N_d$.
	\item \textbf{Object lookup:} find a $d$-simplex in memory given its label and access its properties.
	\item \textbf{Random object selection:} pick a random $d$-simplex (potentially satisfying certain conditions) in the triangulation with uniform probability.
\end{enumerate}
It is precisely the combination of these four operations that makes the implementation nontrivial, and we will see why this is the case in due course. We now proceed to explain the \code{Pool} structure we used to implement operation types 1-3 in our codebase, and treat the \code{Bag} structure that implements operation type 4 separately afterwards.

\subsubsection{The \code{Pool} structure}
It is clear that the deletion of an arbitrary object in a continuous list leaves behind a hole. It is possible to fill this hole by moving the last element \code{obj} of the list to the newly vacant position, but this introduces a new problem: all other objects that refer to \code{obj} should be updated to reflect the change of its position in the list, and this updating operation is potentially of $O(N)$ complexity. An alternate solution is to include hole formation in our strategy, and to work with a list of fixed length. This fixed length imposes an upper bound on the number of simplices that can be created, but memory capacity is typically not the bottleneck for CDT simulations. A computing node with 4GB of memory can easily store on the order of millions of simplices, and at these system sizes long thermalization times are often the primary issue.

The \code{Pool} data structure implements the idea of a list with holes. A \code{Pool} is instantiated with a user-defined capacity $C$, corresponding to the upper bound for the number of objects that can be stored in it. At the time of instantiation, the \code{Pool} allocates $C$ objects in memory and collects the pointers to these objects in the list \code{Pool::elements}. These objects all persist in memory throughout the simulation, and are only deallocated at the end of the simulation run. However, from the point of view of the simulation, these objects can be either ``alive'' or ``dead'' at distinct times, even though they remain present in memory. It is the \code{Pool} structure that tracks which of these objects are alive or dead. Initially, all objects are marked dead. Objects in the \code{Pool} are identified by an integer label that can be both nonnegative and negative. A nonnegative label means that the object is alive; a negative label means that the object is dead. The user can furthermore define custom member variables for objects in the \code{Pool}.

The \code{Pool} uses a stack-like structure to keep track of the indices of dead objects in the list \code{Pool::elements}. These sites in the list are candidates for objects that can be switched to the alive status. Before any objects have been made alive, the stack contains all integers in the range $[0, C-1]$, with the value 0 at the top of the stack and incrementing by 1 with each step downwards.

In order to create an object from the point of view of the simulation, we call the method \code{Pool::create()}. This pops the first value from the stack, i.e.\ the index of an object in \code{Pool::elements} that was dead but is now switched to alive. This index \code{i} is also the label of this object, so that we can always retrieve an object by accessing the pointer stored at \code{Pool::elements[i]}. Objects created by the \code{Pool} can be cast to an integer, which represents its label.

Analogously, we can destroy an object with label \code{i} from the point of view of the simulation by calling \code{Pool::destroy(i)}. The object is marked dead and the label \code{i} is pushed onto the stack of indices of dead objects. Before destroying an object, the \code{Pool} checks whether the object is indeed dead --- otherwise an error is thrown, since the simulation attempting to destroy a dead object is likely the signal of a bug in the code.

Accessing an arbitrary object is straightforward: the method \code{Pool::at(i)} returns the object with label \code{i}. We see that the \code{Pool} structure provides us with object creation, deletion, and lookup operations with $O(1)$ complexity. However, the fact that the list \code{Pool::elements} generically contains holes makes random selection of an object in $O(1)$ time impossible. For this, we need an extra data structure that tracks the position of all the alive elements in the \code{Pool}. This is the \code{Bag} class, which we discuss next.

\subsubsection{The \code{Bag} structure}
It is straightforward to pick a random object from a contiguous list: one generates a random integer $i$ in the range $[0, N-1]$, where $N$ is the length of the list, and retrieves the object at index $i$ in the list. However, applying this to the list \code{Pool::elements} does not work, since the index can point to a dead object. Similarly, we cannot iterate over all alive objects (or a subset of them) present in the \code{Pool}. The \code{Bag} class is a structure that can be tied to a \code{Pool} in order to provide this missing functionality.

The \code{Bag} contains two lists of fixed length $C$, with $C$ equal to the associated \code{Pool} capacity. The lists are named \code{Bag::elements} and \code{Bag::indices}. The list \code{Bag::elements} contains the labels of all elements in this \code{Bag} and is \emph{contiguous}, in the sense that it does not have holes. The list \code{Bag::indices} contains the index of each element in the list \code{Bag::indices}, so that \code{Bag::elements[Bag::indices[i]]} returns the object with label \code{i}. This linked structure allows us to add and remove objects to a \code{Bag} in $O(1)$ time: when an object is added using \code{Bag::add(i)}, its label \code{i} is entered at \code{Bag::elements[Bag::size - 1]}, where \code{Bag::size} is the current number of objects in the \code{Bag}. All unused slots in \code{Bag::elements} are marked \code{EMPTY} with the constant $-1$, so that a \code{Bag} with a nonzero number $n$ of objects in it has positive integers (corresponding to the object labels) in the first $n$ elements of \code{Bag::elements} and $-1$ in the remaining tail of $C-n$ elements. Furthermore, the element \code{Bag::indices[i]} is set to \code{size}, where \code{i} is the added object's label, so that this element points to the position of the objects in \code{Bag::elements}. Finally, we increment \code{Bag::size} by one.

Removal of an object with label \code{i} from a \code{Bag} using \code{Bag::remove(i)} is then straightforward: we find the index located at \code{Bag::indices[i]} pointing to the location \code{k} of the object label in \code{Bag::elements}. Subsequently we replace this label by the \emph{last} element of \code{Bag::elements}, so that the hole is plugged, and update the index of this last element in \code{Bag::indices}. Finally, we set \code{Bag::indices[i]} to the \code{EMPTY} marker $-1$.

We can now select a random element of a \code{Bag} by generating a random integer \code{j} in the range $[0, \texttt{\code{Bag::size}}-1]$ and retrieving the label of the object located at \code{Bag::elements[j]}. This is implemented by the function \code{Bag::pick()}. Similarly, we can iterate over all object labels in the \code{Bag} by restricting iteration over \code{Bag::elements} from index 0 to $(\code{Bag::size}-1)$. Note that the \code{Bag} does not contain the objects themselves, but rather is a bookkeeping device for the labels of objects. Once we have obtained an object label from a \code{Bag}, we can retrieve the corresponding object's properties by accessing it in the associated \code{Pool}.

The relation between \code{Pool} and \code{Bag} is one-to-many, in the sense that we can have several \code{Bag}s for a single object type (e.g. \code{Vertex} or \code{Triangle}). This is convenient for the move implementations of bookkeeping-type, since they require us to keep track of simplex objects that satisfy certain conditions. As an example of this, consider the 2D model with the bookkeeping-type move implementations. The \code{add} move requires an arbitrary \code{Triangle} as an input, so we can define a \code{Bag trianglesAll} from which we can draw random triangles using \code{trianglesAll.pick()}. The \code{delete} move takes a \code{Vertex} object of order four, so we can define a \code{Bag verticesFour} containing all vertices of this type. We should take care to keep this \code{Bag} updated throughout the simulation, adding and removing vertices to and from it in order to maintain consistency. Finally, we can maintain a \code{Bag trianglesFlip} containing all \code{Triangle} objects that can be used as an input for the \code{flip} move described in Sec.\ \ref{mc-sec:metropolis-2d}.

Since every \code{Bag} allocates two lists of integers, both of fixed length equal to the total capacity $C$ of the corresponding \code{Pool}, they are accompanied by large memory usage. We should therefore attempt to restrict the number of distinct \code{Bag}s to a minimum. This is also the reason why we cannot simply add a \code{Bag} to every \code{Vertex} object in order to keep track of its variable number of neighbors: the trade-off for providing all necessary operations in $O(1)$ time is a significantly higher memory usage, and the memory requirements of this set-up would scale as $O(C^2)$.

\subsection{Geometry reconstruction}
We have demonstrated how we can efficiently store and access triangulation data in the computer memory. However, we found that this precludes the use of variable-length lists tracking the neighbors of, for example, vertices in the geometry. Some observables require access to this information, so we should be able to reconstruct it when the time arrives for performing a measurement on the geometry. Furthermore, we have seen that the \code{delete} move in 2D takes a \code{Vertex} of order four as its input and subsequently deletes two \code{Triangle} objects, so this \code{Vertex} should contain \emph{some} information on its neighboring simplices. In what follows, we show how keeping track of a certain minimal amount of extra information suffices for our purposes. There is no unique way of accomplishing this, and we restrict attention to the specific approaches we used when writing our simulation code. We treat the cases of two and three dimensions separately.

\subsubsection{Reconstruction in 2D CDT}
\begin{figure}[ht!]
	\centering
	\includegraphics[width=0.8\textwidth]{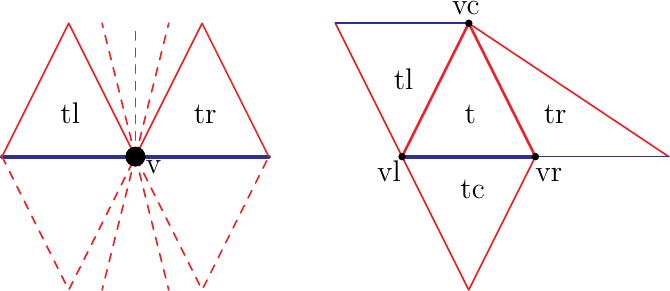}
	\caption{Schematic representation of the data stored in \code{Vertex} (left) and \code{Triangle} (right) objects.}
	\label{mc-fig:neighborhoods-2d}
\end{figure}

There is a simple solution that allows us to reconstruct all topological information about a triangulation, while still preserving $O(1)$ execution time for individual moves. By ``topological'' information we mean information about the connectivity of each individual simplex, e.g. adjacency matrices for vertices in the triangulation, or lists of all the triangles in which a given vertex appears. There are only two fundamental types of simplices that we have to keep track of while performing updates on the geometry: the \code{Vertex} and the \code{Triangle}. These store the following information in each of its instances, of which we show a visual representation in Fig. \ref{mc-fig:neighborhoods-2d}:\footnote{We omit some of the object's member variables that are irrelevant to the current discussion.}
\begin{itemize}
	\item \code{Vertex}
	\begin{itemize}
		\item \code{Triangle tl, tr;} the leftmost and rightmost upwards pointing triangles (i.e.\ 21-simplices) containing the vertex.
	\end{itemize}
	\item \code{Triangle}
	\begin{itemize}
		\item \code{Vertex vl, vr, vc;} the three vertices comprising the triangle. The vertex \code{vl} is the leftmost vertex at the base, \code{vr} is the rightmost vertex at the base (i.e.\ \code{vl} and \code{vr} reside on the same time slice), and \code{vc} is its apex.
		\item \code{Triangle tl, tr, tc;} the three triangles adjacent to the triangle. The triangle \code{tl} is on the left, \code{tr} on the right, and \code{tc} is the vertical neighbor.
	\end{itemize}
\end{itemize}
All moves detailed in Sec.\ \ref{mc-fig:2d-cdt-moves} can be implemented with $O(1)$ execution time using this information alone. Furthermore, we can reconstruct all topological data of the triangulation using this set-up. As an example, consider the following strategy for reconstructing a list of neighbors for a given \code{Vertex v}: start from the \code{Triangle v.tl}, and add all vertices comprising it (except \code{v}) to a collection. We can then step to the right using \code{v.tl.tr} in order to add more neighboring vertices to this list, until we end up at \code{v.tr}. We then step down using \code{v.tr.tc}, and complete the circle by stepping to the left, each time adding new vertices to the collection until we hit \code{v.tl.tc}. Similarly, we can reconstruct a list of all \code{Link} objects in the triangulation by iterating through the list of triangles, and we see that it is superfluous to keep this list up-to-date at every step of the Monte Carlo simulations. We rather choose to perform the full reconstruction at the time we want to take a measurement.

\subsubsection{Reconstruction in 3D CDT}
Similar considerations hold in the 3D version of the model. The fundamental objects that are tracked during geometry updates are the \code{Vertex} and the \code{Tetra} (short for tetrahedron). These store the following geometric data:
\begin{itemize}
	\item \code{Vertex}
	\begin{itemize}
		\item \code{Tetra tetra;} an arbitrary 31-simplex containing this vertex in its base (i.e.\ not as its apex).
		\item \code{int cnum;} the coordination number of the vertex, i.e.\ the number of tetrahedra containing it.
	\end{itemize}
	\item \code{Tetra}
	\begin{itemize}
		\item \code{Vertex[4] vs;} a list of the four vertices comprising the tetrahedron.
		\item \code{Tetra[4] tnbr;} a list of the four tetrahedra neighboring the tetrahedron.
	\end{itemize}
\end{itemize}
There is a specific ordering convention for the two lists in a \code{Tetra} object, which allows one to retrieve specific elements based on their relation to the tetrahedron. The first convention is that for a given tetrahedron \code{t}, we have that \code{t.tnbr[i]} (with $\texttt{\code{i}} \in \{0,1,2,3\}$) is the neighboring tetrahedron of \code{t} \emph{opposite} to the vertex \code{t.vs[i]}. Put differently, the tetrahedron \code{t.tnbr[i]} is the neighbor of \code{t} on the face (a triangle) comprised of all three vertices in \code{t.vs} \emph{not} equal to \code{t.vs[i]}.

The second convention determines the order of the vertices in the list \code{Tetra::vs}, by requiring that the list starts with the vertices with the smallest time slice label (taking into account periodic boundary conditions if applicable). This means that the time label of vertices in this list depends on the tetrahedron type: for a tetrahedron \code{t} in the slab interpolating between the slices labeled $t$ and $(t+1)$, the time labels of the vertices in \code{t.vs} are $[t,t,t,t+1]$. Similarly, the order is $[t,t+1,t+1,t+1]$ for a 13-simplex and $[t,t,t+1,t+1]$ for a 22-simplex. There is no distinguished ordering for vertices with identical time labels.

All moves described in Sec.\ \ref{mc-sec:3d-cdt-moves} can be performed in $O(1)$ time, given the information listed above. This is clear for the \code{add}, \code{shift}, \code{ishift}, and \code{flip} moves. In order to achieve $O(1)$ complexity for the 
\code{delete} move, we keep track of vertex coordination numbers by storing them in \code{Vertex::cnum} throughout the simulation.\footnote{We should mention that continuously updating this information is not free, and incurs a small penalty in execution time. However, high vertex coordination numbers are relatively common in the de Sitter phase of this model, as we have seen in Chapter \ref{ch:slice3d}. Computing the coordination number on the fly using a neighborhood reconstruction is expensive for such vertices, and we therefore consider it likely that the simple operation of updating several integer variables in every move is more efficient.} It is then straightforward to check whether the randomly selected \code{Vertex v} is a candidate for the \code{delete} move, since a vertex qualifies if and only if it is of order six. The tetrahedra involved in the move can be found by a neighborhood reconstruction similar to the one described for the 2D model, where \code{Vertex::tetra} is now taken as a starting point. The same algorithm can be used to reconstruct all information about the connectivity of the triangulation by the time we want to take measurements.

\subsection{Simulation stages}
A single simulation run can generally be divided into the following three stages:
\begin{enumerate}
	\item \textbf{Tuning.} Adjusting the bare cosmological coupling parameter to its pseudocritical value associated with the input values chosen for the remaining coupling constants of the model.
	\item \textbf{Thermalization.} Evolving the system from the initial input geometry to an independent, ``thermalized'' configuration.
	\item \textbf{Measurement.} Collecting measurement data on snapshots of configurations occurring in the random walk.
\end{enumerate}
Separating these stages is a matter of convention, and there are no strict criteria that signal when the next stage of the simulation should be entered. The tuning stage is not relevant in pure 2D CDT, where the exact value of the critical cosmological coupling is known. In models where we require a tuning of the cosmological coupling, this tuning phase partially thermalizes the system. Conversely, in some set-ups the tuning phase is never left, and the coupling is continuously tuned even while measurements are being taken. Finally, there is no binary switch that tells us when the system is ``thermalized'', and deciding when to start the measurement phase is a mix of art and science. We now briefly discuss the global features of the three simulation stages.

\subsubsection{Tuning}
For a general model of random geometry, it is unknown how to analytically compute the critical value of the bare cosmological coupling parameter as a function of the remaining couplings. An exception is pure 2D CDT, where $\lambda_c = \ln 2$ exactly. There are estimates for $\lambda_c$ in the high-temperature and zero-temperature regime in the case of 2D CDT coupled to multiple Ising spins \cite{ambjorn1999new}, but this exhausts the list of CDT models where we have an analytical grip on the problem.

In other cases, like CDT in three and four dimensions, we have to estimate the (pseudo-)critical value of the coupling by manually adjusting it in small steps until we find a range where the system approaches the limit of infinite volume. This is called the \emph{tuning} of the coupling. Roughly speaking, the tuning procedure can be formulated as follows. Denote the bare cosmological coupling $\lambda$, regardless of the dimension of the system, and pick a target volume $\tilde{N}$. We start by guessing an arbitrary initial value for $\lambda_0$, typically of order 1. Now we initialize a Monte Carlo simulation of the system, using $\lambda_0$ as the cosmological parameter when computing acceptance ratios for the moves. We then let the simulation run for some sufficiently large number of moves, while keeping track of the total system volume at set intervals. Subsequently, we compute the averaged value $\langle N \rangle$ of all system volumes recorded during this period. If $\langle N \rangle > \tilde{N}$, adjust $\lambda$ upwards. If $\langle N \rangle < \tilde{N}$, adjust $\lambda$ downwards. Call the resulting value $\lambda_1$, and continue the simulation using this value as the cosmological coupling. Repeating this procedure gives us a sequence $\lambda_i$ that approaches the pseudocritical value $\lambda_c$ associated to the target volume $\tilde{N}$ and the remaining coupling parameters. We stop the tuning procedure once the averaged $\langle N \rangle$ is within a sufficiently small range of the target volume $\tilde{N}$.

It is convenient to make the tuning process \emph{adaptive}, in the sense that we choose larger adjustments to $\lambda$ if the absolute difference $\left|\tilde{N}-\langle N \rangle \right|$ is large, and smaller adjustments if the difference between the target and average simulated volumes grows closer together. Finding appropriate step sizes for such an adaptive procedure is typically done on a trial-and-error basis, since we do not know how to estimate them analytically. A tuning procedure for 3D CDT is implemented in our code in the function \code{Simulation::tune()}.

While tuning the coupling, it is useful to include a volume-fixing term as discussed in Sec.\ \ref{mc-sec:vol-fix}. This reduces the magnitude of fluctuations around the target volume, and thereby prevents the system from shrinking to zero size or exploding to infinity when the initial guess $\lambda_0$ is too far from the true pseudocritical value.

\subsubsection{Thermalization}
Once we have found a suitable estimate of the pseudocritical value for the cosmological constant, the system volume fluctuates around the target $\tilde{N}$ with a typical fluctuation size depending on the prefactor $\epsilon$ of the volume-fixing term. We can now start the thermalization phase, which basically amounts to letting the simulation run for a sufficiently long time so that the random walk has brought us sufficiently far away from our starting geometry. As mentioned earlier, there are no strict definitions of ``sufficiently long'' and ``sufficiently far away'' that are applicable in a general context.

We group the moves during the thermalization and measurement phases in bunches called \emph{sweeps}, where one sweep corresponds to some fixed number of attempted moves. Typically, we set the sweep size to be on the order of 10 or 100 times the target volume $\tilde{N}$. CDT in two dimensions without matter does not require long thermalization times, and it is often sufficient to let the thermalization run for $\sim 100$ sweeps. The higher-dimensional models often thermalize much more slowly, and the thermalization speed depends on the location in the phase space. We therefore prefer to set the length of the thermalization phase (in terms of the number of sweeps) based on an analysis of the measurement data --- for example, by checking whether the value of some simple order parameter has reached a stable average with fluctuations around it.\footnote{Note that this is not appropriate when we investigate the model near a first-order phase transition, where the order parameters can exhibit discontinuous jumps between two distinct values.} When computing ensemble averages of our measurements, we simply throw out the initial region of the measurement phase.

\subsubsection{Measurement}
Once thermalization has been (partially) completed, we enter the measurement phase. This is the stage where we collect the data that allow us to compute estimators of ensemble averages $\langle \mo \rangle$ of observables $\mo$. At every step in the measurement phase, we first let the system evolve for a certain amount of time, and subsequently collect one data point for every observable included in the simulation. These data points are written to external files in order to allow the user to analyze the data before the full simulation run has completed.

Just as for thermalization, appropriate sweep sizes $s$ and the number $k$ of sweeps in between measurements are situation-dependent. The relevant quantity here is the \emph{autocorrelation time} for the observable(s) being investigated. The autocorrelation time is expressed in terms of the number of attempted Monte Carlo moves, and it measures how long the system should thermalize in between two subsequent measurements of a certain observable to be considered statistically independent. We refer to \cite{newman1999monte} for a more detailed discussion of the computation of autocorrelation times in Monte Carlo simulations.

A slightly adapted strategy is needed when we want to perform our measurement at fixed volume, which is typically the case for a finite-size scaling analysis. After $k$ sweeps, the system is generally not exactly at the desired target volume $\tilde{N}$. A simple solution is to complete $k$ sweeps first, and to subsequently continue the random walk until the system hits the size $\tilde{N}$, upon which we perform a measurement. The number $k \cdot s$ is therefore the \emph{minimum} number of attempted moves in between two subsequent measurements, and the actual number of attempted moves is generally larger.

By the time we want to perform a measurement, we should first reconstruct all topological data of the triangulation as detailed in the previous section. As an example, some observables need access to the neighborhoods of vertices in order to construct metric spheres using a breadth-first search algorithm (as will be detailed in the next section). The reconstruction is triggered in our code by calling the function \code{Simulation::prepare()}. After the reconstruction has been completed, all observables that have been registered using \code{Simulation::addObservable()} will be measured in the current state of the geometry.

\subsubsection{Observables}
We can measure properties of the triangulation by defining child classes of the generic parent \code{Observable} class. An instance of such an \code{Observable} can be registered by calling \code{Simulation::addObservable()}, which ensures that the observable is computed on the current state of the system at the end of each measurement step, and its output is written to a file.

Reconstruction of the geometry takes place before measurements are taken, and the results of the reconstruction are stored in indexed lists. As an example, we can retrieve the collection of vertices neighboring a given vertex with label \code{i} by calling \code{Universe::vertexNeighbors.at(i)}. We emphasize again that this information is not directly available during the Monte Carlo updating, and one should keep this in mind when accessing such lists at intermediate stages.

The result of a measurement should be encoded as a text string, and stored in the variable \code{Observable::output}. The contents of this variable are automatically appended to an external data file after the measurement has been completed.

Measuring distances on the lattice can be achieved through the use of a breadth-first search (BFS) algorithm. Such an algorithm starts from a given simplex \code{i} in the triangulation, and enumerates all its neighbors recursively until it arrives at the target simplex \code{f}. Since distance-finding is often necessary when investigating CDT geometries, the methods \code{Observable::distance(i, f)} and \code{Observable::distanceDual(i, f)} implement this procedure on both the direct and dual lattices, where \code{i} and \code{f} represent the initial and target simplex, respectively. Similarly, metric spheres of radius \code{r} centered at a simplex \code{c} can be constructed (also using a BFS algorithm) by calling \code{Observable::sphere(c, r)} and \code{Observable::sphereDual(c, r)}. The implementation of the BFS was chosen to be heavy on memory usage, with the advantage of improved speed. The reasons are similar to those made when implementing the \code{Bag} structure.

Several observables are provided in the open-sourced code, which the user can take as guiding examples when implementing their own custom observables. Both the 2D and 3D codebases include the volume profile, the shell volumes used for determining the Hausdorff dimension, and the average sphere distance used in the context of the quantum Ricci curvature. Additionally, the codebase for the three-dimensional model includes an implementation for measuring the distribution of minimal neck baby universes on spatial slices that was used in Chapter \ref{ch:slice3d} for investigating the entropy exponent of this system.

\section{Summary and conclusion}
We have discussed general aspects of Monte Carlo methods in the context of random geometry. More specifically, we have explained how such methods can be put into practice for two- and three-dimensional CDT quantum gravity. Computer simulations have proved to be an invaluable asset when investigating the properties of CDT ensembles, and it is likely that the role of these numerical methods will only grow over time. Technological developments in computer hardware are sure to expand the scope of CDT research by allowing one to simulate ever larger systems, leading to increased capabilities for studying continuum physics.

Additionally, there may be interesting avenues to pursue on the software side of CDT Monte Carlo simulations. For example, further algorithmic improvements may still be found by investigating the bookkeeping-type methods as discussed in Sec.\ \ref{mc-sec:mcmrg} of this chapter, which could prove especially useful in certain regions of the phase space (for example, near phase transition lines) that are currently difficult to study numerically. 

Another tantalizing possibility is to take inspiration from recent advances in the field of artificial intelligence. Increased usage and availability of graphical processing units (GPUs) has propelled research in machine learning, and this raises the question whether CDT research could similarly benefit from these developments. Unfortunately, it is so far difficult to see how the Markov chain Monte Carlo methods used for CDT simulations can be adapted to make efficient use of GPU architectures. A different and, perhaps, simpler option may be to employ GPUs for the measurement of computationally expensive observables, like the average sphere distance.

\chapter{Conclusion and outlook}
In this thesis, we have investigated discretized models of quantum gravity from a variety of perspectives. We identified three main themes in this body of research. 

Firstly, we have studied a recently formulated prescription for defining curvature in the context of quantum gravity, the so-called quantum Ricci curvature \cite{klitgaard2018introducing}. As a part of this investigation, we built upon the definition of the quantum Ricci curvature to construct a bona fide quantum gravity observable called the curvature profile. 
Secondly, we have taken a close look at two-dimensional spatial hypersurfaces in (2+1)-dimensional CDT quantum gravity, and compared their behavior to that of known universality classes of two-dimensional random geometry. Thirdly, and finally, we have studied and expanded upon parts of the technical toolbox for quantum gravity research. 

In Part \ref{part:curv}, we focused on the topic of curvature observables in quantum gravity. We defined the notion of a curvature profile, a real-valued function of a scale variable that can be interpreted as a global signature of the average curvature of a system. The curvature profile, by virtue of being based on the quantum Ricci curvature and the associated average sphere distance, can in principle be implemented on a wide variety of geometric systems where a notion of distance is available.
This includes the simplicial manifolds of CDT, which are endowed with a natural notion of discrete distance between simplices. Studying the curvature profile in this setting may give us insight into the curvature properties of the corresponding quantum gravity theory described by CDT.


We discussed the results of curvature profile measurements on the ensemble of (1+1)-dimensional CDT quantum gravity with torus topology in Chapter \ref{ch:ricci2d}. Our main goal was to study features of the curvature profile for lattice distances $\delta$ in the regime $\delta_0 \ll \delta \ll L$, where $\delta_0$ is a cutoff scale for discretization artifacts, and $L$ is the typical linear size of the system. For $\delta$ near or beyond this linear size $L$, topological effects can have a strong influence on the behavior of the curvature profile due to the compactified structure of the torus. While avoiding such topological effects in the time direction of the geometry is straightforward, it turned out that periodicity in the spatial direction posed a more difficult problem for our analysis. This is due to quantum volume fluctuations in the one-dimensional spatial slices. Topological shortcuts arising due to such fluctuations can affect the average distance between two metric circles, which leaves an imprint on the curvature profile of the system. Therefore, we devised a method for identifying the presence of such topological shortcuts in an average sphere distance measurement. This allowed us to classify a range of scales --- as a function of the target system volume and number of time slices --- at which effects of the global topology on the curvature profile measurements become negligible.

Restricting our attention to the small scales where only local effects play a role, we observed the emergence of a single universal curvature profile, independent of the size of the system. This universal curvature profile exhibits a slight monotonic rising behavior, which would ordinarily be interpreted as signalling the presence of a small negative average curvature. However, due to the apparent absence of a dynamically generated scale corresponding to an effective curvature radius, we dismissed this interpretation. Instead, we concluded that this system of quantum geometry exhibits a new type of curvature behavior, which we dubbed \emph{quantum flatness}. Our findings indicate that this ``quantum torus'' seems to differ from a flat, classical torus from the point of view of the curvature profile.

In Chapter \ref{ch:defects}, we developed machinery to construct geodesics on two-dimensional almost-everywhere flat surfaces with isolated curvature defects. This enabled us to compute partial curvature profiles of the surfaces of the five Platonic solids. These compact surfaces are topologically identical to the smooth two-sphere, and as a consequence have the same amount of integrated scalar curvature due to the Gauss-Bonnet theorem. The simplest of the Platonic solids, the tetrahedron, has the special property that geodesics on its surface can be represented by geodesics in the tessellated plane, allowing us to compute its full curvature profile. Viewing the Platonic solids as polyhedral approximations to a smooth two-sphere, we see that the tetrahedron can be seen as the coarsest approximation, with the total curvature concentrated in just four singular points. This led us to compare the curvature profile of the tetrahedron to that of a smooth two-sphere, in which we found a close resemblance between the two. We interpret this as evidence for the coarse-graining capabilities of the curvature profile. In the context of nonperturbative quantum gravity, this is a desirable feature of a curvature observable candidate, since we can expect geometry to become highly singular near the Planck scale.

In Part \ref{part:cdt3d} of this thesis, we turned to (2+1)-dimensional CDT quantum gravity. We employed computer simulations in order to measure four geometric properties of the spatial slices in both phases of the (2+1)-dimensional CDT model: vertex coordination number distributions, minimal neck baby universe sizes, the local and global Hausdorff dimension, and curvature profiles. Next, we compared our findings to results known from the literature for two-dimensional \emph{Euclidean} DT quantum gravity. We found that the results we obtained for spatial hypersurfaces in the degenerate phase of (2+1)-dimensional CDT can be matched with good accuracy to numerical results known from the literature for 2D DT. On the other hand, our measurements performed in the de Sitter phase of the model do not resemble numerical results from any system of two-dimensional random geometry that we know of. It is possible that these discrepancies can be ascribed to discretization artifacts, meaning that they could disappear when probing larger systems. However, we did not see improved scaling behavior even for the largest systems we used in our simulations. Therefore, our results may indeed indicate the existence of a previously unknown type of two-dimensional quantum geometry, the dynamics of which are affected by its embedding in the surrounding three-dimensional system.

Part \ref{part:mat} of this thesis focused on technical aspects of discrete quantum gravity. Firstly, in Chapter \ref{ch:diff-mat} we discussed random matrix theory, a framework that finds applications in many areas of physics and mathematics, one of which is two-dimensional triangulated quantum gravity. We showed that the perturbative expansion of integrals over one- and two-matrix ensembles can be matched to an expression containing only a single matrix variable and a conjugate differential operator. Subsequently, we explored the details of this differential reformulation, showing how to perform the diagonalization procedure in the differential language by making use of the Harish-Chandra--Itzykson--Zuber formula. We then demonstrated how the differential formulation can be rewritten in terms of Slater determinants, which leads to a decomposition in terms of (bi-)orthogonal polynomials. Finally, we put forward a proposal for how the Slater determinant expression of a two-matrix model may potentially be used in order to construct a solution of the model in terms of \emph{orthogonal} polynomials as opposed to the usual \emph{biorthogonal} polynomials. Certain two-dimensional models of quantum gravity can be represented as a two-matrix model, and finding the associated (bi-)orthogonal families of polynomials allows for the exact computation of correlation functions in the model. Finding such families is typically much more difficult in the biorthogonal case, so that an orthogonal polynomial solution might open up new possibilities for obtaining exact results in these two-dimensional models of quantum gravity.

In Chapter \ref{ch:monte-carlo}, we presented a hands-on introduction to the implementation of Monte Carlo simulations of systems of random geometry. We discussed methods known from the literature for setting up detailed balance equations in the context of dynamical lattices, which allows for random sampling of geometric ensembles. We then explained in detail how these methods can be implemented for two- and three-dimensional CDT, highlighting the practical considerations one encounters when attempting to set up the simulations in an efficient manner. The simulation implementations of these models were constructed (in collaboration with A. G\"orlich and D. N\'emeth) as a part of the research contained in this thesis, and the lessons learned during this process may prove to be a valuable resource for those interested in the numerical simulation of nonperturbative quantum gravity.

\subsection*{Outlook}
I started this thesis by recalling the state of theoretical physics research at the end of the nineteenth century: many physicists felt that the field was nearing a stage of conceptual completion, and that one would need to look elsewhere in science for major new discoveries. It is difficult to overstate how differently the situation has developed. Between then and now, numerous new challenges have appeared on the road to improving our fundamental understanding of Nature. Many of those challenges have, in turn, been met by brilliant minds, and were subsequently resolved. On the other hand, some have given birth to further generations of puzzling problems that repeatedly forced researchers back to the drawing board.

One such problem is finding a quantum theory of gravity, and this has developed into one of the most sought-after goals in theoretical physics. It would have been impossible for the aforementioned nineteenth-century physicist to imagine our current predicament, if only because both quantum mechanics and general relativity had yet to be discovered. While there is no clear path ahead to a complete understanding of quantum gravity, many valuable lessons have been learned over the past century, leading to a wide variety of promising candidate approaches.

One such candidate is Causal Dynamical Triangulations. It is a relatively young theory, having originated in the late 1990s. In the subsequent years, it has produced many encouraging results that have established it as a strong contender for a quantum theory of gravity. Furthermore, it has sparked new lines of investigation in other approaches to quantum gravity. It will be exciting to see how CDT matures into its thirties and beyond, and I speculate here on interesting potential future research directions for the theory.

The quantum Ricci curvature has formed a major theme of this thesis research, and as a natural consequence I am especially interested in seeing further exploration of this new prescription for studying curvature. Determining the curvature profiles --- analytically where possible, and numerically otherwise --- of a wider range of geometries and ensembles of geometry would provide a baseline catalogue that can help in interpreting curvature profiles obtained in unexplored territory. Part of my Ph.D. research has contributed towards this goal, but there is much left to be done. A straightforward extension of existing work would be to investigate curvature profiles of quantum gravitational systems coupled to matter fields. Simple examples are two-dimensional DT and CDT coupled to Ising spin degrees of freedom \footnote{A preliminary investigation of the CDT case was made in \cite{vanderfeltz2021matter}.}, where the addition of sufficiently many matter fields is known to have a backreaction on the geometry. 

As reported on in Chapter \ref{ch:slice3d}, it is difficult to interpret the curvature profiles measured on two-dimensional spatial slices in the de Sitter phase of three-dimensional CDT. Since we conjectured that the spatial hypersurfaces cannot be understood as independent systems of quantum geometry, it may be more appropriate to study curvature profiles of the full three-dimensional system. This means that we no longer determine the average distance between circles confined to the slices, but rather between metric 2-spheres in the three-dimensional triangulations. Further motivation for such an investigation can be found in the fact that analogous measurements have been performed in the de Sitter phase of four-dimensional CDT \cite{klitgaard2020how}, finding evidence that the resulting curvature profile is compatible with that of a continuum four-dimensional Euclidean de Sitter space. It would be interesting to see whether similar behavior is present in the three-dimensional case.

It is worthwile to stress that determining curvature profiles in the context of nonperturbative quantum gravity is, so far, largely a numerical effort. Measurements of the average sphere distance are limited by computational resources in several respects, and increasingly so as the radii of the spheres (and thereby, the coarse-graining scale) is increased. One of the reasons is that larger system sizes are required in order to avoid finite-size effects, leading to longer thermalization times in between subsequent measurements. Another factor in this issue is related to the nature of the curvature profile measurement itself: finding shortest distances between pairs of points on a generic simplicial manifold is an expensive operation, and this process needs to be repeated many times in order to compute the full average distance between a pair of spheres. While it is unlikely that there exist algorithmic improvements that can significantly bring down the computational complexity of the exact average sphere distance measurement on arbitrary simplicial manifolds, it seems reasonable that one can make use of suitable approximation schemes. A simple approach is to employ a Monte Carlo integration scheme, in which one only computes the shortest distance between a randomly sampled subset of all possible pairs of points on the two spheres. It would be interesting to see to what extent this could speed up the data collection process, while still being able to guarantee sufficiently low approximation errors. This can be investigated empirically, although such an analysis must be carried out separately for distinct ensembles and system sizes since we generally do not have prior knowledge about the distribution of distances between pairs of points in the simplicial manifolds. Still, this method could be used to distinguish ``safe'' classes of geometric ensembles, where highly undersampled pairs of points on the spheres can give good estimates of the exact average sphere distance.

Although the average sphere distance on a simplicial manifold is typically computed numerically, it is possible that certain geometric ensembles allow for analytic determination of the curvature profile. We discussed in Chapter \ref{ch:diff-mat} how two-dimensional Euclidean quantum gravity can be modeled by integrals over random matrix ensembles. Such matrix models provide an analytical handle on the corresponding quantum gravity theories, and it is conceivable that these techniques can also be used to compute the average sphere distance exactly, or approximate it in a certain limit. While this is likely a difficult problem to solve, it could form a proof of concept that determining curvature profiles is not exclusively a numerical task.

Finally, I would like to highlight two important open areas of CDT research that I find especially interesting from a physical point of view. While all the work done for this thesis research concerned lower-dimensional pure gravity models, the holy grail is to construct a complete four-dimensional continuum theory of quantum gravity, complete with all the matter fields that we know exist in our Universe. 

The first of the two research areas that I would like to see come to fruition concerns the continuum limit of four-dimensional pure gravity CDT. While there is evidence for continuous phase transitions in certain regions of the phase space of this model, it is so far unclear how one should proceed to locate potential UV fixed points on this transition line. The existence of such a UV fixed point is required for the model to have a continuum interpretation, and it would be a major breakthrough if one was found. Previous attempts at searching for fixed points have used volume fluctuations of spatial slices to define a notion of correlation length, but it has been suggested that \emph{curvature} correlators may be more appropriate for this purpose. I am eager to see whether the quantum Ricci curvature can indeed prove its value in this context as well.

The second area of research that fascinates me for its potential real-world implications consists in coupling the Standard Model of particle physics to CDT quantum gravity. Whereas solving the problem of quantum gravity even in the absence of matter would be a tremendous achievement, the final goal is to have a theory that includes all known fundamental interactions of Nature. What is more, it may even be the case that including the matter fields of the Standard Model is a necessary condition for constructing a consistent quantum gravity theory. Studying the interplay of the Standard Model and CDT quantum gravity requires one to implement lattice gauge theory on the CDT ensemble. A preliminary investigation of Yang-Mills gauge fields on two-dimensional CDT was made in \cite{candido2021compact}, and work is currently underway to extend this to the four-dimensional case. Readers familiar with standard lattice gauge theory will recognize how ambitious this research programme is, and it is perhaps precisely \emph{because} of the extreme difficulty of this undertaking that I will be keeping a close eye on developments in this direction.

\chapter{Research data management}
The research in this thesis is carried out under the research data management policy of the
Institute for Mathematics, Astrophysics and Particle Physics. The policy can be found at \href{https://www.ru.nl/publish/pages/868512/imapp_rdm_policy.pdf}{\code{https://www.ru.nl/publish/pages/868512/imapp\_rdm\_policy.pdf}}. We provide data specifications for relevant chapters of this thesis below.

\begin{itemize}
	\item \textbf{Chapter \ref{ch:ricci2d}}. All data was generated by use of the simulation code that can be found at \href{https://github.com/JorenB/2d-cdt}{\code{https://github.com/JorenB/2d-cdt}} \cite{brunekreef2021jorenb}, and has been archived privately. The data can be made available by the author on request.
	\item \textbf{Chapter \ref{ch:slice3d}}. All data was generated by use of the simulation code that can be found at \href{https://github.com/JorenB/3d-cdt}{\code{https://github.com/JorenB/3d-cdt}} \cite{brunekreef2022jorenb}, and has been archived privately. The data can be made available by the author on request.
\end{itemize}


\appendix
\chapter{Intersection numbers of closed curves on discretized tori}
\label{app-sec:intersection-numbers}
\begin{figure}[ht!]
	\centering
	\begin{subfigure}[t]{0.31\textwidth}
	\centering
		\includegraphics[width=0.9\textwidth]{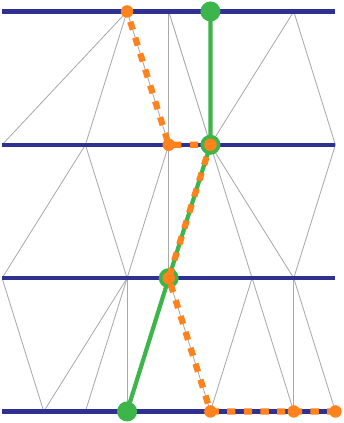}
		\caption{}
	\end{subfigure}
	\hfill
	\begin{subfigure}[t]{0.31\textwidth}
	\centering
		\includegraphics[width=0.9\textwidth]{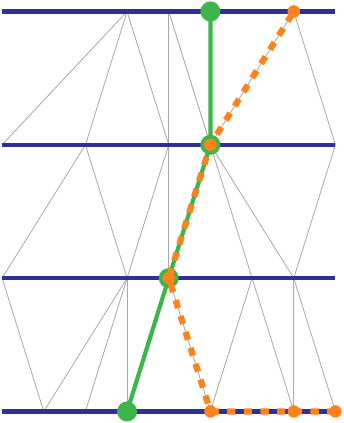}
		\caption{}
	\end{subfigure}
	\hfill
	\begin{subfigure}[t]{0.31\textwidth}
	\centering
		\includegraphics[width=0.9\textwidth]{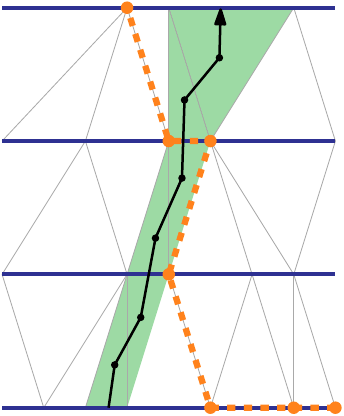}
		\caption{}
	\end{subfigure}
	\caption{
	Left, center: if the reference curve is taken along the links of the triangulation (green), this leads to overlaps with other paths of the same type (orange, dashed) rather than point-like intersections.
	Right: if instead one of the paths is taken along dual links (black, dual to a contiguous ``chain'' of triangles), intersections occur in isolated points only and the
	intersection number introduced in Sec.\ \ref{ric2dcdt-inter} is well defined.}
	\label{app-in-fig:crossing-primal-dual}
\end{figure}

In this appendix we explain how to set up the intersection number analysis introduced in Sec.\ \ref{ric2dcdt-inter} on CDT geometries, leading to the
results presented in Sec.\ \ref{ric2dcdt-measure}.\footnote{We will treat explicitly the case relevant for computing the curvature profile in terms of the link distance. There is
an analogous procedure if the dual link distance is used instead.}
As already mentioned earlier, this analysis is simplified when the reference loop $\gamma$ is constructed 
from dual links. Otherwise we would have to deal with situations where the two curves are ``tangent'' in the sense of sharing one or more links, as illustrated by Fig.\ \ref{app-in-fig:crossing-primal-dual}. If $\gamma$ consists of dual links, as in the black curve in Fig.\ \ref{app-in-fig:crossing-primal-dual}c, 
all intersection points occur between a link and a dual link, which always happens at a nonvanishing angle in an isolated point.

For a given pair of circles on a CDT geometry $T$, we construct the curve $\gamma$ by randomly picking a triangle $\Delta$ outside the region $C$ 
enclosed by the two circles. From this starting point, we perform a breadth-first search (BFS) along dual links or, equivalently, along the triangles of $T$, 
again excluding all triangles located in $C$. The BFS is restricted to move forward in discrete CDT time $t$ or stay at constant time, but not go back in time.
The algorithm is continued until we meet the original triangle $\Delta$ again. We can then trace back our steps to find a shortest path connecting $\Delta$ to itself,
which will serve as the reference curve $\gamma$ (with winding number 1 in the positive time direction) for determining the intersection numbers with
all geodesics contributing to the average sphere distance computation. 
If the BFS exhausts the geometry while moving forward in time, without getting back to the triangle $\Delta$, we have encountered a level-2 violation, 
where the region $C$ wraps around the spatial direction in a full circle.

Once a curve $\gamma$ has been found, we enumerate all pairs of vertices $(q,q')\! \in\! S_p^\delta\! \times\! S_{p'}^\delta$,
and subsequently construct the shortest paths $\phi(q,q')$ connecting these pairs by again performing a BFS. Since we are interested
in their intersection number with $\gamma$, we not only keep track of the length of this shortest path, but also 
of all vertices encountered along the way. With this information, it is straightforward to compute the total intersection numbers $c(\phi,\gamma)$.

\begin{figure}[t]
	\centering
	\includegraphics[width=0.75\textwidth]{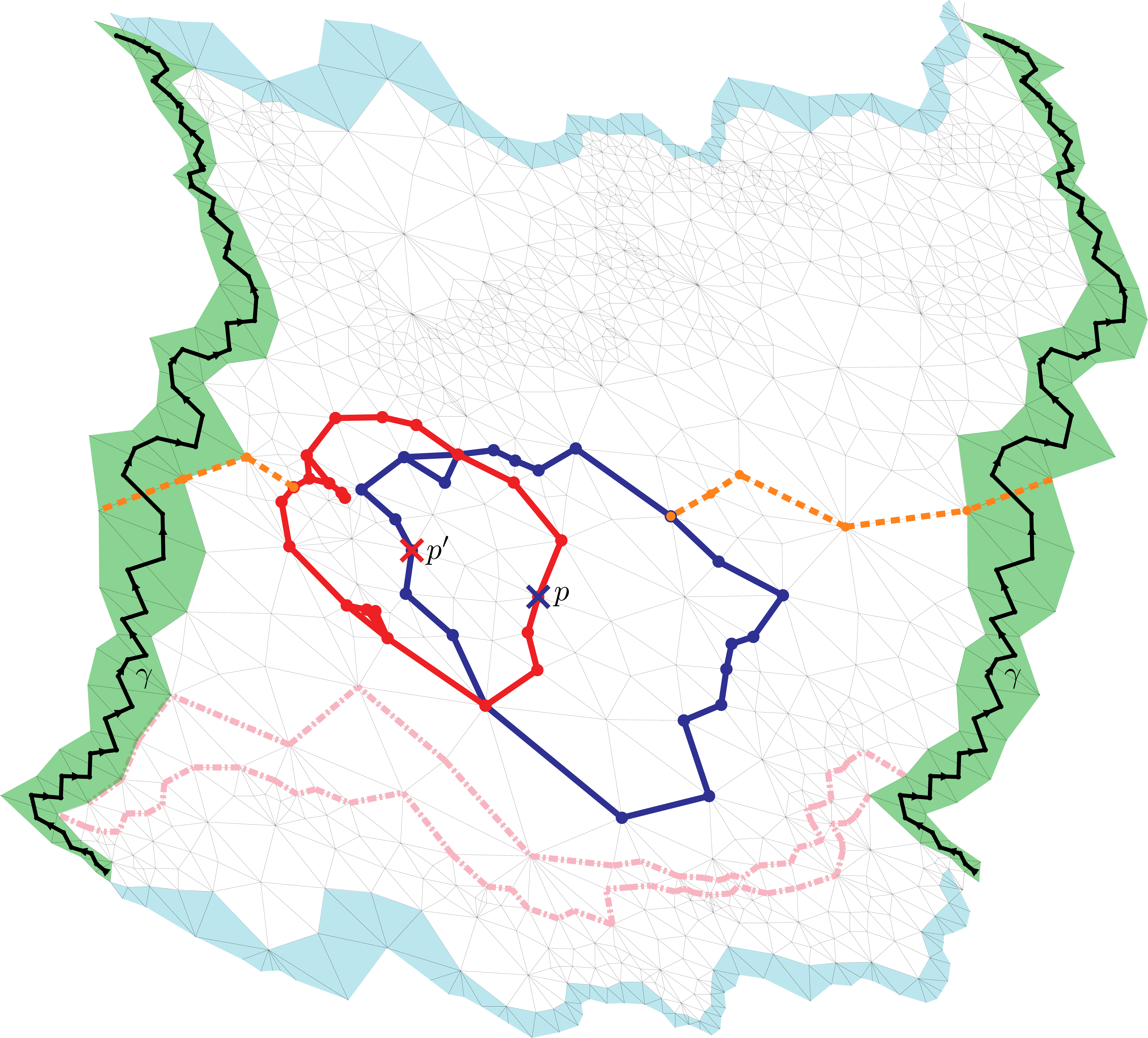}
	\caption{Example of a toroidal CDT triangulation, represented by a harmonic embedding. The centers of the circles $S_p^\delta$ (dark blue) 
	and $S_{p'}^\delta$ (red) of radius $\delta \!= \!3$ are marked by crosses. The presence of a shortest geodesic between two points 
	on the two circles (orange, dashed), which has nonzero intersection number with the reference curve $\gamma$ along dual links, signals a level-1 violation. 
	Opposite boundary strips should be identified, as described in the text. Two neighboring spatial slices of constant time are indicated by 
	dash-dotted red lines.
	}
	\label{app-in-fig:triangulation-crossing-number}
\end{figure}

To illustrate some aspects of the nontrivial geometry of the CDT configurations on which this analysis takes place,  
Fig.\ \ref{app-in-fig:triangulation-crossing-number} depicts a specific example of such a triangulation. It contains a pair of circles 
$(S_p^3, S_{p'}^3)$ of radius 3, together with a pair of points connected by a shortest geodesic $\phi$ 
of nonvanishing intersection number. 
We have chosen to represent the triangulation by its harmonic embedding \cite{ambjorn2012roaming} in the plane, 
which is based on two loops along its links that generate the fundamental group of the torus. They are a spacelike loop at constant time (bottom boundary in 
Fig.\ \ref{app-in-fig:triangulation-crossing-number}) and a timelike loop (left boundary in Fig.\ \ref{app-in-fig:triangulation-crossing-number}), which we have chosen to lie immediately to the left of the reference curve $\gamma$ used to compute the intersection numbers.
Opposite boundaries of this embedding should be identified. We have repeated the triangulated boundary strips (light blue and green) along opposite sides, which
makes it easier to see which boundary links should be identified pairwise to form the torus.

In addition to the two spheres and the topological shortcut $\phi$, which intersects the curve $\gamma$ exactly once, 
we have for reference also marked two neighboring spatial slices (red, dash-dotted).
Unlike in the standard way of depicting two-dimensional CDT geometries (Fig.\ \ref{ric2dcdt-fig:cdt-sample}), in this representation the spatial slices
cannot be identified just by inspection. 
It is not coincidental that the circle configuration with a level-1 violation lies in a region that is relatively sparsely populated by triangles. 
Recall that (after the Wick rotation) all triangles of a CDT configuration are equilateral, a property that cannot be preserved faithfully by any
planar representation. The nature of the harmonic embedding we have chosen is such that regions with small spatial slices are stretched out, whereas regions with large spatial slices are squeezed. 
It implies that regions with lower triangle density are associated with smaller slice volumes, which are exactly the regions where we expect level-1 
violations to occur more frequently. 
\chapter{Estimating the entropy exponent}
\label{app:entropy}
This appendix describes the procedure used in Sec.\ 2.1 of \cite{ambjorn1993baby} to determine best fit values for the entropy exponent $\gamma$ 
with associated error bars, together with the results of this analysis applied to our data. One introduces a subleading correction to the right-hand 
side of \eqref{eq:minbu-dist} by replacing
\begin{equation}
	B^{\gamma-2} \to B^{\gamma-2}\left(1+\frac{c}{B}+O\left(\frac{1}{B^2}\right)\right),
\end{equation}
which allows for a better fit in the regime of small $B$, without significantly affecting the behavior of the function at intermediate and large $B$. 
One then takes the logarithm on both sides of \eqref{eq:minbu-dist} to obtain
\begin{equation}
\label{eq:log-minbu-cor}
	\log(\bar{n}_{N_2}(B)) = a + (\gamma-2)\log\left(B(N_2-B)\right) + \frac{c}{B},
\end{equation}
where the best fit parameters $a$, $\gamma$ and $c$ should now be determined from the observed baby universe distributions shown in 
Fig.\ \ref{fig:minbu-dist}.\footnote{Note that the argument in the logarithm on the right-hand side of this expression differs from Eq.\ (2.2) in \cite{ambjorn1993baby} by a multiplicative factor $N_2$, which is absorbed by the fit parameter $a$.} 
The goodness of fit is defined through the $\chi^2$-statistic, described in greater detail in App.\ \ref{app:ci} below.

\begin{figure}[h!]
	\centering
	\includegraphics[width=0.8\textwidth]{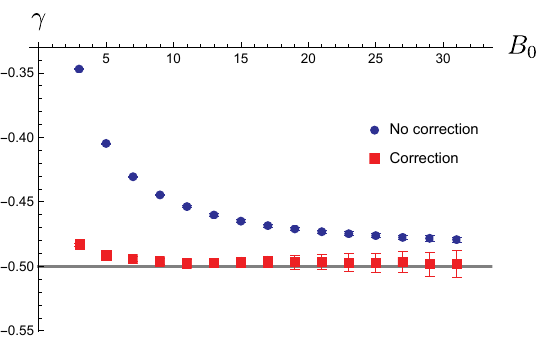}
	\caption{Fitted values of the entropy exponent $\gamma$ in the degenerate phase for slice volume $N_2\! =\! 1.000$ and different lower cut-offs $B_0$. 
	We show the best fit values with and without the correction term appearing in \eqref{eq:log-minbu-cor}.}
	\label{fig:minbu-gamma-fits}
\end{figure}

As mentioned before, the prediction \eqref{eq:minbu-dist} is only expected to hold for sufficiently large baby universes, where discretization effects are negligible. 
We therefore introduce a lower cut-off $B_0$ on $B$ on the data before extracting the best fit parameters. The resulting values of the entropy
exponent $\gamma$ as a function of the cut-off $B_0$ are shown in Fig.\ \ref{fig:minbu-gamma-fits} for $N_2\!\equiv V_2\! =\! 1.000$, both with and without the correction term $c/B$ in \eqref{eq:log-minbu-cor}. Our figure resembles Fig.\ 2 of reference \cite{ambjorn1993baby} extremely closely, including the magnitude of the error bars. 
The authors of \cite{ambjorn1993baby} subsequently obtain an estimate for $\gamma$ by fitting an exponential of the form 
\begin{equation}
	\gamma(B_0) = \gamma - c_1\, e^{-c_2\, B_0}.
\end{equation}
The resulting values for $\gamma$ shown in Table 1 of \cite{ambjorn1993baby} are (within statistical error) identical to the ones we found from our data, using the same method.
\pagebreak
\null
\vfill
\chapter{Determining confidence intervals for best fit parameters}
\label{app:ci}
While researching the literature on numerical estimates of quantities like the Hausdorff dimension and the entropy exponent, we found that previous work 
is often not very explicit about the methodology used to determine the uncertainty in the best fit parameters, if such margins of error are provided at all. 
Since our goal was to investigate whether the behavior of the spatial slices is consistent with that of known models of two-dimensional random geometry, 
we considered it important to attach a degree of confidence to the results obtained. This led us to a more detailed study of the statistical methods available for performing such an analysis. Here we summarize our findings, in the hope that others may find them useful. We also comment on the assumptions required to make use of these methods, and to what extent they apply in the present context. Our main aim is to motivate the choices made in computing the confidence intervals; we do not aim for full mathematical or statistical rigor.

The generic starting point consists of a set of $N$ data points $y_n \in \mathbb{R}$ measured at positions $\vec{x}_n$, to which we want to fit a function $f(\vec{x},\vthet)$ 
with parameters $\vthet$. Firstly, we require a measure of the goodness-of-fit which allows us to find the set of best fit parameters $\vthetm$ that minimizes this measure. Secondly, since the measurements are inherently noisy, we are interested in specifying a range for the fit parameters in which we expect to find the ``true'' values with a certain degree of confidence. In the method of least squares, the goodness-of-fit is defined through the sum of squares of the residuals, and the optimal fit is obtained when this sum is minimized. However, the standard unweighted sum of least squares assumes equal variances on all the data points, 
which typically is not the case for the measurements we perform in lattice quantum gravity. In what follows, we will show how the $\chi^2$-distribution can be used in the context of a linear regression model to define a goodness-of-fit and corresponding confidence intervals for the parameters $\vthet$. The models we fit in Chapter \ref{ch:slice3d} do not fall into this class of linear models, so we subsequently discuss how the analysis is affected when the condition of linearity is relaxed.

\section{Best fit parameter estimation}
A linear model takes the form
\begin{equation}
	f\left(\vec{x}; \vec{\theta}\right) = \sum_{i=1}^p \theta_i f_i(\vec{x}),
\end{equation}
where the $f_i$ are functions of the independent variables. These functions are allowed to be nonlinear in the $x_i$ --- the linearity condition applies to the fit parameters $\theta_i$ only. Suppose the ``correct'' model has fit parameters $\vthet_0$. If the errors $\sigma_n$ on the $N$ measurement outcomes are Gaussian, we can consider the $y_n$ to be normally distributed random variables with mean $f(\vec{x}_n; \vthet_0)$ and standard deviation $\sigma_n$. We can turn these into $N$ standard normals $Z_n=\frac{y_n-f(\vec{x}_n;\vthet_0)}{\sigma_n}$. Let us define the \emph{$\chi^2$-statistic} of a choice of fit parameters $\vthet$ as the sum of the squares of the $Z_n$,
\begin{equation}
\label{eq:chisq}
	\chi^2\left(\vec{\theta}\right) = \sum_{n=1}^N \left(\frac{y_n-f\left(\vec{x}_n; \vthet\right)}{\sigma_n\,}\right)^2.
\end{equation}
The reason we call this the $\chi^2$-statistic is that the sum of $k$ independent standard normals follows a so-called $\chi^2$-distribution with $k$ degrees of freedom. Such a distribution has expectation value $k$ and variance $2k$. The quantity $\chi^2\left(\vthet\right)$ should be minimized to find a \emph{maximum likelihood estimator} for $\vthet_0$, corresponding to the set of best fit parameters.

In the current situation, where we are trying to fit a model to the data points, we must take into account that the individual $y_n$ are explicitly \emph{not} independent --- after all, we have hypothesized the existence of a model $f(\vec{x}; \vthet)$ that can predict the outcomes of our measurements. The sum of $N$ squared standard normal distributions therefore follows a $\chi^2$-distribution with a certain number $k\! <\!  N$ of degrees of freedom. For a linear model without priors (i.e.\ restrictions on the fit parameters), we have $k\! =\! N-p$, where $p$ is the number of fit parameters.

Therefore, when fitting a linear model $f(\vec{x}; \vthet)$ with $p$ fit parameters to $N$ data points $y_n$ with Gaussian errors $\sigma_n$, we expect the $\chi^2$-statistic to follow a $\chi^2$-distribution with $k=N-p$ degrees of freedom. As mentioned earlier, the best fit parameters $\vthetm$ are determined by finding the minimum possible value $\chi^2_\textrm{min}$ of the $\chi^2$-statistic. The fit is considered good when $\chi^2_\textrm{min}\! \approx\! k$, since this is the expectation value of the corresponding $\chi^2$-distribution. Significantly larger values of $\chi^2_\textrm{min}$ indicate that no reasonable fit could be found (e.g.\ our assumptions about the model may be wrong), whereas a $\chi^2_\textrm{min}$ significantly lower than $k$ could mean we are overfitting the data or overestimating the measurement errors $\sigma_n$.

\section{Confidence intervals}
With the best fit parameters $\vthetm$ at our disposal, we can now turn to determining \emph{confidence intervals} on these parameters. After all, slightly varying the parameters around their best fit values should produce approximately equal $\chi^2$-statistics. Moreover, the measurement data we are using to compute 
$\chi^2$ is inherently noisy, which implies that we have merely obtained an estimate of the ``true'' best fits. To specify a range in which we expect to find the true values with a certain degree of confidence, we can use the properties of the $\chi^2$-distribution. 

Confidence intervals (CIs) are computed at a certain \emph{confidence level}, specified by a percentage (a common choice is the 95\% CI). Alternatively, we can specify a significance level $\alpha$, corresponding to a $(1\!-\! \alpha)\%$ confidence level. Given $\alpha$, we can determine the \emph{critical value} $\chi^2_{\textrm{crit},\alpha}$ of the $\chi^2$-statistic for our fitted model containing $p$ parameters by solving $P(\chi^2\! \leq \!\chi^2_{\textrm{crit},\alpha})\! = \! (1-\!\alpha)$, where $P(\chi^2\! \leq\! x)$ is the cumulative distribution function for a $\chi^2$-distribution with $k\! =\! p$ degrees of freedom.\footnote{When estimating the best fit entropy exponent and local Hausdorff dimension, we were only interested in one out the $p$ fit parameters. In this case, we should match to a $\chi^2$-distribution with one degree of freedom.} We can determine $\chi^2_{\textrm{crit},\alpha}$ by the use of quantile functions or lookup tables. The $(1\! -\! \alpha)\%$ confidence interval for the fit parameters $\vthet$ is then defined \cite{avni1976energy} as the region for which 
\begin{equation}
	\chi^2(\vthet) - \chi^2_\textrm{min} < \chi^2_{\textrm{crit},\alpha}.
    \label{eq:ellipsoid-ci}
\end{equation}
Typically, this region is an ellipsoid in $\vthet$-space, which can be determined numerically by performing a grid search around $\vthetm$. As an example, when determining the 95\% confidence inter\-vals for a linear model with two parameters, we find $\chi^2_{\textrm{crit},0.05}\!\approx\! 5.991$, and the joint confidence intervals of the two fit parameters are the region in $\mathbb{R}^2$ for which \eqref{eq:ellipsoid-ci} holds.

\section{Potential caveats}
We have used the procedure just described to determine the confidence intervals for the best fit parameters in Chapter \ref{ch:slice3d} of this thesis. However, as pointed out earlier, the analysis rests on several assumptions that do not necessarily apply to our models and measurements. An important prerequisite for using the $\chi^2$-distribution is that the measurement errors are Gaussian, otherwise the sum \eqref{eq:chisq} is not a sum of squares of standard normals. We often found a slight degree of skewness in the distribution of our measurement results, potentially invalidating the use of $\chi^2$-methods. However, the skewness factors were always near zero, so that we may still consider the computed bounds of the confidence intervals to be good approximations to their true values.

A second potential issue is that the models we fit in Chapter \ref{ch:slice3d} are not linear. Both for the minbu sizes and the microscopic Hausdorff dimension, one of the fit parameters appears in the exponent of an independent variable. Although the best fit parameters for such models can still be obtained by minimizing \eqref{eq:chisq}, determining the correct number of degrees of freedom is known to be difficult \cite{andrae2010dos}. This means that we do not know the proper expectation value of \eqref{eq:chisq}, and therefore do not have a reference point to compare our $\chi^2_\textrm{min}$ to. However, not knowing the true number of degrees of freedom has more serious consequences for computing confidence intervals. The number $p$ of fit parameters in our models is small, $p\!  <\!  4$, and choosing a different number of degrees of freedom near zero has a strong effect on the resulting $\chi^2_{\textrm{crit},\alpha}$. This can significantly alter the width of the corresponding confidence interval. For our purposes, we do not consider this to be a major issue. The main goal of our analysis was to check whether our results are consistent with previously known results from the literature, requiring an order-of-magnitude estimate of the confidence interval. This order of magnitude is not affected if we over- or underestimate the degree-of-freedom count by a few units. 
Furthermore, since all confidence intervals in Chapter \ref{ch:slice3d} of this thesis were obtained by using the same methods, any comparison \emph{among} our confidence intervals is likely to 
still be meaningful.
\chapter{Normalizations of the differential formulations}
\label{dm-app:normalizations}

\section{Normalization of the differential formulation of diagonalized one-matrix models}
\label{dm-par:norm-1}

The normalization $e_N$ in \eqref{dm-eq:diff-diagonal-usual-herm} is computed developing the Vandermonde determinants as in Sec.~\ref{dm-subsub:det-form-usual}:
\be
\begin{split}
\frac 1 {e_N} &= \left[ \e^{\frac 1 {2N}(\frac{\pd}{\pd \x})^2}\vd^{2}(\x)\right]_{\x=0} \\
&= N! \det_{i,j} \left\{ \left[ \e^{\frac 1 {2N} \dxi{2}} x^{i+j} \right]_{x=0} \right\}  \\
&= N! \det_{i,j}\Biggl\{ \sum_{n\ge 0}\frac {(2N)^{-n}} {n!} \left[ \dxi{2n} x^{i+j} \right]_{x=0} \Biggr\},
\end{split}
\ee
where the indices in the determinants range from 0 to $N-1$. The bracket in the rightmost expression is non-vanishing only if $2n=i+j$, in which case $\left[ \dxi{2n} x^{i+j} \right]_{x=0} = (i+j)!$. Therefore, 
$
1/ {e_N} = N!  \det(R) (2N)^{-N(N-1)/2},
$
where the $N\times N$ matrix $R$ is such that $R_{ij}=0$ for $i+j$ odd, and $R_{ij} = (i+j)!/[(i+j)/2]!$ for $i+j$ even. This determinant can be computed to be $\det(R) = 2^{N(N-1)/2} \prod_{j=1}^{N-1} j!$, so that
$
e_N = N^{N(N-1)/2} / \prod_{j=1}^N j!.
$

\section{Normalization of the differential formulation of diagonalized two-matrix models}
\label{dm-par:norm-2}

The normalization is the same as for the one-matrix model, since:
\begin{align}
\begin{split}
\left[ \e^{\frac 1 N \sum_{i=1 }^N\frac{\pd}{\pd a_i}\frac{\pd}{\pd b_i}}\, \right.& \Delta(\ba)\Delta(\bb) \bigg]_{\ba=\bb=0} \\
& = N! \det_{i,j} \left\{ \left[ \e^{\frac 1 N \sum_{i=1 }^N\frac{\pd}{\pd a}\frac{\pd}{\pd b}}\, a^{i}b^{j} \right]_{a=b=0} \right\}  
\\ &= N! \det_{i,j}\Biggl\{ \sum_{n\ge 0}\frac {(N)^{-n}} {n!} \left[ \frac{d^{n}}{d a^{n}} a^i\right]_{a=0}\left[ \frac{d^{n}}{d b^{n}} b^j\right]_{b=0} \Biggr\}
\\ &= N! \det_{i,j}\Biggl\{ \delta_{i,j} i! N^{-i} \Biggr\} = \frac 1 {e_N}.
\end{split}
\end{align}
\chapter{Alternative representations 
using generalized heat kernels}
\label{dm-appendixB}

In Sec.~\ref{dm-subsub:invert-diff-diag-gaussian}, we have detailed how, for one-matrix models,  the Slater determinant form \eqref{dm-eq:slater-1-mat-usual} in the integral formulation can be recovered directly from that in differential formulation \eqref{dm-eq:Slater-diff-onemat} by using the heat kernel formula \eqref{dm-eq:heat-kernel}.

In this appendix, we detail for completeness the analogous computation for two-matrix models, that is,  the integral analogues of expressions \eqref{dm-eq:diff-diagonal-new}, \eqref{dm-eq:Slater-first-expr}, and \eqref{dm-eq:Slater-diff} using generalized heat kernels. Together these form the back lower edge of the diagram shown in Fig. \ref{dm-fig:diagram-two-matrix}.

Our starting point is the differential representation \eqref{dm-eq:diff-diagonal-new}:
\be
\label{dm-eq:diff-app}
Z_{V_1,V_2}=  \left[ \frac 1 {\vd(\x)} \e^{N V_1( \frac 1 {\sqrt N} \frac{\pd}{\pd \x})} \vd(\x) \,\e^{NV_2(\frac 1 {\sqrt N}  \x)} \right]_{\x=0}
\, .\ee
In \eqref{dm-eq:heat-kernel}, we used the heat kernel $K_t(x)$ to transform a differential representation of the one- model to an integral form. For two-matrix models, we can use the following analogous identity (see e.g. \cite{Robinson}):
\be
\label{dm-eq:gen-heat-kernel}
 \e^{ W(\frac{d}{d \x})} F(\x) = \int_{\bR^N} d\y \,K_{W} (\x-\y) F(\y),
\ee
where $W$ is a polynomial, and $K_{W}$ is a generalized heat kernel with the following Fourier representation:
\be
K_{W} (\x-\y) = \frac {1} {(2\pi)^N} \int_{\bR^N} d\bp\,  \e^{\imath \bp \cdot (\x-\y) 
+ W(\imath \bp)}.
\ee
If we set $W(\x) = N V_1\left(\frac{1}{\sqrt{N}} \x\right)$ and $F(\x) = \vd(\x) e^{N V_2\left(\frac{1}{\sqrt{N}} \x\right)}$, we can substitute \eqref{dm-eq:gen-heat-kernel} in \eqref{dm-eq:diff-app} to find
\be
\label{dm-eq:app-c1}
Z_{V_1,V_2}= \frac {1} {(2\pi)^N} \left[ \frac{1}{\vd(\x)} \int_{\bR^N} \d\y \int_{\bR^N} \d\bp\, \e^{\imath \bp \cdot(\x-\y)+NV_1(\frac 1 {\sqrt{N}}\imath \bp)} \vd(\y) \e^{N V_2(\frac 1 {\sqrt N}\y)}\right]_{\x=0}.
\ee
This is the integral analogue of \eqref{dm-eq:diff-diagonal-new}. Subsequently, we can rewrite this expression as a Slater determinant by expanding the Vandermonde determinant, just like we did when rewriting \eqref{dm-eq:diff-diagonal-new} to \eqref{dm-eq:Slater-first-expr}, thus obtaining:
\be
\label{dm-eq:app-c2}
Z_{V_1,V_2}= \frac {1} {(2\pi)^N}  \left[\frac{1}{\vd(\x)} \det_{i,j}\left\{ \int_{\bR} \d y \int_{\bR} \d p\,\e^{\imath p(x_i-y)+NV_1(\frac 1 {\sqrt{N}}\imath p)} y^{j} \e^{N V_2(\frac 1 {\sqrt N} y)}\right\} \right]_{\x=0},
\ee
where $0\le i,j\le N-1$. Finally, we can use identity \eqref{dm-eq:identity-diff-det} to obtain
\be
\label{dm-eq:app-c3}
Z_{V_1,V_2}= \frac {f_N} {(2\pi)^N}   \det_{i,j}\left\{ \int_{\bR} \d y   \int_{\bR} \d p\, \left[\frac{\partial^{i}}{\partial x^{i}}\e^{\imath p(x-y)}\right]_{x=0} \e^{NV_1(\frac 1 {\sqrt{N}}\imath p)} y^{j} \e^{N V_2(\frac 1 {\sqrt N} y)} \right\},
\ee
where the bracket evaluating $x$ at 0 could now be moved inside the determinant and the integral. This equation mirrors \eqref{dm-eq:Slater-diff} using the generalized heat kernel. 
Carrying out the differentiation and evaluating $x$ at 0: 
\be
Z_{V_1,V_2}= \frac {f_N} {(2\pi)^N}   \det_{i,j}\left\{ \int_{\bR} \d y   \int_{\bR} \d p\, (\imath p)^i\e^{-\imath py} \e^{NV_1(\frac 1 {\sqrt{N}}\imath p)} y^{j} \e^{N V_2(\frac 1 {\sqrt N} y)} \right\}.
\ee
To verify that the equation is well-normalized we may use \eqref{dm-eq:Gaussian-mixt} to carry out the integration of the matrix elements. There is a subtlety however: \eqref{dm-eq:Gaussian-mixt-0} and \eqref{dm-eq:Gaussian-mixt}  have to be modified for $\alpha\in \imath \mathbb{R}$.
For $n=m=0$ and  $\alpha\in \imath \mathbb{R}$, \eqref{dm-eq:Gaussian-mixt-0} should be replaced with:
\be
\label{dm-eq:Gaussian-mixt-compl-0}
\int_{\bR^2}\d x\, \d y\, \e^{- \imath \alpha  N  xy } = \frac{2\pi} {\alpha N},
\ee
and for positive $n$ or $m$:
\be 
\int_{\bR^2}\d x\, \d y\, \e^{- \imath \alpha N  xy }\, x^n y^m = \frac {\delta_{n,m}  }{(-\imath N)^n} \frac{\partial^{n}}{\partial \alpha^{n}} \int_{\bR^2}\d x\, \d y\, \e^{- \imath \alpha N  xy } =  \frac {\delta_{n,m}  } {(-\imath N)^n} \frac{\partial^{n}}{\partial \alpha^{n}} \frac{2\pi}{\alpha N} 
,
\ee
so \eqref{dm-eq:Gaussian-mixt} has to be modified for:
\be
\label{dm-eq:Gaussian-mixt-compl}
\int_{\bR^2}\d x\, \d y\, \e^{- \imath \alpha N  xy }\, x^n y^m = \delta_{n,m}\,  2\pi \imath \,  \left(\frac 1 {\imath \alpha N}\right)^{n+1} n!.
\ee
These expressions allow showing that 
\begin{align}
\begin{split}
\label{dm-eq:change-variables-pure-complex}
\int_{\bR^2} \d y\, \d p  \, \e^{-\imath py}  (\imath p)^i & \e^{NV_1(\frac 1 {\sqrt{N}}\imath p)} y^{j} \e^{N V_2(\frac 1 {\sqrt N} y)} \\ 
&= \imath  \int_{\bR^2} \d y\, \d x  \, \e^{-xy} x^i  \e^{NV_1(\frac 1 {\sqrt{N}}x)} y^{j} \e^{N V_2(\frac 1 {\sqrt N} y)},
\end{split}
\end{align}
in the sense that their perturbative expansions match, where \eqref{dm-eq:Gaussian-mixt-compl} has to be used to expand the left-hand side and \eqref{dm-eq:Gaussian-mixt} for the right-hand side, which explains the factor $\imath$, while $\imath^{-1}$ would naively be expected as a Jacobian by directly changing variables for $x=\imath p$ in the integral. Finally, changing variables for $\sqrt N x'=x$ and $\sqrt N y'=y$, we  recover the integral Slater determinant form \eqref{dm-eq:slater-2-mat-usual}, since $\frac{f_N}{(2\pi)^N} N^{N(N+1)/2} \imath^N  = \frac 1 {d_N}$.

\addchap{Samenvatting}
Stel je voor dat je zweeft in een luchtballon, hoog boven een enorme oceaan. Gewapend met een dure camera maak je foto's van het uitgestrekte wateroppervlak. In eerste instantie ziet elk beeld er bijna hetzelfde uit: welk stuk van de oceaan je ook fotografeert, het resultaat is steeds weer hetzelfde effen blauwe vlak. Gelukkig heeft de camera ook een dure zoomlens, waarmee je kleinere stukjes van de oceaan in groter detail kunt vastleggen. Naarmate je verder inzoomt, zie je steeds meer onderscheid tussen de verschillende foto's. Eerst zie je vage kleurverschillen door de grootschalige stromingen, daarna begin je de deining van het water te herkennen. Door nog verder aan de zoomknop te draaien, ontwaar je ook de losse golven. Je blijft uitvergroten: nu zie je zelfs de individuele druppels die opspatten uit het water, en het wilde schuim dat op de golven drijft. 

In zekere zin is het bovenstaande gedachte-experiment de kern waar dit proefschrift om draait. In plaats van de oceaan zoomen we in op het universum en proberen we te begrijpen hoe \emph{ruimte} en \emph{tijd} zich op hele kleine schaal gedragen. We zijn gewend om over de ruimte om ons heen te denken als iets ongrijpbaars, iets statisch, iets passiefs, iets waar we ons simpelweg doorheen bewegen met het verstrijken van de tijd. Maar zien vissen de oceaan waarin ze zwemmen niet ook simpelweg als het `universum' waar ze zich doorheen bewegen? In de rest van deze samenvatting beschrijven we in de eerste plaats de natuurkundige idee\"en achter deze analogie. Daarna gebruiken we de analogie om een beeld te schetsen van de methoden en vraagstukken die behandeld worden in dit manuscript.

De Algemene Relativiteitstheorie van Albert Einstein vertelt ons dat de ruimte om ons heen in veel opzichten meer weg heeft van een dynamische oceaan dan van een statische en regelmatige achtergrond waarop de werkelijkheid zich afspeelt. Een belangrijk aspect van de theorie is dat `ruimte' en `tijd' onlosmakelijk met elkaar zijn verbonden --- de snelheid waarmee de tijd verstrijkt hangt af van de snelheid waarmee je je voortbeweegt door de ruimte. We spreken daarom doorgaans over een samengevoegd concept dat \emph{ruimtetijd} genoemd wordt. De ruimtetijd waar we in leven is vierdimensionaal, met drie voor de ruimte en \'e\'en voor de tijd. Wij, mensen, zijn helaas slecht in staat om vier dimensies te visualiseren, maar gelukkig is dit ook niet nodig. Ons voorstellingsvermogen in twee en drie dimensies is meer dan genoeg voor de voorbeelden die in het onderstaande aan bod komen.

De grote ontdekking van Einstein was dat we dit concept van ruimtetijd kunnen beschouwen als een verklaring voor de \emph{zwaartekracht}. Sinds Isaac Newton weten we dat de zwaartekracht een onderlinge aantrekking is tussen alle massieve objecten. Deze aantrekking zorgt er onder andere voor dat de maan rond de aarde draait en dat we allemaal met beide benen op de grond blijven staan. Einstein liet zien dat we de zwaartekracht kunnen beschouwen als gevolg van de \emph{kromming} van de ruimtetijd. Massieve objecten zoals sterren en planeten veroorzaken een kromming in de ruimtetijd om zich heen. Die resulterende kromming bepaalt vervolgens hoe materie zich door de ruimtetijd voortbeweegt, wat wij ervaren als zwaartekracht. Gedeeltelijk kunnen we dit vergelijken met een grote draaikolk in de oceaan. Een boot die langs de randen van de draaikolk rechtdoor probeert te varen, wordt afgebogen door de kromming van het wateroppervlak.

Tot zover zegt onze analogie dus dat we de \emph{ruimtetijd} waar ons universum uit bestaat kunnen beschouwen als een soort enorme oceaan, die vervormd wordt door objecten die zich in die oceaan bevinden. Zwaartekracht is een gevolg van deze vervorming. Het inzoomen op de ruimtetijd (waar dit proefschrift naar vernoemd is) komt dus overeen met het bestuderen hoe zwaartekracht werkt op minuscule schaal. Dit onderzoeksgebied noemen we \emph{kwantumzwaartekracht}. De term `kwantum' verwijst naar de kwantummechanica, de theorie die beschrijft hoe deeltjes zich gedragen op microscopisch niveau. Grof gezegd modelleert de zwaartekracht (in de vorm van de Algemene Relativiteitstheorie) het universum in het groot, en de kwantummechanica het universum in het klein. Beide theorie\"en zijn enorm succesvol binnen deze afgebakende domeinen; het probleem ontstaat wanneer we het grensvlak tussen de twee proberen op te zoeken. Nog altijd is er geen complete theorie die verklaart hoe zwaartekracht werkt op kwantumniveau, ondanks alle inspanningen van natuurkundigen gedurende ruim honderd jaar. Anders gezegd: we zijn nog altijd op zoek naar een volledig begrip van \emph{kwantumruimtetijd}.

Er bestaan vermoedens dat ruimtetijd op kwantumniveau in een constante staat van fluctuatie is. Wij merken hier in het alledaagse leven niets van, gewoonweg omdat al deze fluctuaties zich op zo'n absurd kleine schaal afspelen. Denk terug aan de fotograaf in de luchtballon: van veraf lijkt de oceaan \'e\'en groot blauw geheel, terwijl van dichtbij de waterdruppels en het schuim om onze oren vliegen.

Door de jaren heen zijn er verscheidene methoden ontwikkeld om de kwantumruimtetijd beter te begrijpen. We behandelen hier \'e\'en van die methoden, genaamd \emph{Causale Dynamische Triangulaties} (CDT). We kunnen CDT beschouwen als een raamwerk waarmee we de wetten van de kwantummechanica kunnen toepassen op de ruimtetijd. Op zichzelf biedt dit nog geen garantie op een consistente theorie die een correcte beschrijving geeft van onze werkelijkheid. Door verschillende aspecten van de theorie nauwkeurig te onderzoeken, kunnen we toetsen of CDT inderdaad een geschikte kandidaat is om dit doel te bereiken.

Het fundament van CDT ligt in de toepassing van Richard Feynmans \emph{padintegraalmethode} op de structuur van ruimtetijd zelf. De padintegraalmethode werd oorspronkelijk ontwikkeld om het gedrag van een deeltje op kwantumniveau te beschrijven. In essentie stelt de methode dat een deeltje niet slechts \'e\'en bepaald pad volgt van punt A naar punt B, maar dat er een `superpositie' bestaat van alle mogelijke paden tussen die twee punten. Elk van deze paden draagt bij aan de uiteindelijke `kansamplitude' voor het deeltje om van A naar B te gaan. Door alle mogelijke paden bij elkaar op te sommen, kunnen we een volledig beeld krijgen van het waarschijnlijke gedrag van het deeltje. Deze sommatie kan uitgevoerd worden door de paden eerst te \emph{discretiseren}, wat betekent dat we elk pad zien als een aaneenschakeling van een aantal rechte segmenten. Door deze segmenten steeds kleiner te maken en er meer te gebruiken, kunnen we ieder mogelijk pad tot op willekeurige precisie benaderen. Dit zien we ge\"illustreerd in Figuur \ref{intro-fig:qmpi} van de introductie van dit proefschrift.

We breiden het idee van de padintegraal nu uit naar de ruimtetijd. In plaats van alle mogelijke paden die een deeltje kan afleggen, sommeren we nu over alle mogelijke configuraties van de ruimtetijd. Dit is vele malen complexer dan het sommeren over paden, met name wanneer we het proberen uit te voeren in vier dimensies. Voor het gemak beperken we ons daarom nu tot tweedimensionale ruimtetijd, waarin we de methode beter kunnen visualiseren. Het doel is nu om te sommeren over alle mogelijke tweedimensionale oppervlakken. Opnieuw gebruiken we het concept van discretisatie om grip te krijgen op het probleem. De fundamentele bouwsteen is nu niet meer een recht lijnsegment, maar een platte driehoek. Door deze driehoeken langs de zijden aan elkaar te lijmen, construeren we een tweedimensionaal oppervlak. Door de afmeting van deze driehoeken geleidelijk te verkleinen en meer driehoeken te gebruiken, kunnen we elk mogelijk oppervlak tot op willekeurige precisie benaderen --- net als voor de paden die een deeltje kon afleggen tussen twee punten.

Neem weer onze analogie tussen oceaan en ruimtetijd. Als we het oceaanoppervlak van grote afstand bekijken, kunnen we een goed passende voorstelling ervan maken met een paar grote vlakke driehoeken. Van dichterbij zien we golven en de deining van het water, waardoor die paar grote driehoeken een steeds minder goede benadering zijn van hoe het wateroppervlak er daadwerkelijk uitziet. Door meer driehoeken te gebruiken van kleiner formaat, kunnen we weer een nieuwe `triangulatie' maken die ook op deze ingezoomde schaal beter op het wateroppervlak past. Dit proces kunnen we blijven voortzetten tot we zelfs de kleinste details van het schuim op de golven kunnen vangen, door middel van een enorm aantal minuscule platte driehoekjes. Op diezelfde manier kunnen we binnen het raamwerk van CDT een `triangulatie' maken van de ruimtetijd, op basis van eenvoudige driehoekjes als bouwstenen.

Binnen de kwantumzwaartekracht gaat het echter niet zozeer om het modelleren van een bepaalde vorm van de ruimtetijd. Net zoals we eerst moesten sommeren over alle mogelijke paden die een deeltje van A naar B kan afleggen, moeten we nu sommeren over \emph{alle mogelijke vormen} die de ruimtetijd kan aannemen. Dit kan mogelijk leiden tot inzichten in de eigenschappen van \emph{kwantumruimtetijd}. De strategie die CDT gebruikt om deze som te berekenen, is eenvoudig te formuleren: de taak is om alle mogelijke configuraties van aaneengelijmde driehoekige bouwstenen te verkennen. Deze worden vervolgens allemaal bij elkaar opgeteld, net als bij het sommeren over paden. In de praktijk is deze opsomming lastig (of zelfs onmogelijk) om met pen en papier te berekenen. Daarom wordt binnen CDT-onderzoek veelvuldig gebruik gemaakt van \emph{computersimulaties} om dit soort berekeningen nauwkeurig te benaderen.

Het is belangrijk om te benadrukken dat we niet claimen dat de ruimtetijd \emph{daadwerkelijk} uit kleine driehoekjes bestaat. De driehoekjes zijn slechts een instrument om de sommatie over alle mogelijke ruimtetijden helder te defini\"eren. We hadden ook vierkantjes, vijfhoeken of een andere vorm kunnen kiezen --- het voordeel van de driehoeken is dat we er eenvoudiger mee kunnen rekenen dan met complexere vormen.

De triangulaties die voorkomen binnen CDT kunnen allerlei vreemde en bijzondere eigenschappen bezitten. Een voorbeeld hiervan is dat de effectieve \emph{dimensie} van de triangulaties niet altijd dezelfde hoeft te zijn als de dimensie van de bouwstenen! Bepaalde triangulaties die slechts uit tweedimensionale driehoeken bestaan, kunnen in sommige opzichten beter gezien worden als een driedimensionale ruimte wanneer we het geheel op een grotere schaal bekijken. Andersom kan een triangulatie opgebouwd uit drie- of vierdimensionale `driehoeken' vanuit een bepaald oogpunt meer weghebben van een tweedimensionaal oppervlak, of zelfs een eendimensionale lijn. We kunnen ons deze twee situaties als volgt voorstellen. Een vel papier is (bij benadering) tweedimensionaal, maar wanneer we er een prop van maken krijgt het bepaalde eigenschappen die lijken op een driedimensionale bol. In de omgekeerde situatie beelden we ons een heel aantal kleine piramides in, die in een lange ketting aan elkaar gelijmd zijn. Ook al zijn de piramides op zichzelf driedimensionaal, als we deze lange ketting vanaf een afstand bekijken zien we slechts nog \'e\'en dimensie.

Dit proefschrift richt zich hoofdzakelijk op de analyse van \emph{kromming} van de ruimtetijd op kwantumniveau. De gebruikelijke methoden voor het onderzoeken van kromming (zoals de krommingstensor van Riemann) zijn in dit domein slecht toepasbaar vanwege de invloed van kwantumfluctuaties, die kunnen worden vergeleken met het `schuim' op de golven. De zogenoemde \emph{kwantum-Ricci-kromming} is een recent ontwikkelde methode om kromming in deze context te kunnen bestuderen. Een belangrijk aspect van de kwantum-Ricci-kromming is dat deze de mogelijkheid biedt om kromming te berekenen op verschillende lengteschalen. Terugkomend bij de analogie met het oceaanoppervlak: in sommige gevallen zijn we bijvoorbeeld ge\"interesseerd in de kromming van een golf, zonder dat we daarbij alle details van het schuim en de kleine waterdruppels willen meenemen. De kwantum-Ricci-kromming stelt ons in staat om alle kleinschalige details te middelen over een groter gebied.

De kwantum-Ricci-kromming kan in twee dimensies berekend worden door twee overlappende cirkels naast elkaar op een oppervlak te plaatsen en vervolgens de gemiddelde afstand tussen deze cirkels te berekenen. Als een oppervlak gekromd is zoals een voetbal (dit noemen we `positieve' kromming), dan liggen de cirkels gemiddeld genomen iets dichterbij elkaar dan wanneer ze zich op een plat vlak bevinden. Als een oppervlak daarentegen gekromd is zoals een ruiterszadel (`negatieve' kromming), liggen de cirkels gemiddeld genomen juist iets verder van elkaar af. Door de grootte van de cirkels aan te passen kunnen we bepalen op welke schaal we de kwantum-Ricci-kromming willen meten: doordat we de \emph{gemiddelde} afstand tussen de cirkels berekenen, worden alle fijnmazige details in zekere zin uitgesmeerd.

In het eerste deel van dit proefschrift hebben we de kwantum-Ricci-kromming bestudeerd in tweedimensionale context. Als onderdeel hiervan hebben we het zogenoemde \emph{krommingsprofiel} gedefinieerd, een voorschrift waarmee we een soort `handtekening' kunnen berekenen die informatie verschaft over de krommingseigenschappen van een bepaalde ruimte. Met behulp van computersimulaties hebben we het krommingsprofiel van het tweedimensionale CDT-model met de topologie van een torus berekend. Vervolgens hebben we het gedeeltelijke krommingsprofiel bepaald van de vijf Platonische lichamen: convexe veelvlakken waarvan de zijvlakken regelmatige veelhoeken zijn. Deze Platonische lichamen zijn nagenoeg overal vlak, met uitzondering van de hoekpunten. De kromming is geconcentreerd in deze `krommingssingulariteiten'. Uit onze berekeningen is gebleken dat de krommingsprofielen van deze Platonische lichamen veel gelijkenissen vertonen met het krommingsprofiel van een regelmatig boloppervlak. Dit interpreteren we als een aanwijzing dat het krommingsprofiel en de onderliggende kwantum-Ricci-kromming inderdaad geschikte methoden zijn om kleinschalige krommingsdetails uit te middelen.

In het tweede deel lag de focus op CDT in drie dimensies. We hebben tweedimensionale `hyperoppervlakken' binnen dit model onderzocht, waarbij het doel was om te bepalen of het gedrag van deze hyperoppervlakken overeenkomt met dat van bekende \emph{universaliteitsklassen} van tweedimensionale willekeurige oppervlakken. Dit bleek inderdaad het geval voor de hyperoppervlakken in \'e\'en fase van het model, waar onze metingen bevestigen dat het gedrag overeenkomstig is met dat van tweedimensionale (niet-causale) Dynamische Triangulaties (DT). In de andere fase van het model lieten onze metingen zien dat de hyperoppervlakken zeer waarschijnlijk niet binnen een van de mogelijke universaliteitsklassen vallen. Op basis daarvan hebben we het vermoeden geopperd dat de hyperoppervlakken in deze fase van het model niet los gezien kunnen worden van de driedimensionale structuur, en daadwerkelijk een onlosmakelijk onderdeel van het geheel vormen.

Het derde en laatste deel van dit proefschrift betrof wiskundige en computationele technieken voor het rekenen aan verzamelingen van getrianguleerde ruimtes. We bestudeerden eerst ensembles van toevalsmatrices, waarmee analytische berekeningen uitgevoerd kunnen worden aan tweedimensionale willekeurige oppervlakken. We hebben laten zien hoe integralen over twee-matrix-ensembles herschreven kunnen worden in termen van differentiaaloperatoren in \'e\'en matrixvariabele. Deze herformulering is mogelijk inzetbaar om twee-matrixmodellen op te lossen in systemen van orthogonale polynomen, in plaats van de gebruikelijke oplossing in termen van bi-orthogonale polynomen. Als laatste hebben we de implementatie-aspecten van CDT-computersimulaties besproken. Deze simulaties zijn doorgaans tijdrovend en het is daarom van groot belang om deze zo effici\"ent mogelijk op te zetten. Allereerst hebben we de relevante theoretische en praktische factoren toegelicht. Daarna hebben we laten zien hoe deze overwegingen zijn meegenomen in het opzetten van de simulatiecode die gebruikt is voor veel van de resultaten in dit proefschrift.

\addchap{Acknowledgements}
The first time I arrived at the Huygens building as a freshly recruited PhD student, I was carrying a mixed bag of feelings and expectations about the years that would follow. On the one hand, I was immensely excited about everything I would have to learn to strengthen my understanding of fundamental physics. On the other hand, this seemed an insurmountable task, and I therefore felt ill-prepared to do research. After receiving my office key, I went to look for my new supervisor for guidance --- but, as it turns out, she was out on a skiing holiday.

Dear prof. Loll, dear Renate: I'm happy to say that your absence during this first week turned out to be an anomaly. You have managed to help me grow far beyond the person I was back then, while keeping intact that sense of excitement for learning more about the universe at its smallest scales. You often left me free to explore whatever topic I was interested in at the time, even if it was difficult to see how it would fit into our research programme. At the same time, you coached me into finding the right questions that \emph{did} fit into our research programme, freeing up ample time to discuss our projects almost every week. The content of this thesis is the result of our joint efforts to answer these questions, and the very existence of this thesis is the result of the fact that you allowed me to experience physics as a hobby instead of an obligation. Furthermore, I was happy to find a companion in you for nitpicking about textual and typographical details when preparing a paper for publication, and your meticulous comments on my thesis drafts have improved the manuscript by orders of magnitude. And, finally --- as evidenced by my first few days at Radboud --- you showed me that it's perfectly fine for a top-level scientist to take up some holidays every now and then. I owe you many thanks for everything you've taught me and done for me over the past years.

Furthermore, I want to thank the members of the HEP department for their involvement during my time there. Wim, being a teaching assistant for your quantum field theory course has been a major factor for me to feel more confident as theoretical physicist, and your style of teaching has inspired me to take up many more subsequent TA assignments. Jan, it has been great fun to learn quantum geometry from you while I had to explain the subject to others, and I would never have expected that my initial efforts at typesetting your problem sheets would eventually lead to a fully-fledged hardcover textbook on the topic. Timothy, I appreciate all the time you took to share your experience on CDT simulations and discrete exterior calculus --- I managed to bypass many roadblocks by taking up your advice.

Marcus, our project on approximate symmetries has taken many unexpected twists and turns, and at times it seemed to me that we might not be able to arrive at a satisfactory set of conclusions. However, your optimism and perseverance were part of the reason that we finally managed to formulate a coherent message based on our findings. I also very much enjoyed reminiscing about our time in Nijmegen during our unexpected reunion in Alsace last summer. Luca, our collaboration with Johannes not only led to a full chapter of my thesis, but also involved several great Friday pasta\&wine nights with you and Alexandra; certainly one of the most fun ways I've ever done research! Daniel, thank you for getting me on board with your plans for studying the 3D CDT effective action, I learned a lot from your higher-dimensional experience. Also, thank you for bearing with me as I kept complaining about extremely minor details in our draft. Willem, het was erg gezellig en leerzaam om je te mogen begeleiden voor je masterscriptie en we treffen elkaar vast nog eens in Domburg voor een frisse duik.

To all my friends I met through the DRSTP: many thanks for showing me that it \emph{is} possible to have a great night out in S\~{a}o Paolo with a group of theoretical physicists. Greg, my ETH buddy: I'm happy we were able to reunite in Amsterdam and Utrecht, sharing our love of the Gaberdiel lectures we both attended in Z\"urich. Kilian, our club has been missing your support for quite some time now, and I'm looking forward to you answering my questions on D-branes while watching FC Utrecht lose from the Bunnikside again. Jorinde, bedankt voor alle leuke gesprekken over het leven als beginnend natuurkundige --- dat stadium ben jij intussen ruimschoots gepasseerd, en ik ben benieuwd hoe je je daar verder gaat ontwikkelen. Eric, we delen veel meer interesses dan ik in eerste instantie had kunnen denken. Dit heeft een grote rol gespeeld bij mijn keuze voor het NKI, enorm bedankt dat je me daar aan een postdocplaats hebt geholpen! Ik kijk niet alleen uit naar alle projecten die we nog gaan oppakken in de nabije toekomst, maar ook naar alle quasi-wetenschappelijke troep die we onderweg nog gaan tegenkomen. Gelukkig heb ik in jou altijd een medestander in mijn frustratie over onzin in de breedste zin van het woord.

Mirjam, Sasha, and all other members of your research group, many thanks for showing me around in the world of infectious diseases modeling for one great summer at the UMC Utrecht. I had lots of fun during the weekly group meetings and I'm excited to see that our project is finally nearing completion!

Pieter, Maarten, Emma, Jorg, Arie, bedankt voor alle boeken die jullie me hebben laten lezen die ik zelf nooit uitgezocht zou hebben, en iets minder bedankt voor de boeken die ik nu nog steeds niet uit zou zoeken. Theo, Arie, Oscar, bedankt voor al tweederde van ons leven aan vriendschap, en voor alles wat daar nog op volgen gaat. Patrick, Michael, shalosh kef! Chris, Tristan, het Nederlandse ETH-departement is nu helaas wat opgebroken, maar weet vooral dat de grote pan met chili altijd voor jullie klaarstaat. Jasper, Jonathan en Thomas, laten we gauw de zelfgemaakte barbecue in Frankrijk weer in de fik zetten, er moeten dringend nieuwe flunkybalregels opgesteld worden. Nikki, bedankt voor alle middagen en avonden zonder einde, en alle uiteenlopende chatgesprekken op ieder moment van de dag. Hangouts heeft aan ons twee trouwe klanten gehad --- twee van de weinigen, waarschijnlijk.

Aan iedereen met wie ik de afgelopen jaren de Witte Vos heb mogen delen: het was een mooi avontuur om te wonen in dat prachtige huis dat zo nu en dan zijn eigen leven leek te leiden. Voor saaie momenten was geen tijd, mede dankzij de vliegende kerstbomen, ontploffende ovens, nachtelijke weddenschappen en permanente binnentemperatuur van 11 graden. Helaas komt ook daaraan binnenkort voor mij een einde, maar ik ga met veel plezier terugkijken op alles wat we daar in zoveel verschillende samenstellingen hebben beleefd.

Domstad Majella 3: ik had niet gedacht dat ik na mijn 27e nog moest leren om tegen mijn verlies te kunnen. Gelukkig wist Tom me al snel uit te leggen dat het bij ons grotendeels een kwestie is van vaak veel pech hebben. Ons jaar komt nog wel. Tom, Sandor, Wout, Souf, Lex, Thomas, Matthijs, Siderion (x2), Nicko, Jaap, Marijn, Frank, Duco, Daan, Arie, ik had in geen beter team terecht kunnen komen. Joost, ik ben blij dat je me een tweede kans hebt gegeven na de eerste paar weken. Tim, bedankt voor je 24/7 bol.com service via WhatsApp. Leon, de rare snuiter die eigenlijk best lief bleek te zijn. Elbert, als je je toga maar thuis laat. Jens, maakt niet uit. Jesse, bedankt dat je de enige bent die \'echt kan voetballen. Rubbe, het spijt me dat ik je tactische meesterplannen nog steeds niet begrijp. Baba (het blijft raar om je zo te noemen), bedankt dat je mij hier met wat mooie praatjes tussen hebt weten te kletsen zodat we na onze huisgenotentijd als teamgenoot verder konden. Djelloe, je zal het niet snel iemand in je bijzijn horen zeggen, maar je betrouwbaarheid en inzet maken je een onmisbare schakel die we zo nu en dan hard nodig hebben.

Arie en Wouter, waarde leden van het zuurstofgenootschap. Kort nadat Wouter weer eens een nieuwe levensfase in zijn schoot geworpen kreeg en bij ons introk, dwong een klein stukje RNA ons om het grootste deel van onze dagen samen thuis door te brengen. In de maanden die daarop volgden hebben we vele Otto's besteed aan het verkennen van de Midden-Oosterse keuken, het uitbreiden en uitdunnen van onze wijnvoorraad, en het beslechten van verhitte discussies over de optimale strategie voor een hypothetisch spel dat allang niets meer met Catan te maken had. Ook al besef ik dat het voor veel anderen een moeilijke tijd is geweest, het feit dat het ons drie\"en zoveel dichter bij elkaar heeft gebracht is voor mij een zilveren randje aan die periode. Vanaf de Ahrend 230 kan een mens de wereld aan.

Wat ooit begon als een surrogaatjaarclub gevormd uit een paar scheppen los zand, is intussen behoorlijk samengekneed tot een hechte vriendengroep die ik niet meer zomaar uit het oog wil en zal verliezen. Tim, Jordy, Marcel, Aupie, Aad, Vincent, Maarten, Jeroen, Bob, Sebas, Rik, Jelle, het verbaast me iedere keer weer hoe bijna elke mogelijke subgroep van ons dertienen iets met elkaar gemeen lijkt te hebben, en hoe makkelijk het vervolgens is om al die kliekjes op \'e\'en grote hoop te gooien bij het heilige kerstdiner, al dan niet met droogijs, dun ijs of een indoor-vuurwerkshow. Vincent, als we elkaar binnenkort eindelijk met ``Herr Doktor'' mogen aanspreken wordt het hoog tijd om onze quatre-mainsambities weer nieuw leven in te blazen. Jeroen, we moeten nodig eens gaan brainstormen over de automatische vinexwijk-designer --- ooit gaan we \'echt iets tastbaars doen voor de mensheid. Bob, bedankt voor je unieke perspectieven en je 0900-adviesjes door alle jaren heen. Ik kijk al uit naar onze volgende ijsberentocht. Jelle, je aanwezigheid geeft kleur en energie aan iedere gelegenheid, en het aanwezige risico om zo nu en dan belachelijk gemaakt te worden neem ik dan ook graag voor lief. Ik hoop dat ik je vanaf binnenkort nog vaker (gepland danwel ongepland) tegen het lijf loop als mijn buurman in de borrelkapel en zal de BWM-plop dan ook nauwlettend in de gaten houden.

Ruben, Bram, de musketiers: de driehoek van onze lange-afstandsvriendschap heeft sinds de middelbare school al elke denkbare vorm aangenomen. Van \includegraphics[width=0.4cm]{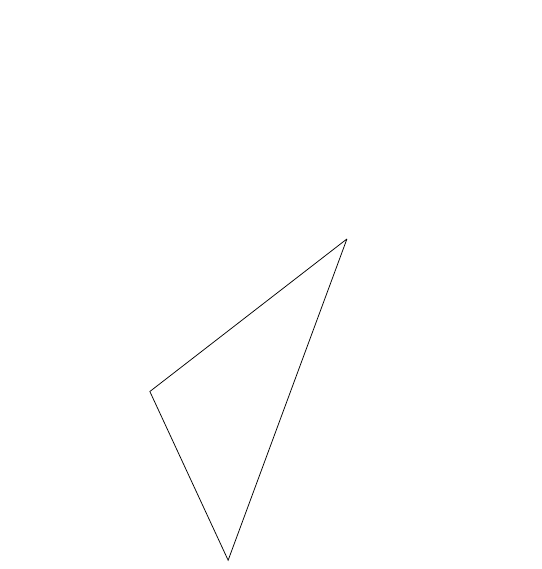} naar \includegraphics[width=0.6cm]{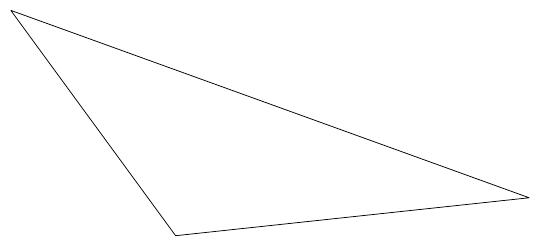} en nu helaas alweer een tijdje \includegraphics[width=0.15cm]{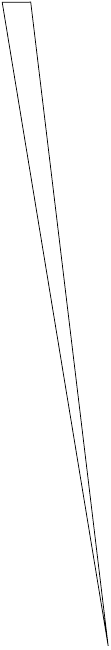}. Gelukkig weten we elkaar nog steeds met enige regelmaat te vinden, of dat nou in Dublin, Amsterdam, Boedapest, Bergamo of Zeeland is. Ik kijk iedere keer weer uit naar de momenten waarop onze driehoek samenkrimpt tot een punt. Ruben, ik geniet van je originaliteit, vrolijkheid, en het aanstekelijke enthousiasme waarmee je al je ontelbare plannen uitvoert. Bram, ik denk dat ik met niemand z\'o veel woorden heb uitgewisseld via digitale kanalen als met jou. Vaak over onzin, maar onze online briefwisselingen hebben er ook voor gezorgd dat je obsessie met Amerikaanse politiek langzaam op mij is overgesprongen. Overigens moet ik eerlijk bekennen dat ik je brakke spraakberichten van zestien minuten zo nu en dan ongeopend heb laten passeren.

En uiteraard, mijn tweede familie: Nico, Maxi, Dani\"el en Noraly, het was geweldig om bij jullie en samen met jullie op te groeien, en jullie hebben nog altijd een bijzondere plek in mijn leven. Intussen hoeven jullie me geen havermout meer te voeren tijdens het oppassen, maar ik ben altijd erg blij als de koffie gewoon weer klaarstaat (of snel gemaakt wordt) als ik onaangekondigd op het raam klop. Dani\"el, de prachtige omslag die je voor dit boekje hebt ontworpen is voor mij een blijvend aandenken aan de mooie tijden die we met zijn allen hebben beleefd.

Lieve opa, lieve oma, wat hadden jullie dit nog graag meegemaakt en wat had ik jullie hier graag bij gehad. Nieuwsgierigheid is wat me op dit pad heeft gebracht en gehouden. Die nieuwsgierigheid heb ik voor een heel groot deel aan jullie te danken. We reisden het hele land af om tentoonstellingen, dierentuinen en musea te bezoeken. Hoeveel boeken heb ik wel niet van jullie gekregen om me verder te verdiepen in een onderwerp dat we samen ontdekt hadden. De eindeloze gesprekken die ik met Opa had op het bankje bij de vijver. De scripties die Oma wist te produceren als ik een vraag had over de stamboom van een figuur uit de Griekse mythologie. Jullie hebben me zo enorm veel geleerd. Ik denk nog z\'o vaak aan jullie.

Voor de laatste vier personen in dit dankwoord bestaat eigenlijk geen passende volgorde. Geheel volgens de conventies voor auteurslijsten binnen de hoge-energiefysica houd ik daarom eenvoudigweg het alfabet aan. 

Mama \& papa, ik had al snel besloten dat twee epidemiologen per gezin ruim genoeg was. Het sprak voor zich dat ik girafoloog zou worden. Uitgebreid hebben we besproken wat ik allemaal zou moeten doen om dat doel te behalen --- en straks sta ik dit proefschrift te verdedigen dat in de verste verte niets met giraffen te maken heeft. Toch is het zaadje geplant tijdens die discussies, en sindsdien hebben jullie me in nog zoveel meer opzichten kunnen helpen met het zoeken van een weg in de academische wereld. Tijdens onze gezellige studieweekjes in Zwitserland en Zeeland kon ik de concentratie vinden die ik zo nu en dan nodig had; andersom waren de zomervakanties met ons vieren een heerlijke afleiding van al het denkwerk, waarna ik altijd weer met hernieuwd enthousiasme in de trein naar Nijmegen stapte. Jullie hebben Thomas en mij altijd onze eigen gang laten gaan, zonder het gevoel te geven dat we in jullie voetsporen moesten treden. Toch heeft het opgroeien in een wetenschappelijk nest duidelijk een stempel op me gedrukt en ik neem aan dat jullie ook wel doorhebben hoe gelukkig mij dat al die jaren heeft gemaakt. Daarbij neem ik graag op de koop toe dat er vroeger wel erg vaak woorden als `cohortstudie' en `infant respiratory infection' over de eettafel heen en weer vlogen. Ik ben ontzettend dankbaar voor alle manieren waarop jullie, zichtbaar en onzichtbaar, gefaciliteerd hebben dat Thomas en ik konden worden wie we nu zijn.

Marte, toen ik je leerde kennen dacht ik toch wel in het laatste staartje van mijn promotietraject te zijn beland. Dit bleek gewoon weer een van mijn veel te optimistische inschattingen, en uiteindelijk heb je nog een aanzienlijk deel van dichtbij meegemaakt. Ik vind het fascinerend om van je te horen hoe het (voor zovelen verborgen) dagelijkse leven in de psychiatrie eruitziet, en daarnaast om van je te leren hoe onderzoek werkt in een totaal ander domein dan waar ik me in bevond. Wonder boven wonder heb je zelfs de hardloper in mij naar boven kunnen halen --- ik kan nog steeds niet zeggen dat het allemaal van harte gaat (leer mij dit aub), maar ik weet zeker dat ik zonder jou niet half zo ver gekomen zou zijn. Onder de streep ben ik ontzettend blij dat ik heb gekozen voor m\'et jou hier in plaats van z\'onder jou in Okinawa en ik kijk nu al uit naar onze tijd samen op jouw postzegeltje met uitzicht op dat mooie, gezellige plein in hartje Utrecht.

Thomas, ik zou het mezelf haast makkelijker maken door hier op te sommen wat we de afgelopen jaren n\'iet samen hebben gedaan. Gelukkig heeft dit dankwoord geen woordenlimiet. Tomaten planten, treinrails leggen, overhooghouden, menu's opstellen, peaty bogs zoeken, trollen slopen, tully's maken, queso's eten, souffl\'es bakken, fisks turven, kerstpakketten rondbrengen, de lijst is eindeloos. Dankjewel voor alle keren dat je bij ons hebt meegevoetbald als de opkomst weer eens laag was --- iets minder bedankt voor de keren dat je me te grazen hebt genomen in de wedstrijden dat we juist tegenover elkaar stonden, ook al deed ik mijn uiterste best om je die dienst terug te bewijzen. Dankjewel voor alle lange kerstvakanties samen, waarin de dagen eigenlijk altijd weer te kort waren voor alle plannen die we maakten. Ik ben ontzettend trots op alle grote stappen die je het afgelopen jaar hebt gezet, ook al wist ik altijd al waar je toe in staat bent. Ik prijs mezelf gelukkig dat een van mijn beste vrienden, van wie ik zoveel heb kunnen leren, toevallig ook mijn broertje is.

\addchap{About the author}
Joren Brunekreef was born on the 9th of April 1990 in Utrecht. He completed bachelor's degrees in physics and mathematics at Utrecht University. For his bachelor's thesis, he researched topological quantum computation at Maynooth University in Ireland under the supervision of Joost Slingerland and Cristiane de Morais Smith. After graduating, he moved to Switzerland to pursue a master's degree in theoretical physics at the Eidgen\"ossische Technische Hochschule (ETH) in Z\"urich. His master's thesis work focused on integrable spin chain models and was supervised by Yunfeng Jiang and Niklas Beisert. He graduated from ETH in the fall of 2016.

Subsequently, he began his Ph.D. research with Renate Loll in the Quantum Gravity group at Radboud University, Nijmegen. Over the next few years, he investigated notions of curvature in nonperturbative Causal Dynamical Triangulations (CDT) quantum gravity. This work eventually culminated in this thesis. He also set up a codebase for performing CDT simulations in two and three dimensions, which has since been released to the public for free use. In addition, he served as a teaching assistant for several master-level courses in physics and earned recognition for teaching excellence from the student body on multiple occasions.

In the summer of 2022, he joined the University Medical Center Utrecht as a temporary junior researcher in the Infectious Diseases Modelling group led by Mirjam Kretzschmar. During this period, he worked on agent-based network models to investigate the effectiveness of lockdown regulations in preventing the spread of infectious diseases.

He has since joined the AI for Oncology group at the Netherlands Cancer Institute (NKI) in Amsterdam as a postdoctoral fellow. Under the supervision of Jan-Jakob Sonke and Jonas Teuwen, he is researching novel Artificial Intelligence (AI) methodologies for cancer diagnostics and treatment, with a special focus on computer vision techniques for medical imaging.

\backmatter

\errorcontextlines=200
\printbibliography

@article{alsing2011simplicial,
  title = {The simplicial {{Ricci}} tensor},
  author = {Alsing, Paul M. and McDonald, Jonathan R. and Miller, Warner A.},
  year = {2011},
  month = jun,
  journal = {Classical and Quantum Gravity},
  volume = {28},
  number = {15},
  eprint = {1107.2458},
  pages = {155007},
  publisher = {{IOP Publishing}},
  issn = {0264-9381},
  doi = {10.1088/0264-9381/28/15/155007},
  urldate = {2022-04-01},
  archiveprefix = {arxiv},
  langid = {english}
}

@article{ambjorn1990summing,
  title = {Summing over all genera for d {$>$} 1: a toy model},
  shorttitle = {Summing over all genera for d {$>$} 1},
  author = {Ambj{\o}rn, Jan and Durhuus, Bergfinnur and Jonsson, Thordur},
  year = {1990},
  month = jul,
  journal = {Physics Letters B},
  volume = {244},
  number = {3},
  pages = {403--412},
  issn = {0370-2693},
  doi = {10.1016/0370-2693(90)90337-6},
  urldate = {2022-04-25},
  langid = {english}
}

@article{ambjorn1993baby,
  title = {Baby universes in 2d quantum gravity},
  author = {Ambj{\o}rn, J. and Jain, S. and Thorleifsson, Gudmar},
  year = {1993},
  month = jun,
  journal = {Physics Letters B},
  volume = {307},
  number = {1-2},
  eprint = {hep-th/9303149},
  pages = {34--39},
  publisher = {{North-Holland}},
  issn = {0370-2693},
  doi = {10.1016/0370-2693(93)90188-N},
  urldate = {2021-10-26},
  archiveprefix = {arxiv}
}

@article{ambjorn1995fractal,
  title = {On the fractal structure of two-dimensional quantum gravity},
  author = {Ambj{\o}rn, J. and Jurkiewicz, J. and Watabiki, Y.},
  year = {1995},
  month = nov,
  journal = {Nuclear Physics B},
  volume = {454},
  number = {1},
  eprint = {hep-lat/9507014},
  pages = {313--342},
  issn = {0550-3213},
  doi = {10.1016/0550-3213(95)00468-8},
  urldate = {2021-12-07},
  archiveprefix = {arxiv},
  langid = {english}
}

@article{ambjorn1995new,
  title = {New critical phenomena in 2d quantum gravity},
  author = {Ambj{\o}rn, Jan and Thorleifsson, Gudmar and Wexler, Mark},
  year = {1995},
  month = apr,
  journal = {Nuclear Physics B},
  volume = {439},
  number = {1-2},
  eprint = {hep-lat/9411034},
  pages = {187--204},
  issn = {05503213},
  doi = {10.1016/0550-3213(95)00014-J},
  urldate = {2022-08-25},
  archiveprefix = {arxiv}
}

@article{ambjorn1995scaling,
  title = {Scaling in quantum gravity},
  author = {Ambj{\o}rn, J. and Watabiki, Y.},
  year = {1995},
  month = jul,
  journal = {Nuclear Physics B},
  volume = {445},
  number = {1},
  eprint = {hep-th/9501049},
  pages = {129--142},
  issn = {05503213},
  doi = {10.1016/0550-3213(95)00154-K},
  urldate = {2022-04-07},
  archiveprefix = {arxiv}
}

@book{ambjorn1997geometry,
  title = {The {{Geometry}} of {{Dynamical Triangulations}}},
  author = {Ambj{\o}rn, Jan and Carfora, Mauro and Marzuoli, Annalisa},
  year = {1997},
  month = sep,
  publisher = {{Springer Science \& Business Media}},
  googlebooks = {9lYHWfzQmS0C},
  isbn = {978-3-540-63330-3},
  langid = {english}
}

@book{ambjorn1997quantumb,
  title = {Quantum {{Geometry}}: {{A Statistical Field Theory Approach}}},
  shorttitle = {Quantum {{Geometry}}},
  author = {Ambj{\o}rn, Jan and Durhuus, Bergfinnur and Jonsson, Thordur},
  year = {1997},
  month = jun,
  publisher = {{Cambridge University Press}},
  doi = {10.1017/CBO9780511524417},
  isbn = {978-0-521-46167-2},
  langid = {english}
}

@article{ambjorn1998nonperturbative,
  title = {Non-perturbative {{Lorentzian}} quantum gravity, causality and topology change},
  author = {Ambj{\o}rn, J. and Loll, R.},
  year = {1998},
  month = dec,
  journal = {Nuclear Physics B},
  volume = {536},
  number = {1},
  eprint = {hep-th/9805108},
  pages = {407--434},
  issn = {0550-3213},
  doi = {10.1016/S0550-3213(98)00692-0},
  urldate = {2022-03-14},
  archiveprefix = {arxiv},
  langid = {english}
}

@article{ambjorn1998quantum,
  title = {The quantum space-time of c=-2 gravity},
  author = {Ambj{\o}rn, Jan and Anagnostopoulos, K. N. and Ichihara, T. and Jensen, L. and Kawamoto, N. and Watabiki, Y. and Yotsuji, K.},
  year = {1998},
  month = feb,
  journal = {Nuclear Physics B},
  volume = {511},
  number = {3},
  eprint = {hep-lat/9706009},
  pages = {673--710},
  issn = {05503213},
  doi = {10.1016/S0550-3213(97)00659-7},
  urldate = {2022-04-07},
  archiveprefix = {arxiv}
}

@article{ambjorn1999euclidean,
  title = {Euclidean and {{Lorentzian}} quantum gravity\textemdash lessons from two dimensions},
  author = {Ambj{\o}rn, J. and Loll, R. and Nielsen, J. L. and Rolf, J.},
  year = {1999},
  month = feb,
  journal = {Chaos, Solitons \& Fractals},
  volume = {10},
  number = {2},
  eprint = {hep-th/9806241},
  pages = {177--195},
  issn = {0960-0779},
  doi = {10.1016/S0960-0779(98)00197-0},
  urldate = {2022-03-14},
  archiveprefix = {arxiv},
  langid = {english}
}

@article{ambjorn1999new,
  title = {A new perspective on matter coupling in 2d quantum gravity},
  author = {Ambj{\o}rn, Jan and Anagnostopoulos, K. N. and Loll, R.},
  year = {1999},
  month = oct,
  journal = {Physical Review D},
  volume = {60},
  number = {10},
  eprint = {hep-th/9904012},
  pages = {104035},
  issn = {0556-2821, 1089-4918},
  doi = {10.1103/PhysRevD.60.104035},
  urldate = {2022-04-07},
  archiveprefix = {arxiv}
}

@article{ambjorn2000crossing,
  title = {Crossing the c=1 barrier in {{2D Lorentzian}} quantum gravity},
  author = {Ambj{\o}rn, Jan and Anagnostopoulos, K. and Loll, R.},
  year = {2000},
  month = jan,
  journal = {Physical Review D},
  volume = {61},
  number = {4},
  eprint = {hep-lat/9909129},
  pages = {044010},
  publisher = {{American Physical Society}},
  doi = {10.1103/PhysRevD.61.044010},
  urldate = {2022-03-14},
  archiveprefix = {arxiv}
}

@incollection{ambjorn2000lorentzian,
  title = {Lorentzian and {{Euclidean}} quantum gravity \textemdash{} analytical and numerical results},
  booktitle = {M-{{Theory}} and {{Quantum Geometry}}},
  author = {Ambj{\o}rn, J. and Jurkiewicz, J. and Loll, R.},
  editor = {Thorlacius, L{\'a}rus and Jonsson, Thordur},
  year = {2000},
  series = {{{NATO Science Series}}},
  pages = {381--450},
  publisher = {{Springer Netherlands}},
  address = {{Dordrecht}},
  doi = {10.1007/978-94-011-4303-5_9},
  urldate = {2022-03-14},
  isbn = {978-94-011-4303-5},
  langid = {english}
}

@article{ambjorn2000nonperturbative,
  title = {A non-perturbative {{Lorentzian}} path integral for gravity},
  author = {Ambj{\o}rn, J. and Jurkiewicz, J. and Loll, R.},
  year = {2000},
  month = jul,
  journal = {Physical Review Letters},
  volume = {85},
  number = {5},
  eprint = {hep-th/0002050},
  pages = {924--927},
  issn = {0031-9007, 1079-7114},
  doi = {10.1103/PhysRevLett.85.924},
  urldate = {2022-08-03},
  archiveprefix = {arxiv}
}

@article{ambjorn2000relation,
  title = {On the relation between {{Euclidean}} and {{Lorentzian 2D}} quantum gravity},
  author = {Ambj{\o}rn, Jan and Correia, J. and Kristjansen, C. and Loll, R.},
  year = {2000},
  month = feb,
  journal = {Physics Letters B},
  volume = {475},
  number = {1-2},
  eprint = {hep-th/9912267},
  pages = {24--32},
  issn = {03702693},
  doi = {10.1016/S0370-2693(00)00058-7},
  urldate = {2022-04-07},
  archiveprefix = {arxiv}
}

@article{ambjorn2001computer,
  title = {Computer simulations of {{3D Lorentzian}} quantum gravity},
  author = {Ambj{\o}rn, J. and Jurkiewicz, J. and Loll, R.},
  year = {2001},
  month = mar,
  journal = {Nuclear Physics B - Proceedings Supplements},
  volume = {94},
  number = {1-3},
  eprint = {hep-lat/0011055},
  pages = {689--692},
  issn = {09205632},
  doi = {10.1016/S0920-5632(01)00878-7},
  urldate = {2022-08-25},
  archiveprefix = {arxiv}
}

@article{ambjorn2001dynamically,
  title = {Dynamically triangulating {{Lorentzian}} quantum gravity},
  author = {Ambj{\o}rn, J. and Jurkiewicz, J. and Loll, R.},
  year = {2001},
  month = sep,
  journal = {Nuclear Physics B},
  volume = {610},
  number = {1},
  eprint = {hep-th/0105267},
  pages = {347--382},
  issn = {0550-3213},
  doi = {10.1016/S0550-3213(01)00297-8},
  urldate = {2022-03-14},
  archiveprefix = {arxiv},
  langid = {english}
}

@article{ambjorn2001lorentzian,
  title = {Lorentzian 3d gravity with wormholes via matrix models},
  author = {Ambj{\o}rn, J. and Jurkiewicz, J. and Loll, R. and Vernizzi, G.},
  year = {2001},
  month = sep,
  journal = {Journal of High Energy Physics},
  volume = {2001},
  number = {09},
  eprint = {hep-th/0106082},
  pages = {022--022},
  issn = {1029-8479},
  doi = {10.1088/1126-6708/2001/09/022},
  urldate = {2022-04-11},
  archiveprefix = {arxiv}
}

@article{ambjorn2001nonperturbative,
  title = {Nonperturbative {{3D Lorentzian}} quantum gravity},
  author = {Ambj{\o}rn, J. and Jurkiewicz, J. and Loll, R.},
  year = {2001},
  month = jul,
  journal = {Physical Review D},
  volume = {64},
  number = {4},
  eprint = {hep-th/0011276},
  pages = {044011},
  publisher = {{American Physical Society}},
  doi = {10.1103/PhysRevD.64.044011},
  urldate = {2021-10-28},
  archiveprefix = {arxiv}
}

@article{ambjorn20023d,
  title = {3d {{Lorentzian}}, dynamically triangulated quantum gravity},
  author = {Ambj{\o}rn, J. and Jurkiewicz, J. and Loll, R.},
  year = {2002},
  month = mar,
  journal = {Nuclear Physics B - Proceedings Supplements},
  volume = {106--107},
  eprint = {hep-lat/0201013},
  pages = {980--982},
  issn = {09205632},
  doi = {10.1016/S0920-5632(01)01904-1},
  urldate = {2022-04-08},
  archiveprefix = {arxiv}
}

@article{ambjorn20033d,
  title = {{{3D Lorentzian}} quantum gravity from the asymmetric {{ABAB}} matrix model},
  author = {Ambj{\o}rn, J. and Jurkiewicz, J. and Loll, R. and Vernizzi, G.},
  year = {2003},
  month = nov,
  eprint = {hep-th/0311072},
  urldate = {2022-04-11},
  archiveprefix = {arxiv}
}

@article{ambjorn2004emergence,
  title = {Emergence of a {{4D}} world from causal quantum gravity},
  author = {Ambj{\o}rn, J. and Jurkiewicz, J. and Loll, R.},
  year = {2004},
  month = sep,
  journal = {Physical Review Letters},
  volume = {93},
  number = {13},
  eprint = {hep-th/0404156},
  pages = {131301},
  publisher = {{American Physical Society}},
  doi = {10.1103/PhysRevLett.93.131301},
  urldate = {2022-03-10},
  archiveprefix = {arxiv}
}

@article{ambjorn2004renormalization,
  title = {Renormalization of 3d quantum gravity from matrix models},
  author = {Ambj{\o}rn, J. and Jurkiewicz, J. and Loll, R.},
  year = {2004},
  month = feb,
  journal = {Physics Letters B},
  volume = {581},
  number = {3-4},
  eprint = {hep-th/0307263},
  pages = {255--262},
  issn = {03702693},
  doi = {10.1016/j.physletb.2003.11.068},
  urldate = {2022-04-11},
  archiveprefix = {arxiv}
}

@article{ambjorn2005reconstructing,
  title = {Reconstructing the universe},
  author = {Ambj{\o}rn, J. and Jurkiewicz, J. and Loll, R.},
  year = {2005},
  month = sep,
  journal = {Physical Review D},
  volume = {72},
  number = {6},
  eprint = {hep-th/0505154},
  pages = {064014},
  publisher = {{American Physical Society}},
  doi = {10.1103/PhysRevD.72.064014},
  urldate = {2022-03-15},
  archiveprefix = {arxiv}
}

@article{ambjorn2005spectral,
  title = {The spectral dimension of the universe is scale dependent},
  author = {Ambj{\o}rn, J. and Jurkiewicz, J. and Loll, R.},
  year = {2005},
  month = oct,
  journal = {Physical Review Letters},
  volume = {95},
  number = {17},
  eprint = {hep-th/0505113},
  pages = {171301},
  publisher = {{American Physical Society}},
  doi = {10.1103/PhysRevLett.95.171301},
  urldate = {2022-03-15},
  archiveprefix = {arxiv}
}

@article{ambjorn2007putting,
  title = {Putting a cap on causality violations in causal dynamical triangulations},
  author = {Ambj{\o}rn, Jan and Loll, Renate and Westra, Willem and Zohren, Stefan},
  year = {2007},
  month = dec,
  journal = {Journal of High Energy Physics},
  volume = {2007},
  number = {12},
  eprint = {0709.2784},
  pages = {017},
  publisher = {{IOP Publishing}},
  issn = {1126-6708},
  doi = {10.1088/1126-6708/2007/12/017},
  urldate = {2022-03-14},
  archiveprefix = {arxiv},
  langid = {english}
}

@article{ambjorn2008nonperturbative,
  title = {The nonperturbative quantum de {{Sitter}} universe},
  author = {Ambj{\o}rn, J. and G{\"o}rlich, A. T. and Jurkiewicz, J. and Loll, R.},
  year = {2008},
  month = sep,
  journal = {Physical Review D},
  volume = {78},
  number = {6},
  eprint = {0807.4481},
  pages = {063544},
  issn = {1550-7998, 1550-2368},
  doi = {10.1103/PhysRevD.78.063544},
  urldate = {2022-04-07},
  archiveprefix = {arxiv}
}

@article{ambjorn2008planckian,
  title = {Planckian birth of the quantum de {{Sitter}} universe},
  author = {Ambj{\o}rn, J. and G{\"o}rlich, A. T. and Jurkiewicz, J. and Loll, R.},
  year = {2008},
  month = mar,
  journal = {Physical Review Letters},
  volume = {100},
  number = {9},
  eprint = {0712.2485},
  pages = {091304},
  issn = {0031-9007, 1079-7114},
  doi = {10.1103/PhysRevLett.100.091304},
  urldate = {2022-04-07},
  archiveprefix = {arxiv}
}

@article{ambjorn2008string,
  title = {A string field theory based on causal dynamical triangulations},
  author = {Ambj{\o}rn, Jan and Loll, Renate and Watabiki, Yoshiyuki and Westra, Willem and Zohren, Stefan},
  year = {2008},
  month = may,
  journal = {Journal of High Energy Physics},
  volume = {2008},
  number = {05},
  eprint = {0802.0719},
  pages = {032--032},
  publisher = {{Springer Science and Business Media LLC}},
  issn = {1126-6708},
  doi = {10.1088/1126-6708/2008/05/032},
  urldate = {2022-03-14},
  archiveprefix = {arxiv},
  langid = {english}
}

@article{ambjorn2009shaken,
  title = {Shaken, but not stirred - {{Potts}} model coupled to quantum gravity},
  author = {Ambj{\o}rn, J. and Anagnostopoulos, K. N. and Loll, R. and Pushkina, I.},
  year = {2009},
  month = jan,
  journal = {Nuclear Physics B},
  volume = {807},
  number = {1-2},
  eprint = {0806.3506},
  pages = {251--264},
  issn = {05503213},
  doi = {10.1016/j.nuclphysb.2008.08.030},
  urldate = {2022-04-07},
  archiveprefix = {arxiv}
}

@article{ambjorn2011secondorder,
  title = {A second-order phase transition in {{CDT}}},
  author = {Ambj{\o}rn, J. and Jordan, S. and Jurkiewicz, J. and Loll, R.},
  year = {2011},
  month = nov,
  journal = {Physical Review Letters},
  volume = {107},
  number = {21},
  eprint = {1108.3932},
  primaryclass = {gr-qc, physics:hep-lat, physics:hep-th},
  pages = {211303},
  issn = {0031-9007, 1079-7114},
  doi = {10.1103/PhysRevLett.107.211303},
  urldate = {2022-08-03},
  archiveprefix = {arxiv}
}

@article{ambjorn2012nonperturbative,
  title = {Nonperturbative quantum gravity},
  author = {Ambj{\o}rn, J. and G{\"o}rlich, A. T. and Jurkiewicz, J. and Loll, R.},
  year = {2012},
  month = oct,
  journal = {Physics Reports},
  volume = {519},
  number = {4-5},
  eprint = {1203.3591},
  pages = {127--210},
  issn = {03701573},
  doi = {10.1016/j.physrep.2012.03.007},
  urldate = {2022-04-07},
  archiveprefix = {arxiv}
}

@article{ambjorn2012pseudotopological,
  title = {Pseudo-topological transitions in {{2D}} gravity models coupled to massless scalar fields},
  author = {Ambj{\o}rn, J. and G{\"o}rlich, A. T. and Jurkiewicz, J. and Zhang, H.-G.},
  year = {2012},
  month = oct,
  journal = {Nuclear Physics B},
  volume = {863},
  number = {2},
  eprint = {1201.1590},
  pages = {421--434},
  issn = {05503213},
  doi = {10.1016/j.nuclphysb.2012.05.024},
  urldate = {2022-04-07},
  archiveprefix = {arxiv}
}

@article{ambjorn2012roaming,
  title = {Roaming moduli space using dynamical triangulations},
  author = {Ambj{\o}rn, Jan and Barkley, J. and Budd, T. G.},
  year = {2012},
  month = may,
  journal = {Nuclear Physics B},
  volume = {858},
  number = {2},
  eprint = {1110.4649},
  pages = {267--292},
  issn = {05503213},
  doi = {10.1016/j.nuclphysb.2012.01.010},
  urldate = {2022-04-07},
  archiveprefix = {arxiv}
}

@article{ambjorn2012second,
  title = {Second- and first-order phase transitions in {{CDT}}},
  author = {Ambj{\o}rn, J. and Jordan, S. and Jurkiewicz, J. and Loll, R.},
  year = {2012},
  month = jun,
  journal = {Physical Review D},
  volume = {85},
  number = {12},
  eprint = {1205.1229},
  pages = {124044},
  issn = {1550-7998, 1550-2368},
  doi = {10.1103/PhysRevD.85.124044},
  urldate = {2022-04-07},
  archiveprefix = {arxiv}
}

@inproceedings{ambjorn2013transfer,
  title = {The transfer matrix method in four-dimensional causal dynamical triangulations},
  booktitle = {{{AIP Conference Proceedings}}},
  author = {Ambj{\o}rn, J. and {Gizbert-Studnicki}, J. and G{\"o}rlich, A. T. and Jurkiewicz, J. and Loll, R.},
  year = {2013},
  volume = {1514},
  eprint = {1302.2210},
  pages = {67--72},
  publisher = {{American Institute of Physics}},
  doi = {10.1063/1.4791727},
  urldate = {2022-04-07},
  archiveprefix = {arxiv}
}

@article{ambjorn2013twodimensional,
  title = {Two-dimensional causal dynamical triangulations with gauge fields},
  author = {Ambj{\o}rn, J. and Ipsen, A.},
  year = {2013},
  month = sep,
  journal = {Physical Review D},
  volume = {88},
  number = {6},
  eprint = {1305.3148},
  pages = {067502},
  publisher = {{American Physical Society}},
  doi = {10.1103/PhysRevD.88.067502},
  urldate = {2022-03-14},
  archiveprefix = {arxiv}
}

@article{ambjorn2013universality,
  title = {Universality of 2d causal dynamical triangulations},
  author = {Ambj{\o}rn, J. and Ipsen, A.},
  year = {2013},
  month = jul,
  journal = {Physics Letters B},
  volume = {724},
  number = {1},
  eprint = {1302.2440},
  pages = {150--154},
  issn = {0370-2693},
  doi = {10.1016/j.physletb.2013.06.005},
  urldate = {2022-03-14},
  archiveprefix = {arxiv},
  langid = {english}
}

@article{ambjorn2014renormalization,
  title = {Renormalization group flow in {{CDT}}},
  author = {Ambj{\o}rn, J. and G{\"o}rlich, A. T. and Jurkiewicz, J. and Kreienb{\"u}hl, A. and Loll, R.},
  year = {2014},
  month = aug,
  journal = {Classical and Quantum Gravity},
  volume = {31},
  number = {16},
  eprint = {1405.4585},
  primaryclass = {gr-qc, physics:hep-lat, physics:hep-th},
  pages = {165003},
  issn = {0264-9381, 1361-6382},
  doi = {10.1088/0264-9381/31/16/165003},
  urldate = {2022-06-03},
  archiveprefix = {arxiv}
}

@article{ambjorn2014restricted,
  title = {A restricted dimer model on a 2-dimensional random causal triangulation},
  author = {Ambj{\o}rn, J. and Durhuus, B. and Wheater, J. F.},
  year = {2014},
  month = sep,
  journal = {Journal of Physics A: Mathematical and Theoretical},
  volume = {47},
  number = {36},
  eprint = {1405.6782},
  pages = {365001},
  issn = {1751-8113, 1751-8121},
  doi = {10.1088/1751-8113/47/36/365001},
  urldate = {2022-04-07},
  archiveprefix = {arxiv}
}

@article{ambjorn2015spectral,
  title = {The spectral dimension in {{2D CDT}} gravity coupled to scalar fields},
  author = {Ambj{\o}rn, Jan and G{\"o}rlich, Andrzej T. and Jurkiewicz, Jerzy and Zhang, Hongguang},
  year = {2015},
  month = apr,
  journal = {Modern Physics Letters A},
  volume = {30},
  number = {13},
  eprint = {1412.3434},
  pages = {1550077},
  issn = {0217-7323, 1793-6632},
  doi = {10.1142/S0217732315500777},
  urldate = {2022-04-07},
  archiveprefix = {arxiv}
}

@article{ambjorn2017new,
  title = {New higher-order transition in causal dynamical triangulations},
  author = {Ambj{\o}rn, J. and Coumbe, D. and {Gizbert-Studnicki}, J. and G{\"o}rlich, A. T. and Jurkiewicz, J.},
  year = {2017},
  month = jun,
  journal = {Physical Review D},
  volume = {95},
  number = {12},
  eprint = {1704.04373},
  pages = {124029},
  issn = {2470-0010, 2470-0029},
  doi = {10.1103/PhysRevD.95.124029},
  urldate = {2022-04-07},
  archiveprefix = {arxiv}
}

@article{ambjorn2020renormalization,
  title = {Renormalization in quantum theories of geometry},
  author = {Ambj{\o}rn, Jan and {Gizbert-Studnicki}, Jakub and G{\"o}rlich, Andrzej and Jurkiewicz, Jerzy and Loll, Renate},
  year = {2020},
  journal = {Frontiers in Physics},
  volume = {8},
  eprint = {2002.01693},
  primaryclass = {gr-qc, physics:hep-lat, physics:hep-th},
  issn = {2296-424X},
  doi = {10.3389/fphy.2020.00247},
  urldate = {2023-02-17},
  archiveprefix = {arxiv}
}

@article{ambjorn2021cdt,
  title = {{{CDT}} quantum toroidal spacetimes: an overview},
  shorttitle = {{{CDT}} quantum toroidal spacetimes},
  author = {Ambj{\o}rn, Jan and Drogosz, Zbigniew and {Gizbert-Studnicki}, Jakub and G{\"o}rlich, Andrzej T. and Jurkiewicz, Jerzy and N{\'e}meth, D{\'a}niel},
  year = {2021},
  month = apr,
  journal = {Universe},
  volume = {7},
  number = {4},
  eprint = {2103.15610},
  pages = {79},
  publisher = {{Multidisciplinary Digital Publishing Institute}},
  issn = {2218-1997},
  doi = {10.3390/universe7040079},
  urldate = {2022-03-14},
  archiveprefix = {arxiv},
  copyright = {http://creativecommons.org/licenses/by/3.0/},
  langid = {english}
}

@article{andersson1998cosmological,
  title = {The cosmological time function},
  author = {Andersson, L. and Galloway, G. J. and Howard, R.},
  year = {1998},
  month = feb,
  journal = {Classical and Quantum Gravity},
  volume = {15},
  number = {2},
  eprint = {gr-qc/9709084},
  pages = {309--322},
  issn = {0264-9381, 1361-6382},
  doi = {10.1088/0264-9381/15/2/006},
  urldate = {2022-08-03},
  archiveprefix = {arxiv}
}

@misc{andrae2010dos,
  title = {Dos and don'ts of reduced chi-squared},
  author = {Andrae, Rene and {Schulze-Hartung}, Tim and Melchior, Peter},
  year = {2010},
  month = dec,
  number = {arXiv:1012.3754},
  eprint = {1012.3754},
  primaryclass = {astro-ph, physics:physics, stat},
  publisher = {{arXiv}},
  doi = {10.48550/arXiv.1012.3754},
  urldate = {2023-02-27},
  archiveprefix = {arxiv}
}

@article{atkin2012analytical,
  title = {An analytical analysis of {{CDT}} coupled to dimer-like matter},
  author = {Atkin, Max R. and Zohren, Stefan},
  year = {2012},
  month = jun,
  journal = {Physics Letters B},
  volume = {712},
  number = {4},
  eprint = {1202.4322},
  pages = {445--450},
  issn = {0370-2693},
  doi = {10.1016/j.physletb.2012.05.017},
  urldate = {2022-03-14},
  archiveprefix = {arxiv},
  langid = {english}
}

@article{avni1976energy,
  title = {Energy spectra of {{X-ray}} clusters of galaxies.},
  author = {Avni, Y.},
  year = {1976},
  month = dec,
  journal = {The Astrophysical Journal},
  volume = {210},
  pages = {642--646},
  issn = {0004-637X},
  doi = {10.1086/154870},
  urldate = {2022-03-22}
}

@article{barkley2019precision,
  title = {Precision measurements of {{Hausdorff}} dimensions in two-dimensional quantum gravity},
  author = {Barkley, Jerome and Budd, T. G.},
  year = {2019},
  month = nov,
  journal = {Classical and Quantum Gravity},
  volume = {36},
  number = {24},
  eprint = {1908.09469},
  pages = {244001},
  publisher = {{IOP Publishing}},
  issn = {0264-9381},
  doi = {10.1088/1361-6382/ab4f21},
  urldate = {2022-03-14},
  archiveprefix = {arxiv},
  langid = {english}
}

@book{bauerschmidt2019introduction,
  title = {Introduction to a renormalisation group method},
  author = {Bauerschmidt, Roland and Brydges, David C. and Slade, Gordon},
  year = {2019},
  volume = {2242},
  eprint = {1907.05474},
  primaryclass = {math-ph},
  doi = {10.1007/978-981-32-9593-3},
  urldate = {2023-02-27},
  archiveprefix = {arxiv}
}

@article{benedetti2007dimensional,
  title = {(2+1)-dimensional quantum gravity as the continuum limit of causal dynamical triangulations},
  author = {Benedetti, D. and Loll, R. and Zamponi, F.},
  year = {2007},
  month = nov,
  journal = {Physical Review D},
  volume = {76},
  number = {10},
  eprint = {0704.3214},
  pages = {104022},
  publisher = {{American Physical Society}},
  doi = {10.1103/PhysRevD.76.104022},
  urldate = {2021-12-20},
  archiveprefix = {arxiv}
}

@article{benedetti2009spectral,
  title = {Spectral geometry as a probe of quantum spacetime},
  author = {Benedetti, Dario and Henson, Joe},
  year = {2009},
  month = dec,
  journal = {Physical Review D},
  volume = {80},
  number = {12},
  eprint = {0911.0401},
  pages = {124036},
  issn = {1550-7998, 1550-2368},
  doi = {10.1103/PhysRevD.80.124036},
  urldate = {2022-04-11},
  archiveprefix = {arxiv}
}

@article{benedetti2015spacetime,
  title = {Spacetime condensation in (2+1)-dimensional {{CDT}} from a {{Ho\v{r}ava-Lifshitz}} minisuperspace model},
  author = {Benedetti, Dario and Henson, Joe},
  year = {2015},
  month = nov,
  journal = {Classical and Quantum Gravity},
  volume = {32},
  number = {21},
  eprint = {1410.0845},
  pages = {215007},
  issn = {0264-9381, 1361-6382},
  doi = {10.1088/0264-9381/32/21/215007},
  urldate = {2022-04-11},
  archiveprefix = {arxiv}
}

@article{benedetti2017capturing,
  title = {Capturing the phase diagram of (2+1)-dimensional {{CDT}} using a balls-in-boxes model},
  author = {Benedetti, Dario and Ryan, James P.},
  year = {2017},
  month = may,
  journal = {Classical and Quantum Gravity},
  volume = {34},
  number = {10},
  eprint = {1612.09533},
  pages = {105012},
  issn = {0264-9381, 1361-6382},
  doi = {10.1088/1361-6382/aa6b5d},
  urldate = {2022-04-11},
  archiveprefix = {arxiv}
}

@article{Bertola:2003cn,
  title = {Mixed correlation functions of the two-matrix model},
  author = {Bertola, M and Eynard, Bertrand},
  year = {2003},
  journal = {Journal of Physics A: Mathematical and General},
  volume = {36},
  number = {28},
  eprint = {hep-th/0303161},
  pages = {7733--7750},
  archiveprefix = {arxiv}
}

@article{Bertola:2007is,
  title = {Biorthogonal polynomials for two-matrix models with semiclassical potentials},
  author = {Bertola, M},
  year = {2007},
  journal = {Journal of Approximation Theory},
  volume = {144},
  number = {2},
  eprint = {nlin/0605008},
  pages = {162--212},
  archiveprefix = {arxiv}
}

@article{binder1997applications,
  title = {Applications of {{Monte Carlo}} methods to statistical physics},
  author = {Binder, K.},
  year = {1997},
  month = may,
  journal = {Reports on Progress in Physics},
  volume = {60},
  number = {5},
  pages = {487--559},
  publisher = {{IOP Publishing}},
  issn = {0034-4885},
  doi = {10.1088/0034-4885/60/5/001},
  urldate = {2022-03-14},
  langid = {english}
}

@article{boulatov1986analytical,
  title = {Analytical and numerical study of a model of dynamically triangulated random surfaces},
  author = {Boulatov, D. V. and Kazakov, V. A. and Kostov, I. K. and Migdal, A. A.},
  year = {1986},
  month = dec,
  journal = {Nuclear Physics B},
  volume = {275},
  number = {4},
  pages = {641--686},
  publisher = {{North-Holland}},
  issn = {0550-3213},
  doi = {10.1016/0550-3213(86)90578-X},
  urldate = {2021-10-26}
}

@article{Bouttier_2002,
  title = {Critical and tricritical hard objects on bicolourable random lattices: exact solutions},
  author = {Bouttier, J and Francesco, P Di and Guitter, E},
  year = {2002},
  month = apr,
  journal = {Journal of Physics A: Mathematical and General},
  volume = {35},
  number = {17},
  eprint = {cond-mat/0201213},
  pages = {3821--3854},
  publisher = {{IOP Publishing}},
  doi = {10.1088/0305-4470/35/17/302},
  archiveprefix = {arxiv}
}

@incollection{BouttierMapEnumeration,
  title = {Enumeration of maps},
  booktitle = {The {{Oxford Handbook}} of {{Random Matrix Theory}}},
  author = {Bouttier, J.},
  editor = {Akemann, Gernot and Baik, Jinho and Francesco, Philippe Di},
  year = {2015},
  pages = {534--560},
  publisher = {{Oxford University Press}},
  address = {{Oxford}},
  chapter = {26}
}

@article{brewin1988riemann,
  title = {The {{Riemann}} and extrinsic curvature tensors in the {{Regge}} calculus},
  author = {Brewin, L.},
  year = {1988},
  month = sep,
  journal = {Classical and Quantum Gravity},
  volume = {5},
  number = {9},
  pages = {1193--1203},
  publisher = {{IOP Publishing}},
  issn = {0264-9381},
  doi = {10.1088/0264-9381/5/9/005},
  urldate = {2022-04-01},
  langid = {english}
}

@article{brunekreef2021approximate,
  title = {Approximate {{Killing}} symmetries in non-perturbative quantum gravity},
  author = {Brunekreef, Joren and Reitz, Marcus},
  year = {2021},
  month = jul,
  journal = {Classical and Quantum Gravity},
  volume = {38},
  number = {13},
  eprint = {2012.14518},
  primaryclass = {gr-qc, physics:hep-th},
  pages = {135009},
  issn = {0264-9381, 1361-6382},
  doi = {10.1088/1361-6382/abf412},
  urldate = {2023-02-10},
  archiveprefix = {arxiv}
}

@article{brunekreef2021curvature,
  title = {Curvature profiles for quantum gravity},
  author = {Brunekreef, J. and Loll, R.},
  year = {2021},
  month = jan,
  journal = {Physical Review D},
  volume = {103},
  number = {2},
  eprint = {2011.10168},
  pages = {026019},
  publisher = {{American Physical Society}},
  doi = {10.1103/PhysRevD.103.026019},
  urldate = {2022-03-07},
  archiveprefix = {arxiv}
}

@misc{brunekreef2021jorenb,
  title = {{{JorenB}}/2d-cdt: {{First}} release},
  shorttitle = {{{JorenB}}/2d-cdt},
  author = {Brunekreef, Joren and G{\"o}rlich, Andrzej T.},
  year = {2021},
  month = oct,
  doi = {10.5281/zenodo.5572409},
  urldate = {2022-03-14},
  howpublished = {Zenodo}
}

@article{brunekreef2021quantum,
  title = {Quantum flatness in two-dimensional quantum gravity},
  author = {Brunekreef, J. and Loll, R.},
  year = {2021},
  month = dec,
  journal = {Physical Review D},
  volume = {104},
  number = {12},
  eprint = {2110.11100},
  pages = {126024},
  publisher = {{American Physical Society}},
  doi = {10.1103/PhysRevD.104.126024},
  urldate = {2022-03-09},
  archiveprefix = {arxiv}
}

@misc{brunekreef2022jorenb,
  title = {{{JorenB}}/3d-cdt: {{First}} release},
  shorttitle = {{{JorenB}}/3d-cdt},
  author = {Brunekreef, Joren and N{\'e}meth, Daniel and G{\"o}rlich, Andrzej T.},
  year = {2022},
  month = jun,
  doi = {10.5281/zenodo.6628721},
  howpublished = {Zenodo}
}

@article{brunekreef2022onematrix,
  title = {One-matrix differential reformulation of two-matrix models},
  author = {Brunekreef, Joren and Lionni, Luca and Th{\"u}rigen, Johannes},
  year = {2022},
  month = sep,
  journal = {Reviews in Mathematical Physics},
  volume = {34},
  number = {08},
  eprint = {2108.00540},
  primaryclass = {math-ph},
  pages = {2250026},
  issn = {0129-055X, 1793-6659},
  doi = {10.1142/S0129055X2250026X},
  urldate = {2023-02-21},
  archiveprefix = {arxiv}
}

@article{brunekreef2022phase,
  title = {The phase structure and effective action of {{3D CDT}} at higher spatial genus},
  author = {Brunekreef, Joren and N{\'e}meth, D{\'a}niel},
  year = {2022},
  month = sep,
  journal = {Journal of High Energy Physics},
  volume = {2022},
  number = {9},
  eprint = {2208.13084},
  primaryclass = {gr-qc, physics:hep-lat, physics:hep-th},
  pages = {212},
  issn = {1029-8479},
  doi = {10.1007/JHEP09(2022)212},
  urldate = {2023-01-03},
  archiveprefix = {arxiv}
}

@article{brunekreef2023nature,
  title = {On the nature of spatial universes in {{3D Lorentzian}} quantum gravity},
  author = {Brunekreef, J. and Loll, R.},
  year = {2023},
  month = jan,
  journal = {Physical Review D},
  volume = {107},
  number = {2},
  eprint = {2208.12718},
  primaryclass = {gr-qc, physics:hep-lat, physics:hep-th},
  pages = {026011},
  issn = {2470-0010, 2470-0029},
  doi = {10.1103/PhysRevD.107.026011},
  urldate = {2023-02-10},
  archiveprefix = {arxiv}
}

@article{brunekreef2023simulating,
  title = {Simulating {{CDT}} quantum gravity (to appear)},
  author = {Brunekreef, Joren and Loll, Renate},
  year = {2023}
}

@article{budd2012effective,
  title = {The effective kinetic term in {{CDT}}},
  author = {Budd, T. G.},
  year = {2012},
  month = may,
  journal = {Journal of Physics: Conference Series},
  volume = {360},
  eprint = {1110.5158},
  primaryclass = {gr-qc, physics:hep-lat},
  pages = {012038},
  issn = {1742-6596},
  doi = {10.1088/1742-6596/360/1/012038},
  urldate = {2022-08-03},
  archiveprefix = {arxiv}
}

@phdthesis{budd2012thesis,
  title = {Non-perturbative quantum gravity: a conformal perspective},
  shorttitle = {Non-perturbative quantum gravity},
  author = {Budd, T. G.},
  year = {2012},
  month = mar,
  journal = {Ph.D. Thesis},
  urldate = {2022-08-03},
  school = {Utrecht University}
}

@article{budd2013exploring,
  title = {Exploring torus universes in causal dynamical triangulations},
  author = {Budd, T. G. and Loll, R.},
  year = {2013},
  month = jul,
  journal = {Physical Review D},
  volume = {88},
  number = {2},
  eprint = {1305.4702},
  pages = {024015},
  publisher = {{American Physical Society}},
  doi = {10.1103/PhysRevD.88.024015},
  urldate = {2021-12-20},
  archiveprefix = {arxiv}
}

@misc{budd2022family,
  title = {A family of triangulated 3-spheres constructed from trees},
  author = {Budd, Timothy and Lionni, Luca},
  year = {2022},
  month = mar,
  number = {arXiv:2203.16105},
  eprint = {2203.16105},
  primaryclass = {gr-qc, physics:math-ph},
  publisher = {{arXiv}},
  doi = {10.48550/arXiv.2203.16105},
  urldate = {2022-08-15},
  archiveprefix = {arxiv}
}

@article{candido2021compact,
  title = {Compact gauge fields on causal dynamical triangulations: a {{2D}} case study},
  shorttitle = {Compact gauge fields on {{Causal Dynamical Triangulations}}},
  author = {Candido, Alessandro and Clemente, Giuseppe and D'Elia, Massimo and Rottoli, Federico},
  year = {2021},
  month = apr,
  journal = {Journal of High Energy Physics},
  volume = {2021},
  number = {4},
  eprint = {2010.15714},
  pages = {184},
  issn = {1029-8479},
  doi = {10.1007/JHEP04(2021)184},
  urldate = {2022-03-14},
  archiveprefix = {arxiv},
  langid = {english}
}

@book{carlip1998quantum,
  title = {Quantum {{Gravity}} in 2+1 {{Dimensions}}},
  author = {Carlip, Steven},
  year = {1998},
  series = {Cambridge {{Monographs}} on {{Mathematical Physics}}},
  publisher = {{Cambridge University Press}},
  address = {{Cambridge}},
  doi = {10/d43dgt},
  urldate = {2022-08-03},
  isbn = {978-0-521-54588-4}
}

@article{carlip2005quantum,
  title = {Quantum gravity in 2+1 dimensions: the case of a closed universe},
  shorttitle = {Quantum gravity in 2+1 dimensions},
  author = {Carlip, S.},
  year = {2005},
  month = dec,
  journal = {Living Reviews in Relativity},
  volume = {8},
  number = {1},
  eprint = {gr-qc/0409039},
  pages = {1},
  issn = {2367-3613, 1433-8351},
  doi = {10.12942/lrr-2005-1},
  urldate = {2022-08-03},
  archiveprefix = {arxiv}
}

@article{carlip2017dimension,
  title = {Dimension and dimensional reduction in quantum gravity},
  author = {Carlip, S.},
  year = {2017},
  month = oct,
  journal = {Classical and Quantum Gravity},
  volume = {34},
  number = {19},
  eprint = {1705.05417},
  primaryclass = {gr-qc, physics:hep-th},
  pages = {193001},
  issn = {0264-9381, 1361-6382},
  doi = {10.1088/1361-6382/aa8535},
  urldate = {2022-08-03},
  archiveprefix = {arxiv}
}

@article{catterall1995scaling,
  title = {Scaling and the fractal geometry of two-dimensional quantum gravity},
  author = {Catterall, S. and Thorleifsson, G. and Bowick, M. and John, V.},
  year = {1995},
  month = jul,
  journal = {Physics Letters B},
  volume = {354},
  number = {1-2},
  eprint = {hep-lat/9504009},
  pages = {58--68},
  publisher = {{North-Holland}},
  issn = {0370-2693},
  doi = {10.1016/0370-2693(95)00623-S},
  urldate = {2021-12-06},
  archiveprefix = {arxiv}
}

@article{cooperman2014first,
  title = {A first look at transition amplitudes in (2+1)-dimensional causal dynamical triangulations},
  author = {Cooperman, Joshua H. and Miller, Jonah},
  year = {2014},
  month = feb,
  journal = {Classical and Quantum Gravity},
  volume = {31},
  number = {3},
  eprint = {1305.2932},
  pages = {035012},
  issn = {0264-9381, 1361-6382},
  doi = {10.1088/0264-9381/31/3/035012},
  urldate = {2022-04-11},
  archiveprefix = {arxiv}
}

@article{cooperman2014scaledependent,
  title = {Scale-dependent homogeneity measures for causal dynamical triangulations},
  author = {Cooperman, Joshua H.},
  year = {2014},
  month = dec,
  journal = {Physical Review D},
  volume = {90},
  number = {12},
  eprint = {1410.0632},
  pages = {124053},
  publisher = {{American Physical Society}},
  doi = {10.1103/PhysRevD.90.124053},
  urldate = {2021-12-20},
  archiveprefix = {arxiv}
}

@article{cooperman2017second,
  title = {A second look at transition amplitudes in (2+1)-dimensional causal dynamical triangulations},
  author = {Cooperman, Joshua H. and Lee, Kyle and Miller, Jonah},
  year = {2017},
  month = jun,
  journal = {Classical and Quantum Gravity},
  volume = {34},
  number = {11},
  eprint = {1610.02408},
  pages = {115008},
  issn = {0264-9381, 1361-6382},
  doi = {10.1088/1361-6382/aa6d38},
  urldate = {2022-04-11},
  archiveprefix = {arxiv}
}

@article{coumbe2016exploring,
  title = {Exploring the new phase transition of {{CDT}}},
  author = {Coumbe, D. N. and {Gizbert-Studnicki}, J. and Jurkiewicz, J.},
  year = {2016},
  month = feb,
  journal = {Journal of High Energy Physics},
  volume = {2016},
  number = {2},
  eprint = {1510.08672},
  pages = {144},
  issn = {1029-8479},
  doi = {10.1007/JHEP02(2016)144},
  urldate = {2022-04-07},
  archiveprefix = {arxiv}
}

@article{curien2020geometric,
  title = {Geometric and spectral properties of causal maps},
  author = {Curien, Nicolas and Hutchcroft, Tom and Nachmias, Asaf},
  year = {2020},
  month = aug,
  journal = {Journal of the European Mathematical Society},
  volume = {22},
  number = {12},
  eprint = {1710.03137},
  pages = {3997--4024},
  issn = {1435-9855},
  doi = {10.4171/jems/1001},
  urldate = {2023-09-22},
  archiveprefix = {arxiv},
  langid = {english}
}

@article{dasgupta2001propertime,
  title = {A proper-time cure for the conformal sickness in~quantum gravity},
  author = {Dasgupta, A. and Loll, R.},
  year = {2001},
  month = jul,
  journal = {Nuclear Physics B},
  volume = {606},
  number = {1},
  eprint = {hep-th/0103186},
  pages = {357--379},
  issn = {0550-3213},
  doi = {10.1016/S0550-3213(01)00227-9},
  urldate = {2021-12-07},
  archiveprefix = {arxiv},
  langid = {english}
}

@article{david1985planar,
  title = {Planar diagrams, two-dimensional lattice gravity and surface models},
  author = {David, F.},
  year = {1985},
  month = jan,
  journal = {Nuclear Physics B},
  volume = {257},
  pages = {45--58},
  issn = {0550-3213},
  doi = {10.1016/0550-3213(85)90335-9},
  urldate = {2022-03-14},
  langid = {english}
}

@book{deutsch2011fabric,
  title = {The {{Fabric}} of {{Reality}}},
  author = {Deutsch, David},
  year = {2011},
  month = apr,
  publisher = {{Penguin UK}},
  googlebooks = {Z7uFxViR19oC},
  isbn = {978-0-14-196961-9},
  langid = {english}
}

@article{DiFrancesco:1998bp,
  title = {Coloring random triangulations},
  author = {Di Francesco, P and Eynard, Bertrand and Guitter, E.},
  year = {1998},
  journal = {Nuclear Physics B},
  volume = {516},
  number = {3},
  eprint = {cond-mat/9711050},
  pages = {543--587},
  archiveprefix = {arxiv}
}

@article{difrancesco19952d,
  title = {{{2D Gravity}} and random matrices},
  author = {Di Francesco, P. and Ginsparg, P. and {Zinn-Justin}, J.},
  year = {1995},
  month = mar,
  journal = {Physics Reports},
  volume = {254},
  number = {1-2},
  eprint = {hep-th/9306153},
  pages = {1--133},
  issn = {03701573},
  doi = {10.1016/0370-1573(94)00084-G},
  urldate = {2023-01-03},
  archiveprefix = {arxiv}
}

@article{Duits:2012vg,
  title = {The hermitian two matrix model with an even quartic potential},
  author = {Duits, Maurice and Kuijlaars, Arno B J and Mo, Man Yue},
  year = {2012},
  journal = {Memoirs of the American Mathematical Society},
  volume = {217},
  eprint = {1010.4282},
  pages = {1--118},
  archiveprefix = {arxiv}
}

@article{durhuus2010spectral,
  title = {On the spectral dimension of causal triangulations},
  author = {Durhuus, Bergfinnur and Jonsson, Thordur and Wheater, John F.},
  year = {2010},
  month = jun,
  journal = {Journal of Statistical Physics},
  volume = {139},
  number = {5},
  eprint = {0908.3643},
  pages = {859--881},
  issn = {1572-9613},
  doi = {10.1007/s10955-010-9968-x},
  urldate = {2022-03-14},
  archiveprefix = {arxiv},
  langid = {english}
}

@article{durhuus2015exponential,
  title = {Exponential bounds on the number of causal triangulations},
  author = {Durhuus, Bergfinnur and Jonsson, Thordur},
  year = {2015},
  month = nov,
  journal = {Communications in Mathematical Physics},
  volume = {340},
  number = {1},
  eprint = {1408.2101},
  primaryclass = {math-ph},
  pages = {105--124},
  issn = {0010-3616, 1432-0916},
  doi = {10.1007/s00220-015-2453-2},
  urldate = {2022-08-03},
  archiveprefix = {arxiv}
}

@article{durhuus2020structure,
  title = {The structure of spatial slices of 3-dimensional causal triangulations},
  author = {Durhuus, Bergfinnur and Jonsson, Thordur},
  year = {2020},
  month = sep,
  journal = {Annales de l'Institut Henri Poincar\'e D},
  volume = {7},
  number = {3},
  eprint = {1712.07502},
  pages = {365--393},
  issn = {2308-5827},
  doi = {10.4171/aihpd/91},
  urldate = {2022-08-25},
  archiveprefix = {arxiv},
  langid = {english}
}

@article{durhuus2021critical,
  title = {Critical behaviour of loop models on causal triangulations},
  author = {Durhuus, Bergfinnur and Poncini, Xavier and Rasmussen, J{\o}rgen and {\"U}nel, Meltem},
  year = {2021},
  month = nov,
  journal = {Journal of Statistical Mechanics: Theory and Experiment},
  volume = {2021},
  number = {11},
  eprint = {2104.14176},
  pages = {113102},
  publisher = {{IOP Publishing}},
  issn = {1742-5468},
  doi = {10.1088/1742-5468/ac2dfa},
  urldate = {2022-03-14},
  archiveprefix = {arxiv},
  langid = {english}
}

@article{einstein1905zur,
  title = {Zur {{Elektrodynamik}} bewegter {{K\"orper}}},
  author = {Einstein, A.},
  year = {1905},
  journal = {Annalen der Physik},
  volume = {322},
  number = {10},
  pages = {891--921},
  issn = {1521-3889},
  doi = {10.1002/andp.19053221004},
  urldate = {2022-07-11},
  langid = {english}
}

@article{einstein1915feldgleichungen,
  title = {Die {{Feldgleichungen}} der {{Gravitation}}},
  author = {Einstein, Albert},
  year = {1915},
  month = jan,
  journal = {Sitzungsberichte der K\"oniglich Preu\ss ischen Akademie der Wissenschaften (Berlin},
  pages = {844--847},
  urldate = {2022-07-11}
}

@book{ellis2012relativistic,
  title = {Relativistic {{Cosmology}}},
  author = {Ellis, George F. R. and Maartens, Roy and MacCallum, Malcolm A. H.},
  year = {2012},
  month = mar,
  publisher = {{Cambridge University Press}},
  googlebooks = {IgkhAwAAQBAJ},
  isbn = {978-1-139-64295-8},
  langid = {english}
}

@article{eynard1997eigenvalue,
  title = {Eigenvalue distribution of large random matrices, from one matrix to several coupled matrices},
  author = {Eynard, B.},
  year = {1997},
  month = dec,
  journal = {Nuclear Physics B},
  volume = {506},
  number = {3},
  eprint = {cond-mat/9707005},
  pages = {633--664},
  issn = {05503213},
  doi = {10.1016/S0550-3213(97)00452-5},
  urldate = {2023-01-03},
  archiveprefix = {arxiv}
}

@article{eynard2003large,
  title = {Large {{N}} expansion of the 2-matrix model},
  author = {Eynard, B.},
  year = {2003},
  month = jan,
  journal = {Journal of High Energy Physics},
  volume = {2003},
  number = {01},
  eprint = {hep-th/0210047},
  pages = {051--051},
  issn = {1029-8479},
  doi = {10.1088/1126-6708/2003/01/051},
  urldate = {2023-01-03},
  archiveprefix = {arxiv}
}

@article{eynard20062matrix,
  title = {2-matrix versus complex matrix model, integrals over the unitary group as triangular integrals},
  author = {Eynard, B. and Ferrer, A. Prats},
  year = {2006},
  month = may,
  journal = {Communications in Mathematical Physics},
  volume = {264},
  number = {1},
  eprint = {hep-th/0502041},
  pages = {115--144},
  issn = {0010-3616, 1432-0916},
  doi = {10.1007/s00220-006-1541-8},
  urldate = {2022-04-07},
  archiveprefix = {arxiv}
}

@book{eynard2016counting,
  title = {Counting {{Surfaces}}},
  author = {Eynard, Bertrand},
  year = {2016},
  series = {Progress in {{Mathematical Physics}}},
  volume = {70},
  publisher = {{Springer}},
  address = {{Basel}},
  doi = {10.1007/978-3-7643-8797-6},
  urldate = {2023-02-27},
  isbn = {978-3-7643-8796-9 978-3-7643-8797-6}
}

@misc{eynard2018random,
  title = {Random matrices},
  author = {Eynard, Bertrand and Kimura, Taro and Ribault, Sylvain},
  year = {2018},
  month = jul,
  number = {arXiv:1510.04430},
  eprint = {1510.04430},
  primaryclass = {cond-mat, physics:hep-th, physics:math-ph},
  publisher = {{arXiv}},
  doi = {10.48550/arXiv.1510.04430},
  urldate = {2023-02-27},
  archiveprefix = {arxiv}
}

@article{fuchs2007closed,
  title = {Closed geodesics on regular polyhedra},
  author = {Fuchs, Dmitry and Fuchs, Ekaterina},
  year = {2007},
  month = jan,
  journal = {Moscow Mathematical Journal},
  volume = {7},
  doi = {10.17323/1609-4514-2007-7-2-265-279}
}

@article{fuchs2014periodic,
  title = {Periodic billiard trajectories in regular polygons and closed geodesics on regular polyhedra},
  author = {Fuchs, Dmitry},
  year = {2014},
  month = jun,
  journal = {Geometriae Dedicata},
  volume = {170},
  number = {1},
  pages = {319--333},
  issn = {1572-9168},
  doi = {10.1007/s10711-013-9883-9},
  urldate = {2022-03-15},
  langid = {english}
}

@book{goldenfeld2019lectures,
  title = {Lectures on {{Phase Transitions}} and the {{Renormalization Group}}},
  author = {Goldenfeld, Nigel},
  year = {2019},
  month = jun,
  publisher = {{CRC Press}},
  address = {{Boca Raton}},
  doi = {10.1201/9780429493492},
  isbn = {978-0-429-49349-2}
}

@article{goldstone2021shortest,
  title = {Shortest paths on cubes},
  author = {Goldstone, Richard and Roca, Rachel and Valli, Robert Suzzi},
  year = {2021},
  month = mar,
  journal = {The College Mathematics Journal},
  volume = {52},
  number = {2},
  pages = {121--132},
  publisher = {{Taylor \& Francis}},
  issn = {0746-8342},
  doi = {10.1080/07468342.2021.1866944},
  urldate = {2022-03-15}
}

@article{goroff1985quantum,
  title = {Quantum gravity at two loops},
  author = {Goroff, Marc H. and Sagnotti, Augusto and Sagnotti, Augusto},
  year = {1985},
  month = oct,
  journal = {Physics Letters B},
  volume = {160},
  number = {1},
  pages = {81--86},
  issn = {0370-2693},
  doi = {10.1016/0370-2693(85)91470-4},
  urldate = {2023-02-02},
  langid = {english}
}

@book{green1988superstring,
  title = {Superstring {{Theory}}: {{Volume}} 1, {{Introduction}}},
  shorttitle = {Superstring {{Theory}}},
  author = {Green, Michael B. and Green, Michael B. and Schwarz, John H. and Witten, Edward},
  year = {1988},
  month = jul,
  publisher = {{Cambridge University Press}},
  googlebooks = {ItVsHqjJo4gC},
  isbn = {978-0-521-35752-4},
  langid = {english}
}

@misc{gurau2014renormalization,
  title = {Renormalization: an advanced overview},
  shorttitle = {Renormalization},
  author = {Gurau, Razvan and Rivasseau, Vincent and Sfondrini, Alessandro},
  year = {2014},
  month = jan,
  number = {arXiv:1401.5003},
  eprint = {1401.5003},
  primaryclass = {hep-th, physics:math-ph},
  publisher = {{arXiv}},
  doi = {10.48550/arXiv.1401.5003},
  urldate = {2023-02-27},
  archiveprefix = {arxiv}
}

@article{gurau2015analyticity,
  title = {Analyticity results for the cumulants in a random matrix model},
  author = {Gurau, Razvan G. and Krajewski, Thomas},
  year = {2015},
  month = may,
  journal = {Annales de l'Institut Henri Poincar\'e D},
  volume = {2},
  number = {2},
  pages = {169--228},
  issn = {2308-5827},
  doi = {10.4171/aihpd/17},
  urldate = {2023-02-27},
  langid = {english}
}

@article{hamber1986simplicial,
  title = {Simplicial quantum gravity with higher derivative terms: {{Formalism}} and numerical results in four dimensions},
  shorttitle = {Simplicial quantum gravity with higher derivative terms},
  author = {Hamber, Herbert W. and Williams, Ruth M.},
  year = {1986},
  month = jun,
  journal = {Nuclear Physics B},
  volume = {269},
  number = {3},
  pages = {712--743},
  issn = {0550-3213},
  doi = {10.1016/0550-3213(86)90518-3},
  urldate = {2022-04-01},
  langid = {english}
}

@article{HarishChandra,
  title = {Differential operators on a semisimple lie algebra},
  author = {{Harish-Chandra}},
  year = {1957},
  journal = {American Journal of Mathematics},
  volume = {79},
  number = {1},
  eprint = {2372387},
  eprinttype = {jstor},
  pages = {87--120},
  publisher = {{Johns Hopkins University Press}},
  issn = {00029327, 10806377}
}

@article{hastings1970monte,
  title = {Monte {{Carlo}} sampling methods using {{Markov}} chains and their applications},
  author = {Hastings, W. K.},
  year = {1970},
  month = apr,
  journal = {Biometrika},
  volume = {57},
  number = {1},
  pages = {97--109},
  issn = {0006-3444},
  doi = {10.1093/biomet/57.1.97},
  urldate = {2022-06-13}
}

@article{heisenberg1927ueber,
  title = {{\"Uber den anschaulichen Inhalt der quantentheoretischen Kinematik und Mechanik}},
  author = {Heisenberg, W.},
  year = {1927},
  month = mar,
  journal = {Zeitschrift f\"ur Physik},
  volume = {43},
  number = {3},
  pages = {172--198},
  issn = {0044-3328},
  doi = {10.1007/BF01397280},
  urldate = {2022-07-11},
  langid = {ngerman}
}

@incollection{hooft1993planar,
  title = {A planar diagram theory for strong interactions},
  booktitle = {The {{Large N Expansion}} in {{Quantum Field Theory}} and {{Statistical Physics}}},
  author = {'t Hooft, G.},
  year = {1993},
  month = aug,
  pages = {80--92},
  publisher = {{World Scientific}},
  doi = {10.1142/9789814365802_0007},
  urldate = {2022-11-13},
  isbn = {978-981-02-0455-6}
}

@book{hua1963harmonic,
  title = {Harmonic {{Analysis}} of {{Functions}} of {{Several Complex Variables}} in the {{Classical Domains}}},
  author = {Hua, Luogeng},
  year = {1963},
  month = dec,
  publisher = {{American Mathematical Soc.}},
  googlebooks = {99wVBAAAQBAJ},
  isbn = {978-0-8218-1556-4},
  langid = {english}
}

@misc{introduction,
  title = {Introduction to {{C}}++ | {{Electrical Engineering}} and {{Computer Science}}},
  journal = {MIT OpenCourseWare},
  urldate = {2023-10-21},
  howpublished = {https://ocw.mit.edu/courses/6-096-introduction-to-c-january-iap-2011/},
  langid = {english}
}

@article{itzykson1980planar,
  title = {The planar approximation. {{II}}},
  author = {Itzykson, C. and Zuber, J.-B.},
  year = {1980},
  month = mar,
  journal = {Journal of Mathematical Physics},
  volume = {21},
  number = {3},
  pages = {411--421},
  publisher = {{American Institute of Physics}},
  issn = {0022-2488},
  doi = {10.1063/1.524438},
  urldate = {2023-02-27}
}

@article{jain1992worldsheet,
  title = {World-sheet geometry and baby universes in {{2D}} quantum gravity},
  author = {Jain, S. and Mathur, Samir D.},
  year = {1992},
  month = jul,
  journal = {Physics Letters B},
  volume = {286},
  number = {3-4},
  eprint = {hep-th/9204017},
  pages = {239--246},
  publisher = {{North-Holland}},
  issn = {0370-2693},
  doi = {10.1016/0370-2693(92)91769-6},
  urldate = {2021-10-26},
  archiveprefix = {arxiv}
}

@book{johnson2002dbranes,
  title = {D-{{Branes}}},
  author = {Johnson, Clifford V.},
  year = {2002},
  series = {Cambridge {{Monographs}} on {{Mathematical Physics}}},
  publisher = {{Cambridge University Press}},
  address = {{Cambridge}},
  doi = {10.1017/CBO9780511606540},
  urldate = {2022-07-13},
  isbn = {978-0-521-80912-2}
}

@article{jordan2013causal,
  title = {Causal dynamical triangulations without preferred foliation},
  author = {Jordan, S. and Loll, R.},
  year = {2013},
  month = jul,
  journal = {Physics Letters B},
  volume = {724},
  number = {1-3},
  eprint = {1305.4582},
  primaryclass = {gr-qc, physics:hep-lat, physics:hep-th},
  pages = {155--159},
  issn = {03702693},
  doi = {10.1016/j.physletb.2013.06.007},
  urldate = {2022-08-03},
  archiveprefix = {arxiv}
}

@article{jordan2013sitter,
  title = {De {{Sitter}} universe from causal dynamical triangulations without preferred foliation},
  author = {Jordan, S. and Loll, R.},
  year = {2013},
  month = aug,
  journal = {Physical Review D},
  volume = {88},
  number = {4},
  eprint = {1307.5469},
  primaryclass = {gr-qc, physics:hep-lat, physics:hep-th},
  pages = {044055},
  issn = {1550-7998, 1550-2368},
  doi = {10.1103/PhysRevD.88.044055},
  urldate = {2022-08-25},
  archiveprefix = {arxiv}
}

@article{kawai1993transfer,
  title = {Transfer matrix formalism for two-dimensional quantum gravity and fractal structures of space-time},
  author = {Kawai, H. and Kawamoto, N. and Mogami, T. and Watabiki, Y.},
  year = {1993},
  month = may,
  journal = {Physics Letters B},
  volume = {306},
  number = {1},
  eprint = {hep-th/9302133},
  pages = {19--26},
  issn = {0370-2693},
  doi = {10.1016/0370-2693(93)91131-6},
  urldate = {2022-03-14},
  archiveprefix = {arxiv},
  langid = {english}
}

@article{KAZAKOV-Ising,
  title = {Ising model on a dynamical planar random lattice: {{Exact}} solution},
  author = {Kazakov, V.A.},
  year = {1986},
  journal = {Physics Letters A},
  volume = {119},
  number = {3},
  pages = {140--144},
  issn = {0375-9601},
  doi = {10.1016/0375-9601(86)90433-0}
}

@article{kazakov1985critical,
  title = {Critical properties of randomly triangulated planar random surfaces},
  author = {Kazakov, V. A. and Kostov, I. K. and Migdal, A. A.},
  year = {1985},
  month = jul,
  journal = {Physics Letters B},
  volume = {157},
  number = {4},
  pages = {295--300},
  issn = {0370-2693},
  doi = {10.1016/0370-2693(85)90669-0},
  urldate = {2022-03-14},
  langid = {english}
}

@article{kelly2019selfassembly,
  title = {Self-assembly of geometric space from random graphs},
  author = {Kelly, Christy and Trugenberger, Carlo A. and Biancalana, Fabio},
  year = {2019},
  month = may,
  journal = {Classical and Quantum Gravity},
  volume = {36},
  number = {12},
  eprint = {1901.09870},
  pages = {125012},
  publisher = {{IOP Publishing}},
  issn = {0264-9381},
  doi = {10.1088/1361-6382/ab1c7d},
  urldate = {2022-03-14},
  archiveprefix = {arxiv},
  langid = {english}
}

@article{kelly2021emergence,
  title = {Emergence of the circle in a statistical model of random cubic graphs},
  author = {Kelly, Christy and Trugenberger, Carlo and Biancalana, Fabio},
  year = {2021},
  month = feb,
  journal = {Classical and Quantum Gravity},
  volume = {38},
  number = {7},
  eprint = {2008.11779},
  pages = {075008},
  publisher = {{IOP Publishing}},
  issn = {0264-9381},
  doi = {10.1088/1361-6382/abe2d8},
  urldate = {2022-03-14},
  archiveprefix = {arxiv},
  langid = {english}
}

@article{klitgaard2018implementing,
  title = {Implementing quantum {{Ricci}} curvature},
  author = {Klitgaard, N. and Loll, R.},
  year = {2018},
  month = may,
  journal = {Physical Review D},
  volume = {97},
  number = {10},
  eprint = {1802.10524},
  pages = {106017},
  publisher = {{American Physical Society}},
  doi = {10.1103/PhysRevD.97.106017},
  urldate = {2022-03-04},
  archiveprefix = {arxiv}
}

@article{klitgaard2018introducing,
  title = {Introducing quantum {{Ricci}} curvature},
  author = {Klitgaard, N. and Loll, R.},
  year = {2018},
  month = feb,
  journal = {Physical Review D},
  volume = {97},
  number = {4},
  eprint = {1712.08847},
  pages = {046008},
  publisher = {{American Physical Society}},
  doi = {10.1103/PhysRevD.97.046008},
  urldate = {2022-03-04},
  archiveprefix = {arxiv}
}

@article{klitgaard2020how,
  title = {How round is the quantum de {{Sitter}} universe?},
  author = {Klitgaard, N. and Loll, R.},
  year = {2020},
  month = oct,
  journal = {The European Physical Journal C},
  volume = {80},
  number = {10},
  eprint = {2006.06263},
  pages = {990},
  issn = {1434-6052},
  doi = {10.1140/epjc/s10052-020-08569-5},
  urldate = {2022-03-14},
  archiveprefix = {arxiv},
  langid = {english}
}

@phdthesis{klitgaard2022new,
  title = {New {{Curvatures}} for {{Quantum Gravity}}},
  author = {Klitgaard, N. F.},
  year = {2022},
  month = jan,
  urldate = {2022-03-14},
  langid = {english},
  school = {Radboud University}
}

@article{knizhnik1988fractal,
  title = {Fractal structure of 2d quantum gravity},
  author = {Knizhnik, V.g. and Polyakov, A.m. and Zamolodchikov, A.b.},
  year = {1988},
  month = jul,
  journal = {Modern Physics Letters A},
  volume = {03},
  number = {08},
  pages = {819--826},
  publisher = {{World Scientific Publishing Co.}},
  issn = {0217-7323},
  doi = {10.1142/S0217732388000982},
  urldate = {2022-04-12}
}

@article{loll1998discrete,
  title = {Discrete approaches to quantum gravity in four dimensions},
  author = {Loll, Renate},
  year = {1998},
  month = dec,
  journal = {Living Reviews in Relativity},
  volume = {1},
  number = {1},
  eprint = {gr-qc/9805049},
  pages = {13},
  issn = {1433-8351},
  doi = {10.12942/lrr-1998-13},
  urldate = {2022-03-14},
  archiveprefix = {arxiv},
  langid = {english}
}

@article{loll2015locally,
  title = {Locally causal dynamical triangulations in two dimensions},
  author = {Loll, R. and Ruijl, B.},
  year = {2015},
  month = oct,
  journal = {Physical Review D},
  volume = {92},
  number = {8},
  eprint = {1507.04566},
  pages = {084002},
  publisher = {{American Physical Society}},
  doi = {10.1103/PhysRevD.92.084002},
  urldate = {2022-03-14},
  archiveprefix = {arxiv}
}

@article{loll2019quantum,
  title = {Quantum gravity from causal dynamical triangulations: a review},
  shorttitle = {Quantum gravity from causal dynamical triangulations},
  author = {Loll, R.},
  year = {2019},
  month = dec,
  journal = {Classical and Quantum Gravity},
  volume = {37},
  number = {1},
  eprint = {1905.08669},
  pages = {013002},
  publisher = {{IOP Publishing}},
  issn = {0264-9381},
  doi = {10.1088/1361-6382/ab57c7},
  urldate = {2022-03-11},
  archiveprefix = {arxiv},
  langid = {english}
}

@article{loll2022quantum,
  title = {Quantum gravity in 30 questions},
  author = {Loll, Renate and Fabiano, Giuseppe and Frattulillo, Domenico and Wagner, Fabian},
  year = {2022},
  journal = {PoS},
  volume = {CORFU2021},
  eprint = {2206.06762},
  primaryclass = {hep-th},
  pages = {316},
  doi = {10.22323/1.406.0316},
  archiveprefix = {arxiv}
}

@article{Mehta:1981jm,
  title = {A method of integration over matrix variables},
  author = {Mehta, M L},
  year = {1981},
  journal = {Communications in Mathematical Physics},
  volume = {79},
  number = {3},
  pages = {327--340}
}

@book{mehta2004random,
  title = {Random {{Matrices}}},
  author = {Mehta, Madan Lal},
  year = {2004},
  month = oct,
  publisher = {{Elsevier}},
  googlebooks = {Kp3Nx03\_gMwC},
  isbn = {978-0-08-047411-3},
  langid = {english}
}

@article{metropolis1953equation,
  title = {Equation of state calculations by fast computing machines},
  author = {Metropolis, Nicholas and Rosenbluth, Arianna W. and Rosenbluth, Marshall N. and Teller, Augusta H. and Teller, Edward},
  year = {1953},
  month = jun,
  journal = {The Journal of Chemical Physics},
  volume = {21},
  number = {6},
  pages = {1087--1092},
  publisher = {{American Institute of Physics}},
  issn = {0021-9606},
  doi = {10.1063/1.1699114},
  urldate = {2022-04-12}
}

@article{michelson1887relative,
  title = {On the relative motion of the {{Earth}} and the luminiferous ether},
  author = {Michelson, A. A. and Morley, E. W.},
  year = {1887},
  month = nov,
  journal = {American Journal of Science},
  volume = {s3-34},
  number = {203},
  pages = {333--345},
  publisher = {{American Journal of Science}},
  issn = {0002-9599, 1945-452X},
  doi = {10.2475/ajs.s3-34.203.333},
  urldate = {2022-07-11},
  chapter = {Extraterrestrial geology},
  copyright = {GeoRef, Copyright 2008, American Geological Institute.},
  langid = {english}
}

@article{miermont2013brownian,
  title = {The {{Brownian}} map is the scaling limit of uniform random plane quadrangulations},
  author = {Miermont, Gr{\'e}gory},
  year = {2013},
  journal = {Acta Mathematica},
  volume = {210},
  number = {2},
  eprint = {1104.1606},
  pages = {319--401},
  urldate = {2022-04-15},
  archiveprefix = {arxiv}
}

@book{montvay1994quantum,
  title = {Quantum {{Fields}} on a {{Lattice}}},
  author = {Montvay, Istvan and M{\"u}nster, Gernot},
  year = {1994},
  series = {Cambridge {{Monographs}} on {{Mathematical Physics}}},
  publisher = {{Cambridge University Press}},
  address = {{Cambridge}},
  doi = {10.1017/CBO9780511470783},
  urldate = {2023-02-14},
  isbn = {978-0-521-59917-7}
}

@book{newman1999monte,
  title = {Monte {{Carlo Methods}} in {{Statistical Physics}}},
  author = {Newman, M. E. J. and Barkema, G. T.},
  year = {1999},
  month = feb,
  publisher = {{Clarendon Press}},
  googlebooks = {HgBREAAAQBAJ},
  isbn = {978-0-19-158986-7},
  langid = {english}
}

@article{ollivier2009ricci,
  title = {Ricci curvature of {{Markov}} chains on metric spaces},
  author = {Ollivier, Yann},
  year = {2009},
  month = feb,
  journal = {Journal of Functional Analysis},
  volume = {256},
  number = {3},
  eprint = {math/0701886},
  pages = {810--864},
  issn = {0022-1236},
  doi = {10.1016/j.jfa.2008.11.001},
  urldate = {2022-03-14},
  archiveprefix = {arxiv},
  langid = {english}
}

@incollection{orantin2015chain,
  title = {Chain of matrices, loop equations, and topological recursion},
  booktitle = {The {{Oxford Handbook}} of {{Random Matrix Theory}}},
  author = {Orantin, Nicolas},
  editor = {Akemann, Gernot and Baik, Jinho and Di Francesco, Philippe},
  year = {2015},
  month = sep,
  eprint = {0911.5089},
  primaryclass = {hep-th, physics:math-ph},
  publisher = {{Oxford University Press}},
  doi = {10.1093/oxfordhb/9780198744191.013.16},
  urldate = {2023-02-27},
  archiveprefix = {arxiv},
  isbn = {978-0-19-874419-1}
}

@book{oriti2009approaches,
  title = {Approaches to {{Quantum Gravity}}: {{Toward}} a {{New Understanding}} of {{Space}}, {{Time}} and {{Matter}}},
  shorttitle = {Approaches to {{Quantum Gravity}}},
  editor = {Oriti, Daniele},
  year = {2009},
  publisher = {{Cambridge University Press}},
  address = {{Cambridge}},
  doi = {10.1017/CBO9780511575549},
  urldate = {2022-07-13},
  isbn = {978-0-521-86045-1}
}

@misc{orlov2002tau,
  title = {Tau {{Functions}} and {{Matrix Integrals}}},
  author = {Orlov, A. Yu},
  year = {2002},
  month = oct,
  number = {arXiv:math-ph/0210012},
  eprint = {math-ph/0210012},
  publisher = {{arXiv}},
  doi = {10.48550/arXiv.math-ph/0210012},
  urldate = {2023-03-10},
  archiveprefix = {arxiv}
}

@misc{raju2021lessons,
  title = {Lessons from the information paradox},
  author = {Raju, Suvrat},
  year = {2021},
  month = jan,
  number = {arXiv:2012.05770},
  eprint = {2012.05770},
  primaryclass = {gr-qc, physics:hep-th},
  publisher = {{arXiv}},
  doi = {10.48550/arXiv.2012.05770},
  urldate = {2022-07-13},
  archiveprefix = {arxiv}
}

@article{regge1961general,
  title = {General relativity without coordinates},
  author = {Regge, T.},
  year = {1961},
  month = feb,
  journal = {Il Nuovo Cimento (1955-1965)},
  volume = {19},
  number = {3},
  pages = {558--571},
  issn = {1827-6121},
  doi = {10.1007/BF02733251},
  urldate = {2022-03-10},
  langid = {english}
}

@book{riordan1987hunting,
  title = {The {{Hunting}} of the {{Quark}}: a {{True Story}} of {{Modern Physics}}},
  shorttitle = {The {{Hunting}} of the {{Quark}}},
  author = {Riordan, Michael},
  year = {1987},
  publisher = {{Simon \& Schuster}},
  googlebooks = {PN\_vAAAAMAAJ},
  isbn = {978-0-671-50466-3},
  langid = {english}
}

@article{Robinson,
  title = {Basic semigroup theory},
  author = {Robinson, Derek},
  year = {1996},
  journal = {Proceedings of the Centre for Mathematics and its Applications},
  pages = {1--34}
}

@book{salmhofer1999renormalization,
  title = {Renormalization},
  author = {Salmhofer, Manfred},
  year = {1999},
  publisher = {{Springer}},
  address = {{Berlin, Heidelberg}},
  doi = {10.1007/978-3-662-03873-4},
  urldate = {2023-02-27},
  isbn = {978-3-642-08430-0 978-3-662-03873-4}
}

@article{schneider1999universality,
  title = {On the universality of matrix models for random surfaces},
  author = {Schneider, Antje and Filk, Thomas},
  year = {1999},
  month = may,
  journal = {The European Physical Journal C},
  volume = {8},
  number = {3},
  eprint = {hep-lat/9809054},
  pages = {523--526},
  issn = {1434-6044, 1434-6052},
  doi = {10.1007/s100529901092},
  urldate = {2022-04-27},
  archiveprefix = {arxiv}
}

@book{schweber1994qed,
  title = {{{QED}} and the {{Men Who Made}} it: {{Dyson}}, {{Feynman}}, {{Schwinger}}, and {{Tomonaga}}},
  shorttitle = {{{QED}} and the {{Men Who Made It}}},
  author = {Schweber, Silvan S.},
  year = {1994},
  volume = {104},
  eprint = {j.ctv10crg18},
  eprinttype = {jstor},
  publisher = {{Princeton University Press}},
  doi = {10.2307/j.ctv10crg18},
  urldate = {2022-07-12},
  isbn = {978-0-691-03685-4}
}

@article{tee2021enhanced,
  title = {Enhanced {{Forman}} curvature and its relation to {{Ollivier}} curvature},
  author = {Tee, Philip and Trugenberger, C. A.},
  year = {2021},
  month = mar,
  journal = {Europhysics Letters},
  volume = {133},
  number = {6},
  eprint = {2102.12329},
  pages = {60006},
  publisher = {{IOP Publishing}},
  issn = {0295-5075},
  doi = {10.1209/0295-5075/133/60006},
  urldate = {2022-03-14},
  archiveprefix = {arxiv},
  langid = {english}
}

@article{teitelboim1983causality,
  title = {Causality versus gauge invariance in quantum gravity and supergravity},
  author = {Teitelboim, Claudio},
  year = {1983},
  month = mar,
  journal = {Physical Review Letters},
  volume = {50},
  number = {10},
  pages = {705--708},
  publisher = {{American Physical Society}},
  doi = {10.1103/PhysRevLett.50.705},
  urldate = {2022-05-10}
}

@article{thurigen2021renormalization,
  title = {Renormalization in combinatorially non-local field theories: the {{BPHZ}} momentum scheme},
  shorttitle = {Renormalization in {{Combinatorially Non-Local Field Theories}}},
  author = {Th{\"u}rigen, Johannes},
  year = {2021},
  month = oct,
  journal = {Symmetry, Integrability and Geometry: Methods and Applications},
  eprint = {2103.01136},
  primaryclass = {gr-qc, physics:hep-th, physics:math-ph},
  issn = {18150659},
  doi = {10.3842/SIGMA.2021.094},
  urldate = {2023-02-27},
  archiveprefix = {arxiv}
}

@article{trugenberger2016random,
  title = {Random holographic ``large worlds'' with emergent dimensions},
  author = {Trugenberger, Carlo A.},
  year = {2016},
  month = nov,
  journal = {Physical Review E},
  volume = {94},
  number = {5},
  eprint = {1610.05339},
  pages = {052305},
  publisher = {{American Physical Society}},
  doi = {10.1103/PhysRevE.94.052305},
  urldate = {2022-03-15},
  archiveprefix = {arxiv}
}

@article{trugenberger2017combinatorial,
  title = {Combinatorial quantum gravity: geometry from random bits},
  shorttitle = {Combinatorial quantum gravity},
  author = {Trugenberger, Carlo A.},
  year = {2017},
  month = sep,
  journal = {Journal of High Energy Physics},
  volume = {2017},
  number = {9},
  eprint = {1610.05934},
  pages = {45},
  issn = {1029-8479},
  doi = {10.1007/JHEP09(2017)045},
  urldate = {2022-03-14},
  archiveprefix = {arxiv},
  langid = {english}
}

@article{trugenberger2022emergent,
  title = {Emergent time, cosmological constant and boundary dimension at infinity in combinatorial quantum gravity},
  author = {Trugenberger, Carlo A.},
  year = {2022},
  month = apr,
  journal = {Journal of High Energy Physics},
  volume = {2022},
  number = {4},
  eprint = {2112.03778},
  primaryclass = {gr-qc, physics:hep-th},
  pages = {19},
  issn = {1029-8479},
  doi = {10.1007/JHEP04(2022)019},
  urldate = {2023-02-10},
  archiveprefix = {arxiv}
}

@phdthesis{vanderfeltz2021matter,
  type = {M.{{Sc}}. thesis},
  title = {Matter coupling to two-dimensional causal dynamical triangulations},
  author = {{van der Feltz}, Willem},
  year = {2021},
  month = nov,
  school = {Radboud University \& University of Cologne}
}

@misc{vanderhoorn2020ollivier,
  title = {Ollivier curvature of random geometric graphs converges to {{Ricci}} curvature of their {{Riemannian}} manifolds},
  author = {{van der Hoorn}, Pim and Lippner, Gabor and Trugenberger, Carlo and Krioukov, Dmitri},
  year = {2020},
  month = sep,
  number = {arXiv:2009.04306},
  eprint = {2009.04306},
  primaryclass = {math},
  publisher = {{arXiv}},
  doi = {10.48550/arXiv.2009.04306},
  urldate = {2023-02-10},
  archiveprefix = {arxiv}
}

@book{wallace2012emergent,
  title = {The {{Emergent Multiverse}}: {{Quantum Theory}} according to the {{Everett Interpretation}}},
  shorttitle = {The {{Emergent Multiverse}}},
  author = {Wallace, David},
  year = {2012},
  publisher = {{Oxford University Press}},
  address = {{Oxford}},
  doi = {10.1093/acprof:oso/9780199546961.001.0001},
  urldate = {2022-05-06},
  isbn = {978-0-19-954696-1},
  langid = {english}
}

@book{zee2010quantum,
  title = {Quantum {{Field Theory}} in a {{Nutshell}}},
  shorttitle = {Quantum {{Field Theory}} in a {{Nutshell}}},
  author = {Zee, A.},
  year = {2010},
  month = feb,
  publisher = {{Princeton University Press}},
  googlebooks = {XrumDwAAQBAJ},
  isbn = {978-0-691-14034-6},
  langid = {english}
}

@article{zhang2021harmonic,
  title = {Harmonic analysis for rank-1 {{Randomised Horn Problems}}},
  author = {Zhang, Jiyuan and Kieburg, Mario and Forrester, Peter J.},
  year = {2021},
  month = aug,
  journal = {Letters in Mathematical Physics},
  volume = {111},
  number = {4},
  eprint = {1911.11316},
  primaryclass = {math-ph},
  pages = {98},
  issn = {0377-9017, 1573-0530},
  doi = {10.1007/s11005-021-01429-7},
  urldate = {2023-02-27},
  archiveprefix = {arxiv}
}

@article{zuber2008large,
  title = {On the large {{N}} limit of matrix integrals over the orthogonal group},
  author = {Zuber, Jean-Bernard},
  year = {2008},
  month = sep,
  journal = {Journal of Physics A: Mathematical and Theoretical},
  volume = {41},
  number = {38},
  eprint = {0805.0315},
  pages = {382001},
  issn = {1751-8113, 1751-8121},
  doi = {10.1088/1751-8113/41/38/382001},
  urldate = {2022-04-07},
  archiveprefix = {arxiv}
}

@article{ZvonkinMapEnumeration,
  title = {Matrix integrals and map enumeration: {{An}} accessible introduction},
  author = {Zvonkin, A.},
  year = {1997},
  journal = {Mathematical and Computer Modelling},
  volume = {26},
  number = {8},
  pages = {281--304},
  issn = {0895-7177},
  doi = {10.1016/S0895-7177(97)00210-0}
}

\end{document}